\documentclass[10pt]{article}
\usepackage{amsmath,amssymb,graphicx}
\usepackage{amsthm}
\usepackage{euscript}
\usepackage{eufrak}

\batchmode

\newtheorem{defn}{Definition}
\newtheorem{lem}{Lemma}
\newtheorem{thm}{Theorem}

\newtheorem{cor}{Corollary}

\newcommand{\pr}{\noindent{\bf Proof}. }
\newcommand{\rem}{\noindent{\bf Remark}. }
\newcommand{\rems}{\noindent{\bf Remarks}. }

\newcommand{\pa}{\partial}
\newcommand{\one}{\cO(1)}

\newcommand{\const}{\textrm{const}}

\newcommand{\supp}{ \mathrm{ supp  }}
\newcommand{\hs}{ \hspace{1cm}}

\newcommand{\Tr}{\textrm{ Tr }}

\newcommand{\un}{\underline}
\newcommand{\loc}{ \textrm{loc}}
\newcommand{\Vol}{\textrm{Vol}}
\newcommand{\nat}{\natural}
\newcommand{\tk}{\bbT^{-k}_{N -k}}
\newcommand{\tz}{\bbT^0_{N -k}}

\newcommand{\im}{\textrm{Im }  }

\newcommand{\B}{\Big}

\newcommand{\be}{\begin{equation}}
\newcommand{\ee}{\end{equation}}
\newcommand{\bs}{\begin{split}}
\newcommand{\es}{\end{split}}

\newcommand{\bp}{\mathbf{p}}
\newcommand{\bq}{\mathbf{q}}
\newcommand{\bx}{\mathbf{x}}
\newcommand{\by}{\mathbf{y}}

\newcommand{\sZ}{\mathsf{Z}}

\newcommand{\sN}{\mathsf{N}}

\newcommand{\sB}{\mathsf{B}}
\newcommand{\sx}{\mathsf{x}}
\newcommand{\sA}{\mathsf{A}}

\newcommand{\al}{\alpha}
\newcommand{\De}{\Delta}
\newcommand{\de}{\delta}
\newcommand{\ga}{\gamma}
\newcommand{\Ga}{\Gamma}
\newcommand{\ka}{\kappa}

\newcommand{\la}{\lambda}
\newcommand{\Om}{\Omega}
\newcommand{\om}{\omega}

\newcommand{\ep}{\epsilon}
\newcommand{\si}{\sigma}
\newcommand{\Si}{\Sigma}
\newcommand{\Up}{\Upsilon}
\newcommand{\vep}{\varepsilon}

\newcommand{\cA}{ \EuScript{A} }
\newcommand{\cB}{{\cal B}}

\newcommand{\cD}{{\cal D}}

\newcommand{\cO}{{\cal O}}

\newcommand{\cH}{{\cal H}}
\newcommand{\cS}{{\cal S}}

\newcommand{\cR}{{\cal R}}

\newcommand{\cK}{{\cal K}}
\newcommand{\cG}{{\cal G}}
\newcommand{\cM}{{\cal M}}
\newcommand{\cN}{{\cal N}}
\newcommand{\cW}{{\cal W}}
\newcommand{\cQ}{{\cal Q}}
\newcommand{\cL}{{\cal L}}

\newcommand{\cZ}{{\cal Z}}

\newcommand{\bbR}{{\mathbb{R}}}
\newcommand{\bbZ}{{\mathbb{Z}}}
\newcommand{\bbC}{{\mathbb{C}}}

\newcommand{\bbT}{{\mathbb{T}}}

\newcommand{\bbI}{{\mathbb{I}}}

\begin{document}

\title{Nonperturbative  renormalization of scalar QED in d=3}
\author{ 
J. Dimock
\thanks{dimock@buffalo.edu}\\
Dept. of Mathematics \\
SUNY at Buffalo \\
Buffalo, NY 14260 }
\maketitle

\begin{abstract}
\end{abstract}

 
 
\section{Introduction}

\subsection{overview}
We  study  scalar  quantum electrodynamics  (QED) as a Euclidean field theory  on a  toroidal lattice in dimension  $d=3$.    Our concern  is  the ultraviolet  problem of controlling the limit  as the lattice spacing goes to zero.   We have nothing to say about   the infinite volume limit,  and   for convenience we  take  unit volume.

The  renormalization problem  is to choose   counter terms   so that   the model remains well-behaved  as the lattice spacing goes to zero.      We  carry this out  in the framework of the renormalization group  (RG)  defined with   block averaging.   Working in  a bounded field approximation
we  study   the  flow of the renormalization  group   transformations   as a problem in discrete dynamical systems.
In this framework we show   that  counter terms  can be chosen  so that  superficially divergent quantities   (for this model   the  vacuum energy and the scalar   mass)   flow to  preassigned values,   and the other parameters  in the model  stay  bounded,   so the model is well-defined.   This is nonperturbative  renormalization:  there are no expansions in the coupling constant  and no Feynman diagrams.

 Our bounded field approximation fits nicely with  the formulation of  the  RG   developed by 
 Balaban     \cite{Bal82a} - \cite{Bal83b}  who also studies scalar  QED in $d=3$.   In this approach  at each stage of the iteration  the field  space is split into large and small (=bounded) fields.   The renormalization problem  is   confined to the small field region which we consider here.     This   is    supplemented   by   a  treatment of the large field region which gives  tiny corrections to the analysis.     This is  carried out   by  Balaban  and      leads  to  an ultraviolet   stability estimate on    the partition function.

     However  in  the papers   \cite{Bal82a} - \cite{Bal83b}  renormalization   is accomplished  by picking specific counter terms   suggested by perturbation theory  and  then exhibiting  the cancellations.   The final result  is non-perturbative,  but in intermediate steps  one  is obliged to  consider  Feymann diagrams of rather high  order.    In this respect the present paper is an improvement.

 Another feature of    \cite{Bal82a} - \cite{Bal83b} inviting improvement  is that the gauge field is taken to be massive.   
  Here  the analysis is   carried out with the  more physical    massless gauge field. 
   
 A third   feature of     \cite{Bal82a} - \cite{Bal83b}  that wants improvement  is  that the full flow of the RG is not developed.  Instead estimates above and below are taken   after each transformation.    This makes in awkward to extend results beyond the partition function, for example   to control   the   correlation functions. Balaban fixes this in subsequent papers  on other models,   but it remains  undeveloped for scalar  QED.

 Closest to the present  paper  is  the  series of paper  \cite{Dim11}, \cite{Dim12},  \cite{Dim13}   by the author.   These are  on the $\phi^4$  model  in $d=3$,        essentially  the  present model  without the gauge field.   The first   paper  in the series    introduced   the non-perturbative   renormalization  technique   employed here.    The remaining papers  completed the analysis
 of the large field region,   developed the full flow of the RG,     and obtained a stability bound.  This  could be a model for the
 completion of the program for  scalar  QED.   but it is not undertaken here.
 
 Another important source of   ideas for the present work   are the papers   \cite{BIJ85},  \cite{BIJ88}  by      Balaban,  Imbrie,  and Jaffe.    
 They     study    the abelian Higgs model    which is     scalar QED  with a special 
scalar   potential.   
 We   also mention    earlier work      by         Brydges, Fr\"{o}hlich, and Seiler   \cite{BFS79},  \cite{BFS80}, \cite{BFS81} who   treat  scalar  QED in $d=2$.   
 
 In this paper our nonperturbative renormalization   method  is   applied to  a model that is super-renormalizable  with   no coupling constant 
renormalization.  However  there is no obstacle  in principle to applying it to renormalizable   models,  or
possibly even some nonrenormalizable models.

\subsection{the model}
 The model is defined    as follows.   Let $L$  be  a    (large)  positive odd   integer.        We  work on three dimensional  lattices 
 \be   \bbT^{-N}_{M}  =    ( L^{-N}  \bbZ/  L^M \bbZ)^3  \ee
  with lattice spacing  $ \ep  =  L^{-N}$  and linear size  $L^M$.
  At  first   we  take   a  fine  lattice with unit volume   $\bbT^{-N} _0$.  
  On this lattice we consider  scalar fields $\phi:  \bbT^{-N}_0  \to  \bbR^2$.  The field   $\phi=  (\phi_1, \phi_2)$    is often regarded as a  complex valued field   $\phi  = \phi_1 + i \phi_2$,  but  not here.
(Later  we  allow complex valued fields,  but  then each component  will be separately complex; 
with this  formulation the action will be analytic in the fields.)  The gauge group is  $SO(2)$ with Lie algebra   the real numbers  $\bbR$.
Elements of the group can be written  $e^{q \theta}$  where      $\theta  \in \bbR$ and 
\be   q = \left[   \begin{array}{cc}   0  &  -1 \\   1 & 0  \end{array}  \right]
\ee  

There  is  also   an   abelian gauge field  (electromagnetic potential,  connection)   $\cA:   \{\textrm{bonds in }   \bbT^{-N}_0  \}  \to  \bbR$.   A bond from $x$ to a nearest neighbor  $x'$   is the ordered pair    $b= [x,x']$.   We  require  that  
  $\cA(b) =  \cA(x,x')  = -\cA(x',x)$.    A   covariant derivative  on scalar fields  
is    defined on   bonds    by   
\be
 ( \pa \cA)(b) =   (\pa_{\cA}  \phi)(x,x')   =  \B( e^{qe\ep\cA(x,x')}  \phi(x')  -  \phi(x)  \B)  \ep^{-1} 
\ee 
where $e$ is the scalar charge. 
 If  $\{ e_{\mu} \}$ is the standard  basis  then  oriented bonds   have the form
$[x,x'] =  [x,  x+ \ep e_{\mu}]$,    we  write   $A_{\mu}(x)  =  A(x, x+  \ep  e_{\mu})$ 
and   define  
\be
(\pa_{\cA, \mu}  \phi)(x)   =   (\pa_{\cA}  \phi)(x,x+ \ep  e_{\mu})    =   \B( e^{qe\ep\cA_{\mu}(x)}  \phi(x+ \ep  e_{\mu})  -  \phi(x)\B) \ep^{-1} 
\ee
  The ordinary derivative  $\pa_{\mu} \phi$   has  $\cA =0$.   
The gauge field   $\cA$ has field strength $d\cA$ defined on plaquettes  (squares)  by  
\be          d\cA(p) = \sum_{b \in \pa p}    \cA (b)\   \ep^{-1}  \ \ \   \textrm{ or }   \ \ \
(d \cA)_{\mu \nu} (x)   =  d \cA  \B(x, x + \ep e_\mu,     
    x + \ep e_\mu + \ep e_{\nu}, x + \ep e_\nu,  x\B)
\ee

The action  is   
\be
S(\cA, \phi)   =    \frac12   \| d \cA  \|^2 +     \frac  12   \|\pa_{\cA}  \phi \|^2     +  V(\phi) 
\ee   
with potential   
\be    \label{s1}
 V( \phi)    =  
 \vep^N     \Vol ( \bbT^{-N} _0)  +  \frac12 \mu^N  \| \phi  \|^2   +  \frac 14  \la   \int  |\phi (x)|^4  dx  
\ee
Here the   norms  are  $L^2$ norms and  integrals are weighted sums,  for example   
\be 
   \| \phi  \|^2    =  \int | \phi(x)|^2 dx   =    \sum_x  \ep^3 |\phi(x)|^2 
  =    \sum_x  \ep^3   \B(    \phi_1(x)^2  +    \phi_2(x)^2 \B)   
  \ee
 The norms  involving   derivatives  or gauge potentials   are sums over oriented bonds and oriented plaquettes:
 \be
 \begin{split}
   \|\pa_{\cA}  \phi \|^2 = &  \sum_{\mu}  \int   | \pa_{\cA}  \phi(b)|^2 db  \equiv      \sum_{\mu}  \sum_x \ep^3 | \pa_{\cA, \mu}  \phi(x)|^2   \\
   \|   d \cA \|^2  = &  \int | d \cA (p) |^2  dp \equiv     \sum_{\mu < \nu} \sum_x  \ep^3  |  (d \cA)_{\mu \nu} (x) |^2
   \end{split}
      \ee
 In the potential    $\la>0 $ is the scalar coupling constant.   
  The vacuum energy $ \vep^N$   and  the scalar mass    $   \mu^N$  will  be chosen to depend on  $N$.   The $N \to \infty  $  limit   formally   gives the standard continuum theory.   We   are interested   in bounds uniform in  $N$  on  things   like   the partition function
\be    \int    \exp (  -  S(\cA,  \phi)    ) \ D  \cA  \   D  \phi  \ee
where
\be   D  \cA  =  \prod_{b}  d  (A(b) )   \hs   D \phi  =  \prod_x   d  ( \phi(x)  )   \ee
However  the integral will need gauge fixing  to enable convergence.

 The  action  is     gauge invariant.
 For   $\la:  \bbT^{-N}_0  \to  \bbR$   a  gauge transformation is defined by   
 \be   \label{snort}
    \phi^{\la} (x)  =   e^{qe \la(x)}  \phi(x)    \hs    \cA^{\la} (x,x')  =  \cA(x,x')  -  \pa  \la(x,x') 
 \ee  
Then $\pa_{\cA^{\la}}  \phi^{\la}  =  ( \pa_{\cA}  \phi)^{\la} $  and    $| \phi^{\la}| =| \phi|$ and so        
$S( \cA^{\la}, \phi^{\la}) =S(\cA,  \phi) $.

Another  symmetry  is charge conjugation invariance.   We  defined by 
\be  
 C = \left[   \begin{array}{cc}   1  &  0 \\   0 & -1  \end{array}  \right]
\ee  
 Then  $Cq= - qC$  and so    $\pa_{- \cA}  C \phi  = C   \pa_{ \cA}   \phi $.        Since also    $|C \phi|  = | \phi |$   we have   
$ 
S(- \cA,C  \phi)  =S(\cA, \phi) 
$.

\subsection{the scaled model}
We scale up  to  the  large   unit lattice  $\bbT^0_N$,  so the ultraviolet  problem  is recast as in  infrared problem,  the natural home of the renormalization group.       Let $\Phi: \bbT^0_N  \to \bbR^2$  
and  $A:   \{\textrm{ bonds in }  \bbT^0_N   \}  \to  \bbR$ be fields  on this  lattice.  These  scale down  to fields   on the 
original lattice   $\bbT^{-N}_0$   by 
\be
       A_{L^{-N}}(b)  =  L^{N/2} A(L^N b)   \hs  \hs      \Phi_{L^{-N}}(x)  =  L^{N/2} \Phi(L^N x)
\ee
The  action on the  new  lattice $\bbT^0_N$   is   $S_0( A, \Phi)  =   S(A_{ L^{-N}} ,  \Phi_{L^{-N}} ) $
which is  
\be   \label{snow1}
S_0(A, \Phi)   =   \frac12   \|  d A  \|^2     + \frac  12   \|  \pa _{A}  \Phi \|^2   +   V_0(\Phi)  
\ee
where  
\be   \label{snow2}  V_0(\Phi)    = 
   \vep^N_0   \Vol ( \bbT^{0} _N)  + \frac12  \mu^N_0  \| \Phi  \|^2   +  \frac 14  \la^N_0   \sum_x  | \Phi(x)|^4   
\ee
Now norms are defined with unweighted sums,   and derivatives are unit lattice derivatives such as
\be
(\pa_{A}  \Phi)(x,x')   =  e^{qe^N_0A(x,x')}  \Phi(x')  -  \Phi(x)  
\ee 
The   scaled coupling constants are  now tiny and given by   
\be    
   e^N_0  =   L^{-\frac12  N} e  \hs     \la^{N} _0  =  L^{-N}  \la
  \ee
The  scaled  counter terms are
\be
 \vep^N_0 =L^{-3N}  \vep  \hs     \mu^N_0  =  L^{-2N}   \mu  
 \ee 
  In the following we omit the superscript $N$  writing  $e_0,  \la _0 $
and    $\vep_0,      \mu_0$.

As  we  proceed with the RG analysis the volume will shrink back down.  After $k$ steps the torus will be $\tk$.   The coupling constants scale up to    
\be    
    e_k    =   L^{\frac12 k  } e_0  =   L^{-\frac12(N-k)  } e  \hs     \la_k    =  L^{k}  \la_0  = L^{-(N-k)} \la  
  \ee
 The  other  coupling constants  $\vep_k,  \mu_k$  will     evolve in a more complicated manner.
 \bigskip

 \noindent
\textbf{Convention}:    Throughout the paper   the convention is    that   $\one$ is a constant   independent of all parameters.  Also  
$C, \ga$  are  constants   ($C \geq 1,  \ga \leq 1$)    which may depend on $L$ and which    may change from line to line.

\section{RG transformation for scalars}
 
We  explain how  the RG     transformation  is   defined  for scalars, but with a gauge field background.   The discussion follows     \cite{Bal82a},   \cite{Bal82b},  \cite{Bal87},     \cite{Dim11},  \cite{Dim14}.
 
 \subsection{block averages}  \label{singer}
 
 We  start with fields  $A,   \Phi$ on  the unit lattice $\bbT^0_N$.
    We  want to define    a  covariant   block averaging operator  $Q(A)$  taking  $\Phi$ to    
    a  function $Q(A) \Phi$    on the  $L$-lattice   $\bbT^1_N$.    
      In any $3$-dimensional  lattice  let  $B(y)$  be  a cube of $L^3$ sites  ($L$ on a side) centered on  a point $y$.   
 Here the lattice is    $\bbT^0_N$,  and    for     $y \in  \bbT^1_N$    we have    
 \be
  B(y) =  \{x  \in  \bbT^0_N: |x-y| < L/2  \}  \hs    |x-y|  = \sup_{\mu} |x_{\mu}- y_{\mu}|  
   \ee
  The $B(y)$ partition the lattice.  
 For   $x \in B(y)$  let   $\pi$ be a permutation of  $(1,2,3)$ and  let  $\Ga^{\pi}(y,x)$ be that path  from  $y$ to $x$ obtained 
 by  varying each coordinate to its final value in the order  $\pi$.     There are $3!$ of these.
  For  any     path  $\Ga$    let    $ 
  A(\Ga )  =  \sum  _{b \in  \Ga}   A(b) $
 and define an average over the various  paths  from  $y$ to $x$    by 
 \be   \label{wombat} 
  (\tau A)(y,x))  =      \frac{1}{3!}  \sum_{\pi }    A(\Ga^{\pi}(y,x) )
 \ee
   Then we  define the averaging operator
    \be    \label{dice1}
(Q(A) \Phi)(y)    =   L^{-3}   \sum_{x \in B(y)} e^{ qe_0  ( \tau A)(y,x) }  \Phi(x)  \hs   y \in \bbT^1_N
 \ee
   This  is  constructed  to be gauge covariant.   We have  $(\tau A^{\la})(y,x )  = ( \tau A)(y,x  )     - ( \la(x) - \la(y))$ 
    for   $\la:  \bbT^0_N  \to  \bbR$ and so
 \be      \label{study} 
 Q(A^{\la}) \Phi^{\la}  =  (Q(A) \Phi)^{\la^{(1)}}  
 \ee
   where  $\la^{(1)}$  is  $\la  $  restricted to  the  lattice    
 $\bbT^1_{N}$.

 Because we  average over paths,  rather than taking  a fixed path,   our definition is 
   covariant under  symmetries of the lattice   $\bbT_N^1$.  In particular    if  $r$ is     a rotation by a multiple of   $\pi/2$  or  a reflection 
   and   $\Phi_r(y)  =  \Phi(r^{-1} y)$   and   $A_r(b)  =   A(r^{-1} b)$  
 then   
\be
Q(A_r) \Phi_r  =    (Q(A) \Phi)_r      
\ee

    The adjoint operator   (transpose operator)  maps  functions $f$    on   $\bbT^1_N$ to functions on $\bbT^0_N$.  It is computed with
    sums on  $\bbT^1_N$  weighted by  $L^d$   and   is  given by  
 \be
   (Q^T(A)f)(x)   =   e^{ -qe_0(\tau A)(y,x)} f(y)    \hs       x  \in B(y)  
    \ee
Then we  have  
\be      Q(A)  Q^T(A)     = I 
 \ee
and  $Q^T(A)Q(A)$ is an orthogonal   projection.

 \subsection{the  transformation}
 
 Suppose we start with a density  $\rho (\cA,  \phi  ) $ with scalar field   $\phi$ and  background gauge field $\cA$   on 
   $\bbT^{-N}_0$.   It scales up
 to a density   
 \be
    \rho_{0}(A,  \Phi_0)  \equiv   \rho_{L^N}  ( A,  \Phi_0)   \equiv       \rho(A_{L^{-N}}  \Phi_{0, L^{-N}} ) 
 \ee
 where   $A, \Phi_0$  are defined      on  $\bbT^0_N $.    
Starting with  $\rho_0( \cA, \Phi_0)$ we  define a sequence of densities  $\rho_k(  \cA,  \Phi_k)$
 defined for   $\cA$  on   $\tk$ and  $\Phi_k$    on      $ \bbT^0_{N-k}$.      They are defined recursively    
first by 
 \begin{equation}
\begin{split}  \label{kth}
 \tilde  \rho_{k+1} ( \cA,   \Phi_{k+1}) 
=  & \int    \de_G\Big( \Phi_{k+1} -  Q(\cA ) \Phi_k  \Big)  \rho_{k} (\cA,  \Phi_k)   D \Phi_k \\
 \end{split}
\end{equation}
where  $\Phi_{k+1}$ is defined on the coarser lattice  $\bbT^1_{N-k}$.
 The   $\de_G$ is a Gaussian approximation to  
the delta function.
\be  \label{spring2}
\begin{split}
\de_G\Big( \Phi_{k+1} -  Q(A) \Phi_k   \Big)  
 =   &  \Big(\frac{aL}{2\pi} \Big)^{s_{N-k-1}}   \exp  \Big(  - \frac{aL}{2}   |\Phi_{k+1} -  Q(A) \Phi_k |^2  \Big)  \\
  =  &   \Big(\frac{aL}{2\pi} \Big)^{s_{N-k-1}}     \exp  \Big(  - \frac{a}{2L^2}  \|\Phi_{k+1} -  Q(A) \Phi_k \|^2  \Big)\\  
\end{split}
\ee
Here     $ |\Phi_{k+1} -  Q(A) \Phi_k | $  is the $L^2$ norm with   a simple sum over  points  in   $\bbT^1_{\sN}$,  whereas  in   
$ \|\Phi_{k+1} -  Q(A) \Phi_k \| $ it is the $L^2$ norm  with   the sum over points   weighted by the factor  $L^3$   natural for this lattice.    
 The averaging operator  $Q(\cA)$    is  taken to be a modification of  (\ref{dice1}):
  \be    \label{dice2}
(Q(\cA) \Phi_k)(y)    =   L^{-d}   \sum_{x \in B(y)} e^{ qe_kL^{-k} (\tau\cA)(y,x) }  \Phi_k(x)  
 \ee
Here    $(\tau\cA)(y,x)$   is still defined by  (\ref{wombat}),  but  now in $\cA(\Ga)$ the sum is over bonds of length  $L^{-k}$ hence the weighting 
factor  $L^{-k}$ in the exponent.    In general     $ s_N \equiv  L^{3N}  $ is  the  number of sites in a three dimensional
tidal   lattice with $L^N$ sites on a side
 The normalization  factor    $(aL/ 2 \pi)^{s_{N-k-1}}$ in  (\ref{spring2})   is      chosen  so that  $ \int  d \Phi_{k+1} \ \de_G\Big( \Phi_{k+1} -  Q(A) \Phi_k  \Big)  =1  $.
(Recall that there are two components  per site.)   Therefore    
\be     \label{bell}
 \int  \tilde   \rho_{k+1} (\cA,  \Phi_{k+1})\    D \Phi_{k+1}  =     \int     \rho_{k} (\cA,  \Phi_{k} )\   D \Phi_k
\ee

Next     one  scales  back to the unit lattice.   
 If   $\cA$ is a field  on   $\bbT^{-k-1}_{N-k-1}$   and  $ \Phi_{k+1}$ is a field   on $\bbT^{0}_{N-k-1}$   then
   then   
\be
\cA_{L}(b)  = L^{-1/2}   \cA( L^{-1}b) \hs  \Phi_{k+1,L}(x) =  L^{-1/2}\Phi_{k+1} (L^{-1}x)
\ee
are  fields  on  $\tk$   and  $\bbT^1_{N-k}$ respectively,   and we  define     
  \begin{equation}   \label{scaleddensity}
 \rho_{k+1 } ( \cA,  \Phi_{k+1})  =  \tilde  \rho_{k+1} (\cA_L, \Phi_{k+1,L}) L^{s_N  -  s_{N-k-1} }   
 \end{equation}
If   $\Phi'_{k+1} =  \Phi_{k+1,L}$ then             $D  \Phi'_{k+1}  =   L^{-s_{N-k-1} }   D \Phi_{k+1} $ and    we have   by  (\ref{bell})     
 \begin{equation}  \label{preserve}
 \begin{split}
 & \int       \rho_{k+1} (\cA,   \Phi_{k+1})   D \Phi_{k+1}  =  L^{s_N  -  s_{N-k-1} }  \int     \tilde  \rho_{k+1} (\cA_L, \Phi_{k+1,L})     D \Phi_{k+1}   \\ 
 & =  L^{s_N}    \int     \tilde  \rho_{k+1} (\cA_L, \Phi'_{k+1})     D \Phi'_{k+1}
    =   L^{s_N} \int       \rho_{k} ( \cA_L,  \Phi_{k})    D \Phi_k
 \\
\end{split} 
\end{equation}

\begin{lem}   For    $\cA$ on  $\tk$  and  $\Phi_k$ on   $\tz$
\be   \label{lincoln0}
   \int  \rho_{k } ( \cA,  \Phi_k)  D \Phi_k   =      \int   \rho_{0} (\cA_{ L^k}, \phi_{L^k} )   \  D \phi   
\ee
where the integral is over  $\phi$ on   $\tk$. 
\end{lem} 
\bigskip

\rem
In  particular  since      $\rho_{0} (\cA_{ L^N}, \phi_{L^N} ) =\rho (\cA, \phi  )$
 \begin{equation}  \label{preserve2}
\int       \rho_{N} (\cA,   \Phi_{N})   D \Phi_{N}   =   \int   \rho_{0} (\cA_{ L^N}, \phi_{L^N} )   \  D \phi   
=  \int       \rho (\cA,   \phi )   D \phi
 \end{equation}
 and we are back to the integral of our original density.      
 The right side  is the integral over a many dimension space,  but   can be computed as the left   side  which    is the integral over a one dimensional space.    This is the point of the renormalization group approach. 
\bigskip

\pr   It is true for $k=0$;  suppose it is true for  $k$.   If   $\phi  =  \phi'_{L}$    then  $D \phi =  L^{ -s_N} D \phi' $  and so     by  (\ref{preserve}) 
\be  
\begin{split}
&  \int  \rho_{k+1 } ( \cA,  \Phi_{k+1})  D \Phi_{k+1} 
=        L^{s_N} \int       \rho_{k} ( \cA_L,  \Phi_{k})    D \Phi_k  \\
=  &      L^{s_N} \int   \rho_{0} (\cA_{ L^{k+1}}, \phi_{L^k} )   \  D \phi   
=       \int   \rho_{0} (\cA_{ L^{k+1}}, \phi'_{L^{k+1}} )   \  D \phi'    \\
\end{split}
\ee
Hence it is true for  $k+1$.

\subsection{compositions of averaging operators}  To  investigate   the sequence      $  \rho_{k} ( \cA,   \Phi_{k})$     we  first  study how averaging operators compose.
Suppose  we   have  already   defined   a $k$-fold averaging operator $Q_k (\cA)$  depending on     $\cA $  on  $\tk$  and sending
 functions   on    $\bbT^{-k}_{N-k} $  to  functions  on  $\bbT^0_{N-k}$.  We  we  define  the same for  $k+1$ as follows.   First
 define   for  the same  $\cA$   an  operator  
 \be
   Q_{k+1} (\cA)  =  Q(\cA) Q_k(\cA)
 \ee
 which   maps   functions   on    $\bbT^{-k}_{N-k} $  to  functions  on  $\bbT^1_{N-k}$.
Then for $\cA,  f $  on     $ \bbT^{-k-1}_{N-k-1} $   define  
\be   \label{dingdong}
   Q_{k+1} (\cA)f  =   (  Q_{k+1} (\cA_L)f_L)_{L^{-1}}
    \ee
which maps    functions on   $ \bbT^{-k-1}_{N-k-1} $  to functions on  $ \bbT^0_{N-k-1} $.

We need an  explicit representation for  $Q_k(\cA)$.    For any lattice  let     $B_k(y)$  be a block  with  $L^{dk}$ sites  ($L^k$ on a side)
centered on  $y$.     
  Suppose    $x \in \tk$  and  $y  \in  \bbT^0_{N-k}$  satisfy  $x \in B_k(y)$, which is the same as   
  $|x-y| < \frac12$.    There is an associated sequence
$x=y_0,  y_1, y_2,  \dots  y_k = y$ such that  $y_j  \in   \bbT^{-k+j}_{N-k}$  and   $x  \in  B_j(y_j)$. Define
\be    \label{sum}
(\tau_k\cA)(y,x)  =  \sum_{j=0}^{k-1}   (\tau \cA) (y_{j+1}, y_{j}  )  
\ee

\begin{lem}   For  $\cA,  f$ on   $\tk$    
\be
\begin{split}
(Q_k(\cA)  f  )  (y)  =  &   \int_{ |x-y| <  \frac12}   e^{ qe_k L^{-k} (\tau_k\cA)(y,x) }f(x) \ dx \\
 (Q_k^T(\cA)f)(x)  =   &     e^{-q e_k L^{-k}(\tau_k\cA)(y,x) }   f(y)    \hs   x \in B_k(y)
  \\  
\end{split}
\ee
\end{lem}
\bigskip

\pr   The proof is by induction on  $k$.   Assuming it is true for $k$ we have with  $\eta = L^{-k}$
\be 
\begin{split}   ( Q_{k+1} (\cA) f) (y')  =  &  \Big(Q(\cA) Q_k(\cA)f  \Big)(y') \\
 =   &   L^{-3}   \sum_{y \in B(y')}  e^{ qe_k \eta   (\tau\cA)(y',y) }      \int_{ |x-y| <  \frac12}   e^{ qe_k\eta (\tau_k\cA)(y,x)   }f(x) \ dx
 \\
  =   &   L^{-3}      \int_{ |x-y'| <  L/2}   e^{ qe_k  \eta (\tilde    \tau_k\cA)(y',x)    }f(x) \ dx   \\
\end{split}
\ee
Here we  have defined
\be  
(\tilde    \tau_k\cA)(y',x) =  (\tau\cA)(y',y) +  (\tau_k\cA)(y,x)  =     \sum_{j=0}^{k} (\tau  \cA)(y_{j+1}, y_{j}   )
\ee
where  $y_{k+1} = y',  y_k =y,  y_0 = x$.
Now we   scale by (\ref{dingdong})   and get  for  $y'   \in   \bbT^0_{N-k-1} $    and  $x'  \in  \bbT^{-k-1}_{N-k-1} $
\be 
\begin{split}   ( Q_{k+1} (\cA) f) (y')    =   &   L^{-3}      \int_{ |x-Ly'| <  L/2}  e^{ qe_k  \eta(\tilde  \tau_{k+1}\cA_L) (Ly',x)   }f(x/L) \ dx
 \\
 =   &        \int_{ |x' -y'| <  \frac{1}{2}}   e^{qe_k \eta(\tilde  \tau_{k+1}\cA_L) (Ly',Lx')   }f(x')  \ dx'
 \\
\end{split}
\ee
Taking into account  that  $\cA_L (\Ga)  = L^{-\frac12}  A(L^{-1}\Ga  ) $
we have  
\be
\begin{split}
e_k \eta(\tilde  \tau_{k+1}\cA_L ) (Ly',Lx')    =  &e_k   \eta\sum_{j=0}^{k} (\tau \cA_L)(y_{j+1}, y_{j}  ) \Big |_{y_{k+1} = Ly', y_0 = Lx'}  \\
=  &e_k   \eta     L^{-   \frac12} \sum_{j=0}^{k} (\tau  \cA)(L^{-1}y_{j+1},L^{-1}  y_{j} )   \Big |_{y_{k+1} = Ly', y_0 = Lx'} \\ 
 = & e_{k+1} L^{-k-1}( \tau_{k+1}  \cA)  (y',x')  \\
\end{split}
\ee
This gives the first   result.  The expression for the adjoint is a short calculation.

\begin{lem}     For   $\cA$  on   $\tk$  and  $\Phi_{k}$  on $\bbT^0_{N-k}$ the density
$\rho_{k, \cA}(\Phi_k)$ can be written  
  \begin{equation}    
 \label{second}
 \rho_{k,\cA} ( \Phi_k)  
 =  \left(  \frac{a_k }{2 \pi}  \right)^{s_{N-k} }  \ \int \exp \left(-  \frac{a_k}{2} 
\|\Phi_{k} -Q_k(\cA) \phi \|^2 \right)    \rho_{0,  \cA_{L^k}} ( \phi_{L^k}) \ \  D  \phi   
\end{equation}
where     $\phi, \cA$ are on    $\bbT^{-k}_{N-k}$    and 
\begin{equation}
a_k  =   a \frac{1-L^{-2}}{1- L^{-2k}}
\end{equation}
\end{lem}
\bigskip

\pr    The proof is by induction.  For $k=1$ it follows from    (\ref{kth}),(\ref{scaleddensity}).  
   Assuming it is true for $k$  we compute
\begin{equation}    \label{queen}
\begin{split}
&  \tilde  \rho_{k+1, \cA} ( \Phi_{k+1})  \\
& = \const
\int \exp \left(- \frac12   \frac{a}{L^2}
\|\Phi_{k+1}-Q(\cA)\Phi_k\|^2 -  \frac{a_k}{2} 
\|\Phi_{k} -Q_k(\cA) \phi \|^2 \right)    \rho_{0, \cA_{L^k}} ( \phi_{L^k}) \ \ D \phi\ \   D \Phi_k  \\
\end{split}
\end{equation}
The   expression inside the exponential has a minimum  in $\Phi_k$ when 
\begin{equation}
\Big(a_k  +  \frac{a}{L^2}    Q(\cA)^TQ(\cA)  \Big)\Phi_k   =    a_k Q_{k}(\cA) \phi    + \frac{a}{L^2} Q^T(\cA)  \Phi_{k+1}   
\end{equation}
This has the solution $\Phi_k  =  \Phi_k^{\min}(\cA)= \Phi_k^{\min}(\cA;  \Phi_{k+1}, \phi) $  where
\begin{equation}  \label{kingmaker0}
\begin{split}
 \Phi_k^{\min}(\cA)  =  & Q_{k}(\cA)  \phi  
- \frac{aL^{-2}}{a_k+ aL^{-2}}Q^T(\cA)  Q_{k+1}(\cA)  \phi  +  \frac{aL^{-2}}{a_k+ aL^{-2}}Q^T(\cA)   \Phi_{k+1}   \\
\end{split}
\end{equation}  
We   compute  the value at the minimum   using  $Q(\cA) Q^T(\cA) =1$  and $a_{k+1}   = aa_k/(a_k+ aL^{-2})$
and find  (see  \cite{Dim11} for details)  
\begin{equation}  \label{fifty}
 \frac{a}{2L^2} \|\Phi_{k+1}-Q(\cA)  \Phi_k^{\min}(\cA)\|^2 +  \frac12    a_k \| \Phi_k^{\min}(\cA) -   Q_{k} (\cA)  \phi \|^2
=  \frac{a_{k+1}}{2L^2}      \|  \Phi_{k+1}   -   Q_{k+1} (\cA) \phi  \|^2  
\end{equation}

Now  in the integral  (\ref{queen})  expand around the minimizer.  We write  $\Phi_k =  \Phi_k^{\min}(\cA) +Z$ and integrate over
$Z$.  The term with no $Z$'s is   (\ref{fifty}). The linear terms in $Z$ vanish  and the terms quadratic  in $Z$  when integrated over $Z$ yield a constant.
Thus we have  
\begin{equation}    \label{queen2}
 \tilde  \rho_{k+1,\cA} ( \Phi_{k+1}) 
= \const
\int \exp \left(-   \frac{a_{k+1}}{2L^2}      \|  \Phi_{k+1}   -   Q_{k+1}(\cA) \phi  \|^2         \right) 
    \rho_{0,  \cA_{L^k}} ( \phi_{L^k})  \ D    \phi  
\end{equation}
Scaling by  (\ref{scaleddensity})  we  have  for    $ \cA, \phi'$  on   $ \bbT^{-k-1}_{N-k-1}  $  and  $\Phi_{k+1}$ on  $\bbT^0_{N -k-1} $
\begin{equation}    \label{queen3}
\begin{split}
   \rho_{k+1,\cA} ( \Phi_{k+1}) =  & \const
\int \exp \left(-   \frac{a_{k+1}}{2}      \|  \Phi_{k+1}   -   Q_{k+1}(\cA) \phi'  \|^2         \right)  
 \rho_{0,  \cA_{L^{k+1}}} ( \phi'_{L^{k+1}})  \ D  \phi'    
\\
\end{split}
\end{equation}
The constant must   be   $( a_{k+1}  / 2 \pi )^{s_{N-k-1} }$ to preserve  the identity  (\ref{lincoln0}).
This completes the proof.
\bigskip   

Hereafter we abbreviate the normalization factors      in  (\ref{spring2}) and    (\ref{second})  by    
\be 
  N_k  =    \left(\frac{aL}{2\pi} \right)^{s_{N-k-1}}     \hs   \cN_k =     \left(  \frac{a_k }{2 \pi}  \right)^{s_{N-k} } 
\ee

 \subsection{free flow}  \label{minmizers1}

Now consider   an initial density which is a perturbation of the free action:    
   \be 
    \rho_{0}(A,   \Phi_0)  = F_0( \Phi_0)    \exp \B(  -  \frac12 \| \pa_A \Phi_0 \|^2\B)
\ee
     Insert this in (\ref{second}) 
and use    for   $\cA, \phi  $   on   $\bbT^{-k}_{N - k}$ 
\be 
   \frac12 \|  \pa_{\cA_{L^k} }  \phi_{L^k}  \|^2   = \frac12 \|  \pa_{\cA}  \phi \|^2     =         \frac12  <\phi, ( -\De_{\cA}   ) \phi>  
   \ee
where       $-\De_{\cA}  \equiv    \pa_{\cA}^T  \pa_{\cA}$ is defined with 
covariant derivatives  containing the coupling constant  $e_k$.
Then   with    $ F_{0, L^{-k}}( \phi ) =   F_{0}( \phi_{L^k} )   $ we have from (\ref{second})
 \be
 \label{something}
 \rho_{k} (\cA,  \Phi_k)    = 
\cN_k  \int     F_{0, L^{-k}}( \phi ) 
  \exp  \Big( -   \frac{a_k}{2}  \|    \Phi_k -  Q_k(\cA) \phi   \|^2      -    \frac12  <\phi, ( -\De_{\cA}  ) \phi>     \Big)    D  \phi  
\ee
The minimizer in $\phi$ of the expression in the exponential  is  
\be 
         \cH_k(\cA )  \Phi_k    
\equiv      a_k  G_k(\cA) Q^T_k(\cA)\Phi_k   \ee
where $G_k(\cA)$ is the Green's function 
\be  \label{Gk}
   G_k(\cA)   =   \Big(- \De_{\cA}   +  a_k  Q^T_k(\cA) Q_k(\cA)  \Big)^{-1} 
\ee
The inverse exists since this is a strictly positive operator. 

Expanding the exponential around the minimizer with  $\phi      =   \cH_k(\cA )  \Phi_k    +  \cZ$         we   find       
 \be   \label{spinit}
   \rho_{k,\cA} ( \Phi_k)    =   \cN_k    \sZ_k(\cA)  F_k \B(  \cH_k(\cA )  \Phi_{k}  \B)   
      \exp  \Big( -   \frac12 < \Phi_k,  \De_k(\cA)  \Phi_k>    \Big)
  \ee
where
\be  \label{route66}
\begin{split}
 < \Phi_k,  \De_k(\cA)  \Phi_k>     =  &      \frac{a_k}{2}  \|    \Phi_k -  Q_k(\cA)    \cH_k(\cA )  \Phi_k    \|^2      +    \frac12  <   \cH_k(\cA )  \Phi_k , ( -\De_{\cA}  )   \cH_k(\cA )  \Phi_k >
\\
= &  < \Phi_k , \B(   a_k -  a_k^2 Q^T_k(\cA)G_k(\cA)Q_k(\cA)  \B) \Phi_k>     \\
   F_k \B(  \cH_k(\cA )  \Phi_{k}  \B) 
    =  &   \sZ_k(\cA)  ^{-1}  \int      F_{0, L^{-k}}\B(  \cH_k(\cA )  \Phi_{k}     +  \cZ \B)  \\
  &    \exp \B(   -    \frac12  <\cZ, \B( -\De_{\cA}       +  a_k  Q^T_k(\cA) Q_k(\cA)   \B)  \cZ >  \B)  \\
 \sZ_k(\cA)   =  & \int  \exp \B(   -    \frac12  <\cZ, \big( -\De_{\cA}       +  a_k  Q^T_k(\cA) Q_k(\cA)  \B) \cZ > \Big)   D  \cZ   \\
 \end{split}
 \ee
 
 \subsection{the next step}   \label{single}

If   we  start  with  the   expression  (\ref{spinit})  for  $\rho_{k, \cA}$   and apply another renormalization transformation we 
again        get  $\rho_{k+1, \cA}$.   Working out the details  will give us some useful  identities.   
We have first   
\begin{equation}   \label{manx}
\begin{split}
&   \tilde  \rho_{k+1 } (\cA,  \Phi_{k+1}) =   \cN_k   N_{k} \sZ_k(\cA) \\  
&   \int    F_k \B(  \cH_k(\cA ) \Phi_{k+1} \B)      \exp \left(-  \frac{a}{2L^2} 
\|\Phi_{k+1}-Q(\cA)\Phi_k\|^2 -    \frac12 < \Phi_k,  \De_k(\cA)  \Phi_k>  \right) D   \Phi_k  D \phi \\
\end{split}
\end{equation} 
Here   $\Phi_{k+1}, \Phi_k  $     are fields on    $\bbT^{1}_{N-k},\bbT^{0}_{N-k}$   respectively.
The minimizer  of the expression in the exponential  in    $\Phi_k$   is  
\be   \label{min1}
  H_{k} (\cA) \Phi_{k+1} \equiv   \frac{a}{L^2}  C_k (\cA)  Q^{T}(\cA) \Phi_{k+1}  
\ee
where
 \be
C_k (\cA)  =  \Big(\De_{k}(\cA) +  \frac{a}{L^2} Q^T(\cA)Q (\cA) \Big)^{-1}
\ee
Expanding around the  minimizer  with  $\Phi_k   =    H_{k} (\cA) \Phi_{k+1} + Z    $
we  obtain      
 \be   \label{spinit2}
 \begin{split}
&  \tilde    \rho_{k+1} (\cA,  \Phi_k) 
  =     \cN_k   N_{k}
   \sZ_k(\cA) \sZ^f_k(\cA)   \\   & F^*_k \B(  \cH_k(\cA ) H_k(\cA)  \Phi_{k+1}    \B)    
     \exp  \Big( -       \frac12 < H_k(\cA)\Phi_{k+1},  \De_k(\cA)  H_k (\cA)  \Phi_{k+1}>   \Big) \\
\end{split}
  \ee
Here 
\be
\sZ^f _k(\cA)   =   \int   \exp \B(   -    \frac12  <Z,  \B(  \De_k  +  \frac{a}{L^2}  Q^{T}(\cA)Q( \cA) \B) Z>       \B)  D  Z   
\ee
and      
\be  \label{route66a}
\begin{split}
     F^*_{k} \B( \cH_k(\cA ) H_k(\cA)  \Phi_{k+1}   \B) 
    =  &   \sZ^f_k(\cA)  ^{-1}  \int    F_k\B(  \cH_k(\cA ) H_k(\cA)  \Phi_{k+1}     +    \cH_k(\cA )  Z \B)  \\
  &    \exp \B(   -    \frac12  <Z,  \B( \De_k  +  \frac{a}{L^2}  Q^T(\cA)Q( \cA)\B)  Z>       \B)  \\
  = &       \int    F_k\B(  \cH_k(\cA ) H_k(\cA)  \Phi_{k+1}     +    \cH_k(\cA )  Z \B)   d \mu_{C_k(\cA)} (Z)  \\
  \end{split}     
 \ee
 where  $\mu_{C_k(\cA)  }$ is      the Gaussian measure    with covariance  $C_k(\cA) $

Next we  scale by (\ref{scaleddensity})    and get
 \be   
 \begin{split}
  \rho_{k+1} (\cA,  \Phi_{k+1})  
   =  &     \cN_k   N_{k}
   \sZ_k(\cA_L) \sZ^f_k(\cA_L)   \  L^{s_N   -s_{N-k-1}} F^*_k \B(  \cH_k(\cA_L ) H_k(\cA_L)  \Phi_{k+1,L}    \B)   \\
  &    \exp  \Big( -       \frac12 < H_k(\cA_L)\Phi_{k+1,L},  \De_k(\cA_L)  H_k (\cA_L)  \Phi_{k+1,L }>   \Big) 
  \Big) \\
\end{split}   
 \ee

Taking  $F_0 =1$  we have  $F_k = 1 $ and $F^*_k =1$.    Then taking  $\Phi_{k+1}  = 0$   and comparing
this  expression     with   (\ref{spinit}) for  $k+1$ we find 
\be    \label{z}
\cN_{k+1}\sZ_{k+1} (\cA)   =  \cN_k   N_{k}     \sZ_k(\cA_L) \sZ^f_k(\cA_L)   \  L^{s_N   -s_{N-k-1}}
\ee
Furthermore  the   exponential must be       $  \exp  \Big( -       \frac12 < \Phi_{k+1},  \De_{k+1}(\cA)  \Phi_{k+1   }>   \Big)$.
Thus in general  
 \be  
  \rho_{k+1 ,\cA} ( \Phi_{k+1})    =\cN_{k+1} \sZ_{k+1} (\cA)     F^*_k \B(  \cH_k(\cA_L ) H_k(\cA_L)  \Phi_{k+1,L}    \B) 
       \exp  \Big( -       \frac12 < \Phi_{k+1},  \De_{k+1}(\cA)  \Phi_{k+1   }>   \Big)
         \ee
  Comparing this with   (\ref{spinit}) for  $k+1$ we find 
 \be    \label{slumming}
   F^*_k \B(  \cH_k(\cA_L ) H_k(\cA_L)  \Phi_{k+1,L}    \B)  =    F_{k+1}\B(  \cH_{k+1}(\cA) \Phi_{k+1}   \B) 
 \ee 
  Next take    $F_0(\Phi_0)  = <\Phi_0, f>$.   Then   $F_k(\phi)  =  <\phi_{L^k}, f>$ for all $k$    and    $F^*_k(\phi)  =  <\phi_{L^k}, f>$ for all $k$,   
  and (\ref{slumming})  
  says  
  \be   <  \B( \cH_k(\cA_L ) H_k(\cA_L)  \Phi_{k+1,L} \B)_{L^k} ,f  >   =  <\B( \cH_{k+1}(\cA) \Phi_{k+1}\B)_{L^{k+1}}, f>  \ee
  and so    
\be   \label{h}
  \cH_k(\cA_L ) H_k(\cA_L)  \Phi_{k+1,L}     =   (\cH_{k+1}(\cA)\Phi_{k+1})_L
\ee
Now    (\ref{route66a})  evaluated at  $\Phi_{k+1, L}  $     can be written 
\be
  F_{k+1} (\cH_{k+1}(\cA)\Phi_{k+1}) 
    =   \int    F_k\B(  (\cH_{k+1}(\cA)\Phi_{k+1})_L    +    \cH_k(\cA )  Z\B) d \mu_{C_k(\cA) }(Z)      \\
\ee
More generally  for any    $\phi$ on $\bbT^{-k-1}_{N-k-1} $   one can   define the fluctuation integral    
\be  \label{f}
  F_{k+1} (\phi ) 
    =   \int    F_k\B(  \phi_L      +    \cH_k(\cA )  Z\B) d \mu_{C_k(\cA) }(Z)      \\
\ee
The identities    (\ref{z}), (\ref{h}), (\ref{f}) are what we were after.

 \section{Greens functions}

\label{green}
 
 We  study     the Green's function  $G_k(\cA)= 
    \B( - \De_{\cA}   +  a_k   (Q_k^TQ_k )(\cA) \B)^{-1}$,      an operator   on functions on    $\tk$ defined for   a background field    $\cA$ on $\tk$.  These  results are mostly   due to Balaban 
  \cite{Bal83b},  but there are some minor differences.
  
    \subsection{basic  properties}

 We collect  some  general    facts.     As  before    the Laplacian is        $-   \De_{\cA}  = \pa^T_{\cA} \pa_{\cA}$ 
where   with  $\eta  = L^{-k}$
\be  \label{pa}
 \begin{split}
 ( \pa_{\cA, \mu}    f )(x)   =& \Big ( e^{qe_k\eta  \cA_{\mu}(x)} f(x + \eta e_{\mu}) -f(x)\Big) \eta^{-1} \\
 (\pa_{\cA, \mu}^T f)(x)  =&  \Big(e^{-qe_k\eta \cA_{\mu}(x-\eta e_{\mu})}  f( x-\eta e_{\mu})  - f(x)   \Big) \eta^{-1} \\
 \end{split}
\ee
Note that these   differ  by a phase factor  for we have
\be  \label{relation0}
\begin{split}
 (\pa_{\cA, \mu}^T f)(x)   =  &  - e^{-qe_k\eta \cA_{\mu}(x-\eta e_{\mu})} ( \pa_{\cA, \mu}  f)( x-\eta e_{\mu})     \\
 \end{split}
\ee
Explicitly   
\be   \label{lap}   (- \De_{\cA} f ) (x)  =  \sum_{\mu}
 \Big( -  e^{qe_k\eta  \cA_{\mu} (x)} f(x+ \eta e_{\mu} )  + 2f(x)
   -   e^{-qe_k\eta  \cA_{\mu} (x- \eta  e_{\mu} )} f(x- \eta e_{\mu})   \Big)/ \eta^2
\ee
We  note for later reference the product rules:
\be   \label{product}
\begin{split}
\pa_{\cA, \mu} ( hf )  =&    h( \cdot + \eta e_{\mu}) \pa_{\cA, \mu}  f    +    (\pa_{\mu}h )  f   \\
\pa^T_{\cA, \mu} ( hf )  =&    h( \cdot - \eta e_{\mu}) \pa^T_{\cA, \mu}  f    +    (\pa^T_{\mu}h )  f   \\
\end{split}
\end{equation}

We  also   record the symmetries of the Green's functions .   The Laplacian  $ \De_{\cA}$  is covariant under  $\tk$  lattice symmetries and
$ (Q_k^TQ_k )(\cA)$  is   covariant under  $\bbT^0_{N-k}$ lattice symmetries.     Hence   $G_k(\cA)$ are covariant under  $\bbT^0_{N-k}$ lattice symmetries  which means
\be
G_k (\cA_r) f_r  =   (G_k(\cA)f )_r
\ee

 With gauge transformation    $\la$    on  $\tk$    defined as in  (\ref{snort})  we have   
\be   \label{buddy}
   \De_{\cA^{\la}}   = e^{qe_k \la} \De_{\cA}e^{-qe_k \la}  \hs   
  Q_k(\cA^{\la})   =   e^{qe_k \la^{(0)}}  Q_k(\cA) e^{-qe_k \la}
\ee
where   $\la^{(0)} $ is the restriction to the unit lattice $\tz$.  
It follows that  
\be   \label{sync}
  G_{k} (   \cA^{\la})   =    e^{qe_k \la}   G_{k}(  \cA)   e^{-qe_k \la} 
\hs      \cH_{k} (   \cA^{\la})   =    e^{qe_k \la}   \cH_{k}(  \cA)   e^{-qe_k \la^{(0)}  } 
\ee
Similarly we     have the charge conjugation invariance    
\be 
   G_k( - \cA)  =  CG_k(\cA) C  
    \ee

 We  also consider   the Green's function for a region  $\Om \subset  \tk$.    This has the form   
\be
G_k(\Om,  \cA))   =    \B( - \De^N_{\cA}   +  a_k    (Q_k^TQ_k )(\cA) \B)^{-1}_{\Om}
\ee
  The notation   $( - \De^N_{\cA} )_{\Om}$
denotes the Laplacian    with Neumann boundary conditions,  i.e.  as a quadratic form, 
  $ <f,( - \De^N_{\cA} )_{\Om}f> =  \| \pa_{\cA} f \|^2_{\Om}$,   only bonds contained in   $\Om$ contribute.
  Thus     $    (\pa_{\cA, \mu}f)(x)$  is  given by     (\ref{pa})   if  $x, x+ \eta e_{\mu}  \in  \Om$  and is zero otherwise.
  We  still have  $( - \De^N_{\cA} )_{\Om}  =    \pa^T_{\cA} \pa_{\cA}$   but now 
     $    (\pa^T_{\cA_0, \mu}f)(x)$  is  given by     (\ref{pa})   if  $x-\eta e_{\mu}, x  \in  \Om$  and is zero otherwise.
The expression  (\ref{lap}) for the Laplacian must be modified near the boundary.

The operator  $G_k(\Om,  \cA)$  has the same symmetry  properties  as  $G_k(\Om)$,  provided  $\Om$ is  
transformed as well for lattice symmetries.

\subsection{changes in background field}

On   $\tk$ we  consider  changing from a background field $\cA$   to a  background field $\cA + \cA'$  by studying 
 \be
    U_k(  \cA,  \cA')  \equiv   \Big(- \De_{\cA+ \cA'}  +  a_k (Q_k^TQ_k)(\cA + \cA')  \Big) -  \Big(- \De_{\cA}  +  a_k  (Q_k^TQ_k)(\cA)  \Big)
\ee
Define
\be ( F_{ \mu}(\cA ) ) (x)     =  \Big( e^{q  e_k  \eta \cA_{\mu} (x)}  -1 \Big)  \eta ^{-1}     
 \ee
Then we have 
\be  \label{f1}
  \Big( \pa_{\cA  +  \cA', \mu}  f  \Big)(x)  =  e^{q   e_k  \eta  \cA'_{\mu}(x)  } \Big( (\pa_{\cA, \mu}  f)(x)
  - (  F_{ \mu} ( - \cA'))(x)  f(x)  \Big)     
\ee
We   also  define
\be
(F^q( \cA, \cA')f)(y)  =      \int_{ |x-y| <  \frac12} \Big(  e^{ qe_k \eta  ( \tau_k \cA' ) (y,x)  }  -1   \Big) 
e^{ qe_k  \eta ( \tau_k    \cA)  (y,x)  }f(x) \ dx  
\ee
and then   
\be   \label{f2}
 Q_k( \cA  + \cA')    =   Q_k( \cA )   +   F^q( \cA, \cA')
\ee
Expanding  $\pa_{\cA + \cA'}$ and $\pa^T_{\cA + \cA'} $
by  (\ref{f1}) and  $Q_k(\cA + \cA')$  by   (\ref{f2})  we find
\be     \label{urep1}
\begin{split}
  U_k(  \cA,  \cA')  = &  -   F^T( - \cA') \cdot  \pa_{\cA}-  \pa^T_{\cA} \cdot   F( - \cA')  + F^T( - \cA')\cdot   F( - \cA')  \\
 &  +a_k F^{q,T}(\cA, \cA') Q_k(\cA )    +a_k  Q^T_k(\cA )   F^q(\cA, \cA')   
   +a_k F^{q,T} (\cA, \cA') F^q(\cA, \cA')    \\
\end{split}
\end{equation}
 On a function $f$ the second term is  by  (\ref{product})
\be      \label{urep2}
\sum_{\mu} \Big(     \pa^T_{\cA, \mu}    F_{\mu}( - \cA') f \Big)(x)  
= \sum_{\mu}   ( F_{ \mu}( - \cA'))(x - \eta e_{\mu}  ) (\pa^T_{\cA, \mu}f)(x)  
  +\sum_{\mu} \Big (  \pa^T_{\mu} F_{ \mu}(-\cA')\Big)(x) f(x) 
\ee
The  pair    (\ref{urep1}),   (\ref{urep2})  gives our final representation of   $ U_k(  \cA,  \cA') f$.
\bigskip

For the next  results  let  $\De = \De_y  \subset  \tk$  be a unit cube centered on a unit lattice point $y$ and let  $\tilde \De$
be the  enlargement to  a  cube  with  three unit cubes on a side.

\begin{lem} \label{snoring}
Let    $\cA, \cA', f$ be complex valued fields on $\tk$  satisfying    $e_k|\im  \cA |,   e_k|\im  \cA' |    \leq 1$.
Then for  $x  \in  \De = \De_y$
\be   \label{urep3}
\begin{split}
 \|  U_k(  \cA,  \cA')f\|_{\De, \infty}   \leq &  \one  e_k  \Big(  \| \cA'\|_{\tilde  \De,  \infty}   +  \| \pa \cA'\|_{\tilde  \De,  \infty}    \Big)  
    \Big  ( \|f \|_{\tilde\De,  \infty}     +  \|  \pa_{\cA} f \|_{\tilde\De,  \infty}       \Big)  \\
  \|  U_k(  \cA,  \cA')f\|_{\De, 2}   \leq   &   \one  e_k  \Big(  \| \cA'\|_{\tilde  \De,  \infty}   +  \| \pa \cA'\|_{\tilde  \De,  \infty}    \Big) 
      \Big  ( \|f \|_{\tilde\De, 2}     +  \|  \pa_{\cA} f \|_{\tilde\De,  2}       \Big)  \\
\end{split} 
\ee
where    $ \| \pa \cA'\|_{\tilde  \De,  \infty}    =  \sup_{\mu, \nu} \sup_{x \in  \tilde \De}  | \pa_{\nu}  \cA'_{\mu} (x)  |  $,  etc.
\end{lem}
\bigskip

\pr  We  give the proof  for the $L^{\infty}$ norm,  the proof for the   $L^2$ norm is very similar.   Consider the various terms  in  (\ref{urep1}).    We  write for $x  \in \De$
\be    \label{torts}  (F_{ \mu  }( - \cA') )(x)   =      \Big( e^{-q e_k \eta  \cA'_{\mu} (x)}  -1 \Big)  \eta ^{-1}    
= - \int_0^1  dt \   e^{- tq   e_k  \eta  \cA'_{\mu} (x)}    q  e_k \cA'_{\mu} (x)
\ee
For  $v \in \bbC^2$   
 \be
| e^{-tqe_k\eta      \cA_{\mu}(x) }v |  =   |  e^{-tqe_k\eta  \im   \cA_{\mu}(x)}v|  \leq    e^{e_k\eta   |  \im   \cA_{\mu}(x)|} |v|
   \leq     \one  |v|
 \ee
and   this  gives the estimate 
\be   \label{125}
 |( F_{ \mu  }( - \cA') )(x) v  |  \leq  \one  e_k  |  \cA'_{\mu}  (x)||v|   \leq   \one  e_k   \|\cA'\|_{   \De,  \infty}  |v| 
\ee
The same holds for the transpose  $F^T_{1}( - \cA')$.

Next we note that since    $(\tau \cA)(y_{j+1}, y_{j}  )$  is  an average over  paths of length  less than   $\frac12 L^{-(k-j-1)}$ we have 
\be  \label{slippery}
e_k \eta   |\im  (  \tau_k\cA)(y,x) )|  \leq     \sum_{j=0}^{k-1}   e_k  \eta |  \im  (\tau \cA)(y_{j+1}, y_{j})|   \leq  \frac12  \sum_{j=0}^{k-1} L^{-(k-j-1)}  \leq  1
\ee
It follows that    
\be   |(Q_k (\cA)f)(y) |  \leq  \one   \|f \|_{\De, \infty}
\ee
The   adjoint      $Q^T(\cA) $    satisfies the  same   bound. 
 We also have 
$
e_k  \eta   |(\tau_k\cA')(y,x)|  \leq   e_k \| \cA'  \|_{\De, \infty }
$
and this gives      
\be     |( F^q( \cA,  \cA'   )f )(y)     |   \leq    \one     e_k     \|\cA'_{\mu}\|_{\De,  \infty}   \|f \|_{\De, \infty}
\ee
and  similarly for the adjoint.

Now consider the terms  in  (\ref{urep2}).  These  terms involve points just outside  $\De$  which we accommodate by enlarging to $\tilde \De$.   In  particular for the first term in    (\ref{urep2}) we have       by   (\ref{relation0})  
\be 
\begin{split}
| ( F_{ \mu}( - \cA'))(x - \eta e_{\mu}  )v|  \leq  &  \one  e_k  | \cA'(x - \eta e_{\mu}  )v|  \leq      \one  e_k  \| \cA' \|_{\tilde \De, \infty}|v|  \\
| (\pa^T_{\cA, \mu}f)(x)  |  \leq & \one | (\pa_{\cA, \mu}f)(x-  \eta e_{\mu})|  \leq   \one  \| \pa_{\cA, \mu}f \|_{\tilde \De, \infty}  \\ 
\end{split}
\ee
Finally  for the second term in  (\ref{urep2})   
\be  \label{131}
\begin{split}
|(\pa^T_{\mu}  F_{ \mu} (-\cA')) (x)v  |    =& \B | \B ( F_{ \mu}  (-\cA', x- \eta e_{\mu})     -  F_{ \mu}  (-\cA', x)    \B) v\B | \eta^{-1}    \\      
 =& \B |  \B (  e^{-qe_k \eta \cA'_{\mu}(x- \eta e_{\mu})}  -     e^{-qe_k \eta \cA'_{\mu}(x) }  \B)v \B | \eta^{-2} \\
 \leq   & \one     e_k   | (\cA'_{\mu}(x- \eta e_{\mu})  -   \cA'_{\mu}(x))v|  \eta^{-1}  \\
 \leq    & \one     e_k  \| \pa  \cA' \|_{\tilde \De, \infty}   |v|      \\
 \end{split}
 \ee

Now all the terms in   $( U_k(\Om,  \cA,  \cA')f)(x)$   can be estimated
and      we have the result.
\bigskip  

\rem
 Let  $\Om  \subset  \tk$ which is a union of unit cubes.   Consider   the difference with
  Neuman boundary conditions on  $\Om$.
\be
    U_k( \Om,  \cA,  \cA')  =   \Big[- \De^N_{\cA+ \cA'}   +  a_k  (Q_k^TQ_k)(\cA + \cA')  \Big]_{\Om}
     -  \Big[- \De^N_{\cA}    +  a_k   (Q_k^TQ_k)(\cA)  \Big]_{\Om}
\ee
The representation   (\ref{urep1}),   (\ref{urep2})  still  holds but now everything is restricted to $\Om$. 
The estimate  (\ref{urep3})  still holds,   but  the enlargement  $\tilde \De$  only adds cubes in $\Om$.

\subsection{local   estimates }

Partition the  lattice    $\tk$  into  large   cubes $\square$  of linear  size  $M=L^m$ centered on  points in   $\bbT^m_{N-k}$ for some integer $m>1$.    Let  $\tilde \square$ be cube of linear  size $3M$  centered on  the same points.

We   quote some estimates on the  local   Green's functions  $G_k(\tilde \square, \cA)$.    We  want to bound   
$G_k(\tilde \square, \cA)$,   $\pa_{\cA} G_k(\tilde \square,  \cA )$,   and a certain Holder derivative   
 $ \de_{\al, \cA }\pa_{\cA}   G_k(\tilde \square, \cA  ) $.
The    Holder derivative for  $0< \al < 1$  is defined    by    
\be 
  (\de_{\al,\cA} f) (x,y)   =   \frac{e^{qe_k \eta \cA(\Ga_{xy} )} f(y)  -  f(x) }{ |x-y|^{\al} }   
\ee
where  again     $\Ga_{xy}$ is one of the standard paths from $x$ to $y$.  There is an associated  
 norm  
\be  
 \|  \de_{\al, \cA} f  \|_{\infty}  =  \sup_{|x-y|  \leq  1}    | (\de_{\al, \cA} f) (x,y) |
  \ee

\begin{lem}  \label{sweet2} 
 Let  $e_k$ be sufficiently small  depending on  $L,M$.   Let   $\cA$ on $\tilde \square$ be real-valued and  gauge equivalent to  a field $\cA'$  ($\cA  \sim \cA'$)  satisfying      $| \cA'|,|  \pa    \cA'| <    e_k^{-1 + \ep}$  for some small positive constant  $\ep$.
\begin{enumerate}
\item With Holder derivative  $  \de_{\al, \cA }$  of order  $\al<1$ \begin{equation}  \label{sycamore1}
| G_k(\tilde \square, \cA) f |, \ 
| \pa_{\cA} G_k(\tilde \square,  \cA ) f |,  \ 
\|  \de_{\al, \cA }\pa_{\cA}   G_k(\tilde \square, \cA  ) f \|_{\infty}  
 \leq       C  \|f\|_{\infty} 
 \end{equation}
\item
Let    $  \De_y,  \De_{y'}$  be unit squares centered on unit lattice points     $   y,y'  \in  \tilde \square$  and  let 
 $\zeta_y$  be   a smooth partition on unity with $\supp\  \zeta_y \subset  \tilde  \De_{y'} $.   
 Then for a constants  $C,\ga $ 
  \begin{equation}  \label{sycamore2}
\begin{split}
&|1_{\De_y} G_k(\tilde \square, \cA)1_{\De_{y'}} f  |, \ 
|1_{\De_y} \pa_{\cA} G_k(\tilde \square,  \cA )1_{\De_{y'}} f |,    \
\| \de_{\al, \cA }   \zeta_y \pa_{\cA} G_k(\tilde \square, \cA  )1_{\De_{y'}} f \|_{\infty}  \\
& \hs   \leq      C    e^{  -   \ga  d(y,y') } \|f\|_{\infty}    \\
\end{split}
 \end{equation}
 \item   The same bounds hold with  the  $L^2$ norm replacing the $L^{\infty}$ norm.
\end{enumerate}
 \end{lem}
 \bigskip

 \pr     The result holds for    $\cA =0$    see   \cite{Bal83b}, \cite{Dim11}.    
 The $L^2$   result for  $\cA=0$ is actually an input for the $L^{\infty}$ result.   The $L^2$ result can be found for example as a special case 
 of lemma 34 in \cite{Dim11}.

 For  the  general case   if       $   \cA'  =    \cA  - \pa    \la  $
 then         
 \be
 \begin{split}  
 G_{k}(\tilde \square,   \cA)      = & e^{-qe_k \la } G_{k} (\tilde \square, \cA') e^{qe_k \la }    \\
  \pa_{\cA} G_k(\tilde \square,  \cA )  = &   e^{-qe_k \la }   \pa G_k(\tilde \square,  \cA')  e^{qe_k \la }   \\   
  \de_{\al, \cA }\pa_{\cA}   G_k(\tilde \square, \cA  )  =  &   e^{-qe_k \la }  \de_{\al} \pa G_k(\tilde \square, \cA' )  e^{qe_k \la } \\ 
  \end{split} 
   \ee     
Thus it suffices  to prove the result  with  $\cA'$. 

 The  Green's function  $ G_k(\tilde \square,   \cA' )  $ satisfies    
 \be    G_k(\tilde \square,   \cA' )    =    G_k(\tilde \square,  0 )   -    G_k(\tilde \square,  0)   U_k(0, \cA')    G_k(\tilde \square,    \cA') 
 \ee
and so  is given  by 
\be    \label{expand}
    G_{k}(\tilde \square,   \cA)      =      G_{k}(\tilde \square,   0)     \sum_{n=0}^{\infty}   
\Big( - U_k(\tilde \square, 0,  \cA')   G_{k}(\tilde \square,0)    \Big)^n
\ee
provided the   the series converges,  which  we  now establish.
It follows from   (\ref{urep3})  and  our hypotheses on $\cA'$     that 
\be 
 | U_k(\tilde \square,  0,  \cA')f |  \leq   \one    e_k^{\ep} \Big  ( \|f \|_{\infty}  +  \|  \pa f \|_{\infty}   \Big)
\ee
Then  by the result for  $ G_{k}(\tilde \square, 0)    $
 \be    \label{sugar}
\begin{split}
 | U_k( \tilde \square, 0,  \cA') G_k(\tilde \square, 0) f | 
& \leq   \one    e_k^{\ep} \Big  ( \| G_k(\tilde \square, 0) f \|_{\infty}  +  \| ( \pa  G_k(\tilde \square, 0) f \| _{\infty}     \Big)  
 \leq   C    e_k^{\ep} \| f \|_{\infty} \\
\end{split}
\ee
Since $  \sum_n  (C     e_k^{\ep} )^n $  converges  for $e_k$ small,   
this is sufficient to establish the convergence of  (\ref{expand}) and give   (\ref{sycamore1}).

Next   using   the  estimate on    $U_k(  0,  \cA')$   and    the local estimate on   $G_k(\tilde \square, 0)$
we can establish a local version of  (\ref{sugar})
\be  \label{sugar2}
\begin{split}
&  | 1_{\De_y}U_k( \tilde \square,  0,  \cA') G_k(\tilde \square, 0)1_{\De_{y'} }  f | \\  \leq   &
\one   e_k^{\ep}\Big(  \|   1_{  \tilde   \De_y} G_k(\tilde \square, 0)1_{\De_{y'} }  f \|_{\infty}    +     \|   1_{\tilde \De_y}\pa G_k(\tilde \square, 0)1_{\De_{y'} }  f \|_{\infty}  \Big)  \\
\leq    &  C   e_k^{\ep}   \sum_{|y''-y| \leq  1}  e^{ - \ga  d(y'',y')  } \| f \|_{\infty}   \\
\leq    &  C   e_k^{\ep}     e^{ - \ga  d(y,y')  } \| f \|_{\infty}   \\
\end{split}
\ee
Now    we  have 
\be 
\begin{split}
   G_{k}(\tilde \square,   \cA)          =  & \sum_{n=0}^{\infty}   \sum_{y_1, \dots,  y_{n-1} }G_{k}(\tilde \square,     0)
1_{\De_{y_1}}  \Big(- U_k(\tilde \square,  0,  \cA')   G_{k}(\tilde \square,   0)    \Big)  \\
&   \cdots  \ 1_{\De_{y_{n-1}}}
  \Big( -U_k(\tilde \square,  0,  \cA')   G_{k}(\tilde \square,  0)  \Big)\\
\end{split}
\ee
Then     with  $y_n = y'$   the estimate   (\ref{sugar2})  gives
\be  
\begin{split}
 |1_{\De_y} G_k(\tilde \square, \cA)1_{\De_{y'}} f  |      \leq  &   \sum_{n=0}^{\infty}    \sum_{y_1, \dots,  y_{n-1} }    Ce^{ - \ga  d(y,y_0)  }  \prod_{j=1}^n 
   e_k^{\ep} C e^{ - \ga  d(y_{n-1},y_n)  }  
 \|f\|_{\infty}      \\
 \leq    &
  C   e^{  - \frac12   \ga  d(y,y') }     \sum_{n=0}^{\infty}  (C     e_k^{\ep} )^n \|f\|_{\infty}  
    \leq     C   e^{  - \frac12   \ga  d(y,y') }      \|f\|_{\infty}     \\
 \end{split}
  \ee
  Thus the bound holds with a new  $\ga$.
The estimate on derivatives  is similar.  This completes the proof.
\bigskip

Next we  extend the previous result  to  complex fields   $\cA$   on $\tk$  of the form  
 \be  \label{listless} 
 \begin{split} 
   & \cA=  \cA_0  + \cA_1   \\
    & \cA_0  \textrm{ is  real  and  on each }  \tilde \square 
  \textrm{ admits   }   \cA'_0 \sim \cA_0
 \textrm{  satisfying   }    |  \cA'_0 |,     |\pa  \cA'_0     |  <    e_k^{-1 + \ep},   \\ 
&   \cA_1 \textrm{ is complex and  satisfies  }     |  \cA_1 |,     |\pa  \cA_1     |  <    e_k^{-1 + \ep}      \\
\end{split}  
\ee  
This is  an open set  in  some $\bbC^n$  and  we  can  consider   functions analytic in this domain.  
\begin{lem}  \label{sweet2.5}
 Under the same hypotheses  
 $G_k(\tilde \square, \cA)$  has  an analytic  extension to   the region   (\ref{listless}),  and for such fields     $G_k(\tilde \square, \cA)$ 
 again   satisfies bounds of the form  (\ref{sycamore1}), (\ref{sycamore2}).
 \end{lem}
 \bigskip

  \pr   We  again   have    
 \be    G_{k}(\tilde \square,   \cA )      =      G_{k}(\tilde \square,    \cA_0)     \sum_{n=0}^{\infty}   
\Big(-  U_k(\tilde \square,   \cA_0, \cA_1)   G_{k}(\tilde \square,   \cA_0)    \Big)^n
\ee
and by      lemma  \ref{snoring}   and  lemma  \ref{sweet2}  
\be
 | U_k( \tilde \square,  \cA_0,  \cA_1) G_k(\tilde \square, \cA_0) f |
 \leq   C    e_k^{\ep} \| f \|_{\infty} 
 \ee
which   gives   (\ref{sycamore1}).    The bound   (\ref{sugar2})  also  holds,   and  the local version   (\ref{sycamore2})   follows as before.

 \subsection{random walk expansion}  \label{random}
 We  study the global   Green's functions    $G_k(\cA)$ by      random walk expansions.

Again  partition the  lattice    $\tk$  into cubes $\square$  of linear  size  $M= L^m$.   We write    $\tk =  \bigcup_z   \square_z$   where
 $z$  is a point  on the $M$ lattice  $\bbT^m_{N-k}$ and $\square_z$  is the $M$ cube centered on  $z$.
   Let  $\tilde \square_z$  be the      $3M$ cube centered on  $z$.   The random walk expansion is  based on   the operators  
      $G_k(\tilde   \square_z, \cA)$,  discussed previously.   We assume that  $\cA$ is in the domain  (\ref{listless}) so  that  these have good estimates by lemma  \ref{sweet2.5} .

Let   $h^2_z$  be  a partition of unity with  $\sum_z h_z^2 =1$  and   $\supp\  h_z$ well inside  $ \tilde \square_z$.  We  define a parametrix
\be  G_k^*(\cA)  =   \sum_z  h_z  G_k(\tilde   \square_z, \cA) h_z  \ee
On   $\supp \  h_z$ there is no   distinction between  $-\De_{\cA}$
and  $[- \De^N_{\cA}]_{\tilde \square_z} $   and  so we can compute
\be     
  \B( - \De_{\cA}    +  a_k    (Q_k^TQ_k)(\cA)\B) G_k^*(\cA)=    I  -  \sum_z   K_z(\cA) G_k( \tilde \square_z, \cA) h_z  \equiv  I -K
  \ee
  where
  \be    K_z(\cA)  =   -\B[  \B( - \De_{\cA}  +  a_k   (Q_k^TQ_k)(\cA)\B) , h_z\B]  \ee
 Then   
  \be
  G  _k(\cA)  =  G_k^*(\cA) ( I - K)^{-1}   =   G_k^*(\cA) \sum_{n=0}^{\infty}  K^n
  \ee    
  if it converges.  
  This can be written   as    the random walk expansion 
   \begin{equation}  \label{g1}
 G_{k}(\cA) =    \sum_{\om }  G_{k,\om}(\cA)
 \end{equation}
where   a   path  $\om$ is a sequence of points $   \om   =  (\om_0, \om_1, \dots,   \om_n )$
 in  $\bbT^m_{N-k}$
such that  $\om_i, \om_{i+1}$ are nearest neighbors  (in a sup metric),   and  
\be  G_{k, \om} (\cA)= h_{\om_0} G_k(\tilde \square_{\om_0},\cA )h_{\om_0}
 K_{\om_1}(\cA) G_k(\tilde \square_{\om_1},\cA)h_{\om_1}\cdots    K_{\om_n}(\cA) G_k(\tilde \square_{\om_n},\cA)h_{\om_n}
\ee
 Note  that    $ G_{k,\om}(\cA)$  only depends on  $\cA$ in the set  $ \bigcup_{i=0}^n  \tilde \square_{\om_i} $

 \begin{lem}   \label{sweet3}  Let   $M$  be sufficiently large (depending on  $L$),  
  and  $e_k$  sufficiently small (depending on  $L,M$),   and let   $\cA $  be in the domain  (\ref{listless}).  Then
\begin{enumerate}
\item 
The random walk  expansion (\ref{g1})  for $G_k(\cA)$  converges to a function analytic in $\cA$   which satisfies  
 \begin{equation}  \label{sycamore3}
| G_k( \cA) f  |, \ 
| \pa_{\cA} G_k(  \cA ) f|,   
\|  \de_{\al, \cA }\pa_{\cA}   G_k( \cA  ) f\|_{\infty}  
 \leq       C  \|f\|_{\infty} 
 \end{equation}
\item
Let    $  \De_y,  \De_{y'}$  be unit squares centered on unit lattice points     $   y,y'  \in  \bbT^0_{N-k}$  and let
 $\zeta_y$   be a smooth partition on unity with $\supp\  \zeta_y \subset  \tilde  \De_{y'} $. 
 Then there are   constants  $C, \ga$      so   
  \begin{equation}  \label{sycamore4}
\begin{split}
&|1_{\De_y} G_k( \cA)1_{\De_{y'}} f |, \ 
|1_{\De_y} \pa_{\cA} G_k( \cA )1_{\De_{y'}} f |,   
\| \de_{\al, \cA }    \zeta_y  \pa_{\cA}   G_k( \cA  )1_{\De_{y'}} f \|_{\infty}  \\
& \hs   \leq      C    e^{  -\ga  d(y,y') } \|f\|_{\infty}    \\
\end{split}
 \end{equation}
 \item    The same bounds hold with  the  $L^2$ norm replacing the $L^{\infty}$ norm.
\end{enumerate}
 \end{lem}
\bigskip

\rem  The same bounds hold for  $\cH_k(\cA)$,  for example   
 \begin{equation}  \label{slavic}
| \cH_k( \cA) f  |, \ 
| \pa_{\cA} \cH_k(  \cA ) f|,   
\|  \de_{\al, \cA }\pa_{\cA}   \cH_k( \cA  ) f\|_{\infty}  
 \leq       C  \|f\|_{\infty} 
 \end{equation}
\bigskip

\pr  \cite{Bal83b}.   We give the proof for the $L^{\infty}$ norm.
We  compute  using  (\ref{product})
 \be  \label{no1}
  \B([  - \De_{\cA},  h_z ] f\B) (x ) 
=  ( \pa^T h)(x + \eta e_{\mu}) \cdot  \pa_{\cA}f(x)
+    (\pa h_z)(x - \eta e_{\mu}) \cdot  \pa^T_{\cA}f(x)
+   (-\De h_z)(x) f(x)  
\ee
and    with  $x \in \De_y$
\be  \label{no2}
  \B( [ (Q_k^TQ_k)(\cA),   h_z ] f\B)(x)     =  \int_{|x'-y| < \frac12}  e^{-qe_k \eta (\tau_k \cA)(y,x)} e^{qe_k \eta (\tau_k \cA)(y,x')}
\B(h(x') - h(x)  \B)  f(x')   dx'
\ee
The functions   $\{ h_z\}$ can be chosen so that   $| \pa h_z | \leq   \one M^{-1}$  and  $| \pa \pa h |  \leq   \one M^{-2}$.
Then    the representations   (\ref{no1}),  (\ref{no2})   lead to the bound
\be   | K_z(\cA) f | \   \leq    \one  M^{-1}( \|f\|_{\infty}   +   \|\pa_{\cA} f  \|_{\infty} )   \ee
and   therefore  by (\ref{sycamore1})   
\be   \label{spitfire1} | K_{z}(\cA) G_k(\tilde \square_{z}, \cA)f|   \leq   CM^{-1}   \|f \|_{\infty}
\ee
These  imply  that  if  $|\om|  = n$ then  
\be       \label{night}
  |G_{k, \om}(\cA) f|  \leq  C (CM^{-1} )^n \|f\|_{\infty} 
 \ee
This is sufficient to establish the convergence  of  the expansion for $M$ large,   since the number of paths with a fixed length $n$ is bounded
by  $(3d)^n= 9^n$.   The bounds on derivatives follow as well.

For the local estimates  use  the locality of $K_z(\cA)$ and  (\ref{sycamore2})  to obtain
\be      \label{spitfire2} |1_{\De_y}  K_{z}(\cA) G_k(\cA,\tilde \square_{z})1_{\De_{y'}}f|   \leq   CM^{-1}   e^{  -  \ga  d(y,y') }  \|f \|_{\infty}
\ee
Proceeding   as in  lemma  \ref{sweet2} we have the result with a new  $\ga$.
 \bigskip

\rem  We     introduce weakening parameters  $\{ s_{\square} \}$  with  $0 \leq  s_{\square} \leq  1$  and    define 
\be  
s_{\om}  =  \prod_{\square \subset X_{\om}}  s_{\square}  \hs   
  X_{\om}   \equiv  \bigcup_{i=1}^n  \tilde \square_{\om_i} 
\ee
If  $\om = \om_0$ is a single point  then  $|\om| =0$ and  $X_{\om} = \emptyset$.  In this case  we define  $s_{\om} = 1$.
  
    Weakened  propagators are defined by  
\begin{equation}  \label{again}
   G_{k}(s, \cA) =    \sum_{\om} s_{\om}  G_{k,\om}(\cA)
\end{equation}
If  $s_{\square}$  is  small  then the coupling through  $\square$ is reduced.    The  $ G_{k}(s,\cA)$  interpolate between
$ G_{k}(\cA)= G_{k}(1,\cA)$   and a strictly local  operator  $ G_{k}(0,\cA)$.

   The  results of  lemma  \ref{sweet3}   hold  for  the weakened Green's functions   $   G_{k}(s,\cA)$.   In  fact   we  can allow  
 complex  $s_{\square}$   satisfying   $ |s_{\square}|  \leq  M^{\frac12}$ and still get estimates of the same form. 
 Also  $    G_{k}(s,\cA)$ has  the   analyticity and symmetries  of  $G_k ( \cA)$.

 \subsection{more  random walk expansions}

 We also need a random walk expansion  for  $C_k(\cA) =  \B( \De_k(\cA) + aL^{-2} (Q^T Q)(\cA)  \B)^{-1}   $ or  even better for     $C_k^{\frac12}(\cA)$.  
 These are  treated for  $\cA =0$ in   Balaban and    \cite{Dim11} and the treatment is similar here.  
 The operator $C_k(\cA)  $ has  a  simple expression in terms of  $G_{k+1}(\cA) $   (see  (\ref{lamp})),    and this gives the expansion.  
 The analysis  for       $C_k^{\frac12}(\cA)$  is based on the representation   
\be   \label{half} 
C_k^{1/2}(\cA)  =    \frac{1}{\pi}  \int_0^{\infty}  \frac{dx}{\sqrt{x}} \B( \De_k(\cA)  + aL^{-2} (Q^T Q)(\cA)  + x \B)^{-1}   
\ee
As in  appendix  C  in    \cite{Dim11}  one can show that  
\begin{equation}   \label{swipe}
\B( \De_k(\cA) + aL^{-2} (Q^T Q)( \cA)   + x \B)^{-1}   =    \sA_{k,x} (\cA) +   a_k^2  \Big(\sA_{k,x} Q_k  G_{k,x} Q_k^T  \sA_{k,x}\Big)(\cA)  
\end{equation}
  where 
\begin{equation}  \label{three}
\begin{split}
\sA_{k,x}(\cA)   =&   \frac{1}{a_k+x}  (I -( Q^TQ)(\cA))   +   \frac{1}{ a_k + aL^{-2}  +x}   (Q^T Q)(\cA)  \\
G_{k,x}(\cA)  = &  \Big[ -\De_{\cA}  +  \frac{a_k x}{a_k +x}  (Q_{k}^TQ_{k})(\cA)  + 
 \frac{a_k^2aL^{-2}}{(a_k+x)(a_k + aL^{-2} +x)}(Q_{k+1}^T Q_{k+1})(\cA)\Big]^{-1}
  \\
 \end{split}
 \end{equation}

Since all the other  operators   are local  it suffices to establish a random walk expansion for   $G_{k,x}(\cA)$,   and it turns out 
that  an   $L^2$  expansion suffices.  
The expansion    follows  from good local estimates  on  the   local  operator   $G_{k,x}(\tilde \square, \cA)$   defined just  as   $G_{k,x}(\cA)$
but restricting the operator  to  $\tilde \square$  before taking the inverse.
We   claim  that  if  real   $\cA$ is gauge equivalent  to $\cA'$   satisfying  $|\cA' |, | \pa \cA'|  < e_k^{-1+ \ep}$
 \begin{equation}    \label{usher}
 \begin{split}
\| 1_{\De_y}  G_{k,x}(\cA, \tilde \square) 1_{\De_{y'} } f\|_2     \leq  &  C  e^{- \ga  d(y, y')}  \|f \|_2 \\
\| 1_{\De_y} \pa_{\cA}  G_{k,x}(\cA, \tilde \square) 1_{\De_{y'}} f\|_2     \leq  &  C  e^{- \ga  d(y, y')}  \|f \|_2\\
\end{split}
\end{equation}
As before  it suffices to  prove  the result   for $\cA'$.
This is known for  $\cA=0$,  see Appendix E in  \cite{Dim11}.    
For the general case we expand     
\be   
    G_{k,x}(\tilde \square,   \cA')      =      G_{k,x}(\tilde \square,    \cA')     \sum_{n=0}^{\infty}   
\Big( - U_{k,x}(\tilde \square, 0,  \cA')   G_{k,x}(\tilde \square, 0)    \Big)^n
\ee
where    now  $
 U_{k,x}(\tilde \square, 0,  \cA')  =   G_{k,x}(\tilde \square, \cA' ) ^{-1}  -  G_{k,x}(\tilde \square,0 )^{-1}   
$.
As in  (\ref{urep3}) one    establishes   
\be   \label{urep4}
\begin{split}
 \| U_{k,x}(\tilde \square,   0,  \cA')f\|_2   \leq &  \one  e_k  \Big(  \| \cA'\|_{\tilde  \De,  \infty}   +  \| \pa \cA'\|_{\tilde  \De,  \infty}    \Big)     \Big  ( \|f \|_{\tilde\De, 2}     +  \|  \pa_{\cA_0} f \|_{\tilde\De, 2}       \Big) \\
 \leq &  \one  e_k^{-\ep}     \Big  ( \|f \|_{\tilde\De, 2}     +  \|  \pa_{\cA_0} f \|_{\tilde\De, 2}       \Big)  \\
 \end{split}  
\ee
 This gives the convergence of the series  and  the estimate (\ref{usher}).   We  can also extend the result to 
 $\cA$ in the  complex   domain  (\ref{listless}).

 As  in  (\ref{g1})  the control  over   $G_{k,x}(\tilde \square,  \cA)$  leads to  a random walk expansion 
 \be     \label{rw2}
   G_{k,x}(\cA)  =   \sum_{\om}   G_{k,x, \om }(\cA)
  \ee
  and  $L^2$  bounds  like  (\ref{sycamore4}) for   $  G_{k,x}(\cA) $  follow.    By    (\ref{swipe})  we get a random walk 
  expansion for    $C_k^{\frac12} (\cA) $.     This also gives the bound.  
\be    \label{slinky0}
  | C_k^{\frac12} (\cA) f |  \leq   C  \|f \|_{\infty} 
\ee

\section{RG transformations for  gauge fields }

\subsection{axial gauge}
For gauge fields  we  more or  less    follow the  treatment of Balaban       
 \cite{Bal84a}, \cite{Bal84b},  \cite{Bal85b}    and   Balaban, Imbrie, and Jaffe  
 \cite{BIJ85},  \cite{BIJ88},   \cite{BaJa86},  \cite{Imb86}.    This  differs from   
 the  treatment  of   the scalar  field  in that  we  need to employ  gauge fixing   and 
 we use   a different  definition of the averaging operators.     Even so    the gauge fixing  here
 is not exactly  the axial gauge  employed  in the above references,   but a covariant axial gauge
 introduced  by  Balaban    in  \cite{Bal87}, \cite{Bal88a}  and further developed in  
 \cite{Dim14}  to  which we refer for more details.

Start with an integral   over fields    on    $\bbT^{-N}_0$  
 of  the form   
\be     \label{formal3}   
 \int   f(A)   \exp \B( - \frac12 \| dA \|^2  \B)  DA        \hs       DA  =  \prod_{b} d(  A(b)  )
 \ee
 This generally does not converge since $dA$ has a large null space;  we proceed formally.  
 We  scale  up to the lattice  $\bbT_N^0 $.     Let   $\rho_0 $  be  the function   $  f(A)   \exp ( - \frac12 \| dA \|^2  )$
scaled up.   For  $A_0$     on    $\bbT_N^0 $    it is    
\be   
 \rho_0 (   A_0 )   =    F_0(A_0)    \exp \B( - \frac12 \| dA_0 \|^2  \B)  \hs  
\ee
where  $  F_0(A_0)  =    f_{L^N}(  A_0 )  =   f(  A_{0, L^{-N }}   ) $.

 On this lattice       we  define  an  averaged field     on   oriented bonds   in      $\bbT^{1}_{N} $ by     (for reverse oriented bonds take minus this)  
\be   
\begin{split}
(\cQ A) (y,  y + L e_{\mu} )  
= & \sum_{x \in B(y)  } L^{-4}    A(  \Ga_{x,  x +  L e_{\mu} } ) \\
\end{split}
\ee
where   $  \Ga_{x,  x +  L e_{\mu} }$ is the straight line between the indicated points.   
   Note however that   $\cQ^T \cQ$ is not a projection operator.  
The  means that an   exponential RG transformation   cannot be treated as they were in  scalar case.    Instead we use a delta function  RG   transformation  which has other advantages and difficulties.

We  would  like to define  a sequence of  densities  $\rho_0,  \rho_1,  \dots, \rho_N    $ 
 with $\rho_k(A_k) $  defined for  $A_k$  on  $\bbT_{N-k}^0$.  First consider
\be    \label{funny0} 
\tilde   \rho_{k+1} (A_{k+1})   = \int   \de ( A_{k+1} - \cQ A_k)   \rho_k(A_k)  \  DA_k   
\ee
For convergence we introduce  an axial gauge fixing function    (justified by a Fadeev-Popov argument)
\be 
\de  (\tau A_k)    =  \prod_{y  \in   \bbT^{1}_{N-k}  }  \prod_{x  \in B(y),  x \neq  y}    \de \B((\tau   A_k)(y, x) \B)   
\ee
where    $(\tau A_k)(y,x)$ is defined in  (\ref{wombat}). 
Instead  of  (\ref{funny0}) we        define  $\tilde  \rho_{k+1} (A_{k+1})$    for     $A_{k+1}$ on     $ \bbT^1_{N-k}$   
by  
\be    \label{funny1} 
\tilde   \rho_{k+1} (A_{k+1})   = \int   \de ( A_{k+1} - \cQ A_k)      \de ( \tau A_k  )  \rho_k(A_k)  \  DA_k   
\ee
and   then     
 $   \rho_{k+1}(A_{k+1}) $  for     $A_{k+1}$ on    $ \bbT^0_{N-k-1}$        
by 
\be  
    \rho_{k+1}(A_{k+1}  )  =     \tilde  \rho_{k+1} (A_{k+1,L})  L^{\frac12 (b_N- b_{N-k-1}) }  L^{-\frac12 (s_N- s_{N-k-1}) }
\ee
Here   $b_n =  3L^{3N}$ is the number of bonds in a three dimensional   toroidal lattice with $L^N$ sites on a side, and 
$s_N = L^{3N}$ is the number of sites.

The   delta function  averaging operators compose nicely  and we have   
\be  \label{four} 
  \rho_k (A_k)  =      \int    \de (A_k -   \cQ_k  \cA ) \de^{\sx}_k(\cA)  \rho_{0,L^{-k}} ( \cA)    D\cA 
\ee
where now   $\cA$ is defined on bonds in     $\tk$    and   the $k$-fold averaging operator is defined by     $\cQ_k  =   \cQ  \circ  \cdots  \circ  \cQ$.  Then       $ \cQ_k \cA$ is   given   on    oriented bonds  in  $\bbT^0_{N-k}$  by     
\be   
(  \cQ_k \cA) (y,  y +  e_{\mu} )  =  \int_{|x -y|  < \frac12  }  L^{-k}   \cA  ( \Ga_{x,  x +   e_{\mu}})\ dx
\ee
and the   gauge fixing function is   now   
 \be 
  \de^{\sx}_k  (\cA)   =      \prod_{j=0}^{k-1}    \de  (\tau  \cQ_j \cA )   
 \ee
One  can show  that   $ \rho_k (A_k)$    is well-defined     \cite{BaJa86}, \cite{Dim14}.

The integral  of  final density  $\rho_N (A_N) $   gives back the original  integral  (\ref{formal3}) 
but now  with a hierarchical axial gauge fixing function which enables convergence.    See  \cite{Dim14} for details.

For future reference we note the identity
\be
\label{lockdown}
( d \cQ_k \cA) \B(y,  y +  e_{\mu}, y +  e_{\mu}+ e_{\nu} ,y +  e_{\nu}  \B)  =
  \int_{|x -y|  < \frac12  }  L^{-2k}  d \cA  \B( \Si_{x,  x +  e_{\mu}, x +  e_{\mu}+ e_{\nu} ,x +  e_{\nu}}\B) \  dx
\ee
Here  $\Si_{x,  x +  e_{\mu}, \dots  }$ is the square with the indicated corners,   and  in general  $\cA (\Si)  = \sum_{p \in \Si}  
d \cA (p) $.

\subsection{free flow}   \label{minimizers2}

Scaling the $\rho_0$  we can also write  (\ref{four}) as 
\be  \label{fourfour} 
  \rho_k (A_k)  =      \int    \de (A_k -   \cQ_k  \cA ) \de^{\sx}_k(\cA)   F_{0,L^{-k}}  (\cA)    \exp   \B( - \frac 12 \|  d \cA \|^2    \B)    D\cA 
\ee
which we analyze further.

Let   $\cH^{\sx}_{ k}A_k  $  be the minimizer of    $ \| d \cA  \|^2 $  subject to the constraints of the delta functions in  (\ref{four}). 
We  give    an        explicit   representation  later on  in  (\ref{lounge}). 
It has the property that it preserves gauge equivalence:  if  $A_k \sim A'_k$ then  $ \cH^{\sx}_{ k}A_k  \sim  \cH^{\sx}_{ k}A'_k $.

 Expanding around the minimizer by  $\cA  =  \cH^{\sx}_kA_k   + \cZ$        we find  
\be  \label{sex} 
      \rho_k(A_{k})   =    \sZ_k      F_k (\cH^{\sx}_kA_k)     \exp \B(   - \frac12 <A_k,  \De_k  A_k >   \B)   
 \ee
where for    $\cH^{\sx}_kA_k$ and $\cZ$   defined on  $\tk$.
\be
\begin{split}
    <A_k,  \De_k  A_k >   = &    \|  d \cH^{\sx}_{k} A_k  \|^2  \\    
 F_k (\cH^{\sx}_kA_k )   
   = & \sZ_k^{-1}    \int    \de (  \cQ_k \cZ  )  \de^{\sx}_k(\cZ)  F_{0,L^{-k}}  (\cH^{\sx}_kA_k   + \cZ)    \exp   \B( - \frac 12 \|  d \cZ \|^2    \B)  D \cZ
    \\   
\sZ_{k}   =   &   \int    \de (  \cQ_k \cZ  )  \de^{\sx}_k(\cZ)     \exp  \B ( - \frac 12 \|  d \cZ \|^2   \B )  D \cZ \\
\end{split}
\ee

\subsection{the next step}    
     Suppose we are starting with the expression  (\ref{sex}) for  $ \rho_k(A_k) $.   In the next step   generated by  (\ref{funny1})  we  have
\be  \label{six}
\begin{split}
\tilde \rho_{k+1} (A_{k+1}) 
 =   & \sZ_{k}      \int    \de (A_{k+1} -    \cQ A_k  )\   \de(  \tau  A_k)  \     F_k (\cH^{\sx}_k   A_k)     \exp \B(   - \frac12 <A_k,  \De_k  A_k >   \B)   
 DA_k  \\
 \end{split}
\ee
Let   $ H^{\sx}_{k} A_{k+1}$  be  the minimizer for   $ \frac12   <A_k, \De_k  A_k  >  $  subject to the constraints.       Expanding around the minimizer  with   $A_k =  H^{\sx} _{k} A_{k+1} + Z$  
we   again get the representation 
\be  \label{spotless} 
      \rho_{k+1}(A_{k})   =    \sZ_{k+1}       F_{k+1}  (\cH^{\sx}_{k+1}A_{k+1} )     \exp \B(   - \frac12 <A_{k+1},  \De_{k+1}  A_{k+1} >   \B)   
 \ee
But now with the  identifications   
\be    \label{sammy}
\begin{split}
 \sZ_{k+1}  = &  \sZ_k \sZ_k^f     L^{\frac12 (b_N- b_{N-k-1}) }  L^{-\frac12 (s_N- s_{N-k-1}) }       \\
( \cH^{\sx}_{k+1}A_{k+1})_L 
=   & \cH^{\sx}_{k}   H^{\sx}_{k} A_{k+1,L}     \\
F _{k+1}  (\cH^{\sx}_{k+1}A_{k+1}   ) 
   = & (\sZ^f_k)^{-1}    \int     \de (  \cQ   Z  )   \de(  \tau Z)  \    F_k \B( ( \cH^{\sx}_{k+1}A_{k+1} )_L    + \cH^{\sx}_kZ\B)     \exp \B(   - \frac12 <Z,  \De_k Z >   \B)  D Z  \\
     \sZ^f_{k}   =   &    \int    \de (  \cQ   Z  )   \de(  \tau Z)  \       \exp \B(   - \frac12 <Z,  \De_k Z >   \B)  D Z \\ 
 \end{split}
\ee
See   \cite{Dim14} for details. 
More generally we define for  any  $\cA$ on   $\bbT^{-k-1}_{N-k-1}$  
\be    \label{sequentially}
  F _{k+1}  (  \cA   ) 
   =  (\sZ^f_k)^{-1}    \int     \de (  \cQ   Z  )   \de(  \tau Z)  \    F_k \B( \cA_L    + \cH^{\sx}_kZ\B)     \exp \B(   - \frac12 <Z,  \De_k Z >   \B)  D Z  
\ee
Note  that if  $F_0$ is gauge invariant then   $F_k$ is gauge invariant for any $k$.

\subsection{other gauges}

Restrict now to the  case   $F_0=1$ so  $\rho_0 =  \exp( - \frac12 \|dA \|^2 )$.
Instead of  (\ref{four})  the density  $  \rho_k (A_k) $  can     be expressed in the  
 modified Feynman gauge  for any  $\al >0$ by   \cite{Bal84a},  \cite{Dim14} 
\be       \label{int5} 
 \rho_k (A_k)    = \const       \int      \de (A_{k  } -    \cQ_k \cA  )\  \ 
         \exp \B(  -  \frac{1}{2 }     \|  d  \cA \|^2  -   \frac{1}{2 \al} < \de \cA,  R_k\   \de \cA>  \B)     D\cA        
\ee
where  $\de  = d^T$ on 1-forms   (functions on bonds)   is the adjoint of 
$d = \pa$ on scalars,  and    $R_k  $  is  the projection onto the subspace  $\De ( \ker  Q_k  )  $.      It is 
explicitly given by
\be   \label{leg} 
R_k   =     I   -  G_k Q_k^T  (  Q_k G_k^2  Q_k^T)^{-1}  Q_k G_k  
\ee
where  $  G_k  =  ( -\De + a Q_k^TQ_k  ) ^{-1}$    for any   $a \geq  0$  (essentially  the same  as  $G_k(0)$ in (\ref{Gk})).
This  includes  the Landau gauge at   $\al =  0$  in which case 
\be       \label{int6} 
 \rho_k (A_k)    = \const       \int      \de (A_{k  } -    \cQ_k \cA  )\      \de_{R_k}  ( R_k \de  \cA    )      
         \exp \B(  -  \frac{1}{2 }     \|  d  \cA \|^2   \B)     D\cA        
\ee

Let   $\cH_kA_k $  be the minimizer  of    $  \| d \cA  \|^2  + \al^{-1} < \de \cA,  R_k\   \de \cA>  $    subject to the constraint  
$\cQ_k \cA  = A_k$   imposed    in  (\ref{int5}).   An explicit expression  for the minimizer  can  be given using   Green's function for this gauge  defined     by    
\be
\cG_{k}   =   \B(      \de   d   +  \frac{1}{2 \al}  d R_k \de     + a \cQ^T_k \cQ_k \B) ^{-1}  
\ee
 Then   one   can show   \cite{Bal84a}     that  $\cQ_k  \cG_k \cQ_k^T $ is invertible and  for any  $a>0$
\be   \label{lamb} 
\cH_k   =  \cG_k   \cQ_k^T   ( \cQ_k  \cG_k \cQ_k^T )^{-1}   
\ee
It  turns out  that     $\cH_k$ is independent of $\al$ and is also the minimizer for  Landau gauge.
Furthermore    
\be  
\label{relation}  \cH^{\sx}_{k}  =    \cH_{k}  +    \pa  D_k  \ee
   for some operator  $D_k$.
 This means that  in  gauge invariant  expression we  can  can   replace  $ \cH^{\sx}_{k} $ by     $ \cH_{k}$.   In
 particular we  can make this replacement in  the fluctuation integral  (\ref{sequentially}).  This is useful because  $\cH_k$ is more regular than  the axial  $\cH^{\sx}_k$. 

 The  relation  (\ref{relation}) also shows that   $\De_k$  can be expressed in the Landau gauge as
 \be    \label{zee}
 <    Z,     \De_k   Z>   =   \|  d \cH_k  Z \|^2  
 \ee
 Using this   one  can show  \cite{Bal84b}, \cite{Dim14}  that   there  are   constants  $C_{\pm} $  depending only on $L$     
 such that   on   the subspace  $\cQ \cZ =0,  \tau \cZ  =0$
\be     \label{snuff}
 C_{ - } \|  Z \|^2 \    \leq    \    <  Z,     \De_k  Z> \    \leq\   C_+  \| Z \|^2 
  \ee

\subsection{parametrization of the  fluctuation integral}    \label{naughty}

We  parametrize the fluctuation    integral  (\ref{sequentially})    as  in  \cite{Bal84b},   \cite{Dim14}. 
Let   $Z = (Z_1,Z_2)$   where   $Z_1$ is defined on bonds that lie in some $B(y)$  and  $Z_2$ is defined on bonds
joining neighboring cubes  $B(y), B(y')$. The delta function   $\de (\tau Z ) =  \de (\tau Z_1 )   $ is fulfilled 
by taking    $Z_1 = \tilde Z_1  \in \ker \tau $.  Let   $\tilde Z_2$ be defined on bonds joining   $B(y), B(y')$,  but not
the central bond denoted  $b(y,y')$
The  delta function   $\de (\cQ Z )$  selects   $b(y,y') =  S ( \tilde Z_1,\tilde    Z_2)$ for some local linear operator  $S$.   See
(\ref{lignon2}) in the appendix for  the explicit formula.   
     Then    the integral is parametrized by   $Z  = (  \tilde Z_1,  \tilde Z_2, S( \tilde Z_1, \tilde Z_2) $.  Or  if  we let 
 $ \tilde Z = ( \tilde Z_1, \tilde  Z_2)$   then it is parametrized
by  
\be    Z   =  C  \tilde  Z   \equiv   ( \tilde Z, S \tilde Z   )
\ee  
The   fluctuation  integral     (\ref{sequentially})    
can now be written   
\be   \label{springgarden4}
 F_{k+1} ( \cA )  =      \int     F_k\B(\cA_L  + \cH_k  C   \tilde Z   \B)      \exp \B(  - \frac12   <C  \tilde  Z , \De_k C \tilde  Z   >   \B) 
    \    D    \tilde   Z  
   \Big/  \{  F_k =1 \}        
\ee
If  we  define   
\be  
  C_{k}  =   (  C^T     \De_k  C )^{-1}     
\ee
then   the  integral can be expressed with  the Gaussian measure  $\mu_{C_k}$  with covariance  $C_k$   as      
\be    \label{sometimes2} 
\begin{split}
  F_{k+1}  (\cA)     =   &     \int           F_k \B( \cA_L +  \cH_k  C  \tilde   Z        \B)    d \mu_{  C_{k}}   ( \tilde  Z )   
   =        \int           F_k \B( \cA_L +  \cH_k  C  C_k^{\frac12}  \tilde W        \B)    d \mu_I   (   \tilde  W )   \\
\end{split}
\ee      
By   (\ref{snuff})    $  C^T   \De_k  C$    is  uniformly bounded  above and  below.   Hence     the same is true of the inverse  $C_k$
and
\be
\| C_k^{  \pm  \frac 12}  Z  \|   \leq  C  \| Z  \|  
\ee    
These   are      basic    facts for controlling the integrals   (\ref{sometimes2}), but we will    still need more.   

We  note   also     that the integral can  be written      
\be    \label{sometimes3} 
  F_{k+1}  (\cA)     =        \int           F_k \B( \cA_L +  \cH_k    Z        \B)    d \mu_{  C'_k}   (   Z )   \hs  C_k'  \equiv     CC_kC^T
\ee      
where  $C'_k$ is now defined on functions on all of   $\tz$.

\subsection{representation for  $C_k$, $C_k^{\frac12}$ }
  
 We  will need a      representation of  $C_k$    which admits a random walk expansion.   It  is easier  to  treat  $C'_k$
 and   we consider that first.       The following from  \cite{Dim14}   is a simpler version of an analysis by Balaban  \cite{Bal85b}.

 For  $\la, A$ on $\tz$ let  
  $\la = \cM A $ be    the solution of the 
 equations
 \be    ( \tau  ( A + d \la  ) ) (y,x)  =  0   \ \ \   x \neq y    \hs   Q \la (y)   = 0   \hs   x \in B(y)    \ee       
 This is 
  \be  
 \la(x)  =  \cM  \cA(x)   =
- ( \tau  A  ) (y,x)   +    L^{-3} \sum_{x' \neq y }( \tau  A  ) (y,x')    
\ee
 Also define  
\be    \label{lorenzo}
\tilde \cG_{k+1}  =   \cG^0_{k+1} -  \cG^0_{k+1} \cQ_{k+1}^T   \B(\cQ_{k+1} \cG^0_{k+1} \cQ_{k+1}^T\B)^{-1}   \cQ_{k+1} \cG^0_{k+1} 
\ee
where for any  $a >0$
\be   
\cG^0_{k+1}  = \B(   \de  d   +    d R_{k+1} \de   + a \cQ^T_{k+1} \cQ_{k+1}   \B)^{-1}
\ee
The     operator $\cG^0_{k+1} $ is defined on functions on     $\tk$.  
Then  the representation is     
   \be 
       C'_k=   \B(I +  \pa \cM\B) \cQ_k \tilde      \cG_{k+1}      \cQ^T_k\B(I  +  \pa  \cM\B)^T
  \ee

 We  also need a better representation of   $C_k^{\frac12}$ or  $(C_k^{\frac12} )' = CC_k^{\frac12}C^T$. 
 We  have    
 \be  
 \label{picnic2}
   C_k^{\frac12}  =      \frac{1}{\pi}  \int_0^{\infty}  \frac{dx}{ \sqrt x  }     C_{k,x}  
\hs   
    C_{k,x}  =    \B( C^T   \De_k C  +x\B)^{-1}  
\ee
So it is sufficient to  find a representation for       $C_{k,x}$  or   $C'_{k,x} =   CC_{k,x}C^T$. 
Define       
\be    \label{swan}
  \tilde       \cG_{k+1,x}   =  
   \cG^0_{k+1,x}-  \cG^0_{k+1,x}\cQ_{k+1}^T   \B(\cQ_{k+1} \cG^0_{k+1,x}\cQ_{k+1}^T\B)^{-1}   \cQ_{k+1} \cG^0_{k+1,x}
\ee
where  for any  $a>0$   
\be   
\cG^0_{k+1,x}  = \B(   \de  d   +    d R_{k+1} \de   + a \cQ^T_{k+1} \cQ_{k+1}    +  x\cQ_k^T   (I + \pa \cM )\chi^* (I + \pa \cM )^T    \cQ_k   \B)^{-1}
\ee
and  $\chi^*$ suppresses the contribution of central bonds $b(y,y')$ joining  $L$-cubes. 
Then the representation is      
\be   \label{moon}
  C'_{k,x}   =     \B(I  + \pa  \cM\B) \cQ_k   \tilde    \cG_{k+1,x} \cQ_k^T  \B(I  + \pa    \cM\B)^T  
\ee

\subsection{random walk expansions}  
\bigskip

We  quote some results about various random walk expansions, almost all due to Balaban.  
\begin{lem}  \cite{Bal84a}   \ \label{random1} 
The Green's function   $\cG_k$  has  a random walk expansion    
 based on blocks  of size $M$, convergent for $M$ sufficiently large.   These yield  the bounds
 for      $  \De_y,  \De_{y'}$    unit squares centered on unit lattice points     $   y,y'  \in  \bbT^0_{N-k}$   and 
 $\zeta_y$     a smooth partition on unity with $\supp\  \zeta_y \subset  \tilde  \De_{y'} $:   
   \begin{equation}  \label{bound1}
|1_{\De_y} \cG_k1_{\De_{y'}} f |, \ 
|1_{\De_y} \pa \cG_k1_{\De_{y'}} f|,   
\| \de_{\al} \zeta_y \pa   \cG_k1_{\De_{y'}} f \|_{\infty}       \leq      C    e^{  -\ga  d(y,y') } \|f\|_{\infty}    
\end{equation}   
\end{lem}
\bigskip

Here  $ (1_{\De_y} f  ) (x, x + \eta e_{\mu} ) =  1_{\De_y}(x)  f  (x, x + \eta e_{\mu} ) $  and  $(\pa_{\nu}f)(x, x+ \eta e_{\mu}) =  ( \pa_{\nu} f_{\mu})(x)$.
The constant for the Holder derivative  $\de_{\al}$ depends on $\al$.  
The statement that  $\cG_k$ has a random walk expansion means    that   
$  \cG_k  = \sum_{\om }  \cG_{k, \om}$, 
and   just as in  (\ref{night})   
\be     \label{night2}
   |\cG_{k, \om}f |, 
      \leq   C (CM^{-1})^{|\om| }  \| f\|_{\infty}
\ee    
and similarly for the derivatives.  
It also means that bounds of the same form hold  for  $\cG_k(s)$ defined with   weakening parameters  $s$ as in  (\ref{again}).       The estimates   (\ref{bound1})   also     have a global version:
 \begin{equation}  \label{bound2}
| \cG_k f  |, \ | \pa \cG_k f|,  \|    \de_{\al}  \pa  \cG_k f \|_{\infty}         \leq       C  \|f\|_{\infty} 
 \end{equation}
 Similar remarks can be added after each of the following lemmas.

\begin{lem}  \label{random2}   \cite{Bal85b}  
The operators $(Q_k \De^{-2}  Q_k^T)^{-1}$  and    $ (\cQ_k  \cG_k \cQ_k^T)^{-1}$     have       random walk expansions    
 based on blocks  of size $M$, convergent for $M$ sufficiently large.   These yield  the bounds
 \be  
 \begin{split}
     |(Q_k   G_k^2  Q_k^T)^{-1}(x,x') |  \leq  &  C  e^{ - \ga d(x,x')  }  \\
      |  (\cQ_k  \cG_k \cQ_k^T)^{-1}(b,b')   |  \leq  &  C  e^{ - \ga d(b,b')  }  \\
 \end{split}
 \ee
 \end{lem}   
    
     These   operators are not inverses of local operators so the expansions  are more complicated.  

\begin{lem}  \label{random3}  \cite{Bal85b}  
The   operators   $R_k$  and    $\cH_k$ have       random walk expansions 
 based on blocks  of size $M$, convergent for $M$ sufficiently large.   This yields the bounds     
 \be    \label{alfie}
 \begin{split}  
   |1_{\De_y} R_k1_{\De_{y'}} f |, \   |1_{\De_y} \pa R_k1_{\De_{y'}} f|,   
\| \de_{\al } \zeta_y \pa  R_k1_{\De_{y'}} f \|_{\infty}       \leq    &  C    e^{  -\ga  d(y,y') } \|f\|_{\infty}     \\
  |1_{\De_y} \cH_k1_{\De_{y'}} f |, \   |1_{\De_y} \pa \cH_k1_{\De_{y'}} f|,   
\| \de_{\al } \zeta_y \pa   \cH_k1_{\De_{y'}} f \|_{\infty}       \leq    &  C    e^{  -\ga  d(y,y') } \|f\|_{\infty}     \\
\end{split}
\ee
\end{lem}

The expansion for  $R_k$   follows from the expansion for  $G_k  = G_k(0)$   in   section \ref{random} 
and the expansion for  $(Q_k   G_k^2  Q_k^T)^{-1}$  and the representation   (\ref{leg}).   The expansion for
$\cH_k $  follows from the expansion for    $\cG_k$  and  the expansion for      $(\cQ_k  \cG_k \cQ_k^T)^{-1}$
and the representation  (\ref{lamb}).    

For future reference we record the global estimate on  $\cH_k$:
\be
\label{slavic2}
 | \cH_k f |, \   | \pa \cH_k f|,   
\| \de_{\al}  \pa  \cH_k f \|_{\infty}       \leq      C     \|f\|_{\infty}  
\ee

Next we  consider operators  like $C_k$ which  act  on functions  of   the type  $\tilde Z  =   (\tilde  Z_1, \tilde Z_2) $ defined in section \ref{naughty}.  For such functions    define  $1_{B(y)} \tilde Z   =   (1_{B(y)}\tilde  Z_1,1_{B(y)} \tilde Z_2)  $   where    $(1_{B(y)} \tilde Z_2)(x, x + e_{\mu}) 
=  1_{B(y)}(x) \tilde Z_2(x, x + e_{\mu})$.   This is again a variable of the same type  and   $\tilde  Z =  \sum_y  1_{B(y)} \tilde Z $.

\begin{lem}    \cite{Bal85b},  \cite{Bal88a}   \label{random4}  
The   operators  $C_k,    C_{k,x},       C_k^{\frac12} $ have       random walk expansions 
 based on blocks  of size $M$, convergent for $M$ sufficiently large.   These yield the bounds for  $y,y'$ on  $\tz$:    
 \be  \label{succinct}
|1_{B(y)} C_k  1_{B(y')} f |,   | 1_{B(y)} C_{k,x}  1_{B(y')} f|,   |   1_{B(y)} C^{\frac12} _k  1_{B(y')} f|  \leq    C  e^{ - \ga d(b,b')  }\| f \|_{\infty}
\ee
\end{lem}
\bigskip

We sketch the proof.
For  $x \geq 0$ the  Green's function  $ \cG^0_{k+1,x} $  has a random walk expansion just  as   for  $\cG_k$,    and  
  $(\cQ_{k+1}\cG^0_{k+1,x}\cQ_{k+1}^T)^{-1}$  has a random walk expansion just as for   $(\cQ_k  \cG_{k} \cQ_k^T)^{-1}$.
  The other operators in   (\ref{swan}) are local so we have an expansion for    $\tilde \cG_{k+1,x}$.   Then the  other operators in   (\ref{moon})  are local so this yields 
  an expansion for    $C'_{k,x}  \equiv     C   C_{k,x}   C^T $.
Next  we  write 
\be      C_{k,x}    = C^{-1} C'_{k,x} (  C^T) ^{-1}
\ee
Since  $C^{-1},(  C^T) ^{-1}$  are    not local this does not immediately give a random walk expansion for  $  C_{k,x}  $.   
However  $C^{-1},(  C^T) ^{-1}$ themselves have     random walk expansions  which we   develop in appendix  \ref{czero}.   Together with the expansion for   $   C'_{k,x}  $  
we get an expansion for     $   C_{k,x}  $  and  $C_k$ is the special case  $x=0$.

We  cannot use  the  expansion for $C_{k,x}$ directly   in  (\ref{picnic2})     unless we can establish  that 
$  C_{k,x}  =  \cO(x^{-1} ) $  to  ensure the convergence of the integral over  $x$.  This  bound which is not readily available.    Instead we use the modified representation.   Break the integral  over  $x$  at  some  $\ga_1$.
Then for $x> \ga_1$  write
\be
   C_{k,x}  =     \B( C^T    \De_k C   +x\B)^{-1}   =  \sum_{n=0}^{\infty}  x^{-(n-1)} (-1)^n (C^T    \De_k C)^n
 \ee  
This coverages for  $\ga_1$  sufficiently large.    Doing the integral over $x$ in the sum yields     
\be     C_k^{\frac12}   =  \frac{1}{\pi}  \int_0^{\ga_1}    \frac{dx}{ \sqrt x  }       C_{k,x}   
+     \sum_{n=0}^{\infty}   \frac{ (-1)^n  }{n+ \frac12}  \ga_1^{-n - \frac12}  (C \De_k  C )^n
\ee
Then for    $    C_{k,x}$   we use the random walk     expansion above,  which is uniformly bounded in $x$.    For  $C^T   \De_k C$ and powers    we can   use the representation  
$\De_k  =  \cH_k^T \de d \cH_k$   from     (\ref{zee})    and the expansion for  $\cH_k$.  
\bigskip

The bound   (\ref{succinct})   also has the global version:
\be   \label{slinky} 
     | C_k^{\frac12} f |  \leq  C \|f\|_{\infty}   
\ee

 \section{Polymer functions}  \label{symmetries}

 \subsection{a preliminary lemma}

Before defining polymer functions 
we   first      show that    every  gauge  potential  $A$    is  locally    equivalent  to  
a field depending only  on   the field  strength  $d A$.  
We  only  need this on a unit lattice. 

\begin{lem}  \label{five}
Let  $A$ be  a  gauge field  on  a unit lattice lattice.   For any reference point  $y $   on any       any cube 
centered on $y$ we have  $A  =  A'  +   \pa \la$   where $A'$ depends only on $dA$  and satisfies
\be   \label{wonder}
   |A'(b)|  \leq   d(b,y) \|d\cA \|_{\infty} 
\ee
\end{lem}    
\bigskip

\pr   We go to an axial gauge.    Let  $\Ga(y,x)$ be the path from  $y$ to $x$ in which coordinates are increased  in the
standard order,  and let    $\la(x)   = A ( \Ga(y,x) )$.   If  $b = [x,x']$ is on one of the paths   $\Ga(y,x)$  then
\be   A(x,x')   =    A ( \Ga(y,x') )  - A(\Ga(y,x) )   =  \pa \la  (x,x')  
\ee
and so  $A'   =  A - \pa \la$  vanishes on such bonds and  hence on the paths  $\Ga(y,x)$.  

Now for any   bond  $b=[x,x']$ we  have that    $\Ga(y,x) + [x,x'] -    \Ga(y,x')$ is a closed  path   which bounds a surface 
$\Sigma_{y,x,x'}$ made up of   at most   $d(b,y)$  unit plaquettes.   Therefore by the lattice   Stoke's theorem  
\be  A'(x,x)  =  A'  \B(  \Ga(y,x) + [x,x'] -    \Ga(y,x')   \B)  = dA'(\Sigma_{y,x,x'} )  = dA(\Sigma_{y,x,x'} ) \ee
and the result follows.

\subsection{a regularity result}

The Landau gauge minimizer     $\cA_k =  \cH_k A_k $  will play  an important  role in the following.   In particular we  want to  use  
it  as a background field  in  the boson Green's function  $G_k(\cA_k)$.   Hence  it must  satisfy the conditions  (\ref{listless}).
However as  we  explain later we only want  to assume  bounds  on $d \cA_k$  not   $\cA_k$  or   general  derivatives  $\pa  \cA_k$.
To obtain the result will require    some gymnastics,       roughly following     \cite{BIJ85}, \cite{BIJ88}.

  Recall that  $\cH_k$  is  gauge equivalent  to  the axial  gauge  $\cH_k^{\sx} $.   The explicit expression  for the latter
 is 
 \be     \label{lounge}
\cH^{\sx}_k   =   \cQ_k^{s,T}  -  \cG^{\sx}_k \de   \cQ_k^{e,T} d
 \ee
 The operator   $\cQ_k^s$   averages over  the faces unit cubes  and   $\cQ^e_k$  averages over plaquettes on the corners of
 unit cubes.    (See  \cite{BIJ85} or   \cite{Imb86} for the exact definition.)   Here  the operator  $\de = d^T$ on two-forms  (functions on plaquettes) is the adjoint  $d$ on one-forms (functions on bonds).
 The operator    $ \cG^{\sx}_k $  on $\tk$  is the axial Green's function  
 defined by  
 \be   
 \exp\B( \frac12  < f,  \cG^{\sx}_k f> \B)  = \sZ_k^{-1}
   \int    \de (   \cQ_k  \cA ) \de^{\sx}_k(\cA)     \exp   \B( - \frac 12 \|  d \cA \|^2  + <f, \cA>   \B)    D\cA 
\ee
(It is not identical with the Green's function  of  \cite{BIJ85} since the  axial gauge fixing is a little different.)

 After   a calculation   using the identity  (\ref{sequentially})   one  finds that the kernel satisfies 
 \be
  \cG^{\sx}_{k+1}(b,b')   =    L \cG^{\sx}_{k}(Lb,Lb')   +   L      ( \cH^{\sx}_k  C'_k   \cH^{\sx,T}   ) ( L b,L  b') 
 \ee
 Iterating this  we see that the     Green's function admits the decomposition    
   \be  \label{yyy}
    \cG^{\sx}_k( b,b')  =  \sum_{j=0}^{k-1} L^{k-j}  ( \cH^{\sx}_j   C'_j  \cH^{\sx,T}_j   ) ( L^{k-j} b,L^{k-j}  b')     
 \ee
In  the  combination   $\cG^{\sx}_k \de  $   we have on   the right   $\cH^{\sx,T}_j  \de   =   (d \cH^{\sx}_j )^T
=     (d \cH_j )^T    =  \cH^{T}_j  \de $,  and on the left we use     $\cH^{\sx}_j   \sim   \cH_j $.  Thus    $
 \cG^{\sx}_k  \de $   is gauge equivalent to  $\cD_k  \de$  where       $\cD_k$ is  defined by the kernel  
 \be   
   \label{yyyy}
    \cD_k( b,b')  =  \sum_{j=0}^{k-1} L^{k-j}  ( \cH_j   C'_j \cH^{T}_j   )  ( L^{k-j} b,L^{k-j}  b')     
 \ee
 and      $\cH_k^{\sx}$ (and hence  $\cH_k$)   is gauge equivalent  to    
\be
  \cH^{\cD}_k    \equiv   \cQ_k^{s,T}  -   \cD_k \de   \cQ_k^{e,T} d
\ee
The only discontinuous part  of     $\cH^{\cD}_k  $  is  $\cQ_k^{s,T}$.    
The operator  $ \cD_k$      has good regularity and decay bounds   as we now show.

We claim that  
 \be  
   \label{lazy1}
 | (\cD_k     f) (  b)   |,   |( \pa   \cD_k  f ) (b)   |,      |   ( \de_{\al} \pa  \cD_k   f)(b)   |     
           \leq    C e^{- \ga   d(b, \supp f) }  \|f  \|_{\infty} 
 \ee
To see this  temporarily   drop the scaling factors and let   $f_L(b)  = f(b/L)$.     Then    with            $ \tilde   C_k    = \cH_k  C'_k \cH_k^T$ 
we have      
\be    \label{lousy} 
  (\cD_k  f  )(b)       =    \sum_{j=0}^{k-1} L^{-2(k-j)} (    \tilde  C_j f_{L^{k-j}})(L^{k-j}b)
\ee    
But  it   can be deduced from     (\ref{alfie})  (\ref{succinct})  that
\be 
 | (\tilde  C_k f)(x)|,    | (\pa   \tilde  C_k f)(b)|,    | ( \de_{\al} \pa   \tilde  C_k   f)(b)|     \leq   Ce^{ - \ga  d(  b,  \supp  f )   }   \|f  \|_{\infty}    
\ee   
Therefore      
\be       \label{longing}  
|  (\cD_k  f  )(b)  |     \leq  C     \sum_{j=0}^{k-1} L^{-2(k-j)}  e^{ - \ga    L^{k-j} d(b,  \supp  f)  }\|f \|_{\infty}
\ee    
this   yields the first bound in  (\ref{lazy1}).      The derivatives  reduce   the   $ L^{-2(k-j)} $ to $ L^{-(k-j)} $ or  $L^{-(1- \al)(k-j)}$,  and we
still have convergence.  Thus     (\ref{lazy1}) is established.    Note that we   cannot allow   two derivatives unless
    $d(b,  \supp  f) \geq  \one  >  0$.

 We also need a local version   of $\cD$.     Again let     $\zeta_{\De_y} $ be a  smooth   partition of unity with 
  $\supp \   \zeta_{\De_y}  \subset    \tilde   \De_y  $
and define  
 \be 
 \begin{split}
   \cD^{\loc}_k  =    &     \sum_{y,y':  d(y,y' )  <   4} \zeta_{\De_y}  \cD   \zeta_{\De_{y'}}   \\
  \cH^{\loc }_k     =  &   \cQ_k^{s,T}  -   \cD^{\loc}_k \de   \cQ_k^{e,T} d     \\
\end{split}
\ee
Then   $ \cD^{\loc}_k   $   again satisfies the bounds    (\ref{lazy1}),  since if a derivative  falls  on  a  $\zeta_{\De_y} $
nothing important is changed    

The  difference   $  \cD_k  -   \cD_k^{\loc}  $ has no short distance singularity  and we  can  allow more  derivatives, 
also on the right.     
We  have instead of  (\ref{lousy})   
\be   
   \label{zzz}
\B( (   \cD_k  -   \cD_k^{\loc} )f \B)(b)   =
  \sum_{y,y':  d(y,y' )  \geq    4}  \zeta_{\De_y} (b)  \sum_{j=0}^{k-1} L^{-2(k-j)}
   \B(   \tilde C_j ( \zeta_{\De_{y'} }  f)_{ L^{k-j} } \B)   (L^{k-j} b)     
 \ee
and since    $d(y,y' )  \geq    4 $ implies  $ d(\tilde \De_y,\tilde \De_{y'} )   \geq    1$  we have instead  of    (\ref{longing}) 
\be 
\begin{split}
   |  \B(  (   \cD_k  -   \cD_k^{\loc} )  f  \B)(b)  |     \leq    &  C     \sum_{j=0}^{k-1} L^{-2(k-j)} 
     \sum_{y,y':  d(y,y' )  \geq    4}     \zeta_{\De_y} (b) e^{ - \ga    L^{k-j} d(\tilde \De_y,\tilde \De_{y'} )  }  \|f \|_{\infty}  \\
   \leq    &  C     \sum_{j=0}^{k-1} L^{-2(k-j)}   e^{ - \frac12 \ga    L^{k-j}   }\|f \|_{\infty}  \\
   \leq   & C   \|f \|_{\infty}  \\
\end{split}    
\ee
Now we can allow  any  number of   extra  derivatives,   each derivative  adds a factor  $L^{k-j}$ to the last estimate but the
factor     $ e^{ - \frac12 \ga    L^{k-j}   }$ still gives convergence.
In particular  we  have      
\be  \label{nachos} 
| (  \cD_k  -   \cD_k^{\loc} ) \de       F    |,    | \pa   (  \cD_k  -   \cD_k^{\loc} ) \de     F    |,   | \de_{\al}  \pa   (  \cD_k  -   \cD_k^{\loc} ) \de     F    |      \leq   C   \|F \|_{\infty} 
\ee

   With these  preliminaries out of the way we can now state  the regularity result.    Let  $\square^{\nat}$  be a cube centered on  $\square$ which which is a union of $M$-cubes with    $\one L $      $M$-cubes on a side.  We have  $\square \subset   \tilde   \square
   \subset  \square^{\nat}$.

  \begin{lem}   \label{licorice}
  $  \cA_k = \cH_k A_k $  has the property   that  in each $\square^{\nat}$      it      is    gauge equivalent  to  some     $\cA' $ satisfying      
  \be
  |\cA'  |,   |\pa   \cA'  |,  | \de_{\al} \pa \cA'|  \leq    CM   \| d \cA_k  \|_{\infty}   
  \ee    
  \end{lem}
  \bigskip
  
 \pr       
 We  write   
 \be   
  \cA_k   =  (\cH_k -  \cH^{\cD}_k)A_k  +    ( \cH^{\cD}_k -  \cH^{\loc}_k )A_k   +     \cH^{\loc}_k  A_k  
 \ee
 and argue that each term has the stated property.     The first  is globally pure gauge and  hence pure gauge on
 any  $\square^{\nat}$. 
 
  The second  is  the same as      $( \cD_k - \cD_k^{loc} )  \de   \cQ_k^{e,T} dA_k$  and  by   (\ref{nachos}) we have globally  
 \be   
   |( \cD_k - \cD_k^{loc} )  \de   \cQ_k^{e,T} dA_k|,   \leq      C   \| d A_k  \|_{\infty}     
 \ee
 and the same for the derivatives.  
 
   For the third  term    note  that
    by lemma  \ref{five}     
we have     $A_k   = A_k'   + \pa \la   $   on a suitable neighborhood of     $\square^{\nat} $  and   
   \be  \label{tingly}
    |A'_k|  \leq    C M   \| d A_k  \|_{\infty}    \leq    C M   \| d \cA_k  \|_{\infty}          
\ee
The last  inequality  follows    
since  $A_k = \cQ_k \cA_k$
 hence by     (\ref{lockdown})    $ | d  A_k  |  \leq    \| \pa \cA_k  \|_{\infty}    $.
Next     $   \cQ_k^{s,T} d \la   =   d  Q^T_k \la$   and so      in  $\square^{\nat} $
\be
  \cH^{\loc}_k A_k   =     \cH^{\loc}_k  A'_k   +     d  Q^T_k \la 
\ee
Thus it suffices to show  that   $     \cH^{\loc}_k A'_k $  is a sum of terms  with the stated properties.    The function
has a good bound, but we have to work harder for the derivative.   

Extend the definition of  $A'_k$ to the whole lattice by defining it to be zero off the neighborhood of  $\square^{\nat} $.  The extension
is  still      bounded by    $C M   \| d \cA_k  \|_{\infty}$,  as are    derivatives since we are on a unit lattice.   Now write on $\square^{\nat} $  
\be
     \cH^{\loc}_k A'_k   =       (  \cH^{\loc}_k -    \cH^{\cD}_k  )A'_k   +    ( \cH^{\cD}_k -  \cH_k)A'_k    +    \cH_k A'_k  
\ee
The first  term    $( \cD_k - \cD_k^{loc} )  \de   \cQ_k^{e,T}  dA'_k$   and its derivatives   are         again bounded by  (\ref{nachos}),  the second term is  again   pure gauge,  and the third  term has  good bounds  
by    (\ref{slavic2}).    This completes the proof.

\subsection{bounded fields}  \label{bdd}

We define some bounded field conditions.   To motivate the definitions 
consider  the minimizers
\be    
\cA_k =  \cH_k A_k   \hs       \phi_k(\cA)= \cH _k(\cA)  \Phi_k  
\ee
As suggested by our discussion to this point,  and as we show in detail,     the    action after $k$  steps  will    have the
leading terms     
\be 
  \frac12   \|d \cA_k \|^2   +   \frac12  \| \Phi_k  -  \cQ_k( \cA_k)  \phi_k(\cA_k ) \|^2 
    + \frac12  \| \pa_{\cA_k} \phi_k(\cA_k) \|^2  +  \la_k \int \phi_k(\cA_k)^4 + \dots  
\ee
We   choose the small field  conditions so that  if  they are violated somewhere,  then some piece of this action  is large and the contribution to the density is suppressed.    To specify   the conditions  let
 \be
 p_k  =  (- \log \la_k )^p
 \ee
 for some positive integer $p$.     We  assume  $\la_k$   is  small   so that  $p_k$ is large.    Further  we  assume
 that $e_k^2 \leq \la_k$   so that  $p_k  \leq   \la_k^{-\ep}   \leq  e_k^{-2 \ep}$.

 \begin{defn}
 The small field domain    $\cS_k$   is all  real-valued  fields  $A_k, \Phi_k$  on  $\bbT^{0}_{N-k}$  such that    
 \be
      | d \cA_k  |  \leq    p_k
 \ee
  and   
 \be  \label{talc}
| \Phi_k -Q_k(\cA_k)  \phi_k(\cA_k)   |  \leq  p_k    \hs  |   \pa_{\cA_k} \phi_k (\cA_k) | \leq  p_k  \hs    |   \phi_k (\cA_k) |  \leq  \la_k^{-\frac14} p_k
 \ee  
 \end{defn}
 The     bounds  on   $\cS_k$  imply the bounds on the fundamental fields
\be  
\label{fund}
  | d  A_k  |  \leq    p_k  \hs     | \pa \Phi_k  |  \leq    3p_k   \hs    | \Phi_k  |  \leq  2\la_k^{-\frac14}  p_k
\ee
The first  follows from  $A_k   = \cQ_k \cA_k$ and the identity (\ref{lockdown}).   The  other two  follow  in a straightforward manner
(see for example  \cite{Dim11}).

We   also want a  larger complex domain  for the polymer functions we are about to introduce.

\begin{defn} Let  $\ep>0$ be a fixed small number  and consider the bounds 
 \be  \label{ding}
  | \cA|     <   e_k^{-1 + \ep}  \hs
   |\pa \cA|  <   e_k^{-1 + 2\ep}    \hs 
   |\de_{\al} \pa \cA|      <   e_k^{-1 +3 \ep}    
\ee 
and   
 \be   \label{dong}
 |   \phi |   <   \la_k^{-\frac14-  \ep}   \hs  |   \pa_{\cA} \phi| <  \la_k^{-\frac16- 2 \ep}  
 \hs    | \de_{\al, \cA}  \pa_{\cA}  \phi |  <  \la_k^{-\frac16- \ep}
 \ee  
 The  small field  domain   $\cR_k$   is all    complex-valued fields  $\cA, \phi$ on $\tk$  such that
 \begin{enumerate}
 \item  $\cA  = \cA_0  + \cA_1$   where  $\cA_0$ is  real and      each $\square^{\nat} $  is    gauge equivalent  to  some    $\cA'_0 $ satisfying      
 (\ref{ding}) with a factor    $\frac12$ on the right and   $\cA_1$ is complex and satisfies  (\ref{ding}) with a factor $\frac 12$   on the right.   
  \item    $\phi$ satisfies  the bounds  (\ref{dong})  
  \end{enumerate} 
\end{defn}

We also say  $\cA  \in \cR_k$ if $\cA$ satisfies condition 1.  
Then   $\cA$ is locally     gauge equivalent  to  a field  $\cA'$    satisfying  (\ref{ding}),  and if  $\phi$ satisfies    (\ref{dong} ) the 
pair    $(\cA, \phi)$ is locally gauge equivalent to  a pair  $(\cA', \phi')$   satisfying      (\ref{ding}), (\ref{dong})
 (The latter since  $|\phi'|  = | \phi|,   |\pa_{\cA'} \phi' |  =   |\pa_{\cA} \phi |$, etc. )   
We also note that if   $\cA  \in \cR_k$  then  
\be
   | d \cA  |  \leq  \one e_k^{-1 + 2 \ep}     \hs   | \im \cA  |  \leq   \one  e_k^{-1+ \ep}   
\ee

These bounds are somewhat  arbitrary.   They must be large enough so that  $\cS_k \subset  \cR_k$,   a fact we establish next.
The conditions  on  $\cA$ are  more restrictive than the domain  (\ref{listless}) and  hence    $G_k (\cA),  \cH_k(\cA)$
and  derivatives  of order less than two  have
good bounds.  The  conditions   also ensure that  the polymer functions do not become too large  and hence erode the convergence of 
our expansions.  
Also it is convenient to have slightly  sharper bounds for higher derivatives.  
\bigskip

\begin{lem}     \label{inclusion}
If   $\al  <   2/3$   then       $A_k, \Phi_k$ in  $\cS_k$  implies      $  \cA_k ,  \phi_k(  \cA_{k} )$   in    $  \frac12   \cR_k$.   
\end{lem}
\bigskip

\pr  By      Lemma   \ref{licorice},       $  \cA_k$  is     gauge equivalent in each $\square^{\nat} $ to  some     $\cA' $ satisfying      
\be   
  | \cA'|, |\pa    \cA'|,  | \de_{\al} \pa \cA' |   <   CM p_k   \leq   e_k^{- \ep}   
  \ee    
Hence  the  bounds  (\ref{ding}) are  easily   satisfied.   
 The     bounds on  $ \phi_k (\cA_k),    \pa_{\cA_k} \phi_k (\cA_k) $   are   also   immediate.  
    For the  last we    write   for  $d(x,y) \leq  1$  and  $\cA  = \cA_k$
\be
\begin{split}  &  \B|\B( \de_{\al, \cA}  \pa_{\cA}  \phi_k(\cA)  \B)(x,y)\B|    
  =   
  \left| \frac{ e^{qe_k \cA(\Ga_{xy})} ( \pa_{\cA}  \phi_k(\cA)  )(y) - \pa_{\cA}  \phi_k(\cA)  )(x)} {  |x-y|^{ \al}   }  \right|  \\
 &    =   
\left|    \frac{ e^{qe_k \cA(\Ga_{xy})} ( \pa_{\cA}  \phi_k(\cA)  )(y) - \pa_{\cA}  \phi_k(\cA)  )(x)} {   |x-y|^{\frac32 \al}  }  \right|^{2/3}       
 \Big| e^{qe_k \cA(\Ga_{xy})}(  \pa_{\cA}  \phi_k(\cA)  )(y) - \pa_{\cA}  \phi_k(\cA)  )(x)   \Big|^{1/3}    \\
 &   \leq  (  C \la_k^{- \frac14}  p_k  )^{2/3}   (  2p_k   )^{1/3}   \leq  \frac12  \la_k^{- \frac16 - \ep} \\    
\end{split}
\ee
Here  we  used    (\ref{slavic}) and  (\ref{fund})  for the first factor and   (\ref{talc}) for the second factor.
This completes the proof.

\bigskip

For future reference we also note the following result

\begin{lem}  \label{sunfish}
If    $\cA,  \phi  \in \cR_{k+1}$,  then in any   $\square^{\nat} $  the pair   $ (\cA_L, \phi_L)$  is gauge equivalent to  
$(\cA', \phi')$ satisfying   
 \be  \label{ding2}
  | \cA'|  <   L^{-1+ \ep} [  e_k^{-1 + \ep}]   \hs     |\pa \cA'|,        <   L^{-2+ \ep}    [ e_k^{-1 + 2\ep} ]  
 \hs        |\de_{\al} \pa \cA'|      <    L^{-2- \al + 2  \ep}      [e_k^{-1 +3 \ep} ]  
\ee 
and   
 \be   \label{dong2}
 |   \phi' |   <  L^{-\frac34 - \ep} [ \la_k^{-\frac14-  \ep}]   \hs  |   \pa_{\cA'} \phi| < L^{-\frac53 - \ep} [ \la_k^{-\frac16- 2 \ep}]  
 \hs    | \de_{\al, \cA'}  \pa_{\cA'}  \phi' |  <     L^{-\frac53 -\al -  \ep} [\la_k^{-\frac16- \ep}]
 \ee     
In  particular    $( \cA', \phi')   \in     L^{-\frac 34 - \ep} \cR_k$. 
\end{lem}
\bigskip

\pr  Choose an  $M$-cube    $\square^{\nat}_0 $  in  $\bbT^{-k-1}_{N-k-1}$ so that  $\square^{\nat}  \subset  L \square^{\nat}_0 $.   We have
  $\cA  = \cA_0  + \cA_1$ with  $\cA_0 \sim \cA'_0$ in   $\square^{\nat}_0 $ and  $\cA'_0, \cA_1$ satisfy the bounds for $k+1$.    Hence   $\cA_L  \sim \cA'_{0,L}  + \cA_{1,L}  \equiv  \cA' $ in  $L \square^{\nat}_0$ and hence in $\square^{\nat} $.    Since      $e_{k+1}  = L^{1/2}  e_k$     we have  in $\square^{\nat} $  
\be  \label{toffee1}
\begin{split}
   |   \cA'_{0,L}|     \leq   & 
     L^{-\frac12} \| \cA'_0  \|_{\infty}  \leq  \frac12   L^{-\frac12}   e_{k+1}^{-1+\ep}     \leq   \frac12  L^{-1  +    \ep  }[ e_k^{-1+ \ep} ]\\
\end{split}
\ee
The bound  for  $\pa \cA'_{0,L} =  L^{-1}  (\pa \cA'_0)_L$ is similar as  is the bound for    $  \de_{\al} \pa \cA'_{0,L}$
The same bounds hold for  $\cA_{1,L}$.
Therefore   $\cA'$ satisfies (\ref{ding2}).        Similarly  for  $\phi' \sim \phi_L$
since  $\la_{k+1}  =   L \la_k $  
\be 
 | \phi' |   = |  \phi_L |  \leq   L^{- \frac12 } \| \phi\|_{\infty}
  \leq    L^{- \frac12 } \la_{k+1}^{-\frac14 - \ep}    \leq      L^{- \frac34 - \ep  } \la_{k}^{-\frac14 - \ep}
\ee   
Since    $\pa_{\cA'} \phi'  \sim       \pa_{\cA_L } \phi_L  =  L^{-1}( \pa_{\cA} \phi)_L$, etc.  the derivatives add extra powers of $L^{-1}$ as indicated.

\subsection{definition of polymer functions}  \label{polymersection}

A   \textit{polymer}   $X$  in  $\tk$  is a  connected union of $M$ cubes,  with the convention that two  cubes are connected   if
they have an entire face in common.  The set of all polymer functions is denoted  $\cD_k$.   Our interaction  terms will be expressed in terms of polymer functions  $E(X, \cA, \phi)$
which  depend on the fields  $\cA,  \phi$  only in $X$  
 
 We  require that    $E(X, \cA, \phi)$ is bounded and analytic  on the domain   $\cR_k$   so  the norm 
 \be    \|E(X) \|_{k}  =  \sup_{   \cA,  \phi  \in \cR_k  }     |E(X, \cA, \phi)|  
 \ee
 is finite.  
 
We  also require  that   $E(X, \cA, \phi)$
be  exponentially decaying in the size of $X$.    Size is measured on the $M$-scale.   define  $d_M(X)$
by  
\begin{equation}   
  M d_M(X)  =  \textrm{length of the shortest  continuum    tree  joining   the $M$-cubes    in     $X$}  .
\end{equation} 
The  requirement  is that     $E(X, \cA, \phi)$     be bounded by    a constant times  $  e^{- \ka d_M(X)}  $
for  some      $\ka   = \one$.    
To  put it another way  the norm  
\be   \label{kk}
   \|E\|_{k, \ka}  =  \sup_{X \in \cD_k}     \|E(X) \|_{k}     e^{ \ka d_M(X)}   
\ee
must be finite.   The space of all polymer functions with finite  norm is    is a Banach space called  $\cK_k$.

We  also note that  if   $|X|_M$ is  the
number of $M$ cubes in  $X$,    then  
\be    \label{chile}
   d_M(X)  \leq   |X|_M  \leq   \one  ( d_M(X)  + 1 )
\ee  
Also  there are constants  $\ka_0, K_0 = \one$ such that  for any  $M$-cube  $\square$ 
\be   \sum_{X \in \cD_k, X  \supset \square}    e^{- \ka_0 d_M(X)}  \leq  K_0  \ee
We assume   $\ka \geq  \ka_0$.

To  scale  polymer functions   we first introduce a blocking operation.    If  $Y$  is  a polymer  in $\tk$  which is a connected
union of    $LM$-cubes
we  define  
\be   (\cB   E)(Y)  =  \sum_{X:   \bar X  = Y}   E(X)    \ee
where  $\bar X$ is the union of all  $LM$-polymers intersecting  $X$.
Then   
\be 
 | (\cB   E)(Y,  \cA, \phi) |  \leq      \one L^3  e^{- L(\ka - \ka_0 -1) d_{LM} (Y)}   \|    E \|_{k,\ka}
\ee
This can be scaled down   to  a   polymer function  $ (\cB   E)_{L^{-1}}$ on  $\bbT^{-k-1}_{N- k-1}$  by   
\be    (\cB   E)_{L^{-1}}(X,  \cA, \phi)  =    (\cB   E)(LX,  \cA_L, \phi_L) 
\ee
and then   
\be   \label{L3}  
  \| (\cB   E)_{L^{-1}}  \|_{k+1,L(\ka-\ka_0 -1)}  \leq  \one L^3     \|    E \|_{k,\ka}
\ee
Note that if   $\ka$ is large enough  then  $L(\ka - \ka_0 -1)  > \ka$ and we can take  $\ka$ on the left.    But the $L^3$ means 
that the   size can   grow.

\subsection{symmetries}
We  consider  polymer functions    $E(X, \cA, \phi)  \in \cK_k$   which   are    invariant under the following symmetries
\begin{enumerate}
\item  (lattice symmetries)  If   $r$ is a $\tz$ unit  lattice symmetry and  $\cA_r,   \phi_r$  are the transformed fields  then
$E(rX,  \cA_r,  \phi_r)  =  E(X, \cA,  \phi)$.
\item  (gauge invariance)
$E(X, \cA^{\la},   \phi^{\la})  =  E(X,\cA,  \phi )$.
\item  (charge conjugation invariance)  $E(X,-\cA,  C  \phi)  =  E(X,\cA,   \phi )$. 
\end{enumerate}

Here are some consequences.    The  $n^{th}$   derivative of   $E(X, \cA,  \phi ) $   in  $\cA$ at
$\phi=0,  \cA=0$ is  
  is the multilinear
functional 
\be  
 \frac{ \de^n E }{\de \cA^n} \Big (X, 0;  f_1,  \dots   ,f_n \Big)
 =   \frac{  \pa^n} { \pa t_1 \dots \pa t_n}  \Big[     E(X,  t_1  f_1    +  \dots  + t_n f_n,  0  ) \Big]_{t=0}
 \ee
If    one of the functions  $f_i =  \pa  \la$  then  by gauge invariance there is no dependence  on   $t_i$
and  the derivative vanishes.   Thus we have the Ward identity
\be   \label{Ward} 
 \frac{ \de^n E }{\de \cA^n} \Big (X, 0;  f_1,  \dots,  \pa \la,  \dots    ,f_n \Big)
 = 0
 \ee
A special   case  of   gauge invariance is  rotation in charge space.   If  $e_k \la = \theta$= constant then  
\be E(X,  \cA,   e^{q  \theta}  \phi)  =  E(X,\cA,  \phi )\ee
 A   rotation by  $\theta  = \pi$ in charge  space  gives 
$E( X, \cA, - \phi) =  E(X,   \cA, \phi)$.   Hence  any odd number of $\phi$  derivatives   at $\phi=0$  gives zero. 
Therefore  
\be  \label{suzy1}
   \frac{\de E}{ \de \phi} \B(X,0 \B)=0   \hs   \frac{\de^2 E}{ \de \phi  \de  \cA} \B(X,0 \B)=  0  \hs
\frac{\de^3 E}{ \de \phi  \de  \cA^2} \B(X,0 \B)=  0  
  \hs   \frac{\de^3 E}{ \de \phi^3 } \B(X,0 \B)=0   
\ee
Charge    conjugation invariance   gives $E(X,-\cA,  0)  =  E(X,\cA,  0 )$  and this  implies   
\be \label{suzy2}
    \frac{\de E}{ \de \cA} \B(X,0 \B)=  0  \hs   \hs      \frac{\de^3 E}{  \de  \cA^3} \B(X,0 \B)=  0 
\ee

\subsection{normalization}

As we iterate the RG transformations    the scaling operation    can increase the size of   the  polymer functions   by as much as $\cO(L^3)$  as is
evident from (\ref{L3}).    We   have  to watch this carefully  and start with a discussion  of what criteria we  need to 
avoid this growth.    The following generalizes the analysis in   \cite{BDH98},  \cite{Dim11}.

\begin{defn}
A polymer function     $E(X,\cA,  \phi ) $ with the stated   symmetries   is said to be {\em normalized}   if in  addition to the   
vanishing  derivatives (\ref{Ward}),    (\ref{suzy1}), (\ref{suzy2})     we   have  for   $1 \leq  i,j  \leq   2$  and some $x_0 \in X$
\begin{equation}  \label{normalization}
\begin{split}
E(X,0)  = 0    \hs  &
\frac{ \de^2 E }{\de \phi^2}  \B(X, 0;  e_i,e_j  \B) = 0
  \hs    \frac{ \de^2 E }{\de \phi^2}   \B(X, 0;  e_i, ( \cdot -x_0)_{\mu} e_j  \B) = 0  \\
 \end{split}
\end{equation}
\end{defn}
\bigskip

    Define a  polymer  $X$  to be  \textit{small} 
if  $d_M(X) <L$ and \textit{large}  if   $d_M(X)  \geq  L$.  The set of all small polymers   in denoted $\cS$.
Next  we show that a  polymer  function  normalized for small polymers  contracts under scaling.

\begin{lem}   
\label{scalinglem}  Let   $E (X, \cA, \phi)   $  be normalized for small polymers.   Then  for   $L$ sufficiently large and  
  $e_k, \la_k$  sufficiently small   (depending on  $L,M$)   and   $\frac {7}{12}  \leq  \al   < \frac23$
\begin{equation}   \label{crude}  
  \|  ( \cB E)_{L^{-1}} \|_{k+1, \ka}  \leq         \cO( 1) L^{-\ep}      \|E\|_{k, \ka} 
\end{equation}
\end{lem}
\bigskip

\pr       This  follows a similar proof in \cite{Dim11},  where one can find more details.
   For  large sets $d_M(X)\geq  L$,  We  can borrow  a factor  $e^{-L}$ from   the decay factor  $e^{ - \ka  d_M(X)}$.   This beats the $L^3$  and    and   gives an estimate  $\one L^{-n}$ for any  $n$.

For small sets  $X$  
  we  will show that   for $\cA,   \phi \in  \cR_{k+1}$   
  \begin{equation}  \label{snort2}
   |E(X,  \cA_L,  \phi_L)|   \leq   \cO(1) L^{-3-\ep}      \| E(X) \|_k
  \end{equation}
  This  improves on the general   bound  $ |E(X,  \cA_L,  \phi_L)|   \leq  \one     \| E(X) \|_k$
which was the input  to  (\ref{L3}). 
  The   extra factor  $L^{-3}$  beats the $L^3$ and yields the result.

  Every small  polymer  $X$ contains    some $M$-cube  $\square$.   By (\ref{chile}) $|X|_M  \leq \one L$  and so  $X$ is contained 
     in some enlargement    $ \square^{\nat}$.  By lemma
     \ref{sunfish}   $( \cA_L,  \phi_L)$  is gauge equivalent in $\square^{\nat}$ to  $(\cA', \phi')$ satisfying the bounds    (\ref{ding2}), (\ref{dong2}).
 Since  $E(X)$ is gauge invariant it suffices  show  that   $E(X,  \cA',  \phi')$ satisfies   (\ref{snort2})  for fields satisfying  (\ref{ding2}), (\ref{dong2}).
 
 We  make a further gauge transformation.   
 Pick a point  $x_0$  in $X$.  Since  the   constant   $\cA_{\mu}'(x_0) = \pa_{\mu}  ( \cA'(x_0)\cdot (x-x_0) ) \equiv   \pa \la$ is pure  (complex) gauge  in  $\square^{\nat}$ we can 
 define   
 \be   
 \begin{split}
 \tilde   \cA(x) = &   \cA'(x) -  \pa  \la   =  \cA'(x) - \cA'(x_0)  \\ 
\tilde   \phi(x)  =    &   e^{qe_k  \la(x)} \phi'(x)   \\
\end{split}
\ee
 We claim that  
 the    new fields satisfy the  bounds      
  \be  \label{ding3}
  |\tilde \cA|  <   L^{-2+ \ep} [  e_k^{-1 + \ep}]   \hs     |\pa \tilde \cA|,        <   L^{-2+ \ep}    [ e_k^{-1 +2 \ep} ]  
 \hs        |\de_{\al} \pa \tilde  \cA|      <    L^{-2- \al +  2 \ep}      [e_k^{-1 +3 \ep} ]  
\ee 
and   
 \be   \label{dong3}
 |  \tilde  \phi |   <  3 L^{-\frac34 - \ep} [ \la_k^{-\frac14-  \ep}]   \hs  |   \pa_{\tilde \cA} \tilde  \phi| <  3  L^{-\frac53 - \ep} [ \la_k^{-\frac16- 2 \ep}]  
 \hs    | \de_{\al,\tilde  \cA}  \pa_{\tilde  \cA} \tilde   \phi |  <  3 L^{-\frac53 -\al -  \ep} [\la_k^{-\frac16- \ep}]
 \ee     
Indeed   since   $X$ is small it has diameter less than   $M|X|_M \leq  \one ML$.    We  assume  $e_k$ is small enough  so     $  \one ML  e_k^{\ep}  \leq  1$.
  Then we have  the improved bound    
 \be    \label{new2}
\begin{split}
|\tilde  \cA|         \leq    &   \one ML   \|  \pa   \cA'    \|_{\infty}     \leq    ( \one ML )    L^{-2  +  \ep }  e_k^{-1 + 2 \ep}  \leq       L^{-2  +  \ep } [ e_k^{-1 +  \ep}] \\
\end{split}
\ee
The bounds on  derivatives stay the  same.  The  gauge function satisfies    $  |\la  |  \leq   \one ML |\cA'(x_0)|  \leq  CMe_k^{-1 + \ep}   \leq  e_k^{-1}$  and so the bounds on the scalar field are only altered by the inconsequential   
  $ | e^{qe_k  \la(x)}|  \leq   e^{e_k | \la(x)| } \leq  e < 3$.
 
 We  now have  $(\tilde   \cA,  \tilde     \phi )  \in    3  L^{-\frac34 - \ep}  \cR_k$  and since
 since   $ E(  \cA'   , \phi')   = E( \tilde \cA,  \tilde \phi )$   it suffices to  prove the bound
  (\ref{snort2})  for fields satisfying  (\ref{ding3}), (\ref{dong3}).

We     make a Taylor expansion of  $t  \to   E(X,t \tilde \cA, t \tilde \phi)$  around $t=0$ and evaluate at  $t=1$. 
For     complex  $t$ with  $|t|   \leq    \frac16   L^{\frac34 + \ep} $   we  have
$(t \tilde \cA, t \tilde \phi)  \in \frac12 \cR_k$.   
 Taking account the vanishing derivatives  and choosing    $r = \frac16 L^{\frac34 + \ep}$  
 the expansion is  then  
 \be    
 \begin{split}
   \label{deft}
 E(X, \tilde \cA,   \tilde \phi )
= &   \frac12  \frac{ \de^2 E }{\de \cA^2} \B(X, 0;  \tilde \cA ,  \tilde \cA \B)  +
\frac12   \frac{ \de^2 E }{\de \phi^2} \B(X, 0;  \tilde \phi ,  \tilde \phi \B)  \\
 &+ \frac12     \frac{ \de^3 E }{\de  \cA  \de \phi^2} \B(X,  0;\tilde \cA,  \tilde \phi ,  \tilde \phi \B) 
+   \frac{1}{2\pi i}   \int_{|t |=  r }    \frac{  E(X, t \tilde \cA,   t\tilde \phi ) } {t^4(t-1)}   dt \\
\end{split}
\end{equation}
Since   $ |E(X, t \tilde \cA,   t\tilde \phi )|  \leq   \|E(X)\|_k$   the  last      term   in   (\ref{deft})    is  bounded  by   $\cO(1)  L^{-3-  4\ep}  \|E(X) \|_k$
which suffices.

 With  $\phi=0$  the first term can be expressed in a larger  analyticity domain    as  
 \be     \frac12  \frac{ \de^2 E }{\de \cA^2} \B(0;  \tilde \cA ,  \tilde \cA \B)  
 =     \frac{1}{2\pi i}  \int_{|t| =     L^{2 - \ep} }  \frac{ dt}{t^3}E (X, t \tilde \cA,   0)
 \ee
Then  this term is bounded by     $\cO(1)  L^{-4  + 2 \ep}  \|E(X) \|_k$ which suffices.
 \bigskip

 Next  consider the term       $( \de^3 E/\de  \cA  \de \phi^2)(0;\tilde \cA,  \tilde \phi ,  \tilde \phi) $  in   (\ref{deft}). 
 Now  $(t,s)  \to   E_k(X,  t\cA,  s \phi)$  is analytic in    $|t|  \leq   L^{2+ \ep}  $
 and  $|s| \leq   L^{\frac34  + \ep}$    and so  
 \be  
 \frac12   \frac{ \de^3 E}{\de  \cA  \de \phi^2} \B(X, 0;\tilde \cA,  \tilde \phi ,  \tilde \phi \B) 
=  \frac{1}{ (2\pi i)^2}   \int_{|t| = L^{2- \ep}  } \frac{dt}{t^2}    \int_{|s| = L^{\frac34- \ep}  } \frac{ds}{s^3}   
  E_k(X,  t\tilde \cA,  s \tilde \phi)
\ee
Then  this term is bounded by     $\cO(1) L^{-2 + \ep}  L^{-\frac32   + 2 \ep}  \|E(X) \|_k  =  \cO(1) L^{-\frac72 + 3\ep}    \|E(X) \|_k  $ which suffices.

For the    analysis of the term   $(  \de^2 E /\de \phi^2)   (  0; \tilde \phi,  \tilde \phi)$  in   (\ref{deft})
we  write
   \begin{equation}   \label{knuckles}
\tilde \phi(x)  =  \tilde \phi(x_0)   +(x-x_0) \cdot \pa \tilde \phi(x_0)   + \De( x, x_0)
\end{equation}
and  expand     taking account the vanishing derivatives
\be
\begin{split}
  \frac{ \de^2 E }{\de \phi^2}  \B(X,  0; \tilde \phi,  \tilde \phi\B)= 
  &  \frac{ \de^2 E }{\de \phi^2}\B(X,0; (   \cdot-x_0) \cdot \pa \tilde \phi(x_0)  , ( \cdot-x_0) \cdot \pa \tilde \phi(x_0)  \B)   \\
+  &2  \frac{ \de^2 E }{\de \phi^2}\B(X,0; ( \cdot-x_0) \cdot \pa \tilde \phi(x_0) ,  \De \B)   \\
+  &  \frac{ \de^2 E }{\de \phi^2}\B(X,0;   \De ,   \De   \B)  
+  2  \frac{ \de^2 E }{\de \phi^2}\B(X,0; \tilde \phi(x_0) , \De \B)   \\
\end{split}
\ee
All these terms can be estimated  by  Cauchy bounds    and  the information that  
\be  \tilde  \phi(x_0)  \in   L^{-\frac34 - \ep  }  \cR_k
\hs   (x-x_0) \cdot \pa \tilde   \phi(x_0)   \in  L^{-\frac53 -2 \ep}\cR_k    \hs  \De  \in     L^{-\frac53-  \al   - \ep}\cR_k   
\ee
See \cite{Dim11}    for  estimates of this form  (where the exponents are a little different).   The first  term  is 
then   $\cO(1) L^{-10/3  - 4 \ep}\|E (X)  \|_k$  which suffices.    
The second  and third terms are even smaller.  The last  term  is    
less than    $\cO(1)L^{-29/12 - \al    - 2 \ep}\|E (X)  \|_k$, which suffices since we are assuming  $\al  \geq  \frac{7}{12}$.

Thus  (\ref{snort2}) is established and the lemma is proved.

\subsection{arranging normalization}

The next result shows   that if we  remove certain relevant terms from     the polymer function,   the  
remainder    is normalized.

Given  $E (X, \cA, \phi)$ on $\tk$  satisfying lattice, gauge, and charge conjugation   symmetries    we  define    $(\cR E)(X,\cA, \phi) $   as  follows. 
If  $X$ is large then  $(\cR E)(X,\cA, \phi)= E(X,\cA, \phi)$.    If   $X$ is small  ($X  \in \cS$) then  $(\cR E)(X)$  is 
defined by 
\begin{equation}  \label{renorm}
\begin{split}
 E( X, \cA,  \phi)  
= &   \al_0(E,X)  \Vol( X )
+  \al_{2}(E,X)   \int_X | \phi |^2        + \sum_{\mu}  \al_{2, \mu}(E,X)  \int_X  \phi \cdot  \nabla_{\cA,  \mu} \phi   
    +   (\cR E)( X,\cA,  \phi)    \\
\end{split}
\end{equation} 
where    
\be    \label{numbers}
\nabla_{\cA,  \mu}  =    \frac12  ( \pa_{\cA,  \mu}-  \pa^T_{\cA,  \mu}) 
  \hs        \Vol_{\mu}(X)   =\sum_{x \in X:  x + \eta e_{\mu} \in X}    \eta^3  
\ee
and   
\begin{equation} 
\begin{split}
\al_0(E, X)   =& \frac{1}{ \Vol (X)}  E(X,0 )  \hs   \al_{2}(E,X)\de_{ij} =      \frac{1}{2 \ \Vol(X)}   
\frac{\de^2 E}{ \de \phi^2 }  \B( X,0; e_i,e_j  \B)\\
   \al_{2, \mu}(E,X) \de_{ij}    =&\frac{1}{\Vol_{\mu}(X)} \left(\frac{\de^2 E}{ \de \phi^2 }   \B( X, 0, ; e_i,(\cdot - x^0)_{ \mu}e_j \B) -   \frac{1 }{ \Vol(X)}  \frac{\de^2 E}{ \de \phi^2 }   \B( X, 0; e_i,e_j \B)\int_X  (x_{\mu} - x^0_{ \mu})dx\right)\\
 \end{split}
\end{equation}

The  expression for  $ \al_{2, \mu}(E,X) $     is independent of  the base point  $x^0$,  which we take to be  in $X$. 
To  see  that   $\de^2 E/\de \phi^2  ( X,0; e_i,e_j )$ is proportional to  $\de_{ij}$ first note  that  charge conjugation
says    $\de^2 E/\de \phi^2  ( X,0; e_1,e_2 )= \de^2 E/\de \phi^2  ( X,0; Ce_1,Ce_2 )$.  But  $Ce_1=e_1$  and  $Ce_2  =  -e_2$
so this is zero.    The identity  $\de^2 E/\de \phi^2  ( X,0; e_1,e_1 )=\de^2 E/\de \phi^2  ( X,0; e_2,e_2 )$  follows by rotation
invariance.    The same argument works for  $\de^2 E/\de \phi^2  ( X,0; e_i,(x - x^0)_{ \mu}e_j )$. 
\bigskip

The term    $ \int_X  \phi \cdot  \nabla_{\cA,  \mu} \phi   $  requires  some additional comment. 
The derivative    $\nabla_{\cA,  \mu}  $    is the average of a forward and a backward derivative,  and we  use it  because transforms like a   vector field  under lattice symmetries  -  see appendix  \ref{special}.   This would not be the case with just the forward derivative  $\pa_{\cA,  \mu}$.      In  an equation like  (\ref{knuckles}) we are  allowed to use  a forward derivative     since  we are estimating something we
already know  to be invariant.    (The substitution  
$\pa_{ \mu}   \to     \nabla_{  \mu}     $  should also be made in   equation (157) in  \cite{Dim11}.)

     In the expression   $ \int_X  \phi \cdot  \nabla_{\cA,  \mu} \phi   $  we only include bonds 
in  $X$.   To accomplish this     write it  as    $ \int_X    \phi_X \cdot  \nabla_{\cA,  \mu}   \phi_X   $ where
\be   
  \phi_X  (x)  =   \begin{cases}   \phi(x)    &  x \in X  \\
\phi(x  \pm  \eta e_{\mu})    &    x \notin  X,    x  \pm \eta e_{\mu} \in X   
\end{cases}
\ee
Then   if  $r$ is    a lattice symmetry   $   (\phi_r)_{rX }  = (   \phi_X )_r$  and so     $ \int_X    \phi_X \cdot  \nabla_{\cA,  \mu}   \phi_X   $   is covariant.  
 We  need  this property to guarantee that   $\cR E $ is covariant under  lattice symmetries.

\begin{lem}    $\cR  E$  is  invariant under lattice, gauge, and charge symmetries.   $\cR  E$  is normalized for small polymers  and satisfies
for    $e_k, \la_k$ sufficiently small
\begin{equation}     \label{sunup}    
    \|    \cR  E \|_{k, \ka}    \leq  \cO(1)    \|   E   \|_{k, \ka}   
\end{equation}
\end{lem}
\bigskip

\pr 
The  invariance follows since everything else in  (\ref{renorm})  is invariant.
The derivatives  in question match on the left and right  except  for  the  term   $\cR E$,   hence its derivatives vanish.
The  bound  holds  since everything else in  (\ref{renorm}) satisfies the bound.
  See  \cite{Dim11}  for more details.

\bigskip

 For   global quantities we   only have to remove  energy and mass terms.

\begin{cor}   
 \begin{equation}  \label{renorm2}
\sum_X E(X) =     -   \vep(E)  \Vol(  \bbT_{ \sN-k} ) -   \frac12  \mu(E)   \|  \phi \|^2
+    \sum_X  \cR E(X)
\end{equation} 
where
\begin{equation}
\begin{split}
\vep(E)  =  &    -   \sum_{X \supset \square,  X \in \cS}   \al_0(E,X) \\
\frac12  \mu(E)  =& - \sum_{X \supset \square,  X \in \cS}    \al_2(E,X)  \\
\end{split}
\end{equation}
Furthermore
\be  \label{sundown}
|\vep(E)|  \leq \one \| E \|_{k, \ka}    \hs     \mu(E)  \leq   \one  \la_k^{\frac12 + 2 \ep}   \| E \|_{k, \ka}
\ee
\end{cor}
\bigskip

\pr    Sum   (\ref{renorm}) over $X$  and rearrange.      
The   $ \phi \cdot  \nabla_{\cA,  \mu} \phi   $  term vanishes   since
\begin{equation}
  \sum_{X \supset \square,  X \in \cS}     \al_{2, \mu}(E,X)    =0 
\end{equation}
This  follows  since    if  $r$ is a reflection in the $\mu$ direction      $ \al_{2, \mu}(E,rX)  =  -  \al_{2, \mu}(E,X)  $.
Take  a reflection through  the center of  $\square$.

The bound on  $\vep(E)$ follows directly,  and the bound on  $\mu(E)$ uses a Cauchy bound.  See   \cite{Dim11}  
for details.

\subsection{localized  Green's functions}   \label{localgreen}

We  can also localize   the  scalar  Green's functions  with polymers  using the random walk expansion (\ref{g1}).    For a walk   $\om = (\om_0, \om_1, \dots,  \om_n)$  define  $ X'_{\om}  = \cup_{i=0}^n  \tilde \square_{\om_i}$.
Then   write  
\be  G_k (\cA )  =  \sum_{X  \in \cD_k}   G_k (X, \cA  )   \ee
where   
\be
  G_k (X, \cA  )  = \sum_{\om:  X'_{\om}  =X  }   G_{k, \om} (\cA) 
   =\sum_{n=0}^{\infty} \ \     \sum_{\om: |\om| = n,   X'_{\om}  =X  }   G_{k, \om} (\cA)  
\ee
Then   $G_k (X, \cA  )$ only depends on $\cA$ in $X$, and   the kernel  $G_k (X, \cA, x,y)$   vanishes unless $x,y \in X$.

Recall that   if  $|\om| = n$
\be
   | G_{k, \om} (\cA)f|   \leq    C(CM^{-1})^n    \| f\|_{\infty}
\ee
But     $  d_M(X)  \leq    |X |_M =  |X'_{\om} |_M   \leq  27(n+1)$   so we  can  make the estimate    
\be       (CM^{-1})^{n/2}   \leq  \one   (CM^{-1})^{d_M(X) / 54}  \leq    \one  e^{-\ka  d_M(X)}\ee
    for  $M$  sufficiently large.    The remaining factor        
$  (CM^{-1})^{n/2}$   still gives the overall convergence of the series.
Thus we have the bound     
\be   
| G_k (X, \cA  ) f|   \leq    C  e^{-\ka  d_M(X) } \| f \|_{\infty  } 
\ee
as  well as bounds on the derivatives and  $L^2$ bounds.

\section{The  main   theorem}

\subsection{the theorem}

 The   starting   density  on $\bbT^0_{\sN}$ from  (\ref{snow1}),(\ref{snow2})    is 
  \be
\rho_0(A_0, \Phi_0)  = \exp\Big( - \frac12 \| dA_0 \|^2  -  \frac12 \| \pa_{A_0} \Phi_0 \|^2         - V_0(\Phi_0) \Big )
\ee
For the full analysis of the model we     define a sequence of densities  $\rho_k(A_k, \Phi_k)     $  for fields  on $\bbT^0_{N-k}$
by successive RG transformations.
First     for fields on  $\bbT^1_{N-k}$ we define as in (\ref{kth})  and   (\ref{funny1})  
\be     \label{basic1}
\begin{split}
&\tilde   \rho_{k+1} (A_{k+1},  \Phi_{k+1} ) 
=      \\   &  \int \ 
  \de\Big( A_{k+1} -  Q A_k \Big)  \   \de( \tau A_{k} )   \de_G\Big( \Phi_{k+1} -  Q(\tilde   \cA_{k +1}) \Phi_k \Big)
\rho_k(A_k, \Phi_k)      D  A_k   D \Phi_k  \\
\end{split}
\ee
We have chosen a background field    $\tilde   \cA_{k +1}$  which is a smeared out version of $A_{k+1}$ and defined precisely later on.   
Then    we scale to  fields  on   $\bbT^0_{N-k-1}$  by  
 \begin{equation}   \label{scaleddensity2}
 \rho_{k+1} ( A_{k+1},   \Phi_{k+1})  =  \tilde  \rho_{k+1} (A_{k+1,L}, \Phi_{k+1,L}) L^{   \frac12 (b_N - b_{N-k-1}) }  
  L^{   \frac12 (s_N - s_{N-k-1}) }  
 \end{equation}

In this paper  we consider a bounded field approximation in which   (\ref{basic1}) is replaced by  
\be     \label{basic2}
\begin{split}
&\tilde   \rho_{k+1} (A_{k+1},  \Phi_{k+1} ) 
=      \\   &  \int \ 
  \chi_k    \chi_k^w  \ 
  \de\Big( A_{k+1} -  Q A_k \Big)  \   \de( \tau A_{k+1} )   \de_G\Big( \Phi_{k+1} -  Q(\tilde  \cA_{k+1}) \Phi_k \Big)
\rho_k(A_k, \Phi_k)      D  A_k   D \Phi_k  \\
\end{split}
\ee
and   scaling is the same.  
New are  the characteristic functions    $ \chi_k    \chi_k^w $ enforcing bounds on the fields..
Here  $\chi_k =   \chi ( (A_k, \Phi_k)  \in \cS_k)$  is the characteristic function of  the small field  region  $\cS_k$ as defined
in  section  \ref{bdd}.  The other  characteristic function   $  \chi_k^w $
restricts      the fluctuation field and is defined   by    
\be  \label{sally}
  \chi^w_k =        \chi^w_k\Big( C_k(\tilde   \cA_{k+1}) ^{-\frac12}(  \Phi_k -  H_k(\tilde  \cA_{k+1})   \Phi_{k+1}) \B) \ 
   \chi^w_k\B(  C_k^{-\frac12 } C^{-1} ( A_k -   H^{\sx}_k A_{k+1} )   \B)
\ee
where  $\chi^w_k(W)$ is the characteristic function of    $|W| \leq  p_{0,k}$   and  $p_{0,k}  =   ( - \log  \la_k )^{p_0}$
for  some  $p_0 < p$.   
    These  restrictions  are natural   
in Balaban's  formulation  of the renormalization group.  Our goal   is to  study the flow of these modified transformations. As   noted   earlier this  is the location of   the  renormalization problem.

We  are going to claim  that after    $k$  steps we have a density  $\rho_k$  defined on the domain  $\cS_k$  essentially   of the form
\be  \label{snuffit}
\begin{split}
&\rho_k(A_k, \Phi_k)  \\& =  \cN_k  \sZ_k  \sZ_k(\cA_k) 
\exp  \Big(   -  \frac12   \|  d \cA_k \|^2  - S_{k, \cA_k}( \Phi_k,  \phi_k( \cA_k))      -  V_k(   \phi_k(\cA_k) ) +    E_k(  \cA_k,  \phi_k(\cA_k))  \Big) \\
\end{split}  
\ee
where   
\be 
\cA_k = \cA_k(A_k)  = \cH_k A_k   \hs   \phi_k(\cA)  = \phi_k(\cA, \Phi_k)  =  \cH_k(\cA) \Phi_k   
\ee
and where
\begin{equation}
\begin{split}
 S_{k, \cA} (\Phi_k, \phi )  
   =  &    \frac{ a_k}{2} \| \Phi_k  -  Q_k(\cA) \phi \|^2   + \frac12   \|  \pa_{\cA}  \phi \|^2     \\
 V_k(   \phi  )    =      &
 \vep_k    \Vol ( \bbT_{N-k})+ \frac12    \mu_k  \| \phi  \|^2   +  \frac 14  \la_k   \int |\phi(x)|^4 dx  \\
 E_k(\cA, \phi)  =&   \sum_{X}   E_k(X, \cA,  \phi)  \\
\end{split}
\end{equation}
Note that this is true  for  $k=0$ with   $  Z_0=  Z_0(\cA) = 1$,  $E_0 = 0$,   and the convention that  $\cA_0 =  A_0$  and  
 and  $\phi_0 (\cA_0 )  =  \Phi_0$ and  $\cQ_0(\cA_0) = I$  so that   $\Phi_0  -  \cQ_0(\cA_0) \phi_0(\cA_0)  = 0$.

We  assume that    $L$  is   sufficiently large, $M$  is   sufficiently large  (depending on $L$),   and   that  $e, \la$  are  sufficiently small   (depending on $L,M$).   For definiteness  we    take  $e  \leq  \la^{1/2}$  and   then   
$e_k  \leq  \la_k^{1/2}$   for all  $k$.

\begin{thm}   \label{lanky}  Under these assumptions  
  suppose    $\rho_k(A_k, \Phi_k)$  has the representation   (\ref{snuffit})     for  $A_k,  \Phi_k \in \cS_k$.
  Suppose the  polymer function $E_k(X, \cA, \phi)$  is defined on    $\cR_k$,   has  all the symmetries  of section \ref{symmetries},  and is normalized for small polymers.
Finally suppose    
   \begin{equation}  
|\mu_k|     \leq   \la_k^{1/2}       \hs   \|E_k\|_{k, \ka}   \leq   1
 \end{equation}
  Then up to a phase shift   $\rho_{k+1}(A_{k+1},\Phi_{k+1}) $  has  a representation of the same form   for  $A_{k+1},  \Phi_{k+1}  \in \cS_{k+1}$,  
  now with   $e_{k+1}=L^{1/2}e_k$  and $\la_{k+1}= L \la_k$.    The  bounds are not the same but we do have  
    \begin{equation}  \label{recursive}
\begin{split}
\vep_{k+1}   =&  L^3 \vep_k  + \cL_1E_k   +  \vep_k^*( \mu_k,  E_k) \\
\mu_{k+1}   =&   L^2 \mu_k  +  \cL_2E_k  + \mu_k^*( \mu_k,  E_k)  \\
E_{k+1}   =&    \cL_3 E_k  +  E^*_k(    \mu_k,  E_k)  \\
 \end{split}
\end{equation}
The    $\cL_i$  are linear   operators    which   satisfy  
\begin{equation}
\begin{split}
| \cL_1 E_k |   \leq &\  \cO(1) L^{-\ep} \|E_k \|_{k, \ka}\\
| \cL_2 E_k |   \leq &\  \cO(1) L^{-\ep}   \la_k^{1/2  + 2 \ep} \|E_k \|_{k, \ka}\\
\| \cL_3 E_k \|_{k+1, \ka}   \leq &\  \cO(1) L^{-\ep} \|E_k \|_{k, \ka}\\
\end{split}
\end{equation}
and we have the bounds  
\begin{equation}
|\vep_k^*| \leq    \cO(1) \la_k^{\frac{1}{12} - 11 \ep}  \hs  |\mu_k^*|  \leq   \cO(1)\la_k^{\frac{7}{12} - 11 \ep}   \hs    \|E^*_k \|_{k+1, \ka}     \leq  \cO(1) \la_k^{\frac{1}{12} - 11 \ep}
\end{equation}
\end{thm}
\bigskip

\rems
\begin{enumerate}

\item  The phrase  "up to a phase shift" means   we actually show  that      $\rho_{k+1}(A_{k+1},e^{q \theta} \Phi_{k+1}) $
has the form  (\ref{snuffit})  for some real function $\theta =  \theta(A_{k+1})$.  Changing it back  to   $\rho_{k+1}(A_{k+1}, \Phi_{k+1}) $       
changes the definition of the   RG transformation,  but does not change  the basic property that  the integral over $\Phi_{k+1}$  is
the same for each $k$.

\item  By   lemma  \ref{inclusion} we have that   $A_k, \Phi_k \in \cS_k$  implies  $\cA_k, \phi_k(\cA_k)  \in  \cR_k$
so that   $E_k(X,  \cA_k, \phi_k(\cA_k) )$  is well-defined.  

\item  The polymer functions  $E_k$  contain  all parts of the interaction not in  $ S_{k, \cA_k} $  or   $V_k$
These  are  growing at a controlled  rate  because  we  have  extracted  corrections  $ \vep_k^*( \mu_k,  E_k) $ to the 
energy density  and  $ \mu_k^*( \mu_k,  E_k)$  to the mass squared.   

The terms  $\cL_i(E_k)$  are  the result of normalizing terms which newly qualify as  small polymers.  (They are \textit{not}  the full
linearization of the mapping.)   The starred terms are the result of the fluctuation integral and include contributions from  both  $E_k$
and  $V_k$.

\item   We  have the weak  bound  $ \|E_k\|_{k, \ka}   \leq   1$   or  $|E_k(X, \cA,  \phi) | \leq  e^{-\ka d_M(X) }$  because we are allowing the fields    to be somewhat   large.
But   $E_k$   is actually small.   For example   if   $\cA, \phi $ and  derivatives   are $\cO(1)$,  then for  $|t| \leq    \la_k^{-\frac14}  \leq   e_k^{-\frac12}$  we have
$t \cA,  t \phi  \in  \cR_k$.   Hence     since  $E_k$  is normalized
\be
E_k(X, \cA, \phi)  =  E_k(X, \cA, \phi)  -E_k(X, 0) =  \frac{1}{2 \pi i}  \int_{|t|  =  \la_k^{-\frac14}   }   \frac{dt}{t(t-1)}    E_k(X, t\cA, t \phi)  \ee
This gives the bound    $|E_k(X, \cA, \phi)|    \leq  \one \la_k^{\frac14 } e^{- \ka  d_M(X)}$.
 \end{enumerate}

\subsection{proof of the theorem}  The proof  follows the broad outlines of   Balaban,  Imbrie, and Jaffe  \cite{BIJ88}, but differs in
many  details.

\subsubsection{preliminaries}

We  define   operators  $\cH^0_{k+1}, \cH^0_{k+1}(\cA)$ on $\tz$  and  fields  $\cA^0_{k+1},  \phi^0_{k+1}(\cA)$  on $\tk$
which are scalings of  $\cA_{k+1},  \phi_{k+1} (\cA) $.   For  $A_{k+1}, \Phi_{k+1}$ on   $\bbT^1_{N-k}$  we define
\be 
\begin{split}
  &   \cA^0_{k+1}(A_{k+1}) = \cH^0_{k+1} A_{k+1} =(\cH_{k+1} A_{k+1,L^{-1}})_L  
=  ( \cA_{k+1} ( A_{k+1,L^{-1}}  ) )_L   \\
   &   \phi^0_{k+1}(\cA, \Phi_{k+1}) = \cH^0_{k+1}(\cA)     \Phi_{k+1} =(\cH_{k+1}(\cA_{L^{-1}})   \Phi_{k+1,L^{-1}}   )_L  
= (  \phi_{k+1} (\cA_{L^{-1}}, \Phi_{k+1,L^{-1}}  ) )_L   \\
  \end{split}
\ee
These scale  to  $\cA_{k+1},  \phi_{k+1}(\cA) $,  for example if  $A_{k+1}$ on   $\bbT^{-k-1}_{N-k-1}$  then  
$ \cA^0_{k+1}(A_{k+1,L}) =  ( \cA_{k+1}(A_{k+1}))_L  $
We  can also write      
\be
  \phi^0_{k+1}(\cA, \Phi_{k+1})   =   \cH_k (\cA)  H_k(\cA )   \Phi_{k+1} 
\ee  
by the identity    (\ref{h}).    But the analogous formula for the gauge field would only hold if we were using the axial gange
at this point.

We  study   $\tilde   \rho_{k+1}  (A_{k+1},   \Phi_{k+1}) $  for   fields  $ A_{k+1},   \Phi_{k+1} $  in  $ \cS_{k+1}^0$,   the   scaled version of $\cS_{k+1}$.   The space  $ \cS_{k+1}^0$  is defined as  all    $A_{k+1}, \Phi_{k+1}$ on   $\bbT^1_{N-k}$
 satisfying       
\be  \label{small3}
   | d \cA^0_{k+1}   |  \leq    L^{-\frac32}  p_{k+1}   
\ee
and   
\be  \label{small2}
\begin{split}
 | \Phi_{k+1}  -Q_{k+1}( \cA^0_{k+1}  )   \phi^0_{k+1}(\cA^0_{k+1})     |  \leq    & L^{-\frac12} p_{k+1}  \\ 
   |\pa   \phi^0_{k+1}(\cA^0_{k+1})   |  \leq  &   L^{-\frac32}   p_{k+1}  \\
    |  \phi^0_{k+1}(\cA^0_{k+1})    |  \leq  &  L^{-\frac12}  p_{k+1}  \la_{k+1}^{-\frac14}  \\
\end{split}
\ee
 Then $A_{k+1,L^{-1}}, \Phi_{k+1, L^{-1}}$ on   $\bbT^0_{N-k-1}$
satisfy the conditions  for  $\cS_{k+1}$     and we conclude      by  lemma   \ref{inclusion} that 
$\cA_{k+1} (A_{k+1,L^{-1}}), \phi_{k+1}  (  \cA_{k+1} (A_{k+1,L^{-1}}) , \Phi_{k+1, L^{-1}})  $     
   satisfy the bounds for  $\frac12  \cR_{k+1} $.   This is the same  as saying   
     $ \B(\cA^0_{k+1}(A_{k+1}) \B)_{L^{-1}},  \B(\phi^0_{k+1}(\cA^0_{k+1}(A_{k+1}), \Phi_{k+1})\B)_{L^{-1}}  $    satisfy the bounds for  $\frac12  \cR_{k+1} $.      Then     
lemma  \ref{sunfish}  says that   $ \cA^0_{k+1}(A_{k+1}), \phi^0_{k+1}(\cA^0_{k+1}(A_{k+1}), \Phi_{k+1})$  satisfies the
bounds  (\ref{ding2}), (\ref{dong2})  and in particular      
 \be   \label{linger}
( \cA^0_{k+1},   \phi^0_{k+1}(\cA^0_{k+1})  )    \in    L^{-\frac34- \ep}  \cR_{k}
 \ee

\subsubsection{gauge field  translation}

Now   for  $A_{k+1},   \Phi_{k+1}    \in   \cS^0_{k+1}$ 
\be   
\begin{split}
& \tilde    \rho_{k+1} (A_{k+1},  \Phi_{k+1} ) 
=   
\int     D \Phi_k \   D  A_k\         \chi_k\      \chi_k^w \  
\de \Big( A_{k+1} -  Q A_k \Big)  \  \de (  \tau A_k )    \de_G\Big( \Phi_{k+1} -  Q(  \tilde  \cA_{k+1}) \Phi_k \Big)
 \\
&\cN_k   \sZ_k \sZ_{k+1}(\cA_k)   \exp  \Big(   -  \frac12   \|  d \cA_k \|^2 -  S_{k, \cA_k}(  \Phi_k,    \phi_k(\cA_k) )  -  V_k(    \phi_k(\cA_k) ) +    E_k(  \cA_k,   \phi_k(\cA_k))  \Big)
\\
\end{split}
\ee

We   translate to the minimum   of     $  S_k (  \cA_k)$  on the surface   $\cQ A_k =A_{k+1}, \tau A_k =0$    as  before.  
Write    $A_k  =  H^{\sx}_k A_{k+1}  + Z$   and integrate over  $Z$  instead of  $A_k$.   
Then  $\cA_k = \cH_k A_k$ becomes     $\tilde \cA_{k+1}    + \cZ_k  $
where
\be 
    \tilde \cA_{k+1} =    \cH_k  H_k^{\sx} A_{k+1}  \hs    \cZ_k  =  \cH_k Z   
\ee
This is the  $ \tilde \cA_{k+1} $  that  appears in   (\ref{basic1}) and in (\ref{sally}).   
  Next    we use  $ \cH^{\sx}_{k}  =     \cH_{k}    +   \pa    D_k$    and  
the scaled version           $\cH^{\sx,0}_{k+1}  =   \cH^0_{k}    +   \pa    D^0_k $  to  change from      the axial gauge   to the Landau gauge.    Using also   (\ref{sammy})   we  obtain   
\be 
\begin{split}
 \tilde \cA_{k+1}  \equiv   &   \cH_k  H_k^{\sx} A_{k+1}  \\ = &   \cH^{\sx}_k   H_k^{\sx} A_{k+1}  -  \pa    D_k   H_k^{\sx} A_{k+1}    \\ 
  = &   \cH^{\sx,0}_{k+1}  A_{k+1}  -  \pa    D_k   H_k^{\sx} A_{k+1}    \\ 
  =  &   \cH^{0}_{k+1}  A_{k+1}  -   \pa \B(    D_k   H_k^{\sx}  -   D^0_{k+1}  \B) A_{k+1}    \\ 
  \equiv   &    \cA^0_{k+1}  - \pa   \om   \\ 
\end{split}
\ee
where the last line defines   $\om= \om(A_{k+1})$.
 As in  section \ref{minimizers2}
 $   \frac12   \| d \cA_k \|^2 $   become $  \frac12   \|  d \cA^0_{k+1} \|^2  +           \frac12 \Big<Z, \De_k  Z  \Big>
$    and since  $\sZ_k (\cA)$ is gauge invariant we have  
\be   \label{snore}
\begin{split}
& \tilde    \rho_{k+1} (A_{k+1},  \Phi_{k+1} ) 
= \cN_k  \sZ_{k}  \exp \Big(  -  \frac12   \|  d \cA^0_{k+1} \|^2 \Big)  \int    D \Phi_k    \   D Z 
\exp\B(  -    \frac12 \Big<Z,  \De_k  Z  \Big>  \B)   \de  (   QZ     )  \  \de (  \tau Z  )   \\
 &     \chi_k\     \chi_k^w \      
\de_G\Big( \Phi_{k+1} -  Q( \cA^0_{k+1} -  \pa  \om ) \Phi_k \Big)
\sZ_{k+1}(  \cA^0_{k+1}  + \cZ_k ) \\
&
 \exp  \Big(   - S_{k,  \cA^0_{k+1}  + \cZ_k - \pa   \om    }    \B( \Phi_k,     \phi_k( \cA^0_{k+1}  + \cZ_k  -  \pa   \om     ) \B)     \Big) \\
&   \exp  \Big(   -  V_k\B(   \phi_k(  \cA^0_{k+1}  + \cZ_k -  \pa   \om   ) \B)
 +    E_k\B(  \cA^0_{k+1}  + \cZ_k -  \pa   \om    ,   \phi_k( \cA^0_{k+1}  + \cZ_k -  \pa   \om   )  \Big)   \B) \\
\end{split}
\ee

As      in   section \ref{minimizers2} we  replace      $Z$   by  $C  \tilde   Z$  and identify        
$ ( \sZ_k^f )^{-1}  \de  (   QZ     )  \  \de (  \tau Z  )   
\exp\B(  -    \frac12 \Big<Z,  \De_k  Z  \Big>  \B) $ as the Gaussian measure   $d \mu_{C_k} (\tilde Z)  $.
We     now understand   $\cZ_k$ as         $\cZ_k  = \cH_k C \tilde Z $.

If      $\om^{(0)}$ the restriction of  $\om$  to the unit lattice  $\tz$ then  by (\ref{sync})
\be 
\begin{split}
      \phi_k( \cA -  \pa   \om   )  =  &    \cH_k (  \cA -  \pa   \om  ) \Phi_{k}    
=     e^{qe_k \om}   \cH_k (  \cA   )   e^{-qe_k \om^{(0)} }  \Phi_{k}    \\
\end{split}
\ee
We  also change variables   by    $\Phi_{k}   \to  e^{qe_k \om^{(0)} }  \Phi_{k}  $.    This  is a rotation so the Jacobian is one.
Then   $\phi_k( \cA -  \pa   \om   ) $  becomes     $e^{qe_k \om}  \phi_k( \cA  )$  and       
\be
S_{k,  \cA-   \pa   \om    }    \B(e^{qe_k \om^{(0)} }  \Phi_{k} ,  e^{qe_k \om}    \phi_k( \cA     ) \B)
=        S_{k,  \cA   }    \B( \Phi_{k+1} ,       \phi_k( \cA     ) \B)   
\ee
The  $\om$ also disappears from the gauge invariant  $V_k,E_k$.

We also note  that      by  (\ref{study})  
\be   \de_G\Big( \Phi_{k+1} -  Q( \cA^0_{k+1} -  \pa \om)   e^{qe_k \om^{(0)} }   \Phi_k \Big)  =   
 \de_G\Big(  \Phi_{k+1} -  e^{qe_k \om^{(1)} }Q( \cA^0_{k+1} )    \Phi_k \Big)   \ee
where          $\om^{(1)}$ is the restriction of $\om$  to  $\bbT^{1}_{N-k}$.
We     replace  $\Phi_{k+1} $ by    $ e^{qe_k \om^{(1)} } \Phi_{k+1}$ so the phase factor here
disappears   as well. 

Similar considerations show that the bounds enforced by the    characteristic function  $\chi_k$  are now    
 \be   \label{lazy}
 \begin{split} 
   &    | d (\cA^0_{k+1} + \cZ_k)  |  \leq    p_k     \hs
| \Phi_k -Q_k(\cA^0_{k+1}+ \cZ_k)  \phi_k(\cA^0_{k+1}+ \cZ_k)   |  \leq  p_k   \\
&    |   \pa_{\cA^0_{k+1}+ \cZ_k} \phi_k (\cA^0_{k+1}+ \cZ_k) | \leq  p_k  \hs    |   \phi_k (\cA^0_{k+1}+ \cZ_k) |  \leq  \la_k^{-\frac14} p_k  \\
\end{split}
 \ee  
From the representation (\ref{half})    we have   
\be    C_k (\tilde  \cA_{k+1} )^{\frac12}  =    C_k (  \cA^0_{k+1}  - \pa \om )^{\frac12}   =    e^{qe_k \om^{(0)} }   
  C_k (  \cA^0_{k+1} )^{\frac12}   e^{-qe_k \om^{(0)} } 
  \ee
  The same holds for    $ C_k (\tilde  \cA_{k+1} )^{-  \frac12}$ and  with the phase shifts on $\Phi_k,  \Phi_{k+1}$ we 
  now have  
\be
  \chi^w_k =        \chi^w_k\Big( C_k(\cA^0_{k+1})^{-\frac12}(   \Phi_k -  H_k(\cA^0_{k+1})    \Phi_{k+1}) \Big) \
    \chi^w_k\B(  C_k^{-\frac12 } \tilde    Z \B)
\ee

With these changes:    
\be   \label{snore2}
\begin{split}
& \tilde    \rho_{k+1} (A_{k+1},  e^{qe_k \om^{(1)} }   \Phi_{k+1} ) 
=  \cN_k  \sZ_k \sZ^f_k    \exp \Big(  - \frac12   \|  d \cA^0_{k+1} \|^2 \Big)  \int    \    d  \mu_{C_k} ( \tilde Z   ) \  D \Phi_k    \ 
  \\
 &     \chi_k\     \chi_k^w \      
\de_G\Big( \Phi_{k+1} -  Q( \cA^0_{k+1} ) \Phi_k \Big)
Z_{k+1}\B(  \cA^0_{k+1}  + \cZ_k   \B)
 \exp  \Big(   -  S_{k,  \cA^0_{k+1}  + \cZ_k   }    \B( \Phi_k,     \phi_k( \cA^0_{k+1}    + \cZ_k   ) \B)     \Big) \\
&   \exp  \Big(   -  V_k\B(   \phi_k(  \cA^0_{k+1}  + \cZ_k) \B)
 +    E_k\B(  \cA^0_{k+1}  + \cZ_k ,   \phi_k( \cA^0_{k+1}  + \cZ_k)  \Big)  \\
\end{split}
\ee

Next we    separate out   leading terms in an expansion in the fluctuation field    $\cZ_k$.  First  for general  $\cZ$ on $\tk$ define
 \be  
  \de \phi_k (\cA, \cZ,  \Phi_k)  \equiv         \phi_k(\cA+  \cZ, \Phi_k )  -      \phi_k(\cA,  \Phi_k)
 \ee  
 Then in  (\ref{snore}) we can make the replacement $  \phi_k( \cA^0_{k+1}  + \cZ_k )  =    \phi_k( \cA^0_{k+1}  ) 
  +  \de \phi_k( \cA^0_{k+1},  \cZ_k )  $.
Next      define        $E^{(2)},E^{(3)},E^{(4)}  $   by  
\be  
\begin{split}
\label{sanibel}
   V_k(  \phi  +  \de  \phi_k(  \cA , \cZ) ) =&   V_k(  \phi ) +  E^{(2)}_k( \cA,   \cZ,  \phi,    \Phi_k )  \\
  E_k(   \cA + \cZ,  \phi +  \de  \phi_k( \cA,  \cZ ) )  
=   &  E_k(   \cA,  \phi   )    +   E^{(3)}_k(  \cA,   \cZ, \phi,     \Phi_k   ) \\
  \sZ_{k+1}(  \cA  + \cZ) = &  \sZ_{k+1}(  \cA )  \exp (  E^{(4)}_k( \cA,  \cZ ) ) \\
\end{split}
\ee

 We  want to do the same thing  with  $ S_{k,  \cA^0_{k+1}  + \cZ_k   }    \B( \Phi_k,     \phi_k( \cA^0_{k+1}  ) 
  +  \de \phi_k( \cA^0_{k+1},  \cZ_k ) \B )   $.
 But  first   we    express $\Phi_k$ in terms of   $\phi_k( \cA^0_{k+1}  )$.    One   has the identity
  \be
    \Phi_k  =  T_k(\cA) \phi_k(\cA)
  \ee
 where  
\be   \label{tk}  
 T_k(\cA)  
  =  a_k^{-1} Q_{k}(\cA) \Big (  - \De_{\cA}   +    a_{k} Q_{k}^T(\cA)Q_{k}(\cA) \Big)  =    a_k^{-1} Q_{k}(\cA) (  - \De_{\cA})    + Q_{k}(\cA) 
\ee
 Use this  in place of  $\Phi_k$  and then  
 \be    S_{k,  \cA  + \cZ   }    ( \Phi_k,     \phi_k( \cA + \cZ    ) )  
  =   S'_{k,  \cA  + \cZ   }    (   \phi_k( \cA  + \cZ    ) )  
 \ee
 where   
 \be    S'_{k, \cA} ( \phi )  =
        \frac{ a_k}{2} \| (T_k(\cA)  -  Q_k(\cA)) \phi \|^2   + \frac12   \|  \pa_{\cA}  \phi \|^2     
\ee         
 Now    define  $E^{(1)}$  by   
 \be  \label{sanibel2}
 S'_{k,  \cA  + \cZ   }    (  \phi  +   \de   \phi_k( \cA, \cZ    ) ) 
 = S'_{k,  \cA  }  (      \phi    )   +     E^{(1)}_{k}(  \cA,   \cZ, \phi,    \Phi_k )     \ee

Now    we  have with  $\hat   E (  \cA,   \cZ, \phi,    \Phi_k )  =  \sum_{i=1}^4  E^{(i)}(  \cA,   \cZ, \phi,    \Phi_k )  $
\be     \label{outlier1}
\begin{split}
& \tilde    \rho_{k+1} (A_{k+1}, e^{qe_k \om^{(1)} } \Phi_{k+1} ) 
= \cN_k   Z_k Z^f_k     Z_{k}(  \cA^0_{k+1} ) \exp \Big(  - \frac12   \| d \cA^0_{k+1} \|^2\Big)
  \int   \    d  \mu_{C_k} ( \tilde Z   ) \  D \Phi_k    \   \\ 
  &     \chi_k\     \chi_k^w \     
 \de_G\Big( \Phi_{k+1} -  Q( \cA^0_{k+1}  )\Phi_k \B)   \exp  \Big(   -  S'_{k,  \cA^0_{k+1}  }   (     \phi_k( \cA^0_{k+1}    ) )     -  V_k(  \phi_k(  \cA^0_{k+1}))     \Big)  \\
&  
\exp \B(  E_k\B(  \cA^0_{k+1} ,   \phi_k( \cA^0_{k+1})  \Big)    +   \hat E_k\B(    \cA^0_{k+1},   \cZ_k, \phi_k(\cA^0_{k+1}),   \Phi_k \B)    \Big)   \\
\end{split}
\ee

\subsubsection{first localization}

We  want to localize  the terms contributing  to   $  \hat  E_k( \cA,   \cZ,  \phi,    \Phi_k )$.     
These  will be treated  in the region 
\be   \label{region}  
\cA,  \phi     \in  \frac12 \cR_k  \hs      |  \cZ|,  |\pa \cZ|,  \|\de_{\al} \pa \cZ \|_{\infty}     \leq    \la_k^{-\ep}      \hs   |\Phi_k|  \leq   \la_k^{-\frac14- 2\ep}   
\ee
 Since   $e_k \leq  \la_k^{\frac12}$ the bounds on $\cZ$ imply  
\be   
 |\cZ|,  |\pa \cZ|,  \|\de_{\al} \pa \cZ \|_{\infty}          \leq   e_k^{- 2 \ep}
\ee   
 Note  that     the characteristic  function  $\chi_k^w(C_k^{-\frac12}\tilde  Z)$  enforces that
 $   |C^{-\frac12}_k \tilde   Z  |  \leq   p_{0,k}  $.  Since  $ |C^{\frac 12}_k f  | \leq   C \| f \|_{\infty} $
by  (\ref{slinky})       it follows that    $|\tilde  Z |  \leq   C p_{0,k} $.
  Then     by the bounds  (\ref{slavic2})   on  $\cH_k$  the fluctuation field  $\cZ_k= \cH_k C \tilde Z$ satisfies 
  $  |\cZ_k|       \leq     C  p_{0,k}    \leq  \la_k^{-\ep}  $  and similarly for the derivative.    Thus   $\cZ_k$  qualifies for
  the domain  (\ref{region}).   We already know    $\cA^0_{k+1} ,   \phi_k( \cA^0_{k+1})  $ qualify.

 In lemma  \ref{snoopy} below we   show that  on the domain  (\ref{region})  
$(\cA + \cZ,  \phi+  \de \phi_k )   \in \cR_k$.   Therefore    the   $E^{(i)}_k( \cA,   \cZ,  \phi,    \Phi_k )$ as given by 
 (\ref{sanibel}),  (\ref{sanibel2})   are well-defined 
on this domain.     

  First some preliminary results:

\begin{lem}  \label{stungun}
In the region (\ref{region}) 
\be    | \de \phi_k| , \   | \pa_{\cA} \de \phi_k| ,\   |\de_{\al, \cA}  \pa_{\cA}\de \phi_k|  \ \leq \ \la_k^{\frac14 - 5 \ep}
\ee
\end{lem}
\bigskip

\pr  
  If  $\cA \in \cR_k$   then   by   the bounds  (\ref{slavic})   on $\cH_k(\cA)$  
 \be  \label{street}  |\phi_k(\cA, \Phi_k)| \leq   C   \|\Phi_k\|_{\infty}     \leq  C \la_k^{-\frac14- 2\ep} 
\ee
We  write   for $r>1$ 
\be   
 \de \phi_k (\cA, \cZ,  \Phi_k) =        \phi_k(\cA+  \cZ, \Phi_k )  -      \phi_k(\cA,  \Phi_k)  =
\frac{1} { 2 \pi i}  \int_{|t| =r}    \frac{dt}{t(t-1)} \phi_k(\cA+ t \cZ, \Phi_k )  
\ee
If   we  take  $|t| =  e_k^{-1 + 5 \ep}   $  then   $|t \cZ  |   \leq   C e_k^{-1 + 5 \ep} e_k^{- 2 \ep}  \leq   \frac12   e_k^{-1  + 3   \ep}  $
with the same bound for the derivatives.    Hence   $\cA+ t \cZ  \in \cR_k$    and  we  can use
(\ref{street})  to get the bound
\be   | \de \phi_k (\cA, \cZ,  \Phi_k) |  \leq  e_k^{1 - 5 \ep}   (  C \la_k^{-\frac14- 2\ep} )
\leq    \la_k^{   \frac14  - 5 \ep  }  
\ee
The derivatives have the same bound.
\bigskip

\rem   We  will also need a version in which the coupling is weakened.     In  $  \phi_k(\cA+  \cZ),    \phi_k(\cA)   $  
replace   $G_k(\cA+ \cZ), G_k(\cA)$ by  weakened versions    $G_k(s,\cA+ \cZ), G_k(s,\cA)$   .  This    gives
 weakened fields  $  \phi_k(s, \cA+  \cZ),    \phi_k(s,   \cA)     $   depending on  $s  =  \{  s_{\square}  \}$,  and hence 
 a weakened    
 $ \de \phi_k (s )=  \de \phi_k (s,  \cA, \cZ,  \Phi_k ) $.    All the above analysis holds and  we   still  have  the same bounds on     $ \de \phi_k(s)$    even for   $s_{\square}$  complex and 
satisfying   $|s_{\square}|  \leq  M^{\frac12}$.

\begin{lem}    
  For  $|\im \cA|,  | \im \cZ |     \leq  e_k^{-1}$
\be   \begin{split}      \label{root}
| ( \pa_{\cA + \cZ}  - \pa_{\cA}) f |   \leq &   e_k \|\cZ\|_{\infty}  \|  f \|_{\infty}  \\
\| ( \de_{\al, \cA + \cZ}  - \de_{\al,   \cA}) f \|_{\infty}   \leq &   e_k \|\cZ\|_{\infty}  \|  f \|_{\infty}  \\
\|  \de_{\al, \cA } f \|_{\infty}   \leq &  \| \pa_{\cA}  f \|_{\infty}  \\
\end{split}
\ee
\end{lem}
\bigskip

\pr   The first   follows from   
\be  \label{orca}
   ( \pa_{\cA + \cZ, \mu }f)(x)  - (\pa_{\cA, \mu }) f)(x)  =   e^{qe_k\eta \cA_{\mu}(x)}
  F_{\mu}(\cZ_k)f(x + \eta e_{\mu} )
\ee
and the bound 
\be  |F_{\mu}(\cZ_k)f| =    \Big| \B( \frac{e^{qe_k\eta \cZ_{k\mu}(x)} -1}{\eta}  \B) f \Big|   \leq   e_k \|\cZ \|_{\infty}  \| f \|_{\infty}  
\ee
(This is essentially  (\ref{125}) again.)
The  second follows from       
\be
 ( \de_{\al, \cA + \cZ} f)(x,y)    -     ( \de_{\al, \cA} f)(x,y) )
 =   e^{qe_k \cA(\Ga_{xy})} \B(  \frac{ e^{qe_k \cZ(\Ga_{xy}) }-1 }{d(x,y)^{\al}}  \B) f(y)
\ee
and the bound   for  $d(x,y) \leq 1$    
\be    | (e^{qe_k \cZ(\Ga_{xy}) }-1 ) f |  \leq   e_k  d(x,y)  \|\cZ\|_{\infty}  \| f \|_{\infty}  
  \leq    d(x,y) ^{\al}   e_k \|\cZ\|_{\infty}  \| f \|_{\infty}  
\ee
The last follows from  the representation  
\be     e^{qe_k \cA(\Ga_{xy} )} f(y)  -   f (x)   
=   \int_{\Ga(x,y)}     e^{qe_k \cA(\Ga_{xz} )} (  \pa_{\cA} f   )    (z)   \cdot  dz 
\ee
which  yields the bound  for  $d(x,y) \leq 1$
\be   |  e^{qe_k \cA(\Ga_{xy} )}  f(y) -f (x) |  \leq
d(x,y)  \| \pa_{\cA}   f     \|_{\infty}   \leq   d(x,y)^{\al}  \| \pa_{\cA}  f   \|_{\infty} 
\ee

\bigskip

\begin{lem}  \label{snoopy}
 In the region  (\ref{region})  and for   $|t| \leq   \la_k^{-\frac{5}{12} + 5\ep}$  we have 
   \be
 \B(  \cA + t \cZ,   \phi  +  t \de \phi_k  \B)  \in  \cR_k 
  \ee 
\end{lem}
\bigskip

\pr  Let     $\phi_t =  \phi  + t  \de  \phi_k$.  
By lemma \ref{stungun}   
\be
  |t \de \phi_k| , \   |t \pa_{\cA} \de \phi_k| ,\   |t\de_{\al, \cA}  \pa_{\cA}\de \phi_k|  
  \leq   \la_k^{-\frac{5}{12} +  5 \ep }  \la_k^{\frac14 - 5 \ep }   =      \la_k^{-\frac16}   <  \frac14  \la_k^{-\frac16- \ep}   
\ee
Hence   $ (  \cA,    t \de \phi_k  )  \in   \frac14   \cR_k$
and  it follows that   
$(  \cA,  \phi_t  )  \in   \frac34   \cR_k $.

The lemma claim that     $ (  \cA + t \cZ,   \phi_t    )  \in  \cR_k$.   For the $\cA$ conditions it suffices to show
that  $t \cZ  \in \frac12 \cR_k$.
 Since $|t|  \leq  e_k^{-\frac56   + 10  \ep} $  this follows from   
 \be  \label{splitfire3}
  | t \cZ|,   |t d  \cZ|,   |t  \de_{\al} d  \cZ|     \leq      e_k^{-\frac56+10 \ep}(e_k^{-2 \ep} )   <   e_k^{-\frac56}  
 \ee
 
For the  $\phi$ conditions we already have    $| \phi_t  |  <  \frac34   \la_k^{-\frac14 - \ep} $.
For the derivatives   use   (\ref{root})   and   $ |t \cZ  |    \leq      \la_k^{-\frac{5}{12} + 4 \ep} $ 
to estimate   
\be   \label{lon}
\begin{split}
    | \pa_{\cA + t \cZ}  \phi_t  |   \leq   & 
| \pa_{\cA}  \phi_t   |   +  |(\pa_{\cA + t \cZ} - \pa_{\cA}) \phi_t |     \\
\leq   &   \frac34  \la_k^{-\frac16 - 2 \ep}     + e_k   \la_k^{-\frac{5}{12} + 4 \ep}   \la_k^{-\frac14 - \ep}\\
\leq   &   \frac34  \la_k^{-\frac16 - 2 \ep}     +   \la_k^{-\frac{1}{6} + 3 \ep} <   \la_k^{-\frac16 - 2 \ep}    \\
 \end{split}  
\ee

Finally we estimate the Holder  derivative    
\be   \label{holder}
\begin{split}
  \de_{\al, \cA+t\cZ}  \pa_{\cA+t\cZ} \phi_t   =   &    
    ( \de_{\al, \cA+t\cZ} -    \de_{\al, \cA} )  \pa_{\cA+t\cZ} \phi_t  \\
 &    +\de_{\al, \cA} ( \pa_{\cA+t\cZ}  -\pa_{\cA}) \phi_t   +  \de_{\al, \cA}  \pa_{\cA} \phi_t    \\
\end{split}
\ee
We know the last term is bounded  by  $ \frac34  \la_k^{-\frac16 -  \ep} $.       For the first   term  we  use
the bounds    (\ref{root})  and    (\ref{lon})     to  obtain  
\be   |  ( \de_{\al, \cA+t\cZ} -    \de_{\al, \cA} )  \pa_{\cA+t\cZ} \phi_t  |  \leq   e_k      \la_k^{-\frac{5}{12} + 4 \ep}   \la_k^{-\frac16 - 2\ep}\
<   \frac 18   \la_k^{-\frac16 - \ep}   \ee
For the second term   in   (\ref{holder})  
we use  the bound from    (\ref{root}) 
\be    \label{lumnar}
\begin{split}
\|\de_{\al, \cA} ( \pa_{\cA+t\cZ}  -\pa_{\cA}) \phi_t  \|_{\infty}
 \leq   &   \|\pa_{ \cA} ( \pa_{\cA+t\cZ}  -\pa_{\cA}) \phi_t  \|_{\infty}  \\
\end{split}
\ee
We  write  $  e^{qe_k \eta  \cA_{\nu}(x)   } \phi_t( x +  \eta e_{\nu} ) =    \eta  \pa_{\cA, \nu}  \phi_t(x)  +  \phi_t(x)$  and then   
(\ref{orca}) says   
 \be
 \begin{split}
  ( \pa_{\cA+t\cZ, \nu}  -\pa_{\cA, \nu}) \phi_t 
=   &   F_{\nu} (t\cZ)  \B(  \eta  \pa_{\cA, \nu}  \phi_t  +  \phi_t \B)
\end{split}
\ee
Then  by  (\ref{product}) 
\be  
\begin{split}
\B(\pa_{\cA, \mu}    ( \pa_{\cA+t\cZ, \nu}  -\pa_{\cA, \nu})   \phi_t  \B)(x)
 =  & \B( F_{\nu} (t\cZ) \B) (x+ \eta e_{\mu} ) \B(  \eta   (\pa_{\cA, \mu}  \pa_{\cA, \nu}  \phi_t)(x)  +   (\pa_{\cA, \mu}  \phi_t)(x) \B) \\
&  +  \B(\pa_{\mu}   F_{\nu} (t\cZ)  \B)(x) \B(  \eta ( \pa_{\cA, \nu}  \phi_t)(x)  +  \phi_t(x) \B)\\
\end{split} 
\ee
Note that    $\eta  | \pa_{\cA, \mu} f  |  \leq   \one \|f \|_{\infty} $.   Using this  and   $ (\cA,  \phi_t) \in \cR_k$  and bounds 
like   (\ref{125})  and  (\ref{131}) on  $F_{\nu}(t\cZ)  $  we have   
\be  
\begin{split}
 \|\pa_{ \cA} ( \pa_{\cA+t\cZ}  -\pa_{\cA}) \phi_t  \|_{\infty} 
    \leq   & \one e_k \B( \| t\cZ\| _{\infty}  \|\pa_{ \cA}  \phi_t  \|_{\infty}  
    +    \|  t  \pa   \cZ\| _{\infty}  \|  \phi_t \|_{\infty}        \B)   \\
\leq   &  \one   \la_k^{\frac12}  \B(  \la_k^{-\frac{5}{12} + 4 \ep}  \la_k^{-\frac16 - 2 \ep}  + \la_k^{-\frac{5}{12} + 4 \ep}   \la_k^{-\frac14 -  \ep}     \B)\\
   <  &   \frac18   \la_k^{-\frac16 - \ep} \\
\end{split}
\ee
This is the bound on the  second term in  (\ref{holder}).   Combined with the bounds on the other two terms it 
 gives the required    $ | \de_{\al, \cA+t\cZ}  \pa_{\cA+t\cZ} \phi_t | <      \la_k^{-\frac16 - \ep} $.

\begin{lem}   \label{snooze1}
$E^{(1)}_k$ has a local expansion   $E^{(1)}_k =  \sum_X  \hat  E^{(1)}_k(X)$
where    $ \hat   E^{(1)}_k(X,       \cA,   \cZ,  \phi,    \Phi_k ) $ depends on these fields   only in  $X$,
is  analytic in   (\ref{region})     and satisfies  
there   
\be   \label{pinka}
\B|  \hat E^{(1)}_k\B(X,       \cA,   \cZ, \phi,    \Phi_k \B) \B|
\leq  \one    \la_k^{\frac{1}{12}  -10 \ep} e^{- ( \ka - \ka_0-1) d_M(X)}
\ee
\end{lem}
\bigskip

\pr   
First  split up   $  S'_{k,  \cA}$    into  $M$-cubes  $\square$ by  
\be    S'_{k,  \cA}( \phi)   = \sum_{\square}   S'_{k,  \cA}(\square,  \phi) 
  \hs   S'_{k \cA} ( \square,   \phi )   =  \frac{ a_k}{2} \| (T_k(\cA)  -  Q_k(\cA)) \phi \|^2_{\square}  
   + \frac12   \|  \pa_{\cA}  \phi \|^2_{\square,*}    
\ee
In  $ \|  \pa_{\cA}  \phi \|^2_{\square,*}    $     the  star indicates that  terms   $|(\pa_\cA \phi)(x,x')|$  for bonds  $(x,x')$   which  cross $M$-cubes  $\square$  have been divided 
between the two    cubes.     Then     $S'_{k,  \cA}(\square, \phi) $ depends on  $\phi$ at sites which neighbor
$\square$ but  are not in $\square$.   Hence we    regard    $S'_{k,  \cA}(\square,  \phi)$  as localized in the $3M$-cube
 $\tilde  \square$ centered on  $\square$.  We  define    
 \be
      S^{\#}_{k,  \cA}( \tilde  \square,  \phi)   =   S'  _{k,  \cA}(\square,   \phi) 
   \ee
 Then  the field is  strictly localized in  $\tilde \square$  and we have
 \be 
    S_{k,  \cA}(  \phi)   = \sum_{\tilde   \square}   S^{\#}_{k,  \cA}(\tilde  \square,   \phi) 
 \ee
 There is a corresponding split    $E_k^{(1)}  =  \sum_{\tilde \square}   E_k^{(1)} (\tilde  \square ) $.
where
\be    \label{syrup}
\begin{split}
 E^{(1)}_{k}\B( \tilde    \square,   \cA,   \cZ,    \phi,   \Phi_k   \B) =    &  
 S^{\#}_{k,  \cA + \cZ   }    (\tilde  \square,       \phi  +  \de \phi_k( \cA,  \cZ, \Phi_k)  ) - S^{\#}_{k,  \cA }   \B (\tilde  \square,    \phi  \B )      \\
\end{split}
\ee   

In   appendix  \ref{B}  we establish   that  
\be   | (T_k(\cA)  -  Q_k(\cA)) \phi |    =        a_k^{-1} |   Q_k(\cA) \De_{\cA} \phi  |   \leq     C    \| \pa_{\cA} \phi  \|_{\infty} 
\ee
For  $  \cA,  \phi     \in   \cR_k  $      we have    $|\pa_{\cA} \phi |   \leq   \la_k^{-\frac16 -2 \ep}$ and so   
\be   \label{sloppy}
    | S^{\#}_{k,  \cA   }    (\tilde  \square,    \phi  )|    \leq  C M^3      \la_k^{- \frac13  - 4 \ep}   \leq   \la_k^{- \frac13  -  5 \ep}
 \ee
According to  lemma  \ref{snoopy}    in  $S^{\#}_{k,  \cA + t\cZ   }    (\tilde  \square,   \phi  + t \de \phi_k( \cA,  \cZ, \Phi_k)  ) $    we can take     $|t|  \leq    \la_k^{-5/12+5\ep } $ and stay in the analyticity   region $\cR_k$.   Hence for   $r=       \la_k^{-5/12+5\ep } $
we have the representation
\be    \label{syrup2}
\begin{split}
 E^{(1)}_{k}\B( \tilde    \square,   \cA,   \cZ,    \phi,   \Phi_k   \B)  =    &     \frac{1}{2 \pi i}
 \int_{|t|  =  r }     \frac{dt}{t(t-1)}   
 S^{\#}_{k,  \cA + t\cZ   }    (\tilde  \square,   \phi  + t \de \phi_k( \cA,  \cZ, \Phi_k)  )   \\
\end{split}
\ee
and     the bound  (\ref{sloppy})  yields  
\be   
 |     E^{(1)}_{k}\B( \tilde    \square,   \cA,   \cZ,    \phi,   \Phi_k   \B) |  \leq    \one   \la_k^{5/12   -5     \ep }     \la_k^{- \frac13  - 5 \ep}  
\leq    \one       \la_k^{1/12  - 10 \ep  }  
\ee
Since  $d_M(\tilde \square) = \one$  we can insert a factor  $e^{- (\ka- \ka_0-1) d_M(\tilde \square) }$.
Hence the result  with  $\hat E^{(1)}_k(X)  =    E^{(1)}_k( \tilde    \square)$ if  $X = \tilde \square$ and zero otherwise. 
\bigskip

\begin{lem}   \label{snooze2}
$E^{(2)}_k, E^{(3)}_k$ have    local expansions   $E^{(i)}_k =  \sum_X    \hat  E^{(i)}_k(X)$
where    $\hat  E^{(i)}_k(X,       \cA,   \cZ,  \phi,    \Phi_k ) $ depends on these fields   only in  $X$,
is  analytic in   (\ref{region})     and satisfies  
there   
\be   \label{pinkb}
\begin{split}
\B| \hat  E^{(i)}_k\B(X,       \cA,   \cZ, \phi,    \Phi_k \B) \B|
\leq   &  \one    \la_k^{\frac{5}{12} - 10 \ep} e^{- ( \ka - \ka_0-1) d_M(X)} \\
\end{split}
\ee
\end{lem}
\bigskip

\pr  
The potential has the local decomposition    $V_k(\phi) =  \sum_{\square} V_k(\square, \phi )$ over  $M$-cubes  $\square$.
Then  $E^{(2)}(\square) =  V_k(\square,  \phi + \de \phi_k)   - V_k(\square,  \phi )$ can be written
\be   E^{(2)}(\square, \cA, \cZ,   \phi,  \Phi_k    )
=   \frac{1}{2 \pi i}
 \int_{|t|  = \la_k^{-5/12+5\ep } }     \frac{dt}{t(t-1)}      V_k\Big( \square,  \phi  +   t  \de \phi_k(\cA, \cZ,  \Phi_k)    \Big)
\ee
Here the circle    $|t| = \la_k^{-5/12+5\ep }     $ is chosen so  inside the  circle  
    $\phi  +   t  \de \phi_k    \in  \cR_k$ by lemma  \ref{snoopy}.
On  $\cR_k$ we have    the bound  ($\vep_k$ is irrelevant here)
\be      |V_k(\square,  \phi)|   \leq    M^3\B(  \mu_k  (\la_k^{-\frac14 - \ep} )^2  +    \la_k (\la_k^{-\frac14 - \ep} )^4 \B) \leq  M^3\la_k^{-4 \ep}    \leq   \la_k^{-5 \ep}    \ee
 and this  implies 
\be    |   E^{( 2)}_k(\square) |  \leq   \one    \la_k^{5/12-5\ep }   \la_k^{-5 \ep}
  \leq       \la_k^{ 5/12    -  10 \ep} 
\ee
 The term  $E^{(3)}_k$   inherits  an  expansion in $X$ from  $E_k$   and  we have
  \be
  E^{(3)}_k\Big(X, \Phi_k,  \cA, \cZ,   \phi,  \Phi_k  \Big)   =   \frac{1}{2 \pi i}
 \int_{|t|  =\la_k^{-5/12+5\ep } }     \frac{dt}{t(t-1)}      E_k\Big( X, \cA + t \cZ,  \phi  +   t  \de \phi_k(\cA, \cZ,  \Phi_k)    \Big)
\ee
where again   $(   \cA + t \cZ,  \phi  +   t  \de \phi_k  )   \in \cR_k$ by lemma   \ref{snoopy}.  
Then    the bound  $|E_k(X, \cA,  \phi)|   \leq    e^{- \ka   d_M(X)}$   on  $\cR_k$ now    implies 
that   
\be  \label{nb}   |   E^{( 3)}_k(X) |  \leq  \one  \la_k^{5/12-5\ep }   e^{-  \ka d_M(X)}
 \ee

We  are not finished  because  $E^{(2)}(X),   E^{(3)}(X)$  depends on fields  outside of $X$  through  $\de \phi_k$.
  Consider $ E^{(3)}(X)$.    We   replace   
$\de \phi_k$   by    $\de \phi_k(s)$  in the above formula  and    define   $E^{(3)}(s,  X)$ (see remark  after lemma \ref{stungun}). 
 This   still satisfies the bound  (\ref{nb}).
Now  in each variable  $s_{\square}$  we interpolate between  $s_{\square} =1$  and   $s_{\square} =0$  by
\be   f(s_{\square} =1  )     =  f(s_{\square} =0  )     +   \int_0^1      d  s_{\square}   \frac{\pa  f}{  \pa  s_{\square}}   \ee
This yields  
\begin{equation}  \label{summertime}
\begin{split}
E^{(3)}_k(X)   
= & \sum_{Y \supset X}        E_k(X,Y) \\
 E_k(X,Y;    \cA, \cZ,   \phi,  \Phi_k     )   
= &  \int   ds_{Y-X} 
 \frac { \pa  }{ \pa s_{Y-X}}   \left[  E_k(X,  \cA,   \cZ,    \phi,  \de \phi_k (s,  \cA, \cZ,  \Phi_k ))   \right]_{s_{Y^c} = 0, s_X=1}\\
 \end{split}
 \end{equation}
 The  latter only   only depends  on    $ \cA, \cZ, \phi,   \Phi_k $   in $Y$   since there is  no coupling through $Y^c$.  Now  we   write 
\begin{equation}   
E^{ (3)}   =  \sum_X    E^{(3)}_k( X)    =   \sum_X      \sum_{Y \supset X}         E_k( X,Y) 
=  \sum_{Y }    \hat  E^{(3)}_k (Y) 
\end{equation}
where   the sum is over connected  polymers  $Y$  and   
\begin{equation}    
   \hat  E^{(3)}_k (Y) 
 =  \sum_{X  \subset  Y}       E_k( X,Y)   
\end{equation}
is strictly local in the  fields.

To  estimate  the  new    function  $ \hat   E^{(3)}_k  (Y) $ we argue as follows,   see  \cite{Dim11}  for more details.    Since    $ \de \phi_k (s,  \cA, \cZ,  \Phi_k )$ is analytic
in  $|s_{\square}| \leq   M^{\frac12}$  we  can use a Cauchy bound to estimate the derivatives in  (\ref{summertime}).   Each derivative
contributes a factor   $M^{-\frac12}$   and   $M^{-\frac12} \leq   e^{- \ka} $ for  $M$ sufficiently large.     Hence in an estimate on  
$E_k(X,Y)$  we gain a factor  $e^{- \ka|Y-X|_M}$.    Using also  (\ref{nb})  yields    
\be     | \hat  E^{(3)}_k (Y) | 
 \leq      \one     \la_k^{ \frac{5}{ 12}  -5 \ep}   \sum_{X  \subset  Y}   e^{- \ka|Y-X|_M -  \ka d_M(X)}
\ee
But  one can show  that 
\be   |Y-X|_M     +    d_M(X)   \geq  d_M(Y)   \ee
Hence one can extract a factor   $ e^{- (\ka  - \ka_0)  d_M(X)}$   leaving   a factor   $ e^{ - \ka_0   d_M(X)}$
for the convergence of the sum over $X$.   The sum is bounded by  $\one  |Y|_M \leq   \one ( d_M(Y)  + 1)$   and so we have  
\be     |  \hat   E^{(3)}_k (Y) | 
 \leq      \one    \la_k^{  \frac{5}{ 12}  -5 \ep}  e^{- (\ka  - \ka_0-1 ) d_M(Y)}
\ee
which is more than enough.  The construction of  $ \hat   E^{(2)}_k  (Y) $  follows the same steps.

\begin{lem}  \label{snooze4}
 In the region   (\ref{region})  we have  the local expansion  $E^{(4)} _k=  \sum_X  \hat  E^{(4)}_k (X)$
\be   \label{pinkc}
\B|  \hat E^{(4)}_k\B(X,       \cA,   \cZ    \B) \B|
\leq  \one    e_k^{1 -  6\ep} e^{-  \ka d_M(X)}  
\ee
\end{lem}
\bigskip

This is the most difficult estimate,  and 
we  postpone the proof  to section  \ref{normfactor}.  
\bigskip

\noindent
\textbf{Summary: }   Combining the  results of  lemma  \ref{snooze1},  lemma \ref{snooze2},   and  lemma   \ref{snooze4}   we  have   
$
\hat   E_k   =   \sum_X  \hat  E_k(X)  $  where   $ \hat   E_k(X) =  \hat E^{(1)}_k(X) + \dots  + \hat  E^{(4)}_k(X)$
has fields strictly localized in $X$  and  satisfies   
 on the domain   (\ref{region})  
\be   \label{shoot}
 |\hat    E_k(X, \cA, \cZ, \phi,   \Phi_k)|  \leq  \one   \la_k^{\frac{1}{12} - 10 \ep}   e^{ - (\ka - \ka_0 -1)  d_M(X)  }
\ee

 \subsubsection{restoration of dressed fields }  
 
We  have    some direct  dependence on   $ \Phi_k $ on  the unit lattice  $\bbT^0_{N-k}$ .  We  would  like to express  this in terms of   the dressed field  
 $\phi_{k}(\cA^0_{k+1})$  on the fine  lattice  $\bbT^{-k}_{N-k}$.
 We  again use  the identity    $\Phi_{k}  =     T_k(\cA)  \phi_k(\cA) $   where   $T_k(\cA)$ is  defined  in  (\ref{tk}).  
 Our  new definition  is   
 \be    
\hat    E_k(X,  \cA,   \cZ,  \phi  )  \equiv    \hat   E_k( X,   \cA,   \cZ,   \phi,    T_k(\cA))  \phi   )
 \ee
 (same symbol, different variables).   Then   in   (\ref{outlier1}) we  can make the replacement
 \be  \hat   E_k( X,  \cA_{k+1}^0,  \cZ_k,  \phi_k(\cA^0_{k+1}),   \Phi_k)   =\hat    E_k( X,  \cA^0_{k+1},   \cZ_k,   \phi_k(\cA^0_{k+1})  ) 
 \ee
Using the estimate   $|T_k(\cA)-Q_k(\cA))\phi |  \leq   C     
  \| \pa_{\cA}  \phi  \|_ {\infty} $    from  appendix  \ref{B}  and the estimate   $|Q_k(\cA)\phi |   \leq     \|   \phi  \|_ {\infty}  $ 
we  have
\be  |T_k(\cA)\phi |  \leq    C   \Big(    \|   \phi  \|_ {\infty}   +
  \| \pa_{\cA}  \phi  \|_ {\infty}  \Big)
\ee
Hence on the domain (\ref{region})      
\be      
 |    T_k(\cA)  \phi |   \leq   C  \la_k^{- \frac14  -   \ep}   \leq   \la_k^{-\frac14 - 2 \ep}
\ee
Thus we are  still    in the analyticity domain for  $ \hat E_k(X)$,    and   the bound   (\ref{shoot})  
still holds.

We are not finished because   $\hat   E_k( X, \cA,   \cZ,  \phi  )   $ depends  on   $\phi$ in  $\tilde  X$  through  $\cQ_k(\cA) \De_{\cA} \phi$.  ($\tilde X$ =  union of $M$ blocks touching $X$).
We  define     
\be
 \hat    E^{'}_k( Y )  =  \sum_{X: \tilde X  = Y} \hat   E_k(   X )
\ee 
Then   $\hat E_k  =  \sum_Y  \hat   E^{'}_k(Y  ) $,  and  $E^{'}_k(Y  )$ is strictly local,   
and   
\be     | \hat  E^{'}_k( Y )|  \leq    \one   \la_k^{1/12 -10 \ep}  \sum_{X: \tilde X  = Y}    e^{-  (\ka- \ka_0 -1) d_M(X)}   
\ee
But  $d_M( \tilde  X)   \leq   d_{M}(X)  + \one  |X|_M$  and   $|X|_M  \leq   \one  (d_M(X)  + 1)$ 
so there is a constant  $c = \one$  such that  
\be  c d_M( \tilde  X)   \leq     (d_M(X)  + 1)
\ee
We  use this to extract   a  factor    $ \one  e^{-  c(\ka- 2\ka_0 -1) d_M(Y)}  $.    This leaves   $ e^{- \ka_0 d_M(X)} $    
for convergence of the sum  which is  bounded by   $ \one  | Y|_M   \leq  \one  (d_M(Y)  + 1)$.  Hence we end with the bound
on the domain  (\ref{region}) (without  the condition on $\Phi_k$)
\be      \label{hat}
    |\hat    E^{'}_k( Y , \cA,   \cZ,  \phi )|  \leq    \one   \la_k^{1/12   -  10  \ep}     e^{- c (\ka- 2\ka_0 -2) d_M(Y)}   
\ee

\subsubsection{scalar field   translation}
Now  in  (\ref{outlier1}),   with  $\cA  =  \cA^0_{k+1}$, we  translate to the minimum  of   
\be   
    \frac12  < \Phi_k,  \De_k(\cA)   \Phi_k>     = \frac{ a }{2 L^2}  \|  \Phi_{k+1} -  Q( \cA ) \Phi_k  \|^2   +
 S_{k,  \cA }    ( \Phi_k,    \phi_k( \cA   ) 
 \ee
 As   in section   \ref{single}  this is   $\Phi_k   = H_k (\cA )\Phi_{k+1}$   and we  write
\be
\Phi_k   = H_k (\cA )\Phi_{k+1}    +   Z'   \hs    \phi_k(\cA)   =  \phi^{0}_{k+1}(\cA)  +  
\cZ_k( \cA)    \hs    \cZ_k(\cA)  \equiv  \cH_k(\cA)  Z' 
 \ee
At the minimum  we  have  
\be   \
   \frac12  <  H_k (\cA )\Phi_{k+1} ,  \De_k(\cA)   H_k (\cA )\Phi_{k+1} >  
 +        \frac12 \Big<Z',  \Big(\De_{k}( \cA) +   \frac{a}{L^2}  (Q^TQ )( \cA)  \Big)  Z'  \Big>
\ee
We know the first term  here    scales to     $ \frac12  < \Phi_{k+1},  \De_{k+1}(\cA)   \Phi_{k+1}>$
so it must be       
\be
   S^{0}_{k+1,  \cA  }    ( \Phi_{k+1},    \phi^0_{k+1}( \cA )   )   \equiv    
 \frac{a_{k+1}}{2L^2}   \| \Phi_{k+1}  - Q_{k+1}(\cA)  \phi^0_{k+1}  \|^2  
     +  \frac12     \| \pa_{\cA}\phi^0_{k+1}  \|^2 
\ee
Hence    (\ref{outlier1})       becomes 
\be   
\begin{split}
& \tilde    \rho_{k+1} (A_{k+1},   e^{qe_k \om^{(1)} }  \Phi_{k+1} ) \\
= & \cN_k N_k   \sZ_{k}    \sZ^f_{k}   \sZ_{k}(  \cA^0_{k+1} )   
 \exp \Big(  - \frac12   \|  d \cA^0_{k+1} \|^2 - S^{0}_{k+1,  \cA^0_{k+1}    }    ( \Phi_{k+1},    \phi^0_{k+1}( \cA^0_{k+1} )   ) \Big) \\
& \int    d \mu_{C_k}(\tilde   Z)   \     D Z'   \   \chi_k^w    \chi_k\ 
 \exp \Big(    -  \frac12 \Big<Z',  \Big(\De_{k}( \cA^0_{k+1}) +  \frac{a}{L^2}( Q^TQ)( \cA^0_{k+1})  \Big)  Z'  \Big>  \B)
\\
&   \exp  \Big(   
    -  V_k\Big(   \phi^{0}_{k+1}(\cA^0_{k+1})  +  
\cZ_k( \cA^0_{k+1})  \Big) 
 +    E_k\Big(  \cA^0_{k+1},   \phi^{0}_{k+1}(\cA^0_{k+1})  +  
\cZ_k( \cA^0_{k+1})   \Big)    \\
&  +   \hat   E_k\B(  \cA^0_{k+1},  \cZ_k,   \phi^0_{k+1}( \cA^0_{k+1})+  \cZ_k( \cA^0_{k+1})    \B) \B)  \\
\end{split}
\ee

Now   identify  the Gaussian measure   $ d \mu_{C_k(\cA^0_{k+1})} (Z')  $  by     
 \be       d \mu_{C_k(\cA)} (Z')   =   \sZ^f_k (\cA)  ^{-1}   
  \exp \B( -   \frac12 \Big<Z',  \B(\De_{k} (\cA)+  \frac{a}{L^2} ( Q^TQ)(\cA)  \B)  Z'  \Big> \B)  D Z'
\ee
We  also  define
\be
  \begin{split}
 V_k(   \phi +  \cZ _k(\cA))  = &  V_k(   \phi )
+  E^{(5)}_k (    \phi,   \cZ_k(\cA) )  \\
 E_k(  \cA,   \phi +  \cZ _k(\cA)) =&    E_k(  \cA,   \phi    ) 
 +   E^{(6)}_k ( \cA,   \phi,    \cZ_k(\cA))  \\
\end{split}
\ee
The  $E^{(5)}_k,   E^{(6)}_k$ inherit local expansions.  
Now we have    
\be     \label{bong}
\begin{split}
& \tilde    \rho_{k+1} (A_{k+1}, e^{qe_k \om^{(1)} }   \Phi_{k+1} )  \\
&= \cN_k  N_k  \sZ_k \sZ^f_{k+1} \sZ_k(  \cA^0_{k+1})    \sZ^f_k( \cA^0_{k+1}) \
  \exp \B(  - \frac12   \|  d \cA^0_{k+1} \|^2- S^0_{k+1,  \cA^0_{k+1}    }    ( \Phi_{k+1},    \phi^0_{k+1}( \cA^0_{k+1} )   ) \B) \\
&   \exp  \B(   
    -  V_k\B(   \phi^{0}_{k+1}(\cA^0_{k+1})   
\B)  
 +    E_k\B( \cA^0_{k+1},   \phi^{0}_{k+1}(\cA^0_{k+1})  
 \B)   \B)    \Xi_k\B(  \cA^0_{k+1},  \phi^0_{k+1}( \cA^0_{k+1} )   \B) \\
\end{split}
\ee
Here we have isolated a fluctuation integral
\be 
\begin{split}
& \Xi_k\B(  \cA^0_{k+1},  \phi^0_{k+1}( \cA^0_{k+1} )   \B)  \\
&=
  \int   d \mu_{C_k}  (\tilde       Z)   d \mu_{C_k( \cA^0_{k+1})} ( Z' )  \   \chi_k^w    \chi_k 
  \exp \Big(  E^{\dagger}_k( \cA^0_{k+1},   \cZ_k,   \phi^0_{k+1}( \cA^0_{k+1} ) ,       \cZ_k(\cA^0_{k+1})     \Big) \\
  \end{split}
\ee 
where  
\be 
\begin{split}   E^{\dagger}_k(  \cA,   \cZ_k,     \phi,  \cZ_k(\cA)   )         
=  &  \hat  E_k( \cA,  \cZ_k,     \phi  +     \cZ_k( \cA) )     
+        E_k^{(5)}\Big (    \phi,   \cZ_k( \cA) \Big)  +    E^{(6)}_k \Big( \cA,  \phi,      \cZ_k( \cA)  \Big)\\
\end{split}
\ee

  We   make  another  change of variables  writing  $\tilde  Z = C_k^{\frac12}\tilde    W$   and   $Z' = C_k(\cA)  ^{\frac12}W$.
  Then   $\cZ_k,  \cZ_k(\cA)$ become     $\cW_k,  \cW_k(\cA)$    
  where
\be   \label{tiki}
\cW_k     =   \cH_kC C_k^{\frac12} \tilde W     \hs 
\cW_k(\cA)       =    \cH_k (\cA)    C_k^{\frac12}(\cA) W  
\ee
The fluctuation integral is then     
\be 
  \Xi_k\B(  \cA^0_{k+1},  \phi^0_{k+1}( \cA^0_{k+1} )   \B) =   \int   d \mu_I (\tilde  W)   d \mu_I(W ) )  \   \chi_k^w    \chi_k 
  \exp \Big(   E^{\dagger}_k( \cA^0_{k+1},   \cW_k,   \phi^0_{k+1}( \cA^0_{k+1} ) ,       \cW_k(\cA^0_{k+1})    \Big)   
\ee
The characteristic  function   $\chi_k^w$  has   simplified  (as it was designed to do) so that  now  
 \be  \label{lincoln}
   \chi_k^w    = \chi^w_k(  \tilde   W )     \chi^w_k( W) \
\ee
These enforce that    $|\tilde  W|,   |W|    \leq  p_{0,k}$.   The bounds (\ref{lazy}) enforced by 
  characteristic function  $\chi_k$  are now  with  $\cA = \cA^0_{k+1}$  
\be   \label{contorted}
\begin{split}  
|d(\cA  + \cW_k ) |  \leq   &   p_k    \\
\B|  \B( H_k(\cA)\Phi_{k+1}+ C_k^{\frac12}(\cA)W \B)  
-   Q_k( \cA+ \cW_k ) \B(  \phi_{k+ 1}^0(\cA+ \cW_k )+   \cW_k(\cA + \cW_k )\B) \B|   \leq  &   p_k   \\
\B|  \pa_{ \cA+ \cW_k }   \B(  \phi_{k+ 1}^0( \cA + \cW_k )+   \cW_k(\cA + \cW_k )\B)  \B|   \leq  &      p_k   \\
\B|    \phi_{k+ 1}^0( \cA+ \cW_k )+   \cW_k( \cA + \cW_k )  \B|   \leq  & \la_k^{-\frac14}  p_k   \\
\end{split}
\ee

\subsubsection{estimates}

We  first note that  for       $\cA  \in \cR_k$    
\be    \label{lunky}
\begin{split}   
|C_k^{\frac12 }\tilde  W|,\   | \cW_k |,\  |\pa \cW_k|,\  | \de_{\al}  \pa  \cW_k|    \leq   &    C \|\tilde   W \|_{\infty}    \\
|C_k ^{\frac12 }(\cA)W|,\   | \cW_k(\cA) |,\  |\pa_{\cA} \cW_k(\cA)|  ,\  | \de_{\al, \cA}  \pa_{\cA}  \cW_k(\cA)|    \leq   &    C \|W \|_{\infty}    \\
\end{split}
\ee
The bounds on  $C_k ^{\frac12 },C_k ^{\frac12 }(\cA)$
were     already established  in (\ref{slinky0}), (\ref{slinky}).      
The others follows by the bounds  (\ref{slavic}),  (\ref{slavic2}) on  $\cH_k, \cH_k(\cA)$.

\begin{lem}   \label{sweet}  Let     $\cA,  \phi   \in   \frac14\cR_k$  and  
 and  $|\tilde  W_k|,   |W_k|    \leq  p_{0,k}$.    Then    $ E_k^{\dagger}= \sum_X E_k^{\dagger}(X )$  where    
\be   \label{pink}
| E_k^{\dagger}\B(X,   \cA,   \cW_k,     \phi,  \cW_k(\cA)    \B)  |
\leq   \one  \la_k^{1/12  - 10    \ep}  e^{- c( \ka - 2\ka_0  -2)  d_M(X)}
\ee
\end{lem}
\bigskip

\pr   We  have      $E_k^{\dagger}  (X) =   \hat  E'_k( X )     
+        E_k^{(5)}(X)  +    E^{(6)}_k(X)  $.
 The bound  on    $\hat E'_k(X, \cA, \cW_k,  \phi + \cW_k(\cA)  )$   follows from      (\ref{hat}).   
 For this we  need the fact  that our assumptions and  the bounds (\ref{lunky})  imply that
 $( \cA, \cW_k,  \phi + \cW_k(\cA) )$ is in the domain  (\ref{region}).

  The  bounds  on $ E_k^{(5)}(X),     E^{(6)}_k(X)$  are 
 very similar to the bounds  on  $ E_k^{(2)}(X),     E^{(3)}_k(X)$   given in  lemma \ref{snooze2}.  For example
 \be       E^{(6)}_k \Big(X,  \cA,  \phi,    \cW_k( \cA) \B)
 =     \frac{1}{2 \pi i}
 \int_{|t|  =\la_k^{-\frac16  } }     \frac{dt}{t(t-1)}      E_k\Big( X,      \cA,  \phi   + t   \cW_k( \cA)     )    \Big)
\ee
By  (\ref{lunky})     we have   for such  $t$   
\be   |t\cW_k(\cA) |,\  |t\pa_{\cA}  \cW_k(\cA)|  ,\  |t \de_{\al, \cA}  \pa_{\cA}   \cW_k(\cA)|   \leq  C p_{0,k}\la_k^{-\frac16  }   \leq  \frac12  \la_k^{-\frac16- \ep}
\ee
 and  so   $( \cA,  t  \cW_k(\cA) )    \in   \frac12  \cR_k$  and  we are in the analyticity region for  $E_k(X)$.   Together with  $|E_k(X)|  \leq    e^{-\ka d_M(X)}$  this    gives the bound
\be     |E^{(6)}_k \Big(X,  \cA,  \phi,    \cW_k( \cA) )|   \leq   \one   \la_k^{\frac16}e^{- \ka d_M(X) }
  \ee
which is sufficient.  The bound on $E^{(5)} (X)$ is a little weaker, but still sufficient.

\subsubsection{adjustments}

We  make two adjustments.   The first is to reblock from   polymers $X$ which are unions of $M$ blocks to polymers  $Y$ which are unions of $LM$ blocks. 
We have as in section 
\ref{polymersection}   
\be
 E^{\dagger}_k =   \sum_{X  }   E^{\dagger}_k(  X) =     \sum_{Y} \cB  E_k^{\dagger}( Y)   \equiv  \cB E_k^{\dagger} 
 \ee
 Then     for    $\cA,  \phi   \in   \frac12\cR_k$   and   
  $|\tilde W|,   |W|    \leq  p_{0,k}$
\be   \label{pink2}
\B|   \cB  E_k^{\dagger}\B(Y,   \cA,   \cW_k, \phi,  \cW_k(\cA) \B)  \B|
\leq  \one L^3   \la_k^{\frac{1}{12} -  10 \ep}  e^{-  L(c\ka - 3\ka_0 -3) d_{LM}(Y)}
\ee
We  do the  same  to  the leading term  $E_k$,  introducing  $\cB E_k$
\bigskip

The second  adjustment involves 
 the characteristic function  $\chi_k$ which     enforces  the conditions   (\ref{contorted})..
The next  lemma shows that  if we  assume    $(A_{k+1}, \Phi_{k+1})$  are in  $\cS_{k+1}^0$  as defined  in 
(\ref{small2})  and if    $|\tilde  W|, |W|  \leq  p_{0,k}$   as enforced by (\ref{lincoln}),  then
 we  can  drop  this characteristic function   entirely,   a key simplification.

\begin{lem}  If   
 $(A_{k+1}, \Phi_{k+1}) \in \cS^0_{k+1}$    and  $|\tilde  W|, |W|    \leq  p_{0,k}$  then   the 
 bounds  (\ref{contorted})    are satisfied and hence  $\chi_k =1$.
 \end{lem}
\bigskip

\pr 
For the gauge field it suffices to show separately that   $| d \cA^0_{k+1}  | \leq  \frac12 p_k$ and  $|d \cW_k |
\leq  \frac12 p_k$.    The first follows by (\ref{small3}) and  $p_{k+1} \leq  p_k$.   For the second we have by 
(\ref{lunky})  $|d \cW_k | \leq  C p_{0,k} $. But for      $\la_k$  sufficiently small    $p_{0,k}  / p_k   =    ( - \log \la_k)^{p_0 - p}$   is    as small as   we  like  since  $p_{0}< p$.  Hence  the result.   

It remains to show   that the   scalar    bounds   in    (\ref{contorted}) are satisfied. 
The bounds with  all the $W$'s  gone  and  with a factor of $\frac12$      follow more or less directly from   
from the assumption  $ (A_{k+1}, \Phi_{k+1}) \in \cS^0_{k+1}$ just as for the gauge field.   Thus it suffices  to show that  the difference between the expression
with and without the  $W$'s  satisfy the indicated bounds with a factor $\frac12$.   We have
for  example
\be   \label{underwater}    \pa_{  \cA_{k+1}^0 + \cW_k }   \phi_{k+ 1}^0( \cA_{k+1}^0 + \cW_k ) - \pa_{  \cA^0_{k+1} }   \phi_{k+ 1}^0( \cA_{k+1}^0  )   
=       \frac{1}{2 \pi i}
 \int_{|t|  =e_k^{-1+ 4  \ep } }     \frac{dt}{t(t-1)}      \pa_{  \cA_{k+1}^0 +t \cW_k }   \phi_{k+ 1}^0( \cA_{k+1}^0 + t\cW_k ) 
\ee
To justify this representation  we need control  over   $\pa_{\cA} \cH^0_{k+1}(\cA)$  for  
$\cA  =   \cA_{k+1}^0 + t \cW_k $  and  $|t| \leq  e_k^{-1+ 4  \ep }$.  
 Since  $\pa_{\cA}  \cH^0_{k+1}(\cA)f  
 =    L^{-1} \B(  \pa_{\cA_{L^{-1}  }  } \cH_{k+1} (\cA_{L^{-1} }) f_{L^{-1} }\B)_L$
  we  need       $\cA_{L^{-1}} \in \cR_{k+1}$  
 and it suffices that      $\cA^0_{k+1, L^{-1}} \in \frac12\cR_{k+1}$
and   $t \cW_{k, L^{-1}}  \in   \frac12 \cR_{k+1}$.   We already know the former.  The latter follows by    
 \be     
  |   t\cW_{k,L^{-1}} |,   |   t \pa \cW_{k,L^{-1}} |,  |   t \de_{\al}  \pa    \cW_{k,L^{-1}} |  \leq     e_k^{-1+ 4\ep}     Cp_{0,k}    \leq  \frac12 e_{k+1}^{-1 + 3\ep }  
  \ee 
Thus   we  are in the region of analyticity for   $\pa_{\cA}  \cH^0_{k+1}(\cA)$  and so   $  \pa_{  \cA }   \phi_{k+ 1}^0( \cA)$.   Then       
$  |  \pa_{\cA}  \cH^0_{k+1}(\cA)  f  |   \leq   C  \| f \|_{\infty}   $   and
\be
| \pa_{  \cA }   \phi_{k+ 1}^0( \cA)|   \leq   C \| \Phi_{k+1}\|_{\infty}   \leq  C p_{k+1} \la_{k+1}^{-\frac14}  
\ee
and then  (\ref{underwater})  gives       
\be   \B|  \pa_{  \cA_{k+1}^0 + \cW_k }   \phi_{k+ 1}^0( \cA_{k+1}^0 + \cW_k ) - \pa_{  \cA^0_{k+1} }   \phi_{k+ 1}^0( \cA_{k+1}^0  ) \B | 
\leq   e_k^{1- 4 \ep}    ( C p_{k+1} \la_{k+1}^{-\frac14} )   \leq   Cp_{k+1}\la_{k+1}^{\frac14- \ep }   \leq  \frac12  p_k
\ee
Similarly     $\cA^0_{k+1} +  \cW_k \in \cR_k$ and so    
\be     |  \pa_{  \cA_{k+1}^0 + \cW_k }  \cW_k ( \cA_{k+1}^0 + \cW_k ) |  \leq   C p_{0,k}    \leq  \frac 12 p_k 
\ee
This  completes the bound for the derivative term in   (\ref{contorted})

The bounds on the other terms in  (\ref{contorted})   are   similar.   Note in particular that $\cA_{k+1}^0   \in \cR_k$     and so
$| C^{\frac12}( \cA_{k+1}^0) W|  \leq  Cp_{0,k} < \frac12 p_k$ by  (\ref{lunky}).  This completes the proof

 \subsubsection{second  localization} 
With the characteristic function gone    the fluctuation integral  is   
$\Xi_k\B(  \cA^0_{k+1},  \phi^0_{k+1}( \cA^0_{k+1} ) \B) $   where now for   any  $\cA,  \phi  \in \frac12 \cR_k$  
\be 
   \Xi_k (\cA,  \phi) =   \int   d \mu_I (\tilde  W_k)   d \mu_I(W_k ) )  \   \chi_k^w   
  \exp \Big(  \cB  E^{\dagger}_k \B(\cA,  \cW_k,  \phi,   \cW_k(\cA) \B)  \Big)   
\ee
As   explained   in section    \ref{random}, the Green's functions   $G_k, G_k(\cA)$   have  random walk expansions based on  $M$-cubes for  $M$ sufficiently large.   We use these expansions but now based on  $LM$ cubes.    With them we    define weakened  Green's functions  
$G_k(s), G_k(s, \cA)$  and so minimizers    $\cH_k(s), \cH_k(s, \cA)$    .  
Similarly we  weaken $ C_k^{1/2} , C_k^{1/2}(\cA) $ to  $ C_k^{1/2}(s) , C_k^{1/2}(s,\cA) $,  now    based on the random walk expansions  of  (\ref{rw2})  and lemma   \ref{random4}   with  $LM$-cubes.      Then   define  instead of (\ref{tiki})  
\be
 \cW_k (s) = \cH_k(s)C  C_k^{1/2}(s) \tilde W   \hs   \cW_k(s,\cA)  = \cH_k( s, \cA)  C_k^{1/2}(s,\cA) W
 \ee
 The  term    $ \cB  E^{\dagger}_k \B(Y, \cA, \phi,  \cW_k,  \cW_k(\cA) \B)  $   is local  in  $( \cA,  \cW_k,   \phi,   \cW_k(\cA))$,  but   not    in  $\tilde  W,  W $  To  remedy   this we  write
\begin{equation}  \label{summer}
\begin{split}
\cB E^{\dagger}_k( Y)   
= & \sum_{Z \supset Y}        \cB E^{\dagger}_k\B( Y,Z\B) \\
 \cB E^{\dagger}_k\B( Y,  Z;   \cA, \tilde    W, \phi,   W)\B)   
= &  \int   ds_{Z-Y} 
 \frac { \pa  }{ \pa s_{Z-Y}}   \left[ \cB E^{\dagger}_k\B(Y,  \cA,  \cW_k(s),     \phi,   \cW_k(s, \cA) \B)   \right]_{s_{Z^c} = 0, s_Y=1}\\
 \end{split}
 \end{equation}
Now  we   write 
\begin{equation}   
\cB E_k^{\dagger}    =  \sum_Y   \cB E^{\dagger} ( Y)    =   \sum_Y      \sum_{Z \supset Y}        \cB E^{\dagger} ( Y,Z) 
=  \sum_{Z }    E_k ^{\loc}(Z)   \equiv   E_k^{\loc}
\end{equation}
where   the sum is over   $LM$-polymers  $Z$  and   
\begin{equation}    
    E_k^{\loc}(Z) =  \sum_{Y  \subset  Z}     \cB E^{\dagger} ( Y,Z)   
\end{equation}
is strictly local $( \cA, \tilde  W, \phi,   W)$.

Now   $\cB E^{\dagger}_k\B(Y,  \cA,     \phi,  \cW_k(s),   \cW_k(s, \cA) \B)$ has a bound of the form  (\ref{pink2}) even for 
$| s _{\square} |  \leq  M^{\frac12} $,   and one  can use Cauchy bounds in $s_{\square}$  to prove the following  (see for example lemma 19   in   \cite{Dim11} 
for details).

   \begin{lem}   \label{snuff2}
 For    $\cA,  \phi  \in  \frac12  \cR_{k}$   and       $|\tilde  W|, |W |    \leq  p_{0,k}$
  \begin{equation}   \label{stinger}
 |   E_k ^{loc}(X,  \cA,  \tilde  W, \phi,   W)|   \leq     \cO(1)L^3    \la_k^{1/12 -10  \ep}  e^{ -L(c\ka - 4\ka_0-4)  d_{LM}(X) } 
 \end{equation}
 \end{lem}
\bigskip

\subsubsection{cluster expansion}

 The     fluctuation integral  is     now  
 \begin{equation}
 \Xi_k(\cA,   \phi)=  
  \int \exp     \Big(    \sum_{Y }     E_k ^{\loc}(Y, \cA, \tilde W,  \phi,W)  \Big) \chi_k(\tilde W)      \chi_k(W)      d\mu_I(\tilde W)    d\mu_I(W)   
\end{equation}
We normalized the measure  introducing 
\be 
 d \mu_k^*(\tilde W)  =   \frac{   \chi_k(\tilde W)      d\mu_I(\tilde W)  } { \int   \chi_k(\tilde W)      d\mu_I(\tilde W)  }
 \hs   d \mu^*_k( W)    =   \frac{   \chi_k( W)      d\mu_I( W)  } { \int   \chi_k( W)      d\mu_I( W)  }
\ee
The normalization  factors contribute     $\exp( - \vep^0_k \Vol(\bbT^0_{N-k}  )   )$  where   $\vep^0_k  = \cO(e^{-p_{0,k}^2/2})$
\cite{Dim11}.    So  now      
 \begin{equation}
 \begin{split}
 \Xi_k(\cA,   \phi)=    &  \exp\B( - \vep^0_k \Vol(\bbT^0_{N-k}  )  \B) \Xi'_k(\cA,   \phi) \\
 \Xi'_k(\cA,   \phi)  =  & \int \exp     \Big(    \sum_{Y }     E_k ^{\loc}(Y, \cA, \tilde W,  \phi ,  W)  \Big)      d\mu_k^*(\tilde W)    d\mu_k ^*(W)   \\
 \end{split}
\end{equation}

The cluster  expansion gives this a local structure.  As  in  \cite{Dim11}  using the bound (\ref{stinger})  we  have  
\begin{lem}  \label{cluster0}(cluster expansion) 
For   $\cA,  \phi  \in   \frac12   \cR_k$  
\begin{equation}  \label{sunshine}
  \Xi_k(\cA,\phi)       = \exp  \Big(  \sum_{Y }      \ E^\#_k(Y,\cA,   \phi)  \Big)
\end{equation}
 where  the sum is over $LM$ polymers  $Y$  and  
\begin{equation}  \label{osprey}
| E^\#_{k}(Y,\cA,  \phi    ) |  \leq   \cO(1)  L^3 \la_k^{1/12 - 10 \ep}
  e^{ - L( c \ka     -7 \kappa_0 -7   )  d_{LM}(Y)}  
\end{equation}
\end{lem}
\bigskip

It is straightforward to check  that  the construction of  $E^\#_k(Y, \cA, \phi)$ preserves all the symmetries.

Now  (\ref{bong}) becomes     
 \be     \label{bing}
\begin{split}
& \tilde    \rho_{k+1} (A_{k+1}, e^{qe_k \om^{(1)} }   \Phi_{k+1} )  
=  \cN_k N_k   \sZ_k \sZ^f_{k+1} \sZ_k(  \cA^0_{k+1})    \sZ^f_k( \cA^0_{k+1}) \\
&  \exp \Big( - \frac12   \|  d \cA^0_{k+1} \|^2 - S^0_{k+1,  \cA^0_{k+1}    }    ( \Phi_{k+1},    \phi^0_{k+1}( \cA^0_{k+1} )   ) \Big) \\
&   \exp  \Big(   
    -  V_k(   \phi^0_{k+1}(\cA^0_{k+1}) )   - \vep_k^0  \Vol(  \bbT^0_{ \sN -k}) 
 +  \cB  E_k\B(  \cA^0_{k+1},   \phi^0_{k+1}(\cA^0_{k+1}) \B) +      \ E^\#_k\B( \cA^0_{k+1},   \phi^0_{k+1}(\cA^0_{k+1}\B)     \Big)\\
\end{split}
\ee

\subsubsection{scaling}

Define a scaled phase shift  $\theta  =   \theta (A_{k+1} )$ on  $\bbT^0_{N-k-1}$ by   
\be 
  \theta (A_{k+1} )  = \B(  \om^{(1)} (A_{k+1,L} ) \B)_{L^{-1}}
\ee
Then   
\be
\rho_{k+1} \B(A_{k+1}, e^{qe_{k+1} \theta }  \Phi_{k+1}\B)  =  \tilde  \rho_{k} \B(
A_{k+1,L},   e^{qe_k \om^{(1)} } \Phi_{k+1,L}\B) L^{\frac12(  b_N - b_{N-k-1})  +\frac12(  s_N - s_{N-k-1})               } 
\ee
and so we make the substitutions  $A_{k+1}  \to  A_{k+1,L}$  and   $\Phi_{k+1}  \to  \Phi_{k+1,L}$ in  (\ref{bing}). 
With this substitution     $\cA^0_{k+1}$  becomes   $ \cA_{k+1,L }$ and we identify   by (\ref{z}), (\ref{sammy}) 
\be
\begin{split}
& \B(   \sZ_k \sZ^f_{k+1} L^{\frac12(  b_N - b_{N-k-1})  - \frac12(  s_N - s_{N-k-1}) } \B)   
  \B(\cN_k  N_k \sZ_k(  \cA_{k+1,L })    \sZ^f_k(  \cA_{k+1,L }) 
   L^{  s_N - s_{N-k-1} }  \B)  \\
& =  \cN_{k+1}   \sZ_{k+1} \sZ_{k+1} ( \cA_{k+1} )   \\
\end{split} 
\ee
We also have that   
  $ \phi^0_{k+1}( \cA^0_{k+1} )   $  becomes   $( \phi_{k+1}( \cA_{k+1} ) )_L$, and
  $  \|  \pa \cA^0_{k+1} \|^2$  becomes   $  \|  \pa \cA_{k+1} \|^2$,
and    $S^0_{k+1,  \cA^0_{k+1}    }    ( \Phi_{k+1},    \phi^0_{k+1}( \cA^0_{k+1} )   ) $ becomes  $S_{k+1,  \cA_{k+1}    }    ( \Phi_{k+1},    \phi_{k+1}( \cA_{k+1} )   ) $.   We have also     
$ \vep^0_{k} \Vol(  \bbT_{N -k}) = L^3  \vep^0_{k}  \Vol(  \bbT_{N -k-1})$. 
The potential  $V_k(   \phi^0_{k+1}(\cA^0_{k+1}) ) $
becomes
\begin{equation}
 L^3 \vep_{k} \Vol(\bbT_{N-k-1} )  
  +  \frac 12  L^2 \mu_{k}   \|  \phi_{k+1}(\cA_{k+1})    \|^2    +  \frac14  L \la_k  \int  \B(\phi_{k+1}(\cA_{k+1})\B)^4 
\end{equation}
The   function  $\cB E_k$  becomes   
 $\cB  E_k\B(  \cA_{k+1,L}  , ( \phi_{k+1}(\cA_{k+1}))_L \B) \equiv (\cB  E_k)_{L^{-1}}( \cA_{k+1}  ,  \phi_{k+1}(\cA_{k+1}))$.
 Then we have    $\cB  E_{k, L^{-1}}  =  \sum_X  \cB  E_{k, L^{-1}}(X)  $   where   
$ \cB  E_{k, L^{-1}}(X,  \cA,  \phi)   = \cB   E_k(LX,  \cA_L,  \phi_L) $.
 Since    $E_k$   is normalized  for small polymers   we have   by  lemma \ref{scalinglem}  
\begin{equation}  \label{bb1}
 \|     (\cB  E_k)_{L^{-1}}\|_{k+1, \ka}     \leq  \cO(1) L^{-\ep}     \|  E_{k}\|_{k, \ka}   
\end{equation}
Similarly     $ E^\#_k $  becomes    $   E^\#_k\B(  \cA_{k+1,L}  , (\phi_{k+1}(\cA_{k+1}))_L \B)  \equiv  E^\#_{k, L^{-1}}(  \cA_{k}  , \phi_{k+1}(\cA_{k+1}) )$.    Then   we have  that $ E^\#_{k, L^{-1}}  =  \sum_X   E^\#_{k, L^{-1}}(X)  $   where   
$   E^\#_{k, L^{-1}}(X,  \cA,  \phi)   =    E^\#_k(LX,  \cA_L,  \phi_L) $.  If  $\cA, \phi  \in  \cR_{k+1}$  then   $\cA_L, \phi_L  \in \frac12 \cR_{k}$  and so we can use the bound (\ref{osprey}).     Since   $d_{LM}( LX) = d_M(X)$
 this gives
\be   \label{stingray}
  | E^\#_{k, L^{-1}}(X, \cA,   \phi) |  \leq    
 \cO(1)  L^3     \la_k^{1/12 -10 \ep}     e^{ - L( c \ka     -7 \kappa_0 -7   )  d_{M}(X)}   
\ee
 But for $L$ sufficiently large   $ L( c \ka     -7 \kappa_0 -7   ) \geq  \ka$,   so  the decay factor can be taken  as  $e^{ - \ka d_{M}(X)} $.
 Then the bound is  
 \be   \label{stingray2}
  \| E^\#_{k, L^{-1}} \|_{k+1, \ka}  \leq    
 \cO(1)  L^3     \la_k^{1/12 -10 \ep}      
\ee

 Altogether  then  
 \be     \label{pink8}
\begin{split}
 &   \rho_{k+1} (A_{k+1}, e^{qe_{k+1} \theta } \Phi_{k+1} )  \\
&= \cN_{k+1}  \sZ_{k+1}  \sZ_{k+1}( \cA_{k+1}) \
  \exp \Big( - \frac12  \|  d \cA_{k+1} \|^2- S_{k+1,  \cA_{k+1}    }    ( \Phi_{k+1},    \phi_{k+1}( \cA_{k+1} )   ) \Big) \\
& \exp 
 \Big(  - L^3 (\vep_k + \vep_k^0 )\Vol(  \bbT_{ \sN -k-1})       - \frac12  L^2 \mu_{k}   \|  \phi_{k+1}(\cA_{k+1})   \|^2   
  -  \frac14      \la_{k+1}  \int \B( \phi_{k+1}(\cA_{k+1}) \B)^4  \Big)  \\
&   \exp  \Big(   
(\cB  E_k)_{L^{-1}}\B(  \cA_{k+1},   \phi_{k+1}(\cA_{k+1}) \B) +      \ E^\#_{k,L^{-1}}\B( \cA_{k+1},   \phi_{k+1}(\cA_{k+1}) \B)    \Big)\\
\end{split}
\ee

\subsubsection{completion of the proof}

Neither   $ (\cB E_k)_{L^{-1} }$    nor     $E^\#_{k,L^{-1} }$  are normalized   for small polymers,   and we
need this feature to complete the proof.
We  remove  energy and mass terms to normalize them.

We  have   by  (\ref{renorm2})
 \begin{equation}  
 ( \cB E_k)_{L^{-1}}(\cA, \phi ) =     -  ( \cL_1E_k)   \Vol(  \bbT_{N-k-1} ) 
 -   \frac12  ( \cL_2E_k)    \|  \phi^2 \| 
+  (\cL_3 E_k) (\cA, \phi )
\end{equation} 
where    
 \begin{equation}
\begin{split}
\cL_1E_k  =&   \vep\Big( ( \cB E_k)_{L^{-1}}\Big)   
    \\
 \cL_2E_k  =&   \mu\Big(( \cB E_k)_{L^{-1}}\Big)          \\
 \cL_3 E_k  =  &     \cR  \Big(( \cB E_k)_{L^{-1}}\Big )      \\ 
\end{split}
\end{equation}
By  (\ref{sundown})   and  (\ref{bb1})   $|\cL_1E_k |  \leq  \one L^{-\ep}\|E_k \|_{k, \ka}$  and  
 $|\cL_2E_k |  \leq  \one  \la_k^{\frac12 + 2 \ep} L^{-\ep}\|E_k \|_{k, \ka}$.  By   
(\ref{sunup})  and  (\ref{bb1})     $\| \cL_3 E_k  \|_{k+1, \ka}  \leq   \one L^{-\ep}\|E_k \|_{k, \ka}$. 
These are the required bounds.

We   also   apply    (\ref{renorm2})
  to   $E^\#_{k,L^{-1} }$   but now tack on the extra term  $\vep_k^0$    
We   have
 \begin{equation}  
E^\#_{k,L^{-1} }(\cA, \phi )   -  L^3  \vep_k^0     \Vol(  \bbT_{N-k-1} ) 
=  -   \vep_k^*  \Vol(  \bbT_{N-k-1} ) 
 -   \frac12  \mu_k^* \|  \phi^2 \| 
+  E_k^*  (\cA, \phi )
\end{equation} 
where
\begin{equation}
\begin{split}
\vep_k^*  =    &   L^3  \vep_k^0      +       \vep( E^\#_{k,L^{-1} }) \\
\mu_k^*  =  &    \mu( E^\#_{k,L^{-1} })   \\
E_k^*   =  &  \cR( E^\#_{k,L^{-1} }  ) \\
\end{split}
\end{equation}
By  (\ref{sundown})   and  (\ref{stingray2})   $|\vep_k^*  |  \leq  \one    L^3     \la_k^{\frac{1}{12} -10 \ep}  \leq     \la_k^{1/12 -11 \ep} $  and  
 $|\mu_k^*  |  \leq  \one L^3  \la_k^{\frac{7}{12}  - 8  \ep} \leq     \la_k^{\frac{7}{12}  -11 \ep}$.  By   
(\ref{sunup})  and  (\ref{stingray2})     $\| E_k^*   \|_{k+1, \ka}     \leq  \one    L^{3- \ep}     \la_k^{\frac{1}{12}  -10 \ep}  \leq     \la_k^{1/12 -11 \ep} $. 
These are the required bounds.

Insert   these  expansions  into   (\ref{pink8})  and   define   as in  (\ref{recursive})
  \begin{equation}  \label{recursive2}
\begin{split}
\vep_{k+1}   =&  L^3 \vep_k  + \cL_1E_k   +  \vep_k^*( \mu_k,  E_k) \\
\mu_{k+1}   =&   L^2 \mu_k  +  \cL_2E_k  + \mu_k^*( \mu_k,  E_k)  \\
E_{k+1}   =&    \cL_3 E_k  +  E^*_k(    \mu_k,  E_k)  \\
 \end{split}
\end{equation}
 This gives    the final form 
 \be     \label{pink9}
\begin{split}
    \rho_{k+1} (A_{k+1}, e^{qe_{k+1} \theta }  \Phi_{k+1} ) 
= &  \sZ_{k+1}  \sZ_{k+1}( \cA_{k+1}) \
  \exp \Big( - \frac12  \|  d \cA_{k+1} \|^2 - S_{k+1,  \cA_{k+1}    }    ( \Phi_{k+1},    \phi_{k+1}( \cA_{k+1} )   ) \Big) \\
& \exp 
 \Big( -  V_{k+1} ( \phi_{k+1}(\cA_{k+1}) )    +      \ E_{k+1}( \cA_{k+1},   \phi_{k+1}(\cA_{k+1}) )    \Big)\\
\end{split}
\ee
where   
\be   V_{k+1} ( \phi)  =     \vep_{k+1} \Vol(  \bbT_{ \sN -k-1})    +    \frac12  \mu_{k+1}   \|  \phi   \|^2   +
    \frac14      \la_{k+1}  \int  |\phi|  ^4 
  \ee 
 This completes the proof of theorem \ref{lanky},   except for lemma  \ref{snooze4}.

\section{Normalization factor}

\label{normfactor}

In this section we    prove the missing lemma   \ref{snooze4}.    We need to    understand how the normalization  factor   $ \cZ_k(\cA) $   changes under  a change in $\cA$.
This is somewhat involved since    $\cZ_k(\cA)$  is nonlocal  and we need to express the answer in a local form.   
In  particular we  want to  write  
\be   
    \frac  {\cZ_k(\cA +\cZ)} {  \cZ_k(\cA) }    =  \exp(    E^z(\cA, \cZ) )
    \ee
with     $ E^z(\cA, \cZ) \equiv      E^{(4)}(\cA, \cZ)   $ given   as a sum of local pieces.

There   are two  ways  to  approach this.    On the one hand  from   (\ref{route66})  we  have   
\be  \label{unbirthday}    \frac  {\cZ_k(\cA +\cZ)} {  \cZ_k(\cA) }  
      =\left[  \frac { \det   G_k( \cA  + \cZ) }    { \det   G_k( \cA ) }   \right]^{\frac12}
 \ee   
On     the  other hand we have  from  the    recursion relation   (\ref{z})  
\be  \label{birthday}
   \frac  {\cZ_k(\cA +\cZ)} {  \cZ_k(\cA) } 
     = \prod_{j=0}^{k-1} \left[   \frac{ \det C_j (\cA_{L^{k-j}}+ \cZ_{L^{k-j}})   }{     \det C_j (\cA_{L^{k-j}}) }\right]^{\frac12}
\ee 
In  the first  representation  we  are working   only  on the  fine   lattice  $\tk$  and have to deal with explicit 
ultraviolet  divergences.    In  the  second   case  we  have a product  over   unit lattice  operators   on  $\bbT^{0}_{N -j}$ with gauge fields on  $\bbT^{-j}_{N-j}$   
followed by scalings  down  to  $\tk$.    In  this  case  we  have no explicit ultraviolet divergences  but  have to 
carefully   track  the scaling behavior.    Either  approach should work in principle.    We  prefer to take the
second approach  which  is more in tune with the rest of the paper.     However  at one point we have  to  revert to the first approach  to make the argument.

\subsection{single scale}
 
 We  need estimates on    $C_k(\cA) =   \B(   \De_k(\cA) + aL^{-2} (Q^TQ)(\cA) \B)^{-1} $ 
 and on    
\be 
\begin{split}
\Up_k  (\cA, \cZ)   \equiv   &  C_k(\cA + \cZ) ^{-1}    -  C_k(\cA ) ^{-1} \\
  =  &    \B(  \De_k(\cA+ \cZ) -  \De_k(\cA) \B)   + aL^{-2} \B( (Q^TQ)(\cA + \cZ) -  (Q^TQ)(\cA) \B)\\
  \end{split}
\ee
which   is defined to satisfy   
\be   
C_k(\cA + \cZ)     -  C_k(\cA )=   C_k(\cA)     \Up_k(\cA,  \cZ)  C_k(\cA + \cZ)  
\ee
These  are   all       unit lattice operators defined on functions  on the lattice  $\tz$.

We study these operators for   $\cA \in \frac12  \cR_k$  and  $\cZ  \in \frac12  \cR_k'$ where $\cR_k'$  is all fields   complex valued vector fields   $\cZ$  on  $\tk$
satisfying   
\be    |\cZ|   <  e_k^{-1+3 \ep}  
      \hs     |\pa \cZ|   < e_k^{-1+ 4\ep} 
     \hs     | \de_{\al}  \pa  \cZ    |      <  e_k^{-1+ 5 \ep}
\ee 
We have   $\cR_k'   \subset  e_k^{2\ep}  \cR_k$.

\begin{lem}     \label{noughat}
 For   $\cA  \in \frac12  \cR_k$ ,   $\cZ  \in \frac12  \cR_k'$  the matrix elements satisfy
\be
\begin{split}
|[C_k  (\cA)]_{yy'} |   \leq    &  C       e^{- \ga d(y,y')  }  \\
|[\Up_k  (\cA, \cZ)]_{yy'} |   \leq   &   C    e_k^{2\ep}    e^{- \ga d(y,y')  }\\
\end{split}
\ee
\end{lem}
\bigskip

\pr
First consider   $\Up_k$.  
Define  
\be 
  D_k(\cA)  =  Q_k(\cA) G_k(\cA)  Q^T_k(\cA) 
 \ee
Then  since    
    $  \De_k(\cA)  =     a_k -  a_k^2 D_k(\cA)   $  we have    
\be
\Up_k  (\cA, \cZ) = a_k^2  \B(  D_k(\cA+ \cZ) -  D_k(\cA) \B)   + aL^{-2} \B( (Q^TQ)(\cA + \cZ) -  (Q^TQ)(\cA) \B)
\ee
For matrix elements we have    
\be
   [D_k  (\cA)]_{yy'} =   < \de_y,   D_k(\cA)  \de_{y'}>  =  < Q_k^T(\cA) \de_y , G_k(\cA)  Q_k^T(\cA) \de_{y'}>
\ee
Since  $\supp ( Q_k^T(\cA)     \de_y  )   \subset  \De_y$   we have  by    (\ref{sycamore4})     with $L^2$ bounds 
\be   \label{cork}
  | [D_k  (\cA)]_{yy'}|  
\leq   C e^{- \ga d(y,y')  }   \|Q^T(\cA) \de_y \|_2 \|Q^T(\cA) \de_{y'} \|_2  \leq     C e^{- \ga d(y,y')}    
\ee
Also consider    $[ (Q^TQ)(\cA)]_{yy'}  = < Q^T(\cA)\de_y, Q^T(\cA) \de_{y'}>$.  This vanishes unless  $y,y'$  are in the
same  $L$-cube and satisfies    $|[ (Q^TQ)(\cA)]_{yy'} |  \leq  \one$. 
Next  we  use the analyticity in the fields to write for  $r \geq  1$
\be
  [ D_k(\cA + \cZ)]_{yy'}  - [ D_k (\cA)]_{yy'}  =  \frac{1}{2\pi i}  \int_{|t| = r}  \frac{ dt } {t(t -1)}     [ D_k(\cA + t\cZ)]_{yy'}    
\ee
Here  we   can  take   $r=    e_k^{-2\ep}$    since then   $t\cZ   \in \frac12  \cR_k$
and we  are  in the domain of analyticity.  
Then  (\ref{cork})  yields  
\be    
\B| [ D_k(\cA + \cZ)]_{yy'}  - [ D_k (\cA)]_{yy'}  \B|    
\leq    C  e_k^{2\ep}e^{- \ga d(y,y')}   
\ee
Similarly one shows that   
\be    \label{stunted}
\B| [ (Q^TQ)(\cA + \cZ)]_{yy'}  - [  (Q^TQ) (\cA)]_{yy'}  \B|     
\leq    C  e_k^{2\ep}  
\ee
This is a local operator so the decay factor is optional here.   The bound on $[\Up_k  (\cA, \cZ)]_{yy'}$ follows.

Now  consider  $C_k(\cA)$.   
We  have the identity     (this is   (\ref{swipe}) at  $x=0$ ):
\be     \label{lamp}
 \begin{split}
 C_k(\cA)  = & \sA_k(\cA)  + a_k^2 \sA_k(\cA)Q_k(\cA)  G^0_{k+1}(\cA) Q_k^T(\cA)\sA_k(\cA)   \\
 \sA_k(\cA)   =  &  \frac{1}{a_k} ( I -   (Q^TQ)(\cA))  + \frac{1}{a_k + aL^{-2}}   (Q^TQ)(\cA)   \\
 G^0_{k+1}(\cA)  =  &   \B(  - \De_{\cA}  + \frac{a_{k+1}}{L^2}   (Q_{k+1}^TQ_{k+1})(\cA)  \B)^{-1} \\
 \end{split}
 \ee
 Note that    $G^0_{k+1}(\cA) $ scales to  $G_{k+1}(\cA)$.
Just as for  $D_k(\cA)$  we have 
\be   
|[Q_k(\cA)  G^0_{k+1}(\cA) Q_k^T(\cA)]_{yy'} |  \leq     C e^{- \ga d(y,y')}  
\ee
 Every other  operator in  (\ref{lamp}) is local so we have the result.    
This completes the proof.
\bigskip

Here is a variation of these results.     
As  noted in section \ref{localgreen}  we  can introduce  a local version of  the  Green's function  $G_k(X, \cA)$  so that  
$G_k(\cA)  = \sum_X G_k(X, \cA)$  and the same is true for   $ G^0_{k+1}(\cA) $.
Using these local Green's function we define local operators 
\be
\begin{split}
 D_k(X, \cA)  =  &  Q_k(\cA) G_k(X, \cA)  Q^T_k(\cA)  \\
\Up_k  (X,\cA, \cZ) = &  a_k^2  \B(  D_k(X,\cA+ \cZ) -  D_k(X,\cA) \B)   + aL^{-2} \B(   (Q^TQ)(\cA + \cZ) -   (Q^TQ)(\cA) \B)\bbI_X  \\
 C_k(X,\cA)  = & \sA_k(\cA) \bbI_X + a_k^2 \sA_k(\cA) Q_k(\cA)  G^0_{k+1}(X,\cA)  Q_k^T(\cA)\sA_k(\cA)   \\
 \end{split}
 \ee 
 Here  $\bbI_X(x) = 1$ if  $|X|_M =1$ and  $x \in X$,  and is zero otherwise.
 Summing  over  $X$ we recover    $D_k(\cA),   \Up_k (\cA, \cZ), C_k(\cA)   $.
Repeat the above  proof using   $\|  G_k(X, \cA)f \|_2  \leq  C e^{- \ka d_M(X)}  \|f\|_2$  and the same for  $  G^0_{k+1}(X,\cA) $.       This yields  for the matrix elements    
\be    \label{ugh}
\begin{split}    
|[D_k  (X,\cA)]_{yy'} |   \leq  &  C    e^{- \ka  d_M(X)}      \\
|[\Up_k  (X, \cA, \cZ)]_{yy'} |   \leq   &   C    e_k^{2\ep}  e^{- \ka  d_M(X) }\\
|[C_k  (X, \cA)]_{yy'} |   \leq    &  C   e^{- \ka  d_M(X)}     \\
\end{split}
\ee
These quantities  vanish unless  $y, y' \in X$  and  only depend on $\cA$ in $X$.

\bigskip

\begin{lem}   Let  $e_k$ be sufficiently small depending on  $L,M$.   
 For   $\cA  \in \frac12  \cR_k$ ,   $\cZ  \in \frac12  \cR_k'$ 
 we
have
\be      \label{unbirthday2}
\left[\frac { \det   C_k( \cA  + \cZ) }    { \det   C_k( \cA ) }   \right]^{\frac12}
=  \exp   \B(   \sum_{X \in \cD_k}   E^c_k(X,  \cA,   \cZ)  \B)  \equiv     \exp   \B(     E^c_k(  \cA,   \cZ)  \B)
\ee     
where   $ E^c_k(X, \cA,  \cZ )$ is analytic in $\cA, \cZ$,
depends on the  fields  only   in   $X$,   satisfies   $E_k^c(X, \cA, 0) =0$  and 
\be   \label{basicbound}
   |  E^c_k(X, \cA,  \cZ ) |
 \leq  \one    e_k^{\ep  }   e^{-( \ka- \ka_0 -3) d_M(X) }
  \ee
\end{lem}
\bigskip

\pr    Since      $  C_k(\cA + \cZ)   = C_k(\cA)   +      C_k(\cA) \Up_k(\cA,  \cZ) C_k(\cA + \cZ)   $
\be 
\begin{split}
\frac { \det   C_k( \cA  + \cZ) }    { \det   C_k( \cA ) }   
=   &   \det   \B(    C_k( \cA )   C_k( \cA  + \cZ) ^{-1} \B)^{-1}\\
=  &  \det  \B (  I   -   C_k(\cA) \Up_k(\cA,  \cZ) \B )^{-1} \\
=  &  \exp \B(- \Tr  \log   \B( I   -     C_k(\cA)\Up_k(\cA,  \cZ)    \B)\B )\\
=  &  \exp  \B(   \sum_{n=1}^{\infty}  \frac{1}{n} \Tr \B(  C_k(\cA)  \Up_k(\cA,  \cZ)  \B)^n  \B)\\
\end{split}
\ee

Now  in the sum    insert    $\Up_k( \cA, \cZ)  =  \sum_X  \Up_k(X, \cA, \cZ)$   and   $C_k( \cA)  =  \sum_Y C_k(Y, \cA)$.
The  $n^{th}$ term  is then a   expressed as  a  sum over  sequences of polymers  $(X_1,Y_1, \dots X_n, Y_n)$.    The polymer
$X_i$ must overlap  $Y_i$ and $Y_{i-1}$  and so the union  is  connected.   We group together terms with the 
same union  and get the representation  (\ref{unbirthday2})  with   
\be    E^c_k(X, \cA,  \cZ )
= \frac12  \sum_{n=1}^{\infty}  \frac{1}{n}   \sum_{X_1, \dots  
X_n,  Y_1, \dots,  Y _n     \to  X   }
    \Tr    \B( \Up_k(X_1,\cA,  \cZ)  C_k(Y_1,\cA)  \cdots     \Up_k(X_n, \cA,  \cZ)  C_k(Y_n,\cA)         \B)
\ee  
Here  $X_1, \dots  
X_n,  Y_1, \dots,  Y _n     \to  X$    means the overlap conditions are satisfied and     $\cup_{i=1}^n (X_i \cup Y_i)  = X$.  The trace is evaluated
as    
\be \sum_{x_1,y_1, \dots x_n, y_n}   [ \Up_k(X_1,\cA,  \cZ)]_{x_1y_1} [ C_k(Y_1,\cA)]_{y_1x_2}
  \cdots    [  \Up_k(X_n, \cA,  \cZ)]_{x_ny_n} [ C_k(Y_n,\cA) ]_{y_nx_1}     
\ee
Bound   the  $ [  \Up_k(X_i, \cA,  \cZ)]_{x_iy_i }$ and  $ [ C_k(Y_i,\cA) ]_{y_ix_{i+1}}$  by   (\ref{ugh})  and 
bound the sums  by  estimates like
\be   \label{lignon}
\sum_{x \subset  X }  1  \leq   \Vol(X) \leq   M^3 |X|_M  \leq   \one M^3  e^{d_M(X)}  
\ee    
Thus the trace   has an overall factor  $(\one e_k^{2 \ep} C^2 M^6)^n  \leq   e_k^{n \ep}  $
and dropping  the $1/n$ we have  
\be  |  E^c_k(X, \cA,  \cZ ) |
\leq \one  \sum_{n=1}^{\infty}     \sum_{X_1, \dots  
X_n,  Y_1, \dots,  Y _n     \to  X   }   e_k^{n \ep}
   \prod_{i=1}^n   e^{ - (\ka-1) d_M(X_i)  }   e^{ -(\ka-1)  d_M(Y_i)  } 
\ee  
Now we use
\be   \sum_{i=1}^n  (  d_M(X_i)    + d_M(Y_i)  )  \geq   d_M(X)
\ee
to extract a factor   $e^{-(\ka - \ka_0-2) d_M(X)  }$  leaving   
\be  |  E^c_k(X, \cA,  \cZ ) |
 \leq  \one e^{-(\ka - \ka_0-2) d_M(X)  } \sum_{n=1}^{\infty}     \sum_{X_1, \dots  
X_n,  Y_1, \dots,  Y _n     \to  X   }   e_k^{n \ep}
   \prod_{i=1}^n   e^{ - (\ka_0 +1) d_M(X_i)  }   e^{ -(\ka_0 +1)   d_M(Y_i)  } 
\ee  
We  drop the condition that the union is   $X$,  retaining only the condition $X_1 \subset X$,  and estimate
\be  
\begin{split} 
\sum_{ Y_n  \cap X_n \neq  \emptyset}    e^{ -(\ka_0 +1)   d_M(Y_n)  }   \leq   &    \one  |X_n|_M  \\
\sum_{ X_n  \cap Y_{n-1}   \neq  \emptyset}  |X_n|_M  e^{ -(\ka_0 +1)   d_M(X_n)  }   \leq   &   \one  
\sum_{ X_n  \cap Y_{n-1}   \neq  \emptyset}    e^{ -\ka_0   d_M(X_n)  }  
\leq     \one  |Y_{n-1} |_M  \\
&   \dots    \\ 
\sum_{X_1 \subset X }   |X_1|_M      e^{ -(\ka_0+1)    d_M(X_1)  }   \leq   &   \one   \sum_{X_1 \subset X }       e^{ -\ka_0   d_M(X_1)  }
\leq      \one   |X|_M   \leq  \one e^{d_M(X) }   \\
\end{split}
\ee
The estimate  is now  
\be     |  E^c_k(X, \cA,  \cZ ) |
 \leq   \one  e^{-(\ka - \ka_0-3) d_M(X)  } \sum_{n=1}^{\infty}  ( \one   e_k^{ \ep}  )^n  \leq    \one e_k^{\ep}    e^{-(\ka - \ka_0-3) d_M(X)  } 
\ee  
to finish the proof.
\bigskip

\rem
Note  that       $E^c_k(X, \cA,  \cZ  )$  has the symmetries
\be      \label{gong}
\begin{split} 
E^c_k(X, -\cA,  -\cZ  )
=  &  E^c_j(X, \cA,  \cZ  )  \\
E^c_k(X, \cA+ \pa \la,  \cZ  )
=  &  E^c_j(X, \cA,  \cZ  )  \\
\end{split}
\ee
These can be deduced from the  gauge covariance and  charge conjugation covariance   of  $ \Up_k(X,\cA, \cZ)$ and  $C_k(Y, \cA)$,  which
in turn  follows  from the same properties for  $G_k(X, \cA)$ and  $ Q_k(\cA)$.
It is not the case  that   $E^c_k(X, \cA,  \cZ  )$  is   gauge invariant in $\cZ$.   But for the global version we  do   have 
\be
E^c_k( \cA,  \cZ + \pa \la )
=    E^c_j( \cA,  \cZ  ) \ee
  Indeed the gauge invariance   of  $  \det  C_k(\cA + \cZ)$
implies  that the exponentials are equal,  hence the identity holds for real fields,  and hence for all fields.

\subsection{improved single scale}    

We   want  to improve the last bound  to show it  is small when the fields are small.  
 Let   $R= L^r$ be a  (variable) multiple of  $L$ and let  $\square_R$  be  a partition  of  $\tk$ into  $MR$ cubes. 
 Also let     $\square^{\nat}_R$   be  a cube centered on  $\square_R$  consisting of   $MR$-cubes with   $\one L$   on a side, hence
consisting of   $M$-cubes  with   $\one LR$ on a side.

   We     define  a new domain  based on the inequalities    
\be   \label{region3} 
| \cA|  <     R^{-1+ \ep } e_k^{-1+ \ep}  
      \hs     |\pa \cA|  <     R^{-2+ \ep } e_k^{-1+ 2\ep} 
     \hs     | \de_{\al}  \pa   \cA  |   <    R^{-2 - \al + 2 \ep}   e_k^{-1+ 3 \ep}  
\ee
We  define   $\cR_k(R)$ to be all    complex-valued      $\cA$ on $\tk$  such that
 $\cA  = \cA_0  + \cA_1$   where  $\cA_0$ is  real and  in      each  $\one LMR$  cube   $ \square^{\nat}_R $  is    gauge equivalent  to  some    $\cA'_0 $ satisfying    the bounds  (\ref{region3})   with a factor  $\frac12 $   and   $\cA_1$ is complex  and satisfies the bounds
 (\ref{region3}) with a factor $\frac12$.   We   also  define  $\cR_k'(R)$ by 
\be    \label{symphony1}
  |\cZ|   <     R^{-1+ 2 \ep} e_k^{-1+ 3\ep}  
      \hs    |\pa  \cZ|   <     R^{-2+ 2 \ep } e_k^{-1+ 4\ep} 
     \hs      | \de_{\al}  \pa   \cZ  |   <     R^{-2 - \al+ 3 \ep  }   e_k^{-1+ 5 \ep}  
\ee
If  $R=1$ these are  the domains  $\cR_k, \cR_k'$ we have been discussing.    Eventually large $R$ will be supplied by scaling.

Now  define   
\be   \hat  E^c_k(\cA, \cZ)   =   E^c_k(\cA, \cZ)     -     \frac12   \frac{ \de^2 E^c_k }{\de \cZ^2} \B( 0;   \cZ,  \cZ \B)
\ee
This inherits a  local  expansion   $\hat  E^c_k(\cA, \cZ)  =  \sum_X \hat E^c_k(X,  \cA,   \cZ)$ from the expansion for 
   $ E^c_k(X,  \cA,   \cZ)$.     We  study    $ \hat  E^c_k(\cA, \cZ)  $ postponing the treatment  of    the second derivative term.

\begin{lem}   \label{exact}
For  $\cA \in \frac12 \cR_k(R)$ and  $\cZ \in  \frac12 \cR'_k(R)$    there is a new localization   
\be
  \hat   E^c_k(\cA, \cZ)   =         \sum_{ X  \in \cD_k}    \tilde E^c_k(X,  \cA,   \cZ)  
      \ee
  where   $\tilde   E^c_k(X, \cA,  \cZ )$ is analytic in $\cA, \cZ$,
depends on the  fields  only   in   $X$,   satisfies   $\tilde  E_k^c(X, \cA, 0) =0$  and for a constant  $c \leq  1$ independent of 
all parameters
\be   \label{symphony2}
   |  \tilde E^c_k(X, \cA,  \cZ ) |      \leq  \one     R^{-  10/3  }   e_k^{\ep}    e^{-c(\ka -2 \ka_0 -4)   d_M(X)}  
  \ee
\end{lem} 
\bigskip

\rem   The  key  point is  that  the negative exponent   $10/3$ is greater than  $d=3$;  the specific value is not important.

\bigskip

\pr
Let  $\de$ be a fixed small positive number, say  $\de = \frac18$.   If  $d_M(X)  \geq   L R^{\de}   $  then   
\be         \label{logs} 
\begin{split}
 | E^c_k(X, \cA,  \cZ ) | \leq &    \one    e_k^{\ep}   e^{- \ka d_M(X) }   =  \one   e_k^{\ep}   e^{-  d_M(X) }     e^{- (\ka-1) d_M(X) } \\
\leq  &   \one   e_k^{\ep}   e^{-  LR^{\de}  }     e^{- (\ka-1) d_M(X) }  \leq    \one     R^{-  10/3  }   e_k^{\ep}    e^{-(\ka -1)   d_M(X)}   \\
\end{split}
\ee
If  $|t|  \leq  R^{1- 2 \ep}$ the   $t \cZ \in \frac12  \cR'_k$ and so      
\be      \frac12   \frac{ \de^2 E^c_k }{\de \cZ^2} \B(X,  0;   \cZ,  \cZ \B)
=     \frac{1}{2\pi i}   \int_{|t |= R^{1- \ep }  }  \frac{ dt } {t^3}  E^c_k(X, 0 ,   t \cZ ) 
\ee
 satisfies  a stronger bound than    (\ref{logs}).    Hence     $ \hat   E^c_k(X, \cA, \cZ)  $  satisfies the bound   (\ref{logs})  and  it qualifies as a contribution to    $ \tilde     E^c_k(X, \cA, \cZ)  $. 
Thus it    suffices to consider  $d_M(X)  <  L R^{\de}$ which we write as     $X  \in \cS(R)$

The first step  is to  regroup into terms with greater symmetry.   Again  let  $\square_z$  be    the  $M$-cubes centered on points  $z$  in  the  $M$-lattice and 
write  
\be   \label{babel}
 \sum_{X  \in \cS(R)}  \hat  E^c_k (X,\cA,   \cZ ) 
=  \sum_z  \sum_{X \in  \cS(R),   X \supset  \square_z}      \frac{1}{|X|_M}  \hat    E^c_k (X,\cA,   \cZ ) \ee
Let  $\cO_z$  be the group of all  lattice symmetries that  leave $z$ fixed.
Each  $X  \supset \square_z$  determines another  polymer  $X^{sum}_z$ which is symmetric  around $z$ by  taking 
\be
   X^{sym}_z   =  \bigcup_{r \in  \cO_z}  rX 
\ee
This has  $d_M(   X^{sym}_z  )  \leq   \one  d_M(X)  \leq   \one   L R^{\de}$.
We  group together   polymers  with the same symmetrization  and write 
\be   \label{babel2}
 \sum_{X  \in \cS(R)} \hat     E^c_k (X,\cA,   \cZ ) 
=  \sum_z  \sum_Y  \sum_{X \in  \cS(R),   X \supset  \square_z, X^{sym}_z = Y}      \frac{1}{|X|_M}  \hat   E^c_k (X,\cA,   \cZ ) 
\ee
Change the order of the outside sums and we get   $\sum_Y   \tilde   E_k(Y)$   where
\be    \label{raspberry}
\tilde    E_k(Y, \cA, \cZ)   =  
 \sum_{z:  rY =Y  \textrm{ for }   r  \in \cO_z}  \left(   \sum_{X \in  \cS(R),   X \supset  \square_z, X^{sym}_z = Y}  
     \frac{1}{|X|_M}  \hat     E^c_k (X,\cA,0,   \cZ )   \right) 
     \ee
This is zero unless  $Y$ is symmetric under some  $\cO_z$.    If $rY = Y$  for  $r \in \cO_z$ for some $z$,  then  $z$ is
unique and we have  $z = z(Y)$.   To see  this   we claim  that   $|Y|z   = \sum_{z' \in Y} z'$.    Indeed   on the one hand  $ \sum_{z' \in Y} z'-  |Y|z   $   is invariant  under   $\cO_z$.   On the other hand  since it can be written   as   $  \sum_{z' \in Y} (z'-z) $     it changes sign  under the  reflection 
  $r(z'-z)  = -(z'-z)$.  Thus it must be zero.    Thus outside sum in  (\ref{raspberry})   selects   $z =z(Y)$  and  we
have      
 \be    \label{ek}
 \tilde   E_k(Y, \cA, \cZ)   =      \sum_{X \in  \Om(Y)}  
     \frac{1}{|X|_M}  \hat  E^c_k (X,\cA,  \cZ ) \ee
where we  abbreviate 
\be   \Om(Y)   =   \{  X  \in \cD_k:    X \in  \cS(R),   X \supset  \square_{z(Y)} , X^{sym}_{z(Y)} = Y   \}   
\ee
For any unit lattice symmetry   $\Om(rY) = r \Om(Y)$  and so   $\tilde    E_k(Y, \cA, \cZ)  $ is still invariant

Pick a fixed symmetric  $Y$.  Since     $d_M( Y)  \leq     \one   L R^{\de} $  we have    $| Y|_M  \leq     \one   L R^{\de}    \leq     \one   L R $
and so  $Y$   is contained  in  some   $\square^{\nat}_R$.    
Hence  in  (\ref{ek})    we  can  replace    $\cA$ by $\cA'$ satisfying the conditions   (\ref{region3}).    
In each  term  $  \hat   E^c_k( X,  \cA',  \cZ ) $ contributing to this sum     expand   around  $\cA'=0,  \cZ=0$ taking account the  the function is even and that   $E^c_k( X,  \cA',  0 )=0  $ and that   the second derivative in $\cZ$ is zero.       We   find  for  $r \geq   1$   
      \be    
   \label{deft2}
\hat   E^c_k( X,  \cA',   \cZ) =          \frac{ \de^2 E^c_k }{\de \cA  \de \cZ} \B(X,  0;   \cA'  ,  \cZ \B)   
 +   \frac{1}{2\pi i}   \int_{|t |= r }    \frac{ dt } {t^4(t-1)}  E^c_k(X, t \cA' ,   t \cZ ) 
\end{equation}
In the last  term we can take    $r=  R$ and  then for  $|t| = R$ we have  that  $t \cA',  t\cZ$ satisfies the  $\cR_k, \cR_k'$ bounds.   Hence we  are in the domain of analyticity for    $ E^c_k(X,\cA, \cZ)$  and the 
the formula holds.     From the bound (\ref{basicbound})  on    $ E^c_k(X,\cA, \cZ)$ we  get that the last  term in  (\ref{deft2})   is bounded by  
$ R^{-4}    e_k^{\ep}     e^{- (\ka- \ka_0 -3) d_M(X) }$.
For  the first term  in  (\ref{deft2}) we have the following:

\begin{lem}  \label{nova}  Under the assumptions of lemma  \ref{exact},
$\B( \de^2 E^c_k /\de \cA  \de \cZ  \B)(X,   0 ;  \cA' ,   \cZ ) $    for  $X \in \Om (Y) $
can  be written as a finite sum of terms  which either do not contribute to the sum over $X$ in (\ref{ek})  or are bounded
  on the domain  (\ref{region3}), (\ref{symphony1})   by  
$\one   R ^{ -  10/3} e_k^{\ep}   e^{- (\ka- \ka_0 -3) d_M(X) }$
\end{lem} 
\bigskip

 Assuming the lemma,  $|\hat  E_k^c(X, \cA', \cZ)|  \leq    \one   R ^{ -  10/3} e_k^{\ep}   e^{- (\ka- \ka_0 -3) d_M(X) }$  and so 
  \be
 |\tilde    E_k(Y) |   \leq    \one  R ^{ -  10/3} e_k^{\ep} 
   \sum_{ X \subset  \Om(Y)  } e^{-  (\ka- \ka_0 -3) d_M(X) }
\ee
But  one can show  that   $d_M(X^{sym}_z)  \leq  | \cO_x | d_M(X)  $  where  $| \cO_x |  = \one$ is the number of elements in 
 $\cO_x$.  Then with  $c =   | \cO_x | ^{-1} $ we have   $d_M(X)  \geq  c  d_M(X^{sym}_z)$.   Hence          we  can extract  
 from the sum  $e^{-  c(\ka-2 \ka_0 -3) d_M(Y) }$ 
and    leave   
 \be
 |\tilde   E_k(Y) |   \leq    \one  R ^{ -  10/3} e_k^{\ep}  e^{-  c(\ka-2 \ka_0 -3) d_M(Y) }
   \sum_{ X \subset   Y  } e^{- \ka_0  d_M(X) }
\ee
The  sum over $X$  is   bounded  by   $\one |Y|_M  \leq  \one   e^{c d_M(Y)}$ and    so 
\be
 |\tilde   E_k(Y) |   \leq   \one     R ^{ -  10/3} e_k^{\ep}  e^{-  c(\ka-2 \ka_0 -4) d_M(Y) }
\ee
This completes the proof of  lemma \ref{exact},   except for 
 lemma \ref{nova}.
\bigskip

\pr  (lemma \ref{nova})
Expand  $\cA$ around  $z= z(Y)$:         
\be
  \cA'_{\nu}(x)    =   \cA'_{\nu}(z)  +  \sum_{\sigma}  (x-z)_{\si} ( \pa_{\si} \cA'_{\nu})(z)   +  \De_{ \nu}  (x,z)  \ee   
 As before the   constant   vector field $ \cA'_{\nu}(z) $  is pure gauge in $X$  and disappears.   Thus we have      
\be  \label{sounds}
\begin{split}
 \frac{\de^2 E^c_k }{  \de  \cA    \de \cZ} \B(X,0 ;     \cA', \cZ \B)
=  &  \frac{\de^2 E^c_k }{  \de  \cA   \de \cZ} \B(X,0 ;    (\cdot-z) \cdot  \pa \cA'(z), \cZ(z)   \B)  \\
+ &  \frac{\de^2 E^c_k }{  \de  \cA    \de \cZ} \B(X,0 ;     (\cdot-z) \cdot  \pa \cA'(z) ,   \cZ- \cZ(z)   \B)  \\
+  &  \frac{\de^2 E^c_k }{ \de  \cA    \de \cZ} \B(X,0 ;    \De(\cdot, z),   \cZ   \B) \\
\end{split}
\ee

We claim that    the first term in  (\ref{sounds})   gives zero when summed over $X$ in (\ref{ek}).
Writing $\cZ(z)  =  \sum_{\mu}    \cZ_{\mu}(z)e_{\mu}$  and  $  (x-z) \cdot  \pa \cA(z)  =
 \sum_{\si}  (x-z)_{\si} \cdot  \pa_{\si, \nu } \cA_{\nu}(z)e_{\nu}$ 
It suffices to show that  for any  $\mu, \nu, \si$ the following   sum vanishes:
 \be   \label{sundry}
  \sum_{X  \in  \Om(Y) }    \frac{1}{|X|_M} \frac{\de^2 E^c_k }{  \de  \cA    \de \cZ} \B(X,0 ;  (\cdot-z)_{\sigma} e_{\nu},   e_{\mu}    \B)   
\ee
Let  $r$  the reflection through the point $z$, so   $r(x-z)  =  - (x-z)$.  
Reflection through a unit lattice  point is a symmetry of the theory so 
\be
 \frac{\de^2 E^c_k }{  \de  \cA    \de \cZ} \B(X,0 ;   f, g  \B)  
=    \frac{\de^2 E^c_k }{  \de  \cA    \de \cZ} \B(rX,0 ;  f_{r}  ,  g_{r}  \B)  
\ee
 where (taking account $r^{-1} = r$)
 \be    
 \begin{split}
    (f_r)_{\mu} (x)  = & f_r ( [x, x + \eta e_{\mu} ])  =   f ( r[x, x + \eta e_{\mu} ]) 
    =   f([rx, rx - \eta e_{\mu}] )\\
      = &   -  f(  [rx - \eta e_{\mu},  rx]  ) 
    =  - f_{\mu} ( rx - \eta e_{\nu}  )   \\
  \end{split}  
  \ee  
Here  under reflection  $e_{\mu}$ goes to  $-e_{\mu}$   and    $(x-z)_{\sigma} e_{\nu} $  goes  to     $((x-z)_{\sigma}  + \eta \de_{\sigma \nu} ) e_{\nu}$.
Since also  $|rX|_M =  |X|_M$,    (\ref{sundry}) can be written
 \be   
 \begin{split} 
& -  \sum_{X  \in  \Om(Y) }  \frac{1}{|rX|_M}  \frac{\de^2 E^c_k }{  \de  \cA    \de \cZ} \B(rX,0 ;    (\cdot-z)_{\sigma} e_{\nu},   e_{\mu}     \B)      
 -  \sum_{X  \in  \Om(Y) }\frac{1}{|rX|_M}  \frac{\de^2 E^c_k }{  \de  \cA    \de \cZ} \B(rX,0 ;    \eta e_{\nu},   e_{\mu}   \B) \de_{\si \nu} \\
\end{split}
\ee
However the second term vanishes  since we have  gauge invariance in the first   slot  (the $\cA$ derivative) and the constant
vector field   $\eta e_{\nu}$ is  pure gauge.
In the first term   since  $r \in \cO_x$ we have  $r \Om(Y) = \Om(Y)$  and    summing    over  $rX$ here is the same as the sum over  $X$.   Hence the  first    term   is exactly minus  (\ref{sundry})
 and therefore  zero.
 
   For the second term   in  (\ref{sounds})  note that   since  $X \in \cS(R)$ 
  it  has a diameter smaller than   $M|X|_M \leq  \one    MLR^{\de}$.
Therefore    for  $x \in X$
  \be
  |\cZ(x)- \cZ(z) | \leq  \one MLR^{\de} \|\pa \cZ \|_{\infty}
   \leq  \one MLR^{\de} (R^{-2+ 2 \ep} e_k^{-1 + 4 \ep} ) \leq  \frac12  R^{-2+ \de + 2 \ep } e_k^{-1 + 3 \ep} 
  \ee
  Together with similar   bounds on the derivatives  this gives     $\cZ- \cZ(z)   \in   \frac12  R^{-2 + \de + 2 \ep } \cR'_k$.
  Also     
  \be     | (x-z) \cdot  \pa \cA'(z)  |  \leq  \one MLR^{\de}  \|\pa \cA' \|_{\infty}   
    \leq      \one MLR^{\de} ( R^{-2+ \ep}  e_k^{-1 + 2\ep} ) \leq  \frac12   R^{-2+ \de+ \ep}  e_k^{-1 + \ep} 
    \ee
  Together with similar bounds on   derivatives this implies that  
       $(\cdot-z) \cdot  \pa \cA  \in   \frac12 R^{-2 + \de +  \ep } \cR_k $.  Then by   the bound  (\ref{basicbound})  on   $E_k^c$ and   a Cauchy bound  
    \be
  \B|\frac{\de^2 E^c_k }{  \de  \cA    \de \cZ} \B(X,0 ;   (\cdot-z) \cdot  \pa \cA(z) ,    \cZ- \cZ(z)    \B)\B|
  \leq   \one    R^{-4  + 2\de  + 3 \ep}  e_k^{\ep}     e^{- (\ka- \ka_0 -3) d_M(X) }
  \ee
    which is more than enough.
  
For the third term   in  (\ref{sounds}) 
 we  write
 \be     \De_{\nu} (x,z)  =     \int_{\Ga}     \B( \pa  \cA_{\nu}(y)  -\pa \cA_{\nu}(z)  \B)  \cdot  dy
\ee
where  $\Ga  \in G(z,x)$ is any of the standard paths from $z$  to $x$.
Then  
\be    | \De_{\nu} (x,z) |  \leq    \one ( MLR^{\de})^{1 + \al}   \|\de_{\al}\pa \cA \|_{\infty}   
    \leq      \one   ( MLR^{\de})^{1 + \al}  ( R^{-2- \al+ 2 \ep}  e_k^{-1 + 3\ep} ) \leq  \frac12   R^{-2- \al+ 2\de + 2 \ep}  e_k^{-1 + \ep} 
 \ee
Together with similar bounds  on the derivatives   this implies that  
   $  \De   \in     \frac12    R ^{ -2 - \al+ 2 \de + 2 \ep}  \cR_k   $.
   Then    by  a 
  Cauchy bound
  \be   |  \frac{\de E^c_k }{  \de  \cA    \de \cZ} \B(X,0 ;   \De( \cdot,  z)  ,   \cZ \B)|  \leq   \one
   R ^{ -3 - \al + 2 \de  + 4 \ep} e_k^{\ep}   e^{- (\ka- \ka_0 -3) d_M(X) } 
     \ee
 This is sufficient since with $\al > \frac{7}{12}$  and $\de = \frac18$  and  $\ep$ sufficiently small  we have    $ -3 - \al + 2 \de + 4 \ep  \geq  - \frac {10}{3}$.

\subsection{resummation}

Combining (\ref{birthday})  and   (\ref{unbirthday2})
we  have   
\be   \label{fun}
  \frac{Z_k (\cA+ \cZ)}{Z_k(\cA)}   =   \exp \B(   E^z_k(\cA, \cZ  ) \B)
\hs     E^z_k(\cA, \cZ  )=       \sum_{j=0}^{k-1}    E^c_j( \cA_{L^{k-j}},  \cZ_{L^{k-j}}  )
\ee

\begin{lem}      \label{pinky}
$E^z_k$ has the  partial     local expansion   
\be       
E^z_k (\cA, \cZ)  =    \frac12     \frac{\de^2  E_k ^z  }{ \de  \cZ^2} \B( 0; \cZ,  \cZ  \B)   +      \sum_{X \in \cD_k}   \tilde E^z_k(X,       \cA,   \cZ )
\ee
where    $ \tilde   E^z_k(X,       \cA,   \cZ ) $ depends on the fields  only in  $X$,
is  analytic in   $\cA  \in \frac12  \cR_k$ and  $\cZ  \in \frac12  \cR_k'$        and satisfies  
there   
\be  \B|   \tilde   E^z_k(X,       \cA,   \cZ ) \B|
\leq  \one    e_k^{\ep} e^{- \ka d_M(X)}  
\ee
\end{lem}
\bigskip

\rem    The term  $\frac12   (\de^2  E_k ^z  / \de  \cZ^2) ( 0; \cZ,  \cZ  )$
is localized in the next section.
 \bigskip

\pr 
  In  (\ref{fun}) we     insert the representation  of  $ E^c_j( \cA,  \cZ  )
$    from     lemma \ref{exact}.    Since       
  \be   
  \begin{split} 
&   \sum_{j=0}^{k-1}    \frac{\de^2  E_j ^c  }{ \de  \cZ^2} \B( 0; \cZ_{L^{k-j}},  \cZ_{L^{k-j}}  \B)   
=      \sum_{j=0}^{k-1}    \B[  \frac{ \pa^2E_j ^c  }{ \pa  t \pa  s}   \B(0, t\cZ_{L^{k-j}} + s  \cZ_{L^{k-j}}  \B)    \B]_{t=s=0} \\
&=      \B[   \frac{ \pa^2  E_k ^z }{ \pa  t \pa  s}  \B(0, t\cZ + s  \cZ  \B)    \B]_{t=s=0}
=       \frac{\de^2  E_k ^z  }{ \de  \cZ^2} \B( 0; \cZ,  \cZ  \B)   \\
\end{split}
\ee  
  this gives      
\be    \label{owen}
E^z_k(\cA, \cZ)     
=     \frac12     \frac{\de^2  E_k ^z  }{ \de  \cZ^2} \B( 0; \cZ,  \cZ  \B) +  \sum_{j=0}^{k-1}   \sum_{X  \in \cD_j} \tilde  E^c_j(X, \cA_{L^{k-j}},  \cZ_{L^{k-j}}  )  
\ee

As  in lemma  \ref{sunfish}    our assumption  $\cA \in \frac 12 \cR_k$  implies  that  in each  $L^{k-j}  \square^{\nat}$ we have      $ \cA_{L^{k-j}} \sim    \cA_{0, L^{k-j}} +    \cA_{1, L^{k-j}} $
 where   $ \cA_{0, L^{k-j}} $ is real and satisfies    
 \be   \label{region4} 
 \begin{split}
&|  \cA_{0, L^{k-j}}|  <   \frac14(  L^{k-j}  )  ^{-1+ \ep } e_k^{-1+ \ep}  
      \hs     |\pa  \cA_{0, L^{k-j}}|  < \frac14   (  L^{k-j}  ) ^{-2+ \ep } e_k^{-1+ 2\ep}   \\
&     \hs     | \de_{\al}  \pa   \cA_{0, L^{k-j}}  |   <   \frac14  (  L^{k-j}  ) ^{-2 - \al + 2 \ep}   e_k^{-1+ 3 \ep}   \\
\end{split}
\ee
and    $ \cA_{1, L^{k-j}} $ is complex and satisfies the same bounds.    
 Therefore  $\cA_{L^{k-j}}    \in  \frac12   \cR_j (R)$  with   $R= L^{k-j} $.        
 Similarly  our assumption that  $\cZ_k \in \frac12 \cR_k'$ implies that     $\cZ_{L^{k-j}}     \in  \frac12 \cR'_j (R)$  with   $R= L^{k-j} $.  
Thus  we  can  apply  lemma  \ref{exact}  and  
 and obtain   (using also   $e_j^{\ep} < e_k^{\ep}$ ) 
    \be  \label{back}
| \tilde  E^c_j(X, \cA_{L^{k-j}},  \cZ_{L^{k-j}}  )|  \leq     L^{- \frac{10}{3}(k-j) }      e_k^{\ep}    e^{ - c( \ka- 2\ka_0-4)  d_M(X)  } 
\ee

Now  we  
 reblock.   The sum  in  (\ref{owen})   is now written  in the required form   $\sum_{Y \in \cD_k}   \tilde  E^z_k  (Y )$   where       
  for  $Y \in \cD_k$
 \be   
 \begin{split}
   \label{spitfire}   \tilde  E^z_k  (Y, \cA,  \cZ  )  =  &  \sum_{j=0}^{k-1}   ( \cB^{(k-j)}  \tilde E^c_j )(Y, \cA,  \cZ  )  \\
( \cB^{(k-j)} \tilde E^c_j)(Y, \cA,  \cZ  )
=  &  \sum_{\bar X^{(k-j)}  = L^{k-j} Y }     \tilde  E^c_j(X, \cA_{L^{k-j}},  \cZ_{L^{k-j}}  ) \\
\end{split}
\ee
Here      $\bar X^{(k-j)}$ is the  union of  all   $L^{k-j}M$ blocks  intersecting  $X$.   A minimal spanning   tree on the $M$ blocks in    $X$  is also  a spanning
tree on  the  $L^{k-j}M$ blocks in      $\bar X^{(k-j)}$.  Therefore    $M d_M(X)  \geq   L^{k-j} M  d_{L^{k-j} M}  ( L^{k-j} Y )$   or
just     $ d_M(X)  \geq   L^{k-j}  d_{ M}  ( Y )$.     Then  the decay factor  in  (\ref{back}) satisfies 
\be  
  e^{ - c( \ka- 2\ka_0-4)  d_M(X)  }  \leq    e^{ - L^{k-j}(c( \ka- 2\ka_0-4) - \ka_0)  d_M(Y)  } e^{ - \ka_0 d_M(X)  }
\leq    e^{-( \ka+1)   d_M(Y)  } e^{ - \ka_0 d_M(X)  }
\ee
the last   since  $k-j \geq 1$   and    for  $L$ sufficiently large  $   L(c( \ka- 2\ka_0-4) - \ka_0)  \geq  \ka +1$.  
The sum  over $X$  in   (\ref{spitfire})    is estimated by  
\be    \sum_{X \subset   L^{k-j} Y}   e^{ - \ka_0 d_M(X)  }   \leq  \one | L^{k-j} Y|_M  =   \one   L^{3(k-j)}| Y|_M 
\leq  \one    L^{3(k-j)} e^{d_M(Y) }  
\ee 
Hence we have   
 \be  
|( \cB^{(k-j)} \tilde E^c_j)(Y, \cA,  \cZ  )|      \leq   \one     L^{- \frac13(k-j)}    e_k^{ \ep}      e^{- \ka d_M(Y)  }  
\ee
The  factor  $ L^{- \frac13(k-j)}$  ensures the convergence of the sum over $j$  and we have the required estimate  
$   |\tilde  E^z_k  (Y, \cA,  \cZ  ) |      \leq  \one    e_k^{ \ep}          e^{- \ka d_M(Y) }  $
\bigskip

\subsection{photon self-energy}

We   treat the term   $ \frac12 (\de^2  E_k ^z  / \de  \cZ^2) (  0,0; \cZ,  \cZ  ) $ omitted until now.  
The background field is now zero so we shorten the notation to  $G_k  \equiv G_k(0)$ and  $U_k(\cZ)  \equiv  U_k(0, \cZ)$
and  $E_k^z(\cZ )   =   E_k^z(0, \cZ)$.  
  Since $ G_k(\cZ)  = G_k  + G_k U_k(\cZ) G_k(\cZ)$ we have  
 \be 
\begin{split}
\frac { \det   G_k(  \cZ) }    { \det   G_k }   
=  &  \det  \B (  I   -   G_k U_k ( \cZ) \B )^{-1} 
=    \exp  \B(   \sum_{n=1}^{\infty}  \frac{1}{n} \Tr \B(  G_k  U_k(  \cZ)  \B)^n  \B)\\  
\end{split}
\ee
The function  $\exp  (   E^z_k (\cZ)   )$ is the square root of the last expression  
so    
\be  \label{ort}
 E^z_k(\cZ)    =  \frac12  \sum_{n=1}^{\infty}  \frac{1}{n} \Tr \B( ( G_k  U_k( \cZ)  )^n  \B)
\ee

The derivative  $\frac12    (\de^2  E_k ^z  / \de  \cZ^2) (  0; \cZ,  \cZ  ) $ is a symmetric quadratic form in $\cZ$.  It is called the   photon self-energy and   denoted $\Pi_k$.   Thus 
\be  
   < \cZ,  \Pi_k  \cZ>  =   \frac12 \frac{ \de^2 E^z_k }{  \de  \cZ^2}( 0  ; \cZ, \cZ)  
\ee
Taking account  that  $U_k(0) =0$  we  compute it from   (\ref{ort}) as  
\be   
\begin{split}
 < \cZ,  \Pi_k  \cZ> =& \frac14   \Tr   \B(  \frac{ \de^2  U_k }{ \de  \cZ^2 }( 0  ; \cZ, \cZ  ) G_k\B)   
+  \frac14   
  \Tr   \B(  \frac{ \de  U_k   }{  \de  \cZ}( 0  ;  \cZ)  G_k\frac{ \de  U_k }{  \de  \cZ}( 0  ;  \cZ)  G_k\B)   \\
\end{split}
\ee

Now        $\det G_k (\cZ)$ is gauge invariant,   and it follows that  
  $E^z_k (\cZ)$ is gauge invariant.   So    $ < f_1,  \Pi_k  f_2>  =    \frac12 (\de^2 E^z_k /  \de  \cZ^2 ) (  0; f_1,f_2 ) $
  is gauge invariant in either variable.   This implies the Ward identity      
\be  
<    \pa \la,  \Pi_k   f> =  \frac12  < f,  \Pi_k    \pa  \la >   = 0   \hs   \textrm{ or  }    \hs     \pa ^T \Pi_k  =  \Pi_k    \pa  =0    
\ee

Our goal is to prove the following  local decomposition  (which is not gauge invariant).

\begin{lem}  \label{smooth}
\be     < \cZ,  \Pi_k  \cZ>   =  \sum_X  E_k^{\pi} (X, \cZ)  \ee
where   $E^{\pi}_k (X, \cZ)$ only depends on  $\cZ$ in  $X$,  is invariant under unit   lattice symmetries,   and  
\be 
  |E_k^{\pi}(X, \cZ)|  \leq  
   e_k^{2- \ep  } \B( \|\cZ \|_{\infty}  +   \| \pa \cZ \|_{\infty}    +   \|\de_{\al}  \pa \cZ \|_{\infty}  \B)^2    e^{- \ka d_M(X) } 
\ee
and so  for   $\cZ  \in  \frac12  \cR_k'$
\be   
 |E_k^{\pi}(X, \cZ)|  \leq  e_k^{5 \ep}    e^{- \ka d_M(X) } 
\ee
\end{lem}

\subsubsection{estimates}
We  collect some estimates we will need.   It is now more convenient to use pointwise estimates than  the local $L^{\infty}$
estimates employed earlier.  
We  define  on  $\tk$
\be
  d'(x,y)   =   \begin{cases}  d(x,y)   &    x \neq  y   \\
L^k  &   x=y    \\
\end{cases}
\ee
This is not a true metric since  $d'(x,x) \neq 0$,   but it does satisfy the triangle inequality. 

\begin{lem}
\be      \label{shock}
\begin{split}
    |G_k(  x,y) |    \leq  &           C  d'(x,y)^{-1} e^{ -  \ga  d(x,y)  } \\
      |  \pa_{\mu}    G_k(  x,y) |    \leq  &           C  d'(x,y)^{-2}e^{ -  \ga  d(x,y)  } \\
        |(  \pa_{\mu}    G_k  \pa^T_{\nu})(  x,y) |    \leq  &           C  d'(x,y)^{-3}e^{ -  \ga  d(x,y)  } \\
 \end{split}    
 \ee
\end{lem}
\bigskip

\pr   
We  start with the representation on $\tk$  (see  \cite{Bal82a}, \cite{Dim11} ) 
 \be  \label{yin}
    G_k(  x,y)  =  \sum_{j=0}^{k-1} L^{k-j} \tilde  C_j ( L^{k-j} x,L^{k-j}  y)     
 \ee
where     on  $\bbT^{-j}_{N-j}$
 \be      \tilde  C_j( x,y)  =     ( \cH_j   C_j \cH^T_j   )  (x,y)
 \ee
 and   $\tilde C_0 =  C_0  =  ( - \De  + aL^{-2}Q^TQ)^{-1}$.   
Now  $C_k,  \cH_j$,  and  $\pa  \cH_j$  all have exponential decay and  no short distance singularity.   They satisfy  (see  Appendix  D in  \cite{Dim11};   $L^2$ estimates  suffice  for $\tilde C_0 = C_0$ )   
 \be 
 \begin{split}
   |  \tilde  C_k(x,y)  |, \   | \pa_{\mu}     \tilde  C_k( x,y)  |, \  | ( \pa_{\mu}  \tilde C_k  \pa^T_{\nu}) (  x,y) |\        \leq   &  C e^{ -  \ga  d(x,y)  }   \\     
\end{split}
\ee    
Thus  we  have  
\be      \label{spun}
| G_k(  x,y) |  \leq    C  \sum_{j=0}^{k-1} L^{k-j}   e^{ -  \ga L^{k-j} d(x,y)  } 
 = C   \sum_{ \ell =1}^k     L^{\ell}   e^{ -  \ga L^{\ell } d(x,y)  } 
\ee
Now we split into  three cases.    For   $x=y$ we have    $| G_k(  x,x)  |  \leq  C L^k =  Cd(x,x)^{-1} $.   
   For    $0 <   d(x,y)  \leq  1$ we  need a bound  $C  d(x,y)^{-1}$.  We  choose  $0 \leq  \ell^* \leq  k-1$  so that  
$    L^{\ell^*}  \leq   d(x,y )^{-1}  \leq  L^{\ell^* +1}  $
and  break the sum (\ref{spun})    into  a sum from $1$ to $\ell^*$  (empty if  $\ell^*=0$)  and a sum from $\ell^* +1$ to $k$.   The first sum is dominated by    
\be
     C   \sum_{ \ell =1}^{\ell^*}    L^{\ell}   \leq   C L^{\ell^*}   \leq   C d(x,y)^{-1}  
\ee
The second sum is dominated by  
\be  
\begin{split}
 & C   \sum_{ \ell =\ell^*+1}^{\infty}    L^{\ell}  e^{ -  \ga L^{\ell }d(x,y)} 
  =  C   \sum_{j=1}^{\infty}   L^{\ell^* +j  }   e^{ -  \ga L^{\ell^*+j }d(x,y)}  \\
 &  \leq   C L^{\ell^*}    \sum_{j=1}^{\infty}   L^{j  }   e^{ -  \ga L^{j-1 }}  \leq     C L^{\ell^*}   \leq  C d(x,y)^{-1}\\
 \end{split}
  \ee
   For   $d(x,y)  \geq  1$ we have  
\be  
\begin{split}
& | d(x,y) G_k(  x,y) |  \leq     C   \sum_{ \ell =1}^k     L^{\ell}  d(x,y) e^{ -  \ga L^{\ell } d(x,y)  }  \\
&  \leq    C   \sum_{ \ell =1}^k      e^{ - \frac12 \ga L^{\ell } d(x,y)  }  \leq   C  e^{ - \frac12 \ga Ld(x,y)  } \sum_{\ell=1}^ke^{-\frac12 L^{\ell}}  \leq   
   C  e^{ - \ga d(x,y)  }   \\
\end{split}
\ee
For the derivatives  we  argue similarly  starting with expressions like
 \be
  \pa_{\mu}   G_k(  x,y)  =  \sum_{j=0}^{k-1} L^{2(k-j)} ( \pa_{\mu}  \tilde  C_j )( L^{k-j} x,L^{k-j}  y)     
 \ee 
This completes the proof.  
\bigskip

Next  consider the operator  $U_k( \cZ)$  which     
 we  divide  as   $U_k(\cZ)   =    U_k^s(\cZ)   +  U_k^q(\cZ) $ where  
\be
\begin{split} 
   U_k^s(\cZ)    =   &  - \De_{\cZ}  +   \De_{0} \\
   U_k^q(\cZ)   = &a_k \B( (Q^T_kQ_k)(\cZ) -   (Q^T_kQ_k)(0)\B)  \\
\end{split}
\ee
Here     $U_k^s $  are the standard   pieces  and $U_k^q$  are the pieces involving   averaging operators  $\cQ_k$.

To analyze the contribution of  $U_k^q$
to   $\Pi_k$  we will need  the following   

\begin{lem}      \label{explicit3}
\be   \begin{split}
\B|  \B( \frac  { \de  U_k^q}{ \de  \cZ } (0; \cZ  )G_k\B)(x,y)   \B|
 \leq    &
 C  e_k \|\cZ \|_{\infty} e^{ -  \ga  d(x,y)  }   \\
 \B|  \B(\frac  { \de ^2 U_k^q}{ \de  \cZ^2 }(0;\cZ,\cZ) G_k\B) (x,y)     \B|
\leq   &
   C  e_k^2  \|\cZ \|_{\infty}  e^{ -  \ga  d(x,y)  }   \\
\end{split}
\ee
\end{lem}
\bigskip

\pr Recall that    $\De_z$ is
the unit cube centered on  $z \in \bbT^0_{N-k}$.   The operator  $(Q^T_kQ_k)(\cZ)$    is     local    and  has the kernel
    \be   
    (Q^T_kQ_k)(\cZ; x,y )  =  \begin{cases}
 \exp \B( - qe_k\eta ( \tau_k\cZ)(z,x)    + qe_k\eta ( \tau_k\cZ)(z,y)  \B)   &    \textit{  if   }   x,y \in \De_z \\
 0   &   \textit{   otherwise}
 \end{cases}
 \ee  
It is    analytic and bounded by  $\one$     for  $\|\cZ\|_{\infty} \leq  e_k^{-1}$.   
Then  the  kernel     $\B((Q^T_kQ_k)(\cZ)G_k\B)(x,y)  $  is  analytic     for  $\|\cZ\|_{\infty} \leq  e_k^{-1}$   and if     $x \in \De_z$ then     
  by   (\ref{shock})
\be   
|(Q^T_kQ_k)(\cZ)G_k)(x,y) |   \leq  C  \int_{ \De_z }    d'(x',y)^{-1}   e^{-\ga d(x',y) }\   dx'    \leq    C  e^{-\ga d(x,y) }  
\ee
See appendix \ref{dprime}  for the integrability of   $ d'(x',y)^{-1} $.
Then  for   $\|\cZ\|_{\infty}   \leq  1  $  we have  
\be   
 \B( \frac  { \de  U_k^q}{ \de  \cZ } (0; \cZ  )G_k\B)(x,y)  = a_k   \B( \frac  { \de   (Q^T_kQ_k)}{ \de  \cZ } (0; \cZ  )G_k\B)(x,y)  
=   \frac{a_k}{2 \pi i}   \int_{|t|  = e_k^{-1} }  \frac {dt}{t^2} \B((Q^T_kQ_k)(t\cZ)G_k\B)(x,y)   
 \ee
This  leads to the bound     for   $\|\cZ\|_{\infty}   \leq  1  $ 
\be    
\left| \B( \frac  { \de  U_k^q}{ \de  \cZ } (0; \cZ  )G_k\B)(x,y) \right|   \leq     Ce_k  e^{-\ga d(x,y) }  
     \ee
This is sufficient since   $ ( \de  U_k^q/\de  \cZ ) (0; \cZ  )G_k)(x,y)$ is linear in $\cZ$.
The proof for the second derivative is similar.

\subsubsection{removal of averaging operators from interaction}     

In the expression  for   $< \cZ,  \Pi_k  \cZ> $ we  insert the decomposition  $U_k =  U^s_k  + U_k^q$.
Let  $\Pi_k^q$  be the part with only    $U_k^q$.  We estimate it first.  It is written 
\be   \label{pause1a}
\begin{split}
< \cZ,  \Pi^q_k  \cZ>   =& \frac14   \Tr   \B(  \frac{ \de^2  U^q_k }{ \de  \cZ^2 }( 0  ; \cZ, \cZ  ) G_k\B)  
+  \frac14   
  \Tr   \B(  \frac{ \de  U^q_k   }{  \de  \cZ}( 0  ;  \cZ)  G_k\frac{ \de  U^q_k }{  \de  \cZ}( 0  ;  \cZ)  G_k\B)   \\    
\end{split}
\ee   
Taking account that the trace over charge indices gives a factor of 2 this can   be written   
\be   \label{pause2}
\begin{split}
< \cZ,  \Pi^q_k  \cZ>   = &  \frac12  \int dx   \B(  \frac{ \de^2  U^q_k }{ \de  \cZ^2 }( 0  ; \cZ, \cZ  ) G_k\B) (x,x)  \\
+&  \frac12   \int dx dy \    
  \B(  \frac{ \de  U^q_k   }{  \de  \cZ}( 0  ;  \cZ) G_k\B)(x,y)\B( \frac{ \de  U^q_k }{  \de  \cZ}( 0  ;  \cZ)    G_k\B) (y,x)   \\    
\end{split}
\ee

\begin{lem}  \label{smoothq}
\be  
   < \cZ,  \Pi^q_k  \cZ>   =  \sum_X  E^q_k (X, \cZ) 
 \ee
where   
\be   |E^q _k(X, \cZ)|  \leq   e_k^{2- \ep  }     \|\cZ \|_{\infty} ^2    e^{- \ka d_M(X) } 
\ee
\end{lem}
\bigskip

\pr   The  estimates  of lemma  \ref{explicit3}   show that there is no short distance singularity, and that
 \be   | < \cZ,  \Pi^q_k  \cZ> |        \leq    \B(  C e_k^2  \int   dx
 +    C e_k^2    \int       e^{ - 2 \ga  d(x,y)  }    dx dy\B)  \|\cZ\|^2_{\infty}
 \ee 
 This    bound is proportional to the volume.

We  need to write $< \cZ,  \Pi^q_k  \cZ> $     as a sum of local pieces,    and  do it a way that preserves  invariance under lattice symmetries.  This is best accomplished by  regarding   $\cZ$ is a function on bonds.   We have   
\be    \label{oliver} 
  < \cZ,  \Pi^q_k  \cZ>=     \int \cZ(b)  \Pi^s (  b,b' )  \cZ(b')   db\ db'\   
\ee
where the integral is over oriented bonds  and      $\int  f(b)  db  \equiv   \sum_{\mu} \int  f\big([x, x+ \eta e_{\mu}]\big)\  dx$. 
Alternatively we  take an extended   definition of     $\cZ(b),   \Pi^s (  b,b' )$ to all bonds   with   $\cZ(x,x') = - \cZ(x',x)$, etc.   
Then  a  representation like  (\ref{oliver}) stills holds, but ranging over all bonds  and with an extra factor of $\frac12$ for 
each integral.    In this representation  the invariance    $ < \cZ_r,  \Pi^q_k  \cZ_r>= < \cZ,  \Pi^q_k  \cZ>$  under $\tz$ lattice symmetries  $r$    implies that    
\be  
   \Pi^q_k ( b,  b'  )   =    \Pi^q_k ( rb, rb'  )
     \ee
Again let  $\De_z$  be the  unit  cube centered on the   unit  lattice point  $z \in  \bbT^0_{N-k}$.
Define   a modified characteristic function $\chi_{z}$   on all bonds by   
\be
\chi_{z}(b)
  =  \begin{cases}   1  &   \textrm{ if }   b \subset  \De_z   \\
\frac12   &    \textrm  {if }     b \cap   \De_z  \neq  \emptyset,   b \cap   \De^c_z  \neq  \emptyset   \\ 
0  &    \textrm  {if }     b   \cap    \De_z =  \emptyset  
\end{cases}
\ee
Then     $\sum_z  \chi_z  = 1$     and     $(\chi_z)_r (b) =  \chi_z(r^{-1}b)  =  \chi_{rz}(b)  $.   
We       make the decomposition
\be    \label{gum}
  < \cZ,  \Pi^q_k  \cZ>=   \sum_{z,w}  <( \chi_z\cZ),  \Pi^q_k  (\chi_w \cZ)> 
\ee
The   characteristic functions are insensitive to orientation,  so we  can evaluated this   with    either  oriented or unoriented bonds.
We   have        $<( \chi_{rz}\cZ_r),  \Pi^q_k  (\chi_{rw} \cZ_r)>  =<( \chi_{z}\cZ),  \Pi^q_k  (\chi_{w} \cZ)>$  and  
\be   \label{pause3}
\begin{split}
 <( \chi_z\cZ),  \Pi^q_k  (\chi_{w} \cZ)>   = & \frac12  \int dx     \B(  \frac{ \de^2  U^q_k }{ \de  \cZ^2 }( 0  ; \chi_z\cZ, \chi_w\cZ ) G_k\B) (x,x)  \\
+&  \frac12   \int dx dy \      \B(  \frac{ \de  U^q_k   }{  \de  \cZ}( 0  ; \chi_z\cZ) G_k\B)(x,y)\B( \frac{ \de  U^q_k }{  \de  \cZ}( 0  ; \chi_w\cZ)    G_k\B) (y,x)     \\    
\end{split}
\ee

Again  we  estimate  using the  bounds   of lemma  \ref{explicit3} .    Because  $U_k(\cZ)$ and its derivatives  are local 
operators  the integrals over $x,y$  are  restricted to  the immediate neighborhood of  $\De_z, \De_w$, denoted  $\De^*_z,  \De^*_w$.
Thus we  have    
\be
\B| <( \chi_z\cZ),  \Pi^q_k  (\chi_{w} \cZ)> \B|    \leq    \B(  C e_k^2  \int_{\De^*_z \cap  \De^*_w}  dx
 +    C e_k^2  \int_{\De^*_z   \times  \De^*_w}       e^{ - 2 \ga  d(x,y)  }   dx dy    \B)    \|\cZ\|^2_{\infty}
 \ee 
 The first term only contributes when $\De_z, \De_w$ touch and in the  second term we use   
 Now we  use  $d(x,y)  \geq   d(z,w)-   \one $.  Hence   
\be   \label{gum2}
   | <( \chi_z\cZ),  \Pi^q_k  (\chi_{w} \cZ)>|  \leq   C e_k^2 \| \cZ \|^2_{\infty}  e^{- 2\ga  d(z,w)   } 
\ee
The expansion   (\ref{gum})     localizes  the expression, but not yet in polymers  since  $\De_z \cup \De_w$ is generally not
connected.

For     any  unit lattice points  $z,w$   let   
\be 
\begin{split}
    X_{zw}   =&   \textrm{  the smallest polymer containing }  \De_x^*  \textrm{ for  all  } x \\
   &  \textrm{   in any of the  paths  } \Ga^{\pi}(z,w)  \textrm{    from  } z   \textrm{ to }  w     \\
\end{split}   
\ee
It is roughly the thickened edges  of a cube with  $zw$ on opposite corners.  
Then we have the required   $  < \cZ,  \Pi^q_k  \cZ>   =  \sum_X  E^q_k (X, \cZ) $
where 
\be  
 E^q_k (X, \cZ)  =      \sum_{z,w:  X_{zw}  =X}  <( \chi_z\cZ),  \Pi^q_k  (\chi_{w} \cZ)> 
 \ee
 This satisfies   $ E^q_k (rX, \cZ_r)  = E^q_k (X, \cZ)  $
 and from  (\ref{gum2})
 \be 
   | E^q_k (X, \cZ) | \leq     C e_k^2   \| \cZ \|^2_{\infty}  \sum_{z,w:  X_{zw}  =X} e^{- 2\ga  d(z,w) }   
 \ee
 But $Md_M(X)  \leq  c  d(z,w)$ for some  $c = \one$.  Hence 
   $2 \ga d(z,w)  \geq   2\ga c^{-1} M d_M(X)  \geq   \ka d_M(X)$ for $M$  sufficiently large.  Also the number of points  $z,w$
with  $X_{zw} =X$ is bounded by   $\one  M^6$.   
Hence for  $e_k$ sufficiently small  
 \be   | E^q_k (X, \cZ) | \leq      CM^6 e_k^2  \| \cZ \|^2_{\infty}    e^{- \ka d_M(X) } 
  \leq    e_k^{2-\ep}  \| \cZ \|^2_{\infty}    e^{- \ka d_M(X) } 
 \ee
This completes the proof of the lemma.
\bigskip

There  is also  a term  one  $U_k^q$  and  one of  $U_k^s$.   It has the form
\be   \label{pause8}
< \cZ,  \Pi^{qs} _k  \cZ>   =   \frac14
  \Tr   \B(  \frac{ \de  U^q_k   }{  \de  \cZ}( 0  ;  \cZ)  G_k\frac{ \de  U^s_k }{  \de  \cZ}( 0  ;  \cZ)  G_k\B)   \\    
\ee   
This has integrable  short distanced   singularities and can be treated using  the estimate   just established   on $( \de  U^q_k  /  \de  \cZ)( 0  ;  \cZ)  G_k$
and   estimates on   $(\de  U^s_k /  \de  \cZ) ( 0  ;  \cZ)  G_k$ from the next section.  We omit the details.

 \subsubsection{an explicit representation} 
 
 Now we  are reduced to an expression with standard potential but still non-standard  propagators.   It is partially 
 standard.   It is 
 \be   \label{pause1}
\begin{split}
  < \cZ,  \Pi^p_k  \cZ> 
  \equiv   & \frac14   \Tr   \B(  \frac{ \de^2  U^s_k }{ \de  \cZ^2 }(  0  ; \cZ, \cZ  )  G_k\B)   
+  \frac14   
  \Tr   \B(  \frac{ \de  U^s_k   }{  \de  \cZ}(  0  ;  \cZ)  G_k\frac{ \de  U^s_k }{  \de  \cZ}(  0  ;  \cZ)  G_k\B)   \\ 
\end{split}
\ee

\begin{lem}    
\be    \label{slick2}   
 < \cZ,  \Pi^p_k  \cZ> =  \sum_{\mu \nu} \int     \cZ_{\mu} (x)   \Pi^p_{k, \mu \nu}(x,y)   \cZ_{\nu}(y) \ dx dy
\ee 
where  
\be     \Pi^p_{k,\mu \nu}  (x,y)   =    \de_{\mu \nu} \de (x-y)    \Pi^{p, (0)}_{k, \mu}(x)   
+       \Pi^{p, (1)} _{k,\mu \nu}  (x,y) 
\ee
and   
\be
\begin{split}
 \Pi^{p, (0)}_{k, \mu}(x)    = &    e_k^2  G_k(x+ \eta e_{\mu},x+  \eta e_{\mu})   -  e_k^2 \eta   \pa_{\mu} G_k(x + \eta e_{\mu},x)   \\
 \Pi^{p, (1)} _{k,\mu \nu}  (x,y) 
=  & - e_k^2   (\pa_{\mu} G_k ) (x,y+ \eta e_{\nu}  )  
 (\pa_{\nu} G_k ) (y,x +  \eta e_{\mu} )     \\
 &   +         e_k^2       (\pa_{\mu} G_k  \pa_{\nu}^T ) (x,y  )  
 G_k  (y + \eta e_{\nu},x +  \eta e_{\mu} )    
\end{split}
\ee
\end{lem} 

\rem
Note that     $ \Pi^{p, (0)}_{k, \mu}(x)  =  \cO(\eta^{-1})$
and   
 \be 
 \label{piestimate}
 |\Pi^{p, (1)} _{k,\mu \nu}  (x,y)|   \leq   C e_k^2 d'(x,y)^{-4} e^{-2\ga d(x,y)}
 \ee
 There is a linear ultraviolet  divergence   which must be canceled.
\bigskip

\pr   
Define   an operator  $\cZ^{(1)}_{\mu}$ by   
\be
 ( \cZ^{(1)}_{\mu} f )(x)
=  \frac{d}{dt} \B[    ( \pa_{t\cZ,  \mu} f)(x) \B ]_{t=0}   =    qe_k  \cZ_{\mu}(x) f (x + \eta e_{\mu} )
\ee
Then       
\be 
\begin{split}
 & <f,    \frac{ \de  U^s_k   }{  \de  \cZ}(  0  ;  \cZ)   f> 
=    \frac{d}{dt} \B[  <f,  U^s_k (  t\cZ  ) f> \B] _{t=0}      \\ 
=  &    \frac{d}{dt} \B[  < \pa_{t\cZ} f,   \pa_{t\cZ}  f> \B] _{t=0}  
=     \sum_{\mu}    <   \cZ^{(1)}_{\mu}  f,   \pa_{\mu}  f>  +   < \pa_{\mu}  f,     \cZ ^{(1)}_{\mu}  f> \\
\end{split}
\ee 
 which is equivalent to      the operator identity 
 \be     \frac{ \de  U^s_k   }{  \de  \cZ}(  0  ;  \cZ)  =  \sum_{\mu}      \cZ^{(1),T}_{\mu}  \pa_{\mu}   +    \pa^T_{\mu}      \cZ ^{(1)}_{\mu}
 \ee
 Also  define   
 \be
( \cZ^{(2)}_{\mu} f )(x)=      \frac{d^2}{dt^2} \B[    ( \pa_{t\cZ,  \mu} f)(x) \B ]_{t=0}   =  
      -   e^2_k  \eta  ( \cZ_{\mu}(x) )^2f (x + \eta e_{\mu} )
\ee
Then  
 \be 
\begin{split}
 & <f,    \frac{ \de^2  U^s_k   }{  \de  \cZ^2}(  0  ;  \cZ, \cZ)   f> 
=     \frac{d^2}{dt^2} \B[  <f,  U^s_k (  t\cZ  ) f> \B] _{t=0}    \\    
=  &    \frac{d^2}{dt^2} \B[  < \pa_{t\cZ} f,   \pa_{t\cZ}  f> \B] _{t=0} 
=    2    \sum_{\mu}    <   \cZ^{(1)}_{\mu}  f,   \cZ^{(1)}_{\mu}   f>     +   < \pa_{\mu}  f,  \cZ^{(2)}_{\mu}  f>  \\
\end{split}
\ee 
or        
 \be 
   \frac{ \de^2  U^s_k   }{  \de  \cZ}(  0  ;  \cZ, \cZ)  
=      2   \sum_{\mu}   \cZ^{(1), T}_{\mu}    \cZ^{(1)}_{\mu}  
+     \pa^T _{\mu}    \cZ^{(2)}_{\mu}  
\ee 
Inserting    these in  (\ref{pause1})   we  find     
\be   \label{pause2a}
\begin{split}
  < \cZ,  \Pi^p_k  \cZ> 
=     &   \frac12  \sum_{\mu}  \Tr   \B(       \cZ^{(1), T}_{\mu}    \cZ^{(1)}_{\mu}  
 G_k\B)  +   \eta \sum_{\mu}   \Tr   \B(      \pa^T _{\mu}    \cZ^{(2)}_{\mu}    G_k  \B) \\
+&     \frac12   \sum_{\mu \nu} 
  \Tr   \B(      \cZ_{\mu}^{(1), T}  (  \pa_{\mu}    G_k  )   \cZ_{\nu}^{(1),T}  (  \pa_{\nu}   G_k )\B)    +    \sum_{\mu \nu} 
  \Tr   \B(       \cZ_{\mu} ^{(1),T} (  \pa_{\mu}  G_k    \pa^T_{\nu}   )   \cZ ^{(1)}_{\nu}   G_k\B)   \\ 
\end{split}
\ee
Evaluating this with  $ ( \cZ^{(1),T}_{\mu}  f  )(x)   =    -q  e_k   \cZ_{\mu}  (x- \eta e_{\mu} )f  (x- \eta e_{\mu} )$
and gaining an extra factor of two from the trace over charge indices gives the result.

\subsubsection{removal of averaging operators from propagators}

   Next we change to more standard propagators  (which have more symmetry)
replacing   the propagator   $G_k  =  ( -\De  +   a_k Q_k^TQ_k)^{-1}$   by  $ G^s_k   =  ( \De  + I  )^{-1}  $
This  satisfies  
 \be   \label{picnic}
 \begin{split}
 |G^s_k(x,y) |  \leq   &   \one  d'(x,y)^{-1}e^{-\ga d(x,y)}   \\
 |(\pa   G^s_k)(x,y) |   \leq   &   \one  d'(x,y)^{-2} e^{-\ga d(x,y)}    \\
 |( \pa   G^s_k \pa^T)(x,y) |  \leq   &   \one d'(x,y)^{-3} e^{-\ga d(x,y)}    \\
 \end{split}
 \end{equation}
 This is probably well-known; nevertheless  we include a proof in appendix \ref{orange}. 
\bigskip

Let   $\Pi^s_k$  be   the  photon self energy with this replacement.    It is given by   
\be   \label{stoop}
\begin{split}
  < \cZ,  \Pi^s_k  \cZ> 
  \equiv   & \frac14   \Tr   \B(  \frac{ \de^2  U^s_k }{ \de  \cZ^2 }( 0  ; \cZ, \cZ  )  G^s_k\B)   
+  \frac14   
  \Tr   \B(  \frac{ \de  U^s_k   }{  \de  \cZ}( 0  ;  \cZ)  G^s_k\frac{ \de  U^s_k }{  \de  \cZ}( 0  ;  \cZ)  G^s_k\B)   \\ 
\end{split}
\ee
or  by    an expression like     (\ref{slick2}) with kernel
where  
\be     \Pi^s_{k,\mu \nu}  (x,y)   =    \de_{\mu \nu} \de (x-y)    \Pi^{s, (0)}_{k, \mu}(x)   
+       \Pi^{s, (1)} _{k,\mu \nu}  (x,y) 
\ee
and
\be
\begin{split}
 \Pi^{s, (0)}_{k, \mu}(x)    = &          e_k^2  G^s_k(x+ \eta e_{\mu},x+  \eta e_{\mu})   -  e_k^2 \eta  ( \pa_{\mu} G^s_k) (x + \eta e_{\mu},x)   \\
 \Pi^{s, (1)} _{k,\mu \nu}  (x,y) 
=  &   -    e_k^2     (\pa_{\mu} G^s_k ) (x,y+ \eta e_{\nu}  )  
 (\pa_{\nu} G^s_k ) (y,x +  \eta e_{\mu} )     \\
&   +             e_k^2       (\pa_{\mu} G^s_k  \pa_{\nu}^T ) (x,y  )  
( G^s_k ) (y + \eta e_{\nu},x +  \eta e_{\mu} )   
\end{split}
\ee
Again we have
 \be 
 \label{piestimate2}
 |\Pi^{s, (1)} _{k,\mu \nu}  (x,y)|   \leq   C e_k^2 d'(x,y)^{-4} e^{-2\ga_1 d(x,y)}
 \ee
and  still     there is an apparent   linear ultraviolet divergence.

Note  also   that   $\Pi^s_k$  can be obtained directly  from       $G^s_k  (\cZ)  \equiv  ( - \De_{\cZ}  + I  )^{-1} $ 
just  as  $\Pi_k$ was obtained from  $G_k(\cZ)$, namely 
\be     \label{lunk1}
<  \cZ,  \Pi^s_k  \cZ >  =      \frac12  \frac{ \de^2 E^s_k }{  \de  \cZ^2}( 0  ; \cZ, \cZ)             
  \hs
  E^z_k ( 0, \cZ)  =   \frac12    \log  \left[   \frac{\det  G^s_k(\cZ)}{\det  G^s_k}   \right]  
\ee
 Hence  just    like  $\Pi_k$  we have that  $  \Pi^s_{k,   \mu \nu}$   satisfies the Ward  identity 
 \be 
    \pa^T  \Pi_k^s =0  \hs    \Pi_k^s  \pa =0
 \ee

For the difference we have the following  two  results:       

\begin{lem}    \label{lego} 
 \be 
 \begin{split}
      | G_k(x,y)  -  G^s_k(x,y)   |   \leq   &\    C   e^{ -  \frac12    \ga  d(x,y) }  \\
    |\pa G_k(x,y)  -  \pa    G^s_k(x,y)   |   \leq   &\  C  d(x,y)^{- \ep}   e^{ -  \frac12    \ga  d(x,y) }   \\
 \end{split}
 \ee
\end{lem}

\pr   
 For  the  difference  we  have 
 \be  G_k  -  G^s_k    =   G_k ( I  -  a_kQ_k^TQ_k  )  G^s_k
 \ee
 We      focus on the term   $G_kG^s_k$;  the other term   $a_kG_k Q_k^TQ_k  G^s_k$  is  less singular.   We  have     
 $  (G_kG^s_k)(x,y)  =  \int  G_k(x,z) G^s_k(z,y)  dz$   and so by (\ref{shock}) and (\ref{picnic}) 
 \be   
  |(G_kG^s_k)(x,y) |  \leq   C \int   d'(x,z)^{-1} d'(z,y)^{-1}  e^{ -   \ga  ( d'(x,z) + d'(z,y))  }  dz  
 \ee
 We  can extract  a factor     $ e^{ -  \frac12  \ga    d'(x,y )  } $ here and still have enough decay left  for convergence in $z$  at large  distances.
 For short distances  we have an integrable singularity, for example   by  a Schwarz inequality.    Hence the first bound.    
 For the second bound  we focus  on      $( \pa   G_kG^s_k)(x,y)  =  \int   \pa G_k(x,z) G_k(z,y) dz$  which has the
 bound   
 \be   
|  (\pa   G_kG^s_k)(x,y) | \leq  C  \int   d'(x,z)^{-2} d'(z,y)^{-1}  e^{ -   \ga  ( d'(x,z) + d'(z,y))  }  dz  
 \ee
Again we extract    a factor     $ e^{ -  \frac12  \ga    d'(x,y )  } $ and have no long distance problem.   
  If  either   $d'(x,z)$ or $d'(z,y)$ is greater  than one we have an integrable singularity and get the result.  
If       both  $d'(x,z)  \leq  1$  and    $d'(z,y)  \leq  1$  then    we use   from appendix  \ref{dprime} 
\be
   \int_{d'(x,z)  \leq  1,     d'(y,x)  \leq  1 }   d'(x,z)^{-2} d'(z,y)^{-1}   \leq    \one d'(x,y)^{-\ep}
\ee
and hence the result.

 \begin{lem}  \label{smoothdif}
\be  
   < \cZ,  \Pi^p_k  \cZ>  -    < \cZ,  \Pi^s_k  \cZ>     =  \sum_X  E^q_k (X, \cZ) 
 \ee
where   
\be   |E^q _k(X, \cZ)|  \leq   e_k^{2- \ep  }     \|\cZ \|_{\infty} ^2    e^{- \ka d_M(X) } 
\ee
\end{lem}
\bigskip

\pr      
For the first term  in     $ \Pi^{p,(1)} _k(x,y) -   \Pi^{s,(1)}_k(x,y)   $  we  use lemma  \ref{lego}   to get estimates like   
\be  
|  e_k^2   (\pa_{\mu} G_k  -  \pa_{\mu}G^s_k ) (x,y+ \eta e_{\nu}  )  
 (\pa_{\nu} G^s_k ) (y,x +  \eta e_{\mu} )  |   \leq     C e_k^2    d'(x,y)^{-2-\ep } e^{ -     \ga  ( d(x,y) } 
\ee
which has   an       integrable singularity.  The second      term  is essentially the same.   
In       $ \Pi^{p,(0 )} _k(x,y) -   \Pi^{s,(0 )}_k(x,y)   $   the second       term   is     estimated by      
\be
|     e_k^2 \eta  (   \pa  G_k(x+ \eta e_{\mu}, x  )  -   \pa  G^s_k(x+ \eta e_{\mu}, x  )   |
\leq   \one   e_k^2    \eta     d'(x+ \eta e_{\mu}, x )^{-\ep}   \leq     C   e_k^2  \eta^{1- \ep}  \leq    C   e_k^2  
\ee
The    first     term is equally easy.

To localize  we  proceed  as  in the proof  of  lemma  \ref{smoothq}
writing 
\be  
  < \cZ,  [ \Pi^p_k-  \Pi^s_k ]    \cZ>=   \sum_{z,w}  <( \chi_z\cZ),  [ \Pi^p_k-  \Pi^s_k ]    (\chi_{w} \cZ)> 
\ee
Then   $ <( \chi_z\cZ),  [ \Pi^p_k-  \Pi^s_k ]    (\chi_{w} \cZ)>   $  is  finite  and estimated by   
\be  |  <( \chi_z\cZ),  [ \Pi^p_k-  \Pi^s_k ]    (\chi_{w} \cZ)>  |  \leq    Ce_k^2 \|\cZ \|^2_{\infty} e^{-  \ga d(z,w) }
\ee
Rewriting the expression  as a sum over polymers  and using   $Ce_k^2 \leq  e_k^{2 - \ep}$ gives the result.

\subsubsection{proof of lemma \ref{smooth}}

Our  standard  photon self-energy has the advantage of  being invariant under the full  $\tk$ lattice symmetries,  not just
$\bbT^0_{N-k}$ symmetries.
It  can be written in two  ways    
\be 
\begin{split}
 \label{sty}
< \cZ,  \Pi^s_k  \cZ>   =   &  \int  dbdb'\   \cZ(b)  \Pi^s (  b,b' )  \cZ(b')      
 =    \sum_{\mu \nu}   \int  dx  dy  \   \cZ_{\mu}(x)  \Pi^s_{k,\mu \nu} (x,y)  \cZ_{\nu}(y)    \\
\end{split} 
\ee
where the integral is over oriented bonds and   the kernels are   related by 
\be    \Pi^s_{k,\mu \nu} (x,y)   =   \Pi^s_k (  [x , x+ \eta e_{\mu}],     [y, y+ \eta e_{\nu}] )
\ee
If  the  first form is extended to all bonds as before then     
$  \Pi^s_k ( b,  b'  )   =    \Pi^s_k ( rb, rb'  ) $.    This symmetry  is more complicated in the other notation.    For example is    $r$ is the complete inversion  $rx = -x$ it says   
\be 
\begin{split}  \label{one}  \Pi^s_{k,\mu \nu} (x,y)      =&   \Pi^s_k \B([x,x+ \eta e_{\mu}]  ,[y,y+ \eta e_{\nu}] \B)  
=       \Pi^s_k \B([-x,-x-\eta e_{\mu}]  ,[-y,-y- \eta e_{\nu}] \B)  \\
=    &   \Pi^s_k \B([-x-\eta e_{\mu},-x]  ,[-y- \eta e_{\nu},-y] \B)  
 =   \Pi^s_{k,\mu \nu} (- x- \eta  e_{\mu}, - y  - \eta  e_{\nu})  \\
 \end{split}
\ee

We  break up   $< \cZ,  \Pi^s_k  \cZ>$  into pieces.   Each  piece should be covariant under lattice  symmetries so at first   we work 
with the  representation  $ \Pi^s (  b,b' )  $.  First   let  $\theta$ be  a  smooth  function  on  $\bbR$ such that  $  0 \leq  \theta  \leq  1$
and    $\theta =1$ on   $[-\frac13,  \frac13] $   and   $\supp \theta \subset  [-\frac23,  \frac23]$.   Let   $d_2(b,b')$ be the Euclidean distance   between $b,b'$.     Then   $\theta (d_2(b,b') )$   does not
depend on orientation and we can make the split
\be
\begin{split}
   \label{ring}
   < \cZ,  \Pi^s_k  \cZ>   =   &  \int  dbdb'\   \cZ(b) \B(1-\theta ( d_2(b,b')) \B) \Pi^s (  b,b' )  \cZ(b')   \\ 
+ &  \int  dbdb'\   \cZ(b) \theta (d_2(b,b') )  \Pi^s (  b,b' )  \cZ(b')   \\
\end{split}
\ee
without spoiling covariance.

In the first term  note that $(1-\theta ( d_2(b,b'))$   vanishes unless   $d_2(b,b')  \geq \frac13$.  Thus there is no ultraviolet divergence
here.  We  localize it  as
\be
\sum_{z,w}    \int  dbdb'\  \chi_z(b)  \cZ(b) \B(1-\theta ( d_2(b,b')) \B) \Pi^s (  b,b' )   \chi_w(b') \cZ(b')    
\ee
By (\ref{piestimate2})  the summand  is   bounded by   $ Ce_k^2 \|\cZ \|^2_{\infty} e^{-   \ga d(z,w) } $
and  the sum  is  bounded  by   $ Ce_k^2 \|\cZ \|^2_{\infty}$.    
As in lemma \ref{smoothq}   we  write the expression  as a sum over polymers  and  
get a contributions     to  $E_k^{\pi}(X) $  
bounded  by    
   $ Ce_k^2 \|\cZ \|^2_{\infty} e^{- \ka d_M(X)}  $   
which suffices.

For    the second term in (\ref{ring})   we  localize in the $b$ variable only write it as    
\be
 \sum_z   \int     dbdb'\   \chi_z(b)   \cZ(b) \theta (d_2(b,b'))  \Pi^s_k (  b,b' )  \cZ(b')  
\ee
Since  $\theta (d_2(b,b'))$  vanishes for      $d_2(b,b')  \geq \frac13$,   
  for each   $z$ the term is localized in the threefold enlargement  $\tilde  \De_z$.  Let   $\De_z^M$ be the smallest polymer containing
    $\tilde \De_z$.  Then   we  get a contribution to 
$E^{\pi}_k(X)$   
of the form     
\be     
 \sum_{z:      \De^M_z =X}      \int     dbdb'\   \chi_z(b)   \cZ(b) \theta (d_2(b,b'))  \Pi^s_k (  b,b' )  \cZ(b')  
\ee
 The term  can also be  written  
\be    \label{loud}
 \sum_{z:      \De^M_z =X}  \left[ \sum_{\mu  \nu}     \int      dxdy\    \chi_z ([x,  x + \eta e_{\mu}])  
  \cZ_{\mu}(x)  \theta \B(  d_2( [x, x+ \eta e_{\mu} ],   [y, y+ \eta e_{\nu} ]  )  \B)
  \Pi^s_{k, \mu \nu} (  x,y )  \cZ_{\nu}(y)  \right]
\ee
We will show that   the bracketed expression is bounded  by        
$C e_k^{2}( \| \cZ \|_{\infty} +   \|\pa  \cZ \|_{\infty}  +  \| \de_{\al} \pa  \cZ \|^2)^2$.
Since  $X$ contains at most  $\one$  $M$-cubes,    the sum over  $z$  contributes a factor  $\one M^3$.
Then using  $CM^2 e_k^2  \leq   e_k^{2 - \ep} $  the expression   (\ref{loud}) is bounded by    
$  e_k^{2- \ep}( \| \cZ \|_{\infty} +   \|\pa  \cZ \|_{\infty}  +  \| \de_{\al} \pa  \cZ \|^2)^2$ as required.  
The term    is invariant under lattice symmetries, but in  estimating it
we   break  it  up  into  pieces that  are not      invariant.

The first step is to replace  
$ \theta \B(  d_2( [x, x+ \eta e_{\mu} ],   [y, y+ \eta e_{\nu} ]  )  \B)
 $   by   $ \theta ( d_2(x,y)  ) $.    So for the difference we must consider  
\be
\sum_{\mu \nu}     \int   dxdy\   \chi_z ([x,  x + \eta e_{\mu}])  
  \cZ_{\mu}(x)\B(  \theta \B( d_2( [x, x+ \eta e_{\mu} ],   [y, y+ \eta e_{\nu} ]  )  \B)
 - \theta( d_2(x,y) )  \B)  \Pi^s_{k, \mu \nu} (  x,y )  \cZ_{\nu}(y)  
\ee
If  $d_2( [x, x+ \eta e_{\mu} ],   [y, y+ \eta e_{\nu} ]  ) \leq \frac13$  and  $d_2(x,y) \leq \frac13$   then    $  \theta \B( d_2( [x, x+ \eta e_{\mu} ],   [y, y+ \eta e_{\nu} ]  )  \B)
 - \theta( d_2(x,y) )= 1-1 =0 $
and the term vanishes.   Thus there is no ultraviolet divergence,   only  $\Pi^{s,(1)}_{k, \mu \nu}$ contributes,   and the term can
be estimated by    $C  e_k^{2} \| \cZ \|_{\infty}^2$ as before.

So  now we  consider  
\be
\sum_{\mu  \nu}     \int   dxdy\   \chi_z ([x,  x + \eta e_{\mu}])  
  \cZ_{\mu}(x) \theta( d_2(x,y) )  \Pi^s_{k, \mu \nu} (  x,y )  \cZ_{\nu}(y)  
\ee
We  generate three terms in this expression     by making the  expansion 
\be   \label{lunk}
    \cZ_{\nu}(y)  =    \cZ_{\nu}(x)  +  \sum_{\si} (y-x)_{\si}   \pa_{\si}   \cZ_{\nu}(x)   + \De_{\nu}(y,x)    
\ee

The first  term  is  
\be
\sum_{\mu  \nu }     \int   dx dy\   \chi_z ([x,  x + \eta e_{\mu}])    \cZ_{\mu}(x)\theta( d_2(x,y) )  \Pi^s_{k, \mu \nu} (  x,y )  \cZ_{\nu}(x)  
\ee
We  write  
\be
      \cZ_{\nu}(x) =    \frac{ \pa  }{  \pa  y_{\nu}  }   \B(  \sum_{\si}   (y-x)_{\si}   \cZ_{\si}(x)  \B)
\ee 
It is pure   gauge  and the  expression would vanish by the Ward identity were it not for the factor  $\theta( d_2(x,y) ) $.
With this factor we  get two terms when we  integrate by parts  in  $y$  (see (\ref{product})).   They are  
\be   
\begin{split}
&\sum_{\mu  \nu }     \int   dxdy\   \chi_z ([x,  x + \eta e_{\mu}])  
  \cZ_{\mu}(x)   \theta( d_2(x,y- \eta e_{\mu} ) )     \left(     \frac{ \pa  }{  \pa  y_{\nu}  } ^T    \Pi^s_{k, \mu \nu} (  x,y ) \right) \B(  \sum_{\si}   (y-x)_{\si}   \cZ_{\si}(x)  \B)  \\
+&\sum_{\mu  \nu }     \int   dxdy\   \chi_z ([x,  x + \eta e_{\mu}])  
  \cZ_{\mu}(x)    \left(     \frac{ \pa  }{  \pa  y_{\nu}  } ^T  \theta( d_2(x,y) )   \right)  \Pi^s_{k, \mu \nu} (  x,y )\B(  \sum_{\si}   (y-x)_{\si}   \cZ_{\si}(x)  \B)  \\
\end{split}
\ee  
Since   $ \Pi^s_{k, \mu \nu} (  x,y )= \Pi^s_{k, \nu \mu} ( y,  x )  $    the first term does indeed vanish  by the Ward identity.   For the second term  we have 
\be
 \left(     \frac{ \pa  }{  \pa  y_{\nu}  } ^T  \theta( d_2(x,y) )    \right)
=  \theta'( d_2(x,y) )     \left(     \frac{ \pa  }{  \pa  y_{\nu}  } ^T   d_2(x,y)    \right) 
\ee
and this is bounded by  $\one$.    Furthermore  
$ \theta'( d_2(x,y) )  $  keeps  $x,y$ separate.   There is  no ultraviolet divergence,     only  $\Pi^{s,(1)}_{k, \mu \nu}$ contributes,  
and the term  is  again    estimated by    $C  e_k^{2} \| \cZ \|_{\infty}^2$.
\bigskip

  The second  term  arising from the expansion  (\ref{lunk}) is    
\be    \label{hunk}
\sum_{\mu  \nu \si}     \int   dxdy\   \chi_z ([x,  x + \eta e_{\mu}])    \cZ_{\mu}(x)  \theta( d_2(x,y) )   \Pi^s_{k, \mu \nu} (  x,y )   (y-x)_{\si}   \pa_{\si}   \cZ_{\nu}(x) 
   \ee
This almost changes sign under  reflection through  $x$,  i.e. under the   change of variables  $(y-x)  \to  - (y-x)$  or  $y  \to   -y +2x$, in which case it would   be zero.
The problem is  that   in spite of all our work to get to this point     $  \Pi^s_{k, \mu \nu} (  x,y ) $  is not quite invariant.  We have 
\be
\begin{split}
& \Pi^s_{k, \mu \nu} (  x,y )   =    \Pi^s_{k, \mu \nu} (  0 ,y-x )   \\  \longrightarrow \   &         \Pi^s_{k, \mu \nu} (  0 ,-(y-x)  )  
=      \Pi^s_{k, \mu \nu} (-x,    - y   )   =   \Pi^s_{k, \mu \nu} (x - \eta e_{\mu},     y - \eta e_{\nu}  ) \\
\end{split}
\ee  the last by  (\ref{one}).
Making the  change of variables       the expression    (\ref{hunk})   is the same as  
\be    \label{hunk2}
-  \sum_{\mu  \nu \si  }     \int   dxdy\   \chi_z ([x,  x + \eta e_{\mu}])  
  \cZ_{\mu}(x)  \theta( d_2(x,y) )   \Pi^s_{k, \mu \nu} (x - \eta e_{\mu},     y - \eta e_{\nu}  )
  (y-x)_{\si}   \pa_{\si}   \cZ_{\nu}(x) 
   \ee
and hence it  is also the average of the two which is       
\be    \label{hunk3}
\frac12     \sum_{\mu  \nu \si }     \int   dxdy\   \chi_z ([x,  x + \eta e_{\mu}])  
  \cZ_{\mu}(x)  \theta( d_2(x,y) ) \B(     \Pi^s_{k, \mu \nu} (  x,y )    -  \Pi^s_{k, \mu \nu} (x - \eta e_{\mu},     y - \eta e_{\nu}  )  \B)
  (y-x)_{\si}   \pa_{\si}   \cZ_{\nu}(x) 
   \ee
 In the second term make the replacement      $ \Pi^s_{k, \mu \nu} (x - \eta e_{\mu},     y - \eta e_{\nu}  ) 
 = \Pi^s_{k, \mu \nu} (x ,     y+ \eta e_{\mu}- \eta e_{\nu}  )  $, followed by   the change of variables   $y  \to     y- \eta e_{\mu}+ \eta e_{\nu} $  which yields    
\be
\begin{split}
    \label{hunk5}
&\frac12       \sum_{\mu  \nu \si }     \int   dxdy\   \chi_z ([x,  x + \eta e_{\mu}])  
  \cZ_{\mu}(x)  \theta\B( d_2(x,y) \B)     \Pi^s_{k, \mu \nu} (  x,y ) 
  (y-x)_{\si}   \pa_{\si}   \cZ_{\nu}(x)  \\
-    &\frac12      \sum_{\mu  \nu \si }     \int   dxdy\   \chi_z ([x,  x + \eta e_{\mu}])  
  \cZ_{\mu}(x)  \theta\B( d_2(x, y- \eta e_{\mu}+ \eta e_{\nu} ) \B)     \Pi^s_{k, \mu \nu} (  x,y )  
  ( y -x- \eta e_{\mu}+ \eta e_{\nu} )_{\si}   \pa_{\si}   \cZ_{\nu}(x)  \\
 \end{split}   
   \ee
In the second term  replace $\theta\B( d_2(x, y- \eta e_{\mu}+ \eta e_{\nu} ) \B)   $ by  $ \theta\B( d_2(x,y) \B)$.  The difference is only
non-zero if  $x,y$ are  well separated.   There is  no  ultraviolet divergence,   only  $\Pi^{s,(1)}_{k, \mu \nu}$ contributes,   and   this term is bounded by     $ C e_k^{2} \| \cZ \|_{\infty}    \|\pa  \cZ \|_{\infty}$.
Since 
$  (y-x)_{\si} -     ( y -x- \eta e_{\mu}+ \eta e_{\nu} )_{\si}  =   \eta \de_{\mu \si }-    \eta \de_{\nu \si} $  
 we  are left with  
\be 
\begin{split}  &    \frac12       \sum_{\mu  \nu }     \int   dx\   dy\   \chi_z ([x,  x + \eta e_{\mu}])  
  \cZ_{\mu}(x)  \theta\B( d_2(x, y ) \B)     \Pi^s_{k, \mu \nu} (  x,y )  
   \eta  \B(  \pa_{\mu}   \cZ_{\nu}(x)  -   \pa_{\nu}   \cZ_{\nu}(x) \B) \\  
\end{split}   
\ee
Now  write  
\be      \B(  \pa_{\mu}   \cZ_{\nu}(x)  -   \pa_{\nu}   \cZ_{\nu}(x) \B) 
=       \frac{ \pa  }{  \pa  y_{\nu}  }    \B(  \sum_{\si}   (y-x)_{\si} (   \pa_{\mu}   \cZ_{\si }(x)  -   \pa_{\si}   \cZ_{\si}(x) )  \B)
\ee
As  before   we integrate by parts to transfer the   $y$-derivative to the other factors.    On the  $  \Pi^s_{k, \mu \nu} (  x,y )  $
we again get zero by the Ward identity.   On the factor  $ \theta\B( d_2(x, y ) \B)    $   it again forces $x,y$ to be  separate and
so destroys the UV divergence,    only  $\Pi^{s,(1)}_{k, \mu \nu}$ contributes and this  term can be estimated by 
      $ C e_k^{2} \eta \| \cZ \|_{\infty}    \|\pa  \cZ \|_{\infty}$.
\bigskip

  The final    term  arising from the expansion  (\ref{lunk}) is    
\be    \label{hunk6}
  \sum_{\mu  \nu }     \int   dxdy\   \chi_z ([x,  x + \eta e_{\mu}])  
  \cZ_{\mu}(x)  \theta( d_2(x,y) )   \Pi^s_{k, \mu \nu} (  x,y )  \De_{\nu}(y,x)    
   \ee
We  have the representation     
\be
     \De_{\nu}(y,x)     =  \sum_{\si}  \int_{\Ga} (\pa_{\si} \cZ_{\nu} (z)  -  \pa_{\si}   \cZ_{\nu}  (x) )  dz_{\si}
\ee
where  $\Ga$ is any one of the standard paths  from $x$ to $y$.   For  $d(y,x) \leq 1$ we use   
\be   |\pa_{\si} \cZ_{\nu} (z)  -  \pa_{\si}   \cZ_{\nu}  (x)  |
\leq     d(z,x)^{\al}   \|  \de_{\al} \pa \cZ  \|_{\infty}
\ee
which yields    
\be   
    | \De_{\nu}(y,x) |   \leq    \one  d(y,x)^{1 + \al}   \|  \de_{\al} \pa \cZ  \|_{\infty}
 \ee
If   $d(y,x) \geq  1$ we have     $  | \De_{\nu}(y,x) |   \leq    \one  d(y,x)^{ \al}   \|  \pa \cZ  \|_{\infty}$.  In either case  there is 
no UV divergence,   only  $\Pi^{s,(1)}_{k, \mu \nu}$ contributes since     $\De_{\nu}(x,x)  =0$,   and  the term  is bounded  by   
     $ C e_k^{2 } \| \cZ \|_{\infty}  (   \| \de_{\al} \pa  \cZ \|_{\infty}  +  \|  \pa \cZ  \|_{\infty})    $.   This completes the analysis of 
      $< \cZ,  \Pi_k^s \cZ>$.

 We have   $\Pi_k  =  \Pi_k^q  +   \Pi_k^{qs}  + ( \Pi_k^{p} -   \Pi_k^s)   +  \Pi_k^s$.   Combining the polymer decompositions and     
  estimates  on each of these   completes the proof of lemma  \ref{smooth}.
\bigskip

\pr   (lemma \ref{snooze4}).    From   lemma  \ref{pinky}     and    lemma  \ref{smooth}    we have
\be   E^{(4)}_k(X, \cA, \cZ)  \equiv    E^{z}_k(X, \cA, \cZ)  =      E^{\pi}_k(X, \cA, \cZ) +  \tilde    E^z_k(X, \cA, \cZ)
\ee
and       this is bounded by   
$ \one    e_k^{\ep} e^{- \ka d_M(X)}   $    on the   domain   $\cA  \in   \frac12 \cR_k, \cZ  \in \frac 12  \cR_k'$.
We   want    to show this is bounded by   $  \one   e_k^{1- 6\ep}  e^{-\ka    d_M(X)} $     
   on the smaller domain    $\cA \in \frac12 \cR_k$   and  $| \cZ|, | \pa \cZ |,   \| \de_{\al} \pa \cZ \|_{\infty}  \leq e_k^{-2\ep}$.
To get the better bound   we use a Cauchy inequality.  
Since   $ E^{z}_k\B(X,       \cA,   \cZ    \B) $     vanishes  at  $\cZ =0$  we have 
\be
 E^{z}_k\B(X,       \cA,   \cZ    \B)  =  \frac{1}{2 \pi i}  \int_{|t|  =  r }   \frac{dt}{t(t-1) }  E^{z}_k\B(X,       \cA,   t\cZ    \B)
\ee
If  we  take  $r = \frac12  e_k^{-1 + 7 \ep}  $  then     $|t\cZ |  \leq   e_k^{-1+ 5 \ep}$  with the same bound for the derivatives  and
so       $t \cZ   \in  \frac12   \cR_k'$.   Then we     can use the above bound to obtain
\be
  | E^{z}_k\B(X,       \cA,   \cZ  )    |   \leq  \one  e_k^{1- 7 \ep}  e_k^{\ep} e^{-  \ka  d_M(X)}   =    \one  e_k^{1 - 6 \ep}   
e^{-  \ka    d_M(X)}  \ee
This   completes the proof of   lemma \ref{snooze4}  and  theorem   \ref{lanky}.

\section{The flow}

We    seek  well-behaved    solutions of   the    RG   equations  (\ref{recursive}) .  
Thus we  study
 \begin{equation}  \label{recursive3}
\begin{split}
\vep_{k+1}   =&  L^3 \vep_k  + \cL_1E_k   +  \vep_k^*( \mu_k,  E_k) \\
\mu_{k+1}   =&   L^2 \mu_k  +  \cL_2E_k  + \mu_k^*(\mu_k,  E_k)  \\
E_{k+1}   =&    \cL_3 E_k  +  E^*_k(\mu_k, E_k) \\
 \end{split}
\end{equation}
Our  goal is to show  that for   any    $N$   we can choose  the  initial   point  so  that the solution exists   for  $k=0,1,  \dots,  N$
and finishes  at   values  $(\vep_N,  \mu_N)   =    (\vep^N_N,  \mu^N_N)       $  independent of  $N$.    (Note that   at  $k=N$  we are on the lattice  $\bbT^{0}_0$
with the dressed fields  back   on the original lattice   $\bbT^{-N}_0 $).  This procedure  is  nonperturbative  renormalization -  the initial  values for  $(\vep_0,  \mu_0)  =   (\vep^N_0,  \mu^N_0)  $  will depend  $N$  and in fact be divergent in $N$.  
The problem  is now formally   exactly  the same  as the pure   scalar   
problem  in  \cite{Dim11}.  The functions   $\vep_k^*, \mu_k^*, E^*_k$   are different,   they now contain all radiative corrections,  but the analysis is essentially    the same   as we explain.  

Arbitrarily  fixing the final values at  zero,    and  starting with  $E_0 =0$ as dictated by the model,   
we look for solutions  $\vep_k,  \mu_k,  E_k$    for    $k = 0,1,2,  \dots  , N$
satisfying  
\begin{equation} \label{bc}
\vep_N  = 0       \hs      \mu_N  = 0     \hs     E_0  =0    
  \end{equation}
At  this point we  temporarily drop the equation for the   energy density  $\vep_k$ and just  study  
 \begin{equation}  \label{recursive4}
\begin{split}
\mu_{k+1}   =&   L^2 \mu_k  +  \cL_2E_k  + \mu_k^*(\mu_k,  E_k)  \\
E_{k+1}   =&    \cL_3 E_k  +  E^*_k(\mu_k, E_k) \\
 \end{split}
\end{equation}

Let    $\xi_k  =  (\mu_k, E_k)$  be  an  element of the real  Banach space
  $\bbR \times  \textrm{Re}\  \cK_k  $  where     $ \textrm{Re}\  \cK_k$ is the real    subspace of  $\cK_k$
   defined in section \ref{polymersection}.          Consider sequences      
\begin{equation}
\underline{ \xi }  =  ( \xi_0,  \dots  ,   \xi_N)
 \end{equation}
 Pick    a fixed  $\beta$  satisfying $
0<  \beta   <  \frac{1}{12} -   11 \ep
$
and  let    $\sB$   be the   real Banach  space of all   such   sequences
 with norm   
  \begin{equation}
\|  \underline{\xi } \|  =  \sup_{0 \leq  k   \leq  N}       \{   \la_k^{-\frac12 - \beta}  | \mu_k|, 
 \la_k^{-  \beta}     \| E_k \|_{k, \ka}   \}  
\end{equation}
Let  $\sB_0$   be the   subset  of all sequences satisfying the  boundary 
 conditions.
 Thus   
 \begin{equation}
 \sB_0  =  \{\underline{ \xi}   \in  \sB:    \mu_N  =  0,        E_0  =0\} 
 \end{equation}
  This is a complete metric space  with distance  
  $\| \underline {  \xi } - \underline {\xi' }  \|$.
 Finally   let  
 \begin{equation}
 \sB_1  = \sB_0 \cap    \{\underline{ \xi}   \in  \sB:    \| \underline{\xi} \| <1  \}   
 \end{equation}

 Next  define an    operator  $\un{\xi' } =T \un{\xi } $    by      
  \begin{equation}  \label{recursive5}
\begin{split}
\mu'_k   =&   L^{-2}( \mu_{k+1} -  \cL_2E_k - \mu_k^*)   \\
E'_k   =&      \cL_3 E_{k-1}  +  E^*_{k-1} )  \\
 \end{split}
\end{equation}
Then   $\underline{  \xi }$ is a solution  of  (\ref{recursive3}) iff  it is a fixed point for  $T$ on 
$\sB_0$.     We    look for  such   fixed points  in $\sB_1$.

\begin{lem}   Let  $\la   $  be sufficiently small.  Then   for all  $N$  
\begin{enumerate}
\item   The  transformation   $T$     maps   the set
  $\sB_1$  to itself.  
\item   There is a unique fixed point   $T\underline{ \xi}  = \underline{ \xi }$  in this  set. 
\end{enumerate}
\end{lem}
\bigskip

\pr  (1.) We   use the bounds of  theorem   \ref{lanky}  for   $\cL_2,   \cL_3$
(replacing  $\cO(1)L^{-\ep}$ by   $1$ )   and  for  $\mu_k^*,  E_k^*$.
To   show the the map sends  $\sB_1$ to itself
we  estimate  for  $L$ sufficiently large  and    $\la_k$ sufficiently small
\begin{equation}  \label{jelly}
\begin{split}
\la_k^{- \frac12  -\beta} |\mu'_k| 
 \leq   &
  \la_{k}^{- \frac12 -\beta} L^{-2}
\Big(   | \mu_{k+1}|  + \la_k^{1/2 + 2 \ep}  \| E_k\|_{k, \ka} +   \one   \la_k^{\frac{7}{12}   - 11 \ep}   \Big)\\
  \leq   &
   L^{\beta-  \frac32}\Big[\la_{k+1}^{- \frac12 -\beta}  | \mu_{k+1}|\Big]   +
    L^{-2} \la_k^{2 \ep}\Big[ \la_k^{-\beta} \| E_k\|_{k, \ka} \Big]  +
 \one    \la_{k}^{   \frac{1}{12}   -\beta- 11 \ep}\\
    \leq  & \frac12 (  \|  \underline  \xi  \|  +   1   )  <  1  \\
\end{split}
\end{equation}
We   also   
have 
\begin{equation}
\begin{split}
\la_k^{-\beta} \|E'_k\|_{k, \ka}  
   \leq   &  \la_{k}^{-\beta}   \Big( \|  E_{k-1} \|_{k-1,\ka}   +      \cO(1) \la_{k-1}^{\frac{1}{12} -  11 \ep}          \Big)\\
   \leq    &    L^{  - \beta}  \Big[  \la_{k-1}^{-\beta} \| E_{k-1} \|_{k-1, \ka} \Big]  +  \cO(1) L^{- \beta}
     \la_{k-1}^{\frac{1}{12} - \beta -11 \ep}\\
       \leq  & \frac12 (  \|  \underline  \xi  \|  +   1   )  <   1  \\
     \end{split}
\end{equation}
 Hence  $ \|  T (\underline{ \xi  } ) \|   <  1$  as  required.
\bigskip

\noindent  (2.)
  It suffices  
 to show   that   the mapping is a contraction.
 We show  that   under our assumptions
\begin{equation}
\| T( \underline{ \xi_1}) -   T( \underline{ \xi_2}) \|  \leq  \frac12
\|  \underline{ \xi_1}- \underline{ \xi_2} \| 
\end{equation}

First consider the  $\mu$  terms.   We have   as above  
   \begin{equation}  \label{sudsy0}
\begin{split}
&  \la_k^{- \frac 12  - \beta}|\mu'_{1,k}-\mu'_{2,k}| \leq  
 L^{\beta-  \frac32}\Big[\la_{k+1}^{- \frac12 -\beta}    | \mu_{1,k+1}-  \mu_{2, k+1} |  \Big]   \\
 + &    L^{-2} \la_k^{2 \ep}\Big[ \la_k^{-\beta}   \|    E_{1,k}  -   E_{2,k}    \|_{k, \ka}  \Big]  
+  L^{-2}  \la_k^{- \frac 12  - \beta} \Big|\mu_k^*(  \mu_{1,k}, E_{1,k}) -  \mu_k^*(  \mu_{2,k}, E_{2,k})\Big|  \\
  \end{split}
\end{equation}
The  first  two terms are bounded by a small constant times   $  \|  \underline{ \xi_1}- \underline{ \xi_2} \|$ .
   For the last  term  we   use the fact that    $\mu_k^*(  \mu_{k}, E_{k})$   is actually an analytic function of    $\mu_{k}, E_{k}$
  on its domain  $|\mu_k|  \leq   \la_k^{\frac12}$ and  $\|E_k \|_{k, \ka}  \leq  1$.   We  write
  \begin{equation}   \label{selwyn}
\begin{split}
&\mu^*_k(\mu_{1,k}, E_{1,k}) -\mu^*_k(\mu_{2,k}, E_{2,k}) \\
=  & \Big( \mu^*_k(\mu_{1,k}, E_{1,k}) - \mu^*_k(\mu_{2,k}, E_{1,k}  ) \Big) +
\Big(\mu^*_k(\mu_{2,k}, E_{1,k}) -\mu^*_k(\mu_{2,k}, E_{2,k}) \Big) \\
 \end{split}
\end{equation}
For the first  term  we  write for  $r>1$   
  \begin{equation}
\begin{split}
 &  \mu^*_k(\mu_{1,k}, E_{1,k}) - \mu^*_k(\mu_{2,k}, E_{1,k}  ) 
=      \frac{1}{2 \pi i}
 \int_{|t|  =r }     \frac{dt}{t(t-1)}      \mu^*_k   \Big(\mu_{2,k} +  t( \mu_{1,k}- \mu_{2,k}), E_{1,k})  \Big)   \\
 \end{split}
\end{equation}
We   use the bound  $|\mu^*|  \leq  \one  \la_k^{\frac{7}{12} - 11 \ep}$ on its domain of analyticity.
We take  $r = 4 \la_k^{\frac12 + \beta} |\mu_{1,k}-\mu_{2,k}|^{-1}$.   This  keeps us in the domain of analyticity and   is greater than one since
$ |\mu_{1,k}-\mu_{2,k}|  \leq   \la_k^{\frac12 + \beta}  \| \xi_1- \xi_2 \|  \leq  2  \la_k^{\frac12 + \beta}$. 
 Hence this term is
bounded by   
\be  \one   \Big(  \la_k^{-\frac12 - \beta} |\mu_{1,k}-\mu_{2,k}|  \Big)  \la_k^{\frac{7}{12} - 11 \ep}      \leq   \one \la_k^{\frac{1}{12}  - \beta -  11 \ep}   
 |\mu_{1,k}-\mu_{2,k}|  
\ee
For the second term in (\ref{selwyn}) we write  for  $r>1$
\begin{equation}
\begin{split}
\Big(\mu^*_k(\mu_{2,k}, E_{1,k}) -\mu^*_k(\mu_{2,k}, E_{2,k}) \Big) 
= &   \frac{1}{2 \pi i}
 \int_{|t|  =r }     \frac{dt}{t(t-1)}      \mu^*_k   \Big(\mu_{2,k},  E_{2,k} +  t( E_{1,k}- E_{2,k})  \Big)   \\
 \end{split}
\end{equation}
Now  we  take   $r=  4 \la_k^{\beta}\|E_{1,k} - E_{2,k} \|^{-1}_{k,\ka}  $.  This keeps us in the domain of analyticity 
and  is greater than one since
$\|E_{1,k} - E_{2,k} \|_{k,\ka}    \leq   \la_k^{ \beta}  \| \xi_1- \xi_2 \|  \leq  2  \la_k^{ \beta}$.   Hence this term is bounded by  
\be 
 \one   \Big(  \la_k^{ - \beta}   \|E_{1,k} - E_{2,k} \|_{k,\ka}   \Big)  \la_k^{\frac{7}{12} - 11 \ep}   
   \leq   \one \la_k^{\frac{7}{12}  - \beta -  11 \ep}     \|E_{1,k} - E_{2,k} \|_{k,\ka}    
\ee
Now for    the last term in  (\ref{sudsy0})  we have
  \begin{equation}  \label{sudsy2}
\begin{split}
&  L^{-2}  \la_k^{- \frac 12  - \beta} \Big|\mu_k^*(  \mu_{1,k}, E_{1,k}) -  \mu_k^*(  \mu_{2,k}, E_{2,k})\Big|  \\
&\leq        \one \la_k^{\frac{1}{12}  - \beta -  11 \ep}   
\Big[ \la_k^{- \frac 12  - \beta} |\mu_{1,k}-\mu_{2,k}|   \Big] 
 +    \one \la_k^{\frac{1}{12}  - \beta -  11 \ep}  \Big[ \la_k^{  - \beta}    \|E_{1,k} - E_{2,k} \|_{k,\ka}    \Big]  \\
&\leq        \one \la_k^{\frac{1}{12}  - \beta -  11 \ep}    \|  \underline{ \xi_1}- \underline{ \xi_2} \|
  \end{split}
\end{equation}
Altogether then   
\be  \label{filet}
  \la_k^{- \frac 12  - \beta}|\mu'_{1,k}-\mu'_{2,k}| \leq    \frac12   \|  \underline{ \xi_1}- \underline{ \xi_2} \|
\ee

Now consider the $E$ terms.  We have  
\begin{equation}
\begin{split}
E'_{1,k}-E'_{2,k}   = &  \cL_3 (E_{1,k-1}  -  E_{2,k-1})  \\
  + &  E^*_{1,k-1}(\mu_{1, k-1}, E_{1, k-1})  - E^*_{2,k-1}(\mu_{2, k-1}, E_{2, k-1})     \\     
\end{split}
\end{equation}
Then
\begin{equation}
\la_k^{  - \beta}\|E'_{1,k}-E'_{2,k}\|_{k, \ka} 
\leq   L^{- \beta }    \la_{k-1}^{  - \beta}
\Big(  \|  E_{1,k-1}  -      E_{2,k-1}\|_{k-1, \ka}     +    \| E^*_{1,k-1}  -   E^*_{2,k-1}\|_{k-1, \ka}          \Big ) 
\end{equation}
The first  term  is  bounded by   $ L^{- \beta }  \|  \underline{ \xi_1}- \underline{ \xi_2}\| $.
For the   second    term  we  write  
\be 
\begin{split}
&   E_{k-1}^*(\mu_{1,k-1}, E_{1,k-1}) - E_{k-1} ^*(\mu_{2,k-1}, E_{2,k-1})   \\
= &   \Big( E_{k-1}^*(\mu_{1,k-1}, E_{1,k-1}) - E_{k-1} ^*(\mu_{2,k-1}, E_{1,k-1}) \Big)
+  \Big( E_{k-1}^*(\mu_{2,k-1}, E_{1,k-1}) - E_{k-1} ^*(\mu_{2,k-1}, E_{2,k-1}) \Big)  \\
\end{split}
\ee
For the first  term  we  write for  $r>1$   
  \begin{equation}
\begin{split}
 &   E_{k-1}^*(\mu_{1,k-1}, E_{1,k-1}) - E_{k-1} ^*(\mu_{2,k-1}, E_{1,k-1})   \\
  & =      \frac{1}{2 \pi i}
 \int_{|t|  =r }     \frac{dt}{t(t-1)}      E^*_{k-1}   \Big(\mu_{2,k-1} +  t( \mu_{1,k-1}- \mu_{2,k-1}), E_{1,k-1})  \Big)   \\
 \end{split}
\end{equation}
We   use the bound  $\|E_{k-1}^*\|_{k-1, \ka}  \leq  \la_{k-1}^{\frac{1}{12} - 11 \ep}$ on its domain,  and      take    $r = 4 \la_{k-1}^{\frac12 + \beta} |\mu_{1,k-1}-\mu_{2,k-1}|^{-1}$.  
 Hence this term is
bounded by   
\be  \one   \Big(  \la_{k-1}^{-\frac12 - \beta} |\mu_{1,k-1}-\mu_{2,k-1}|  \Big)  \la_k^{\frac{1}{12} - 11 \ep}      \leq   \one \la_{k-1}^{- \frac{5}{12}  - \beta -  11 \ep}   
 |\mu_{1,k-1}-\mu_{2,k-1}|  
\ee

For the second term we write  for  $r>1$
\begin{equation}
\begin{split}
&\Big(E^*_{k-1}(\mu_{2,k-1}, E_{1,k-1}) -E^*_{k-1}(\mu_{2,k}, E_{2,k-1}) \Big)  \\
  &=    \frac{1}{2 \pi i}
 \int_{|t|  =r }     \frac{dt}{t(t-1)}      E^*_{k-1}   \Big(\mu_{2,k},  E_{2,k-1} +  t( E_{1,k}- E_{2,k-1})  \Big)   \\
 \end{split}
\end{equation}
Again we  use     $\|E_{k-1}^*\|_{k-1, \ka}  \leq  \la_{k-1}^{\frac{1}{12} - 11 \ep}$ and      take   $r=  4 \la_{k-1}^{\beta}\|E_{1,k-1} - E_{2,k-1} \|^{-1}_{k-1,\ka}  $.    Hence this term is bounded by  
\be 
 \one   \Big(  \la_{k-1}^{ - \beta}   \|E_{1,k-1} - E_{2,k-1} \|_{k-1,\ka}   \Big)  \la_{k-1}^{\frac{1}{12} - 11 \ep}   
   \leq   \one \la_{k-1}^{\frac{1}{12}  - \beta -  11 \ep}     \|E_{1,k-1} - E_{2,k-1} \|_{k-1,\ka}    
\ee
Combining these bounds gives
  \begin{equation}  \label{sudsy3}
\begin{split}
&  L^{-\beta}  \la_{k-1}^{  - \beta} \|E_{k-1}^*(  \mu_{1,k-1}, E_{1,k-1}) -  E_{k-1}^*(  \mu_{2,k-1}, E_{2,k-1})\|_{k-1, \ka}  \\
&\leq        \one \la_{k-1}^{\frac{1}{12}  - \beta -  11 \ep}   
\Big[ \la_{k-1}^{- \frac 12  - \beta} |\mu_{1,k-1}-\mu_{2,k-1}|   \Big]  +    \one \la_{k-1}^{\frac{1}{12}  - \beta -  11 \ep}  \Big[ \la_{k-1}^{  - \beta}    \|E_{1,k-1} - E_{2,k-1} \|_{k-1,\ka}    \Big]  \\
&\leq        \one \la_{k-1}^{\frac{1}{12}  - \beta -  11 \ep}    \|  \underline{ \xi_1}- \underline{ \xi_2} \|  
  \end{split}
\end{equation}
Altogether then  for  $L$ sufficiently large and $\la$ sufficiently small    
\be       \label{filet2}
\la_k^{-\beta}\| E'_{1,k}-E'_{2,k}  \|_{k, \ka} 
\leq   \frac12    \|  \underline{ \xi_1}- \underline{ \xi_2} \|  
\ee
Finally  combining        (\ref{filet})  and  (\ref{filet2} )  yields the result    $    \|  \underline{ \xi'_1}- \underline{ \xi'_2} \|  \leq
 \frac12    \|  \underline{ \xi_1}- \underline{ \xi_2} \|  $
\bigskip

 Now  we  can state:

 \begin{thm}  \label{gsf}  Let   $  \la$  be  sufficiently  small.     Then  for  each    $N $  there is a unique sequence
  $\vep_k,  \mu_k,  E_k$    for    $k = 0,1,2,  \dots  , N$   
satisfying of  the dynamical equation   (\ref{recursive3}),   the boundary conditions  (\ref{bc}),     
and     
\begin{equation}  \label{somewhat}
  | \mu_k|  \leq     \la_k^{\frac12 + \beta}   \hs
      \| E_k \|_{k, \ka}   \leq    \la_k^{\beta} 
\end{equation}
Furthermore  
 \begin{equation}  \label{eg}
|\vep_{k}|   \leq     \cO(1) \la_k^{\beta} 
 \end{equation}
\end{thm}
\bigskip

\pr   This  solution is the fixed point from the previous  lemma and  the bounds  (\ref{somewhat})  are
a consequence.   The  bound on the energy density follows from the others,   see \cite{Dim11}.
\bigskip

\rems     Much remains  to be done on this model.    The large field region region needs to be analyzed along the 
lines   of   \cite{Dim12}, \cite{Dim13}.    Then one could  prove an   ultraviolet stability bound     (proved  in  \cite{Bal83a}
for a massive gauge field).     Next one would want     prove   bounds on   the correlation  functions  uniform in the lattice spacing, 
and then show they have a limit as the lattice spacing goes to zero.

There is  also    the question of the infinite volume limit.
In this connection  we  remark that  although our final mass parameter  $\mu_N$   was tuned to zero we could equally  well have
tuned it   to any sufficiently small value.   If this    analysis  could be extended to allow  $\mu_N  = -1$  we would have the abelian
Higgs model.     Then one   could proceed along the lines  suggested   in      \cite{BIJ85},  \cite{BIJ88}  demonstrating
 mass generation for  the gauge field,  exponentially decaying correlations,   and  a   robust infinite volume limit.

\begin{appendix}

 \section{random walk expansion for  $C^{-1}$}  \label{czero}
 
The unit lattice operator  $C$  defined in section  \ref{naughty}   has the form   $C \tilde  Z  =  (\tilde Z,  S\tilde Z  )$
 and  so
 \be   \| C\tilde Z \|^2  =   \|  \tilde Z\|^2  +   \|S\tilde Z  \|^2
 \ee
 and hence
 \be   C^TC   =  I + S^T S  \ee
   which implies
 \be   C^{-1}  = (  I + S^T S)^{-1} C^T
 \ee
 We will show that    $(  I + S^T S)^{-1}$  has a random walk expansion.  Since $C^T$ is local  
 this gives an expansion for  $C^{-1}$.  
   
   \begin{lem}
    $(  I + S^T S)^{-1}$  has a random walk expansion    based on blocks  of size $M$, convergent for $M$ sufficiently large.   
    For  $y,y'$ in  the $L$-lattice 
     \be  
      | 1_{B(y) }  (  I + S^T S)^{-1}   1_{B(y') }    f      |  \leq    C  e^{ - \ga d(y,y')  }  \|f \|_{\infty}  
 \ee
 \end{lem}   
    
\pr  We follow the proof of lemma \ref{sweet3}.       Let   $A  =    I + S^T S  $  and  let      $A_{ \tilde \square_z}$   be the restriction
to  the  $3M$-cubes     $ \tilde \square_z$ centered on $z$ in   the $M$-lattice.     The quadratic form  $A_{ \tilde \square_z}$ is bounded above and below and
has an    exponentially decaying kernel  (actually a finite range  kernel).
By a lemma of Balaban     \cite{Bal83b}   the same is true  of    $\cG_{ \tilde \square_z} \equiv     A_{ \tilde \square_z}^{-1}    $. 
Since $A$ is naturally localized    in terms  of  the $L$-cubes  $B(y)$ we state it as  
 \be    \label{nuts} 
      | 1_{B(y) } \cG_{ \tilde \square_z} 1_{B(y') }    f      |  \leq    C  e^{ - \ga d(y,y')  }  \|f \|_{\infty}  
 \ee

To create the expansion 
 take the partition of unity   $1 = \sum_z  h_z^2$ as before   (but now defined on bonds)    and define
 \be  \cG^*  =   \sum_z h_z  \cG_{\tilde \square_z  } h_z   \ee
 Then  
 \be   A \cG^*   =      \sum_z h_z  A \cG_{\tilde \square_z  } h_z    +   \sum_z [A, h_z]  \cG_{\tilde \square_z  } h_z   
 \ee
But  on the support of $h_z$ we have  $ A \cG_{\tilde \square  }  =   A_{\tilde \square  }  \cG_{\tilde \square  }  =1$
and so       
\be   A \cG^*   =     I      +   \sum_z [A, h_z]  \cG_{\tilde \square  } h_z   = I -  \sum_z   K_z  =  I - K   
 \ee   
  Hence   
  \be  (   I + S^T S  )^{-1}    =   \cG^*  (I -K)^{-1}   \ee
  Expanding  $(I- K)^{-1}$ yields the random walk expansion.

  For  convergence we must show      $[A, h_z]   =  [S^T S, h_z]  =     \cO(M^{-1} )   $
   and since  $ [S^T S, h_z]  =  [S^T, h_z]  S+   S^T [ S, h_z] $  it suffices to show    $[S, h_z]  =     \cO(M^{-1} )   $.  
 This follows since  $S$ is a strictly local operator.    For the details we need  an explicit representation of $S$. 
 We  write  for  $f$ on  the unit lattice  and  $y'  = y + L e_{\mu} $
 \be 
 (\cQ   f )  ( y, y')
 =   \sum_{x \in B(y)}  L^{-4}  \sum_{b  \in  \Ga (x, x+ L e_{\mu})  }  f(b)   
  =      \sum_{b}  f(b)  L^{-4} n_{\mu} (b)   
 \ee
Here      $ n_{\mu} (b)$  is the number of elements in the set   $\{x \in B(y)  : \Ga( x, x+ L e_{\mu})  \ni b \}$.
 For example if   $b \in B(y,y')$  then $ n_{\mu} (b) = L$ and if  $b$ is not in the direction  $e_{\mu}$ then
 $ n_{\mu} (b) = 0$.   In general   $0 \leq  n_{\mu} (b)  \leq  L $.   
Breaking the sum up by the different categories of bonds we have   
 \be 
 \begin{split}
 (\cQ   f )  ( y, y')
  = &  L^{-3}   f(b(y,y'))   +  L^{-3}   \sum_{b \in B(y,y'), b \neq     b(y,y') }  f(b) n_{\mu} (b)    \\
  +  &   \sum_{b \in B(y) }  f(b) \frac{n_{\mu} (b)}{L^4}     +      \sum_{b \in B(y') }  f(b)   \frac{n_{\mu} (b)}{L^4}   \\
 \end{split}
 \ee
  Thus the equation   $\cQ f  = 0$ is solved by     $ f(b(y,y')) =  (Sf)(b(y,y'))  $
 where
  \be   \label{lignon2}
 \begin{split}
(S  f )  \B(b( y, y' ) \B)
  = &  -   \sum_{b \in B(y,y'), b \neq     b(y,y') }  f(b)    
  -     \sum_{b \in B(y) }  f(b)\frac{ n_{\mu} (b) }{ L} -       \sum_{b \in B(y') }  f(b) \frac{ n_{\mu} (b) }{ L}   \\
 \end{split}
 \ee
 Let's  look at the contribution of    the    second term    
here     to    $\B([S, h_z]f \B) (  b(y,y'  ) $.     It  is 
\be 
  \sum_{b \in B(y)}\B(   h_z(  b(y,y ' ) )    -       h_z(b) \B)  \frac{ n_{\mu} (b) }{ L}   f(b)    
   \ee
 But      
 \be
   \B|   h_z(  b(y,y' ) )    -       h_z(b) \B|    \leq   L \| \pa h_z \|_{\infty}  \leq       CM^{-1}         
 \ee
 so the term is  bounded by   $ CM^{-1} \| f \|_{\infty} $.  The other terms have the same bound and this gives
 the convergence of the expansion.     
 The decay factor   is extracted   using   the  local estimate  (\ref{nuts}) as before.

 \section{a covariant derivative}
  \label{special}

We show that  the forward/backward  covariant derivative   $ \nabla_{\cA} f  =  \frac12 (  \pa_{\cA}f  - \pa^T_{\cA}f) $     transforms like a vector field under lattice symmetries. 
For notational convenience we  work on a unit lattice and  absorb the coupling constant into the gauge potential.

\begin{lem}    Let   $r$  be a unit lattice symmetry  fixing the origin with matrix elements  $r_{\mu \nu}$,   and  let    $f_r(x)  =  f(r^{-1} x )$  and  $\cA_r(x,x') = \cA(r^{-1} x, r^{-1} x')$.
Then   
\be
 (\nabla_{\cA_r } f_r)_{\mu}  (x) =    \sum_{\mu}    r_{ \mu  \nu   }   (\nabla_{\cA} f )_{\nu}  (r^{-1}  x)  
\ee
\end{lem}

\pr  Start with   
\be    
(\nabla_{\cA} f ) _{\mu}(x)    = \frac12 \B(  e^{q\cA( x, x+ e_{\mu} )}f(x + e_{\mu} )   -     e^{q\cA( x, x- e_{\mu} )}f(x - e_{\mu} ) \B)
\ee
Given $\mu$   suppose  $  r^{-1} e_{\mu} =  e_{\rho}   $  for some  $\rho$.   Then      $r_{\mu \nu}  = (r^{-1})_{\nu \mu} =  \de_{  \rho \nu}$
for all $\nu$  and
\be    
\begin{split}
(\nabla_{\cA_r} f_r ) _{\mu}(x)    =  & \frac12 \B( e^{q\cA( r^{-1}x,  r^{-1}x+ e_{\rho} )}f(r^{-1}x + e_{\rho } ) 
  -     e^{q\cA(r^{-1} x, r^{-1}x- e_{\rho} )}f(r^{-1}x - e_{\rho} ) \B) \\
=  &    (\nabla_{\cA} f) _{\rho}(r^{-1} x)    \\
=&      \sum_{\mu}    r_{ \mu  \nu   }   (\nabla_{\cA} f )_{\nu}  (r^{-1}  x)  \end{split}
\ee
The other possibility  is  that   $  r^{-1} e_{\mu} =  - e_{\rho}   $ .   Then      $r_{\mu \nu}   =   -    \de_{  \rho \nu}$  
\be    
\begin{split}
(\nabla_{\cA_r} f_r ) _{\mu}(x)    =  &\frac12 \B(  e^{q\cA( r^{-1}x,  r^{-1}x- e_{\rho} )}f(r^{-1}x - e_{\rho } ) 
  -     e^{q\cA(r^{-1} x, r^{-1}x+ e_{\rho} )}f(r^{-1}x + e_{\rho} )   \B) \\
=  & -   (\nabla_{\cA} f) _{\rho}(r^{-1} x)    \\
=&      \sum_{\mu}    r_{ \mu  \nu   }   (\nabla_{\cA} f )_{\nu}  (r^{-1}  x)    
\end{split}
\ee

  \section{an estimate on $Q_k(\cA)\De_{\cA}$}  \label{B}

First    prove a special case of the divergence theorem on the lattice $\tk$ with spacing  $\eta = L^{-k}$.
 Let   $\De_y$    be the unit cube centered on the unit lattice point  $y$. For    a vector field $f_{\mu}$    let     $ \int_{\pa \De_y}  n \cdot    f$
be  the inward surface integral

\begin{lem}   \label{lumpy}
\be
  \int_{\De_y}   \pa ^T \cdot f  
=   \int_{\pa \De_y   }    n \cdot    f
\ee
\end{lem}

\pr    Take    $y=0$ for simplicity.  We  compute
\be  
\begin{split}
\sum_{\mu}\int_{\De_0}  ( \pa_{\mu}^T  f_{\mu} )(x)   dx
=  &\sum_{\mu}  \sum_{ |x|  < \frac12} \eta^{d-1}     ( f_{\mu}(x -  \eta  e_{\mu} ) -f_{\mu}(x) )    \\
 = &\sum_{\mu}   \sum_{ |x_{\nu}|  < \frac12,  \nu \neq  \mu} \eta^{d-1}     \Big ( [  f_{\mu}(x)]_{ x_\mu =  \frac12 -  \frac12 \eta }
  -   [ f_{\mu}(x)]_{ x_\mu = -  \frac12 -  \frac12 \eta}  \Big)    \\
\end{split}
\ee
 The last expression is identified as     $   \int_{\pa \De_0   }    n \cdot    f $.

  \begin{lem}    \label{louie}
   Let   $|\im  \cA|, | d \cA  |      <   e_k^{-1}$.  
  \begin{enumerate}
  \item  For a vector field  $f_{\mu}$ on $\tk$
\be   \|Q_k(\cA) (\pa_{\cA}^T \cdot   f ) \|_{\infty}  \leq    
C   \| f  \|_ {\infty}  
\ee
 \item  For a scalar $\phi$ on  $\tk$
  \be
\|Q_k(\cA)\De_{\cA}  \phi \|_{\infty}  \leq   C      \| \pa_{\cA}  \phi  \|_ {\infty}  
\ee
\end{enumerate}
 \end{lem}
 \bigskip

 \pr    The second follows from the first   with    $f_{\mu}   \equiv   \pa_{\cA, \mu}  \phi$.    
 For the first  let    $U(\cA,y,x)  =  e^{  qe_k  \eta  (\tau_k \cA )(y,x) } $  with  $( \tau_k \cA )(y,x)$ defined in  (\ref{sum}).
  Then we have  for  $y \in \bbT^0_{N-k}$
     \be    \label{swoop}
 \begin{split} 
  ( Q_{k}(\cA) \pa_{\cA}^T \cdot   f  )(y)   =  &  \sum_{\mu} \int_{\De_y } dx\  U(\cA,y,x)  
   \Big( e^{- qe_k \eta  \cA_{\mu}( x- \eta e_{\mu})}  f_{\mu}(x- \eta e_{\mu})  - f_{\mu}(x)   \Big) \eta^{-1}
    \\
=  &        \sum_{\mu}   \int_{\De_y } dx\  
   \Big( U(\cA,y,x)  e^{- qe_k \eta  \cA_{\mu}( x- \eta e_{\mu})}  -  U(\cA,y,x- \eta e_{\mu})    \Big) f_{\mu}(x- \eta e_{\mu})   \eta^{-1}
    \\
&   +  \sum_{\mu}   \int_{\De_y } dx\   \Big(    U(\cA,y,x- \eta e_{\mu}) f_{\mu}(x- \eta e_{\mu})  -   U(\cA,y,x)   f_{\mu}(x)  \Big) \eta^{-1}
\\
\end{split}
\ee
For the second term   here    we  use the divergence theorem   of lemma \ref{lumpy}   to write it as    
\be   \sum_{\mu}  \   \int_{\De_y } dx\  (\pa/ \pa x_{\mu})^T  \Big( U(\cA,y,x)     f_{\mu}(x)    \Big)    dx  
=   \sum_{\mu}  \    \int_{ \pa  \De_y}   dx \  U(\cA,y,x)\  (n_{\mu}   f_{\mu})  (x)   
\ee
Bounding  $U(\cA ,y,x)$   using  (\ref{slippery}), this term is bounded by         by  $ \one  \| f\|_{\infty} $.
For the first    term in    (\ref{swoop})    it suffices to show
\be     \label{shh}
    \int_{\De_y}     dx\   \eta  \B |  (\tau_k \cA )(y,x)  -   \cA(x -  \eta e_{\mu},x  )   -  (\tau_k \cA )(y,x- \eta e_{\mu})    \B|  
   \leq  
 C  \eta  \|d   \cA\|_{\infty}
\ee
Then the term is bounded by    $  C  e_k     \|d \cA\|_{\infty}  \|  f  \|_{\infty}   \leq   C   \|  f  \|_{\infty}     $
as required.

To  prove  (\ref{shh})   recall that   $ (\tau_k \cA )(y,x) =  \sum_{j=0}^{k-1}  (\tau \cA )(y_{j+1}, y_{j})   $  is  defined by  the unique sequence  $x=y_0, y_1,  \dots  y_k  =  y$  with   $y_j \in \bbT^{-k+j}_{N-k}$  and    $x  \in B_j(y_j)$.   Also     $  (\tau_k \cA )(y,x- \eta e_{\mu})   $  is defined by  a similar sequence    $x- \eta  e_{\mu}  =y'_0, y'_1, y'_2,  \dots  y'_k  =  y$. 
Suppose    $x,  x- \eta  e_{\mu} $ are in the same  $L \eta$  cube  $B(y_1) = B(y'_1)$.    Then  $y_i = y_i'$ for  $i = 1,2, \dots, k$.
Hence 
\be 
\begin{split}
&  \eta   (\tau_k \cA )(y,x)  - \eta  \cA(x -  \eta e_{\mu},x  )   - \eta  (\tau_k \cA )(y,x- \eta e_{\mu})\\
=     & \eta    (\tau \cA )(y_1,x)  -  \eta \cA(x -  \eta e_{\mu},x  )   -  \eta (\tau \cA )(y_1,x- \eta e_{\mu}) \\
=   & \frac{1}{d!}  \sum_{\pi}  \eta \cA( \Ga^{\pi}(y_1,x) )     -    \eta  \cA(x -  \eta e_{\mu},x ) 
  -  \frac{1}{d!} \sum_{\pi} \eta \cA( \Ga^{\pi}(y_1, x- \eta e_{\mu} ) )   \\ 
=   & \frac{1}{d!}  \sum_{\pi}  \eta  \cA\B(  \Ga^{\pi}(y_1,x) )   + [x,x -  \eta e_{\mu} ]  -  \Ga^{\pi}(y_1, x- \eta e_{\mu} )   \B)     \\ 
\end{split}
\ee
But   for each $\pi$  the indicated   path 
 is a closed   and  hence is the boundary of   a surface  $  \Sigma^{\pi} $.
 By    the   lattice Stokes theorem with unweighted sums    we have  
 \be 
  \eta  \cA\B( \Ga^{\pi}(y_1,x) )
       + [x,x -  \eta e_{\mu} ]  -  \Ga^{\pi}(y_1, x- \eta e_{\mu})  \B)   = \eta^2 d \cA(   \Sigma^{\pi})  
     \ee   
The  surface    $  \Sigma^{\pi} $  is    made up of  $\eta$-plaquettes    in  an   $L \eta$-cube.   Hence  the number of plaquettes in  $\Sigma^{\pi}$   is bounded by a constant    and  so  $| \eta^2  d \cA(   \Sigma^{\pi} ) | \leq   C \eta^2    \|d \cA\|_{\infty} $.     Thus the integrand in  (\ref{shh})  for points 
with  $x, x- \eta e_{\mu}$ in the same $L \eta$ cube   is bounded by  $ C \eta^2    \|d \cA\|_{\infty}$ which is better than we need.

This would   take    care of most points in (\ref{shh}),  but not all, and not the most important.    More generally       let      $X_j \subset  \De_y$  be the set of points  $x$
with the property that   $x,  x- \eta  e_{\mu} $  are in the same $L^{j+1} \eta  $ cube  but not  in any smaller cube.  
Equivalently $x,  x- \eta  e_{\mu} $  crosses a $L^j \eta $ face but no larger face. 
Then
$\cup_{j=0}^{k-1} X_j  = \De_y$  and in   (\ref{shh} ) we  write 
\be    \label{nuts1}
\int_{\De_y}  dx [ \cdots ] =  \sum_{x \in \De_y}   \eta^3  [ \cdots ]   =  \sum_{j=0}^{k-1}  \sum_{x \in  X_j}    \eta^3   [ \cdots ] 
\ee
Note  that  the number of points in $X_j$ is bounded by  the number of points $x$   such that   $x,  x- \eta  e_{\mu} $ 
 crosses a $L^j \eta $  face and so  
 \be  \label{nuts2}
   |X_j|  \leq   L^{2k} L^{k-j}  =   L^{3k-j} 
 \ee

Suppose  $x \in X_j$. 
Then      $y_i = y_i'$ for  $i = j+1 ,   \dots, k$  and so   
\be 
\begin{split}
& \eta  (\tau_k \cA )(y,x)  -  \eta  \cA(x -  \eta e_{\mu},x  )   - \eta  (\tau_k \cA )(y,x- \eta e_{\mu}) \\
=  & \sum_{i=0}^{j} \eta \tau \cA( y_{i+1}, y_i) )     -    \eta  \cA(x -  \eta e_{\mu},x )   -   \sum_{i=0}^{j} \eta  \cA(y'_{i+1},  y'_i )\\ 
=   & \frac{1}{d!}  \sum_{\pi}   \eta   \cA   \B( \sum_{i=0}^{j}  \Ga^{\pi} ( y_{i+1}, y_i) + [x,x -  \eta e_{\mu} ]  -  \sum_{i=0}^{j}   \Ga^{\pi} ( y'_{i+1}, y'_i)   \B) \\
\equiv  &    \frac{ 1 }{d!}   \sum_{\pi}    \sum_{i=0}^{j}  \eta   \cA\B(\Ga^{\pi} ( y_{i+1}, y_i) + [y_i, y'_i ]  -  \Ga^{\pi} ( y'_{i+1}, y'_i)  + [ y'_{i+1}, y'_i]  \B)     \\
\end{split}
\ee
The last step follows since the added lines  cancel out,  except for  $i=0$ when $ [   y_0 ,   y'_0 ] = [ x, x- \eta e_{\mu} ]$
and    $i=j$  when  $ [y'_{j+1},   y_{j+1}  ] = \emptyset$.    For  each   $\pi, i$  the indicated path is  closed and so  the boundary  of a surface
$\Si^{\pi}_i$.   Hence    the last expression is    the same as   
\be
      \frac{1}{d!}   \sum_{\pi}    \sum_{i=0}^{j}   \eta^2   d \cA(\Si^{\pi}_i )   
\ee  
For each i    the path is      made up of  $L^i\eta$-segments  in  an  $L^{i+1}  \eta $-cube,  
so   the number  of   $L^{i}\eta$-squares is bounded by a constant   $C$  and so     the number  of   $\eta$-plaquettes is bounded by  $CL^{2i} $.  Therefore
 $| \eta^2   d \cA(\Si^{\pi}_i )  |   \leq   C \eta^2  L^{2i}  \leq     CL^{-k + 2i} \eta $.  Hence the expression is bounded by 
\be    
  \frac{1}{d!}   \sum_{\pi}    \sum_{i=0}^{j} | \eta^2  d \cA(\Si^{\pi}_i )|     
\leq       \sum_{i=0}^{j} CL^{-k + 2i} \eta  \| d \cA \|_{\infty}  
\leq   CL^{-k + 2j}  \eta   \|d \cA \|_{\infty}
\ee
Therefore,  referring to   (\ref{nuts1}), (\ref{nuts2}),   the integral  
  (\ref{shh}) is estimated  by 
\be 
 C  \sum_{j=0}^{k-1}  \sum_{x \in  X_j}    \eta^3  L^{-k  + 2j}  \eta \| d\cA  \|_{\infty}
 \leq     C  \sum_{j=0}^{k-1}      L^{-k+ j}  \eta \| d \cA  \|_{\infty}    \leq  C \eta  \| d \cA  \|_{\infty}  
  \ee
which is the result we want.

 \section{integrals}  \label{dprime}
 In  $\tk$  or  $L^{-k} \bbZ^3$ we  consider integrals   of the form 
 \be   \int    f(x)  dx  =  \sum_x  \eta^3  f(x)  \hs    \eta  = L^{-k}
 \ee
 Recall that  $d'(x,y)  = d(x,y) =  \sup_{\mu}  |x_{\mu}  - y_{\mu}|$ for  $x \neq  y$  and  $d'(x,x)  =  L^k = \eta ^{-1}$.
  
 \begin{lem}
 For  $\al <   3$
 \be   \int_{d'(x,y) \leq  1   }     d'(x,y)^{-\al}  dx  \leq  \one  \ee  
 \end{lem}
 \bigskip
 
 \pr  It suffices to consider  $y=0$.  Isolate the $x=0$  term.    Then   with $r \in  L^{-k} \bbZ$
 \be 
 \begin{split}
 \int_{d'(x,0) \leq  1    }   d'(x,0)^{-\al}  dx   \leq  &\  \eta^{3 - \al}   +     \sum_{ |x|  \leq  \frac12, x \neq 0    } \eta^3 |x |^{-\al}    \\
\leq  &\    1  +     \sum_{ 0 < r  \leq   \frac12 }  \eta  r^{-\al}   \sum_{x:     |x|  =r} \eta^2     \\   
\leq  &\    1  +    \one    \sum_{ 0 < r  \leq   \frac12 }  \eta  r^{2  -\al}  \leq     \one     \\   
 \end{split}
 \ee

 \begin{lem}
\be
\begin{split}
    \int_{d'(x,z) \leq  1, d'(y,z) \leq  1}   d'(x,z)^{-\al}   d'(y,z)^{-\beta}    dz   \leq & \one   \hs   \al + \beta  <3 \\
     \int_{d'(x,z) \leq  1, d'(y,z)\leq  1}   d'(x,z)^{-1}   d'(y,z)^{-2}    dz  \leq  &   \one   d'(x,y)^{-\ep }   \\
       \int_{d'(x,z) \leq  1, d'(y,z) \leq  1}   d'(x,z)^{-2}    d'(y,z)^{-2}    dy   \leq  &   \one   d'(x,y)^{-1-\ep }   \\
 \end{split}
 \ee  
\end{lem}
 \bigskip

 \pr    For the first inequality   consider   separately the regions    $d'(x,z)  \leq   d'(y,z)$   and    $  d'(y,z)  \leq   d'(x,y)$  
 and use the previous result.     For the  second inequality  we  need       
\be     \int_{d'(x,z) \leq  1, d'(y, z) \leq  1}    d'(x,z)^{-1}  d'(y, z)^{-2}  d'(x,y)^{\ep }    dz   \leq     \one   
 \ee
We    take     $ d'(x,y)   \leq  d'(x,z)  +  d'(z,y )$.   In the region  $d'(x,z)  \leq   d'(y,z)$ 
we  have  $ d'(x,y)    \leq   2  d'(y,z)  $  and  so  the integral is dominated by  
 \be  \one      \int_{d'(x,z) \leq  1, d'(y, z) \leq  1}     d'(x,z)^{-1}   d'(y,z)^{-2 +  \ep }     dz     
 \ee
 which is  $\one$ by the previous result.   Simlilarly for the region     $d'(y,z)  \leq  d'(x,z) $. 
For the  third   inequality  we need 
\be     \int_{d'(x,z) \leq  1,d'(y, z) \leq  1}     d'(x,z)^{-2}   d'(y,z)^{-2}  d'(x,y)^{1+  \ep }    dz   \leq     \one   
 \ee
  In the region  $d'(x,z)  \leq   d'(y,z)$ 
we  again    have  $ d'(x,y)    \leq   2  d'(y,z)  $ 
  and  so  the integral is dominated by  
 \be  \one    \int_{d'(x,z) \leq  1, d'(y, z) \leq  1}     d'(x,z)^{-2}   d'(y, z)^{-1 +  \ep }     dz     
 \ee
 which is  $\one$ by the first inequality.      Simlilarly for the region     $d'(y,z)  \leq  d'(x,z) $

\section{Green's functions on a lattice}  \label{orange}

We  study  the standard Greens function   $G^s_k   = ( -\De +I )^{-1}$  defined on  $\tk$  or   $L^{-k} \bbZ^3$.  We  are interested in both short and long distance behavior.

\begin{lem}  \label{shock2}  There  is a constant  $\ga = \one$ such that   
\be 
\begin{split}
    |G^s_k(  x,y) |    \leq  &         \one  d'(x,y)^{-1} e^{ -  \ga   d(x,y)  } \\
      |  \pa_{\mu}   G^s_k(  x,y) |    \leq  &      \one  d'(x,y)^{-2}e^{ -  \ga  d(x,y)  } \\
        |(  \pa_{\mu}    G^s_k  \pa^T_{\nu})(  x,y) |    \leq  &      \one  d'(x,y)^{-3}e^{ - \ga   d(x,y)  } \\
 \end{split}    
 \ee
\end{lem}
\bigskip

\pr
 It suffices to  consider  the infinite lattice $\eta \bbZ^3=  L^{-k}\bbZ^3$   since  toroidal   
 case   can   be obtained by  periodizing. 
  Also if  $x \neq  y$  so   $d'(x,y) =  d(x,y)  = \sup_{\mu} |x_{\mu} - y_{\mu}|$ it suffices to consider the sector 
  where   $d'(x,y)  =   |x_0 - y_0|$.

 On the infinite lattice we have the representation
\be   \label{still}
  G^s_k(x,y)   =   \frac{1}{(2 \pi)^3}   \int_{ |p_\mu| < \eta^{-1} \pi} 
  \frac { e^{ip\cdot  (x-y) }}{1 +   \De ( p )      }   
\ee
where
\be 
   \De(p)   =     \sum_{\mu}  2\eta^{-2}( 1- \cos \eta p_{\mu})   =  \sum_{\mu}  \frac{  \sin^2 ( \frac12 \eta p_{\mu}  )}{ ( \frac12   \eta )^2}  
\ee
For    $p=(p_0,p_1,p_2)$     let   $\bp  = (0,p_1,p_2)$.
The denominator in (\ref{still})  vanishes when  
\be
1 +   \De ( p )      =1 +     2\eta^{-2}( 1- \cos \eta p_0)   +    \De(\bp)     =0
\ee
or
\be
    \cos ( \eta p_0 ) =1  +  \frac12  \eta^2     \B( 1 +  \De(\bp)  \B)  
\ee
 So we find poles at  $p_0=  \pm  i  \om(\bf p)$ 
where    
\be
    \cosh  (\eta \om(\mathbf{  p }))  =1  +  \frac12  \eta^2     \B( 1  +   \De(\bp)   \B)  
\ee
Note that if    $\eta$  is small       $ \De(\bp) \approx   | \bp|^2$  and   comparing power series  gives 
$ \om( \mathbf{  p })  \approx    ( 1  +  \De( \bp )  )^{\frac12}  \approx    ( 1  + | \mathbf{p}|^2 )^{\frac12}    $  as  expected.

Now  deform the   $p_0$  contour   to  a rectangle with large imaginary part.    The sides of the rectangle cancel by periodicity.
and  the  far   piece goes to zero.    We only pick up  the pole   at  $p_0  =  \pm  i  \om(\bf p)$  depending on the  sign of  $x_0 -y_0$.
Compute the residue at the pole  using      
\be
   \frac{\pa}{ \pa  p_0}  \B[   2\eta^{-2}( 1- \cos \eta p_0)  \B]_{  p_0  =    \pm  i  \om(\bp)} 
= [ 2 \eta^{-1} \sin  \eta  p_0  ]_{  p_0  =    \pm  i  \om(\bp)}    =    \pm   2  \eta^{-1}    \sinh  \eta      \om(\bp)   
\ee
and find      
\be 
  G^s_k(x,y)   =   \frac{1}{(2 \pi)^2}   \int_{ | p_k  | < \eta^{-1} \pi}       \       e^{-  \om(\bp)  |x_0 - y_0| }
  \frac { e^{i\bp\cdot  ( \bx-\by ) }}{   2  \eta^{-1}    \sinh  \eta      \om(\bp)       }   \    d \bp 
\ee

To  estimate this   start  with   
\be
    \frac12  \leq      \left|  \frac { \sin x  }  {x}   \right|    \leq   1  \hs     |x|  \leq  \frac{\pi}{2} 
\ee
It follows that      $ \frac12    |\bp|^2      \leq       \De( \bp)   \leq    |\bp|^2        $    and so  
  \be   \label{cosh}
  1  +  \frac12  \eta^2     \B( 1  + \frac12   |\bp|^2    \B)   \leq    \cosh  (\eta \om(\bp ))    \leq      1  +  \frac12  \eta^2     \B( 1  +   |\bp|^2    \B)  
\ee
But     $   \sinh^2 x   =  \cosh^2 x -1       \geq    \cosh x  -1  $
so   
\be   \label{elvis1}
\frac12    \eta^2 \B( 1  +  \frac12 |\bp|^2    \B)  \leq          \sinh^2    (\eta \om(  \bp ))  
\hs  \textrm{ hence }  \hs 
 \frac12    \eta  \sqrt{1  +   |\bp|^2 }   \leq          \sinh   (\eta \om(   \bp ))  
 \ee
 
Next  we  claim that there is a constant  $c = \one$   such that  
\be   \label{elvis2}
  c  \sqrt{  1  +   |\bp|^2 }   \leq    \om(\bp)   \leq      \sqrt{  1  +   |\bp|^2 }    
\ee
For the upper bound note that  (\ref{cosh}) implies that  
$    \cosh  (\eta \om(\bp))  \leq  \cosh  (\eta   \sqrt{ 1  +   |\bp|^2 } ) $
and hence    $ \om( \bp)  \leq    \sqrt{ 1  +   |\bp|^2 }  $.   
For the lower  bound  we first note  that  the upper bound implies    
\be   \eta  \om( \bp)  \leq  \sqrt{ 1  +  \eta^2|\bp|^2 } )   \leq     \sqrt{ 1  +2  \pi^2}   \leq  5
\ee
For  $0 \leq  x \leq  5$    we have  $ \sinh x  \leq  \int_0^x \cosh t  \ dt  \leq  x \cosh 5$.
Hence      by   (\ref{elvis1})
\be    \frac12    \eta  \sqrt{1  +   |\bp|^2 }   \leq    \sinh  \eta  \om( \bp)    \leq    \eta  \om( \bp)  \  \cosh 5 
\ee
which gives the lower bound  with  $c =  ( 2   \cosh 5 )^{-1}$

Using   (\ref{elvis1}) and (\ref{elvis2})   we have   for    $x \neq y$ and  $ |x_0-y_0| =d(x,y) \neq  0$  
\be 
|  G^s_k(x,y) |    \leq  \one   \int           \       e^{-  c \sqrt{  1  +   |\bp|^2 }    
  |x_0 - y_0| }
  \frac {  d \bp  }{    \sqrt{  1  +   |\bp|^2 }     }     \
\ee  
Now with $\ga   = \frac12  c$ we  can extract a factor  $\exp(  - \ga |x_0-y_0| )$  and obtain
\be 
|  G^s_k(x,y) |    \leq  \one      e^{  - \ga |x_0-y_0| } \int           \       e^{-  \frac12  c \sqrt{  1  +   |\bp|^2 }    
  |x_0 - y_0| }
  \frac {  d \bp  }{    \sqrt{  1  +   |\bp|^2 }     }     \
\ee  
Change variables to   $\bq  =  |x_0 - y_0|   \bp $  and  find  that the
integral  here   is
\be 
\begin{split}
&      |x_0 - y_0|^{-1}  
   \int           \       e^{- c  \sqrt{ |x_0 - y_0|^2  +   |\bq|^2 }   } 
  \frac {  d \bq    }{    \sqrt{  |x_0 - y_0|^2  +   |\bq|^2 }     }      d \bq   \\
&  \leq    \one     |x_0 - y_0|^{-1}    \int           \       e^{-  c |\bq|    } \frac { 1 }{        |\bq|     }     \ d \bq   
   =     \one     |x_0 - y_0|^{-1}   \\
\end{split}
\ee
Thus we  get     
\be 
|  G^s_k(x,y) |    \leq  \one     |x_0 - y_0|^{-1}     e^{  - \ga |x_0-y_0| }
 \ee

If  $x=y$    the estimate  comes down to  
\be     |  G^s_k(0,0) |    \leq  \one   \int_{|p_k|  \leq  \eta^{-1}\pi    }      \frac {  d \bp  }{    \sqrt{  1  +   |\bp|^2 }     }     \
\ee  
Enlarge the   integration region  to   $ |\bp|  \leq   2 \eta ^{-1} \pi$ and go to polar coordinates to  get  
 \be        |  G^s_k(0,0) |    \leq  \one   \int_0^{2\eta^{-1}  \pi}           \frac { r dr  }{    \sqrt{  1  + r^2 }   }
=  \one  \eta^{-1}  =  \one L^{k}  =  \one  d'(0,0) ^{-1}          \
\ee  
  
For derivatives  we    note that for  $x_0 > y_0$   
\be   
(x_0 - y_0 )\frac{ \pa }{ \pa x_0}     G^s_k(x,y) 
=     \frac{-1}{(2 \pi)^2}   \int_{ | p_k  | < \eta^{-1} \pi}       \    \om(\bp)  (x_0 - y_0)       e^{-  \om(\bp)  (x_0 - y_0) }
  \frac { e^{i\bp\cdot  (\bx-\by) }}{   2  \eta^{-1}    \sinh  \eta      \om(\bp)       }   \ d \bp 
\ee
Now use   $ |  \om(\bp)  (x_0 - y_0)       e^{- \frac12   \om(\bp)  (x_0 - y_0) }| \leq  \one$  and proceed as before to estimate
the quantity  by  $\one |x_0 - y_0|^{-1}  e^{  - \ga |x_0-y_0| } $.   The same works  for   $y_0 > x_0$  so we have
\be  \label{sudsy}
  \left|  \frac{ \pa }{ \pa x_0}     G^s_k(x,y) \right|    \leq     \one    |x_0 - y_0|^{-2}     e^{  - \ga |x_0-y_0| }
\ee
For  the other      derivatives   we  have  for  $k=1,2$
\be   
(x_0 - y_0 )\frac{ \pa }{ \pa x_k}     G^s_k(x,y) 
=     \frac{1}{(2 \pi)^2}   \int_{ | p_k  | < \eta^{-1} \pi}       \ i p_k   (x_0 - y_0)       e^{-  \om(\bp)  (x_0 - y_0) }
  \frac { e^{i\bp\cdot  (\bx-\by) }}{   2  \eta^{-1}    \sinh  \eta      \om(\bp)       }   \ d \bp 
\ee
and this   again  yields     the  bound  (\ref{sudsy}).    Higher derivatives are similar.

\end{appendix}

\end{document}